\documentclass[preprint]{aastex}
\usepackage{color}

\shorttitle{The Frequency of Planets Around B and A Stars}
\shortauthors{Nielsen et al.}

\begin{document}

\title{The Gemini NICI Planet-Finding Campaign: \\ The Frequency of Giant Planets around Young B and A Stars}

\author{Eric L. Nielsen,\altaffilmark{1}
Michael C. Liu,\altaffilmark{1}
Zahed Wahhaj,\altaffilmark{2}
Beth A. Biller,\altaffilmark{3}
Thomas L. Hayward,\altaffilmark{4}
Laird M. Close,\altaffilmark{5}
Jared R. Males,\altaffilmark{6}
Andrew J. Skemer,\altaffilmark{7}
Mark Chun,\altaffilmark{1}
Christ Ftaclas,\altaffilmark{1}
Silvia H. P. Alencar,\altaffilmark{6}
Pawel Artymowicz,\altaffilmark{7}
Alan Boss,\altaffilmark{8}
Fraser Clarke,\altaffilmark{9}
Elisabete de Gouveia Dal Pino,\altaffilmark{10}
Jane Gregorio-Hetem,\altaffilmark{10}
Markus Hartung,\altaffilmark{4}
Shigeru Ida,\altaffilmark{11}
Marc Kuchner,\altaffilmark{12}
Douglas N. C. Lin,\altaffilmark{13}
I. Neill Reid,\altaffilmark{14}
Evgenya L. Shkolnik,\altaffilmark{15}
Matthias Tecza,\altaffilmark{9}
Niranjan Thatte,\altaffilmark{9}
Douglas W. Toomey\altaffilmark{16}
}

\altaffiltext{1}{Institute for Astronomy, University of Hawaii, 2680 
Woodlawn Drive, Honolulu HI 96822, USA}
\altaffiltext{2}{European Southern Observatory, Alonso de C\'{o}rdova 3107, 
Vitacura, Santiago, Chile}
\altaffiltext{3}{Max-Planck-Institut f\"ur Astronomie, K\"onigstuhl 17, 
69117 Heidelberg, Germany}
\altaffiltext{4}{Gemini Observatory, Southern Operations Center, c/o AURA,
Casilla 603, La Serena, Chile}
\altaffiltext{5}{Steward Observatory, University of Arizona, 933 North Cherry 
Avenue, Tucson, AZ 85721, USA}
\altaffiltext{6}{Departamento de Fisica - ICEx - Universidade Federal de 
Minas Gerais, Av. Antonio Carlos, 6627, 30270-901, Belo Horizonte, MG, 
Brazil}
\altaffiltext{7}{University of Toronto at Scarborough, 1265 Military Trail, 
Toronto, Ontario M1C 1A4, Canada}
\altaffiltext{8}{Department of Terrestrial Magnetism, Carnegie Institution of 
Washington, 5241 Broad Branch Road, N.W., Washington, DC 20015, USA}
\altaffiltext{9}{Department of Astronomy, University of Oxford, DWB, Keble 
Road, Oxford OX1 3RH, UK}
\altaffiltext{10}{Universidade de Sao Paulo, IAG/USP, Departamento de 
Astronomia, Rua do Matao, 1226, 05508-900, Sao Paulo, SP, Brazil}
\altaffiltext{11}{Tokyo Institute of Technology, 2-12-1 Ookayama, 
Meguro-ku, Tokyo 152-8550, Japan}
\altaffiltext{12}{NASA Goddard Space Flight Center, Exoplanets and Stellar 
Astrophysics Laboratory, Greenbelt, MD, 20771, USA}
\altaffiltext{13}{Department of Astronomy and Astrophysics, University of 
California, Santa Cruz, CA, USA}
\altaffiltext{14}{Space Telescope Science Institute, 3700 San Martin Drive, 
Baltimore, MD 21218, USA}
\altaffiltext{15}{Lowell Observatory, 1400 West Mars Road, Flagstaff, AZ 
86001, USA}
\altaffiltext{16}{Mauna Kea Infrared, LLC, 21 Pookela St., Hilo, HI 
96720, USA}

\begin{abstract}
We have carried out high contrast imaging of 70 young, nearby B and A stars to 
search for brown 
dwarf and planetary companions as part of the Gemini NICI 
Planet-Finding Campaign.  
Our survey represents the largest, deepest survey for 
planets around high-mass stars ($\approx$1.5--2.5 M$_\sun$) 
conducted to date and 
includes the planet hosts $\beta$ Pic and Fomalhaut.  
We obtained follow-up astrometry of all candidate companions within 400 AU 
projected separation for stars in uncrowded fields and identified new 
low-mass companions to HD~1160 and HIP~79797.  
We have found that the previously known young brown dwarf companion to 
HIP~79797 is itself a tight (3 AU) binary, composed of 
brown dwarfs with masses 58$^{+21}_{-20}$ M$_{Jup}$ and 
55$^{+20}_{-19}$ M$_{Jup}$, making this system one of the 
rare substellar binaries in orbit around a star.  
Considering the contrast limits of our NICI data 
and the 
fact that we did not detect any planets, we use high-fidelity Monte Carlo 
simulations to show 
that fewer than 20\% of 2 M$_{\sun}$ stars can have giant planets greater 
than 4 M$_{Jup}$ between 59 and 460~AU at 95\% confidence, and fewer than 
10\% of these stars can have a planet more massive than 10 M$_{Jup}$ between 
38 and 650 AU.  Overall, we find that large-separation giant planets are 
not common 
around B and A stars: fewer than 10\% of B and A stars can have an 
analog to the HR~8799~b (7~M$_{Jup}$, 68 AU) 
planet at 95\% confidence.  We also describe a new 
Bayesian technique for 
determining the ages of field B and A stars from photometry and theoretical 
isochrones.  Our method produces more plausible ages for 
high-mass stars than 
previous age-dating techniques, which tend to underestimate stellar ages and 
their uncertainties.
\end{abstract}


\section{Introduction}

In the last two decades, over 800 planets have been detected outside of our 
solar system, the vast majority found by the transit 
(e.g. \citealt{kepler11}) or radial velocity (RV, e.g. \citealt{Marcy08}) 
methods.  These planets tend to be close to their parent star, 
$\lesssim$1 AU for transits and $\lesssim$5 AU for RV, and obtaining 
spectra is difficult for transiting planets and impossible for non-transiting 
RV planets.  Direct imaging of self-luminous planets, by contrast, is most 
sensitive to large-separation ($\gtrsim$10 AU) 
planets which can be followed up with spectroscopy.  Thus far, however, 
only a handful of planetary mass companions have been detected by direct imaging 
(e.g. \citealt{fomalhautb,hr8799b,hr8799e,betapic2}), 
despite several concerted efforts at large telescopes.

One unexpected result from these initial 
direct imaging planet detections is 
the apparent prevalence of planets around high-mass stars: Fomalhaut 
(1 planet, \citealt{fomalhautb}), HR~8799 (4 planets, 
\citealt{hr8799b,hr8799e}), 
and $\beta$ Pic (1 planet, \citealt{betapic2}) are all A stars.  These stars 
also all harbor debris disks, though as yet there is no compelling 
evidence that debris disk hosts are more likely to harbor 
giant planets \citep{debris}.  \citet{bryden09} found 
no significant correlation between presence of RV planets (mainly giant 
planets) and debris disks.  However, studies of hosts of 
lower-mass RV planets (less than the mass of Saturn) by \citet{wyatt12} and 
hosts of Kepler planet candidates by \citet{lawler12} suggest that stars 
hosting lower-mass planets are more likely to also host debris disks.  
High-mass 
stars are intrinsically brighter than later-type stars, so a faint planet 
is much harder to directly detect around an A star than an M star.  The planet 
detections to date suggest either that giant planets 
are more common around A stars, as is the case for close-in RV planets 
(e.g. \citealt{johnson_new}), that wide-separation planets are more 
common around high mass stars, or both.

The planets around $\beta$ Pic and Fomalhaut were found as part of targeted 
studies of these stars specifically, motivated primarily by their bright 
debris disks, making it difficult to place statistical constraints 
on exoplanet populations around high mass stars from these 
two detections.  
The four HR 8799 planets were found as part of the International Deep Planet 
Survey (IDPS), and an 
analysis of the frequency of giant planets around A~stars was presented 
recently by \citet{vigan12}.  Based on their sample of 38 A stars 
and 4 early-F stars, they find that between 6 and 19\% of A stars 
have a giant planet between 5 and 320 AU at 68\% confidence.  However, these 
statistical constraints are based on optimistically young ages for many 
stars in their sample (see Section~\ref{ages_sec}).

The Gemini NICI Planet-Finding Campaign is a 4-year program to detect 
extrasolar giant planets, measure their frequency, determine how 
that frequency depends on stellar mass, and study the spectral energy 
distributions of young exoplanets \citep{liunici}.  The Campaign has thus far 
detected three brown dwarf companions around the young stars PZ Tel 
\citep{pztel}, CD$-$35~2722 \citep{cd35}, and HD 1160 \citep{hd1160} but 
no planetary-mass companions ($\lesssim 13$ M$_{Jup}$).  Here we 
present an analysis of the frequency of giant planets around high 
mass stars based on the 70 young B and A stars observed by the Campaign.  
In companion papers we present an analysis of the 
planet frequency around stars in young moving groups \citep{moving_groups} and 
debris disk hosts \citep{debris}.

\section{Target Selection}

Prior to the start of the Gemini NICI Planet-Finding 
Campaign in December 2008, we assembled a list of 1353 
potential target stars.  Among the many considerations used to construct 
this input list were the radial velocity results of \citet{johnson}, which 
pointed to a correlation between the frequency of short-period giant planets 
and stellar host mass.  We sought to determine if this trend continues to 
longer-period planets by selecting target stars from a range of 
masses, so as to directly measure planet frequency as a function of 
spectral type.  For the B and A stars in the sample, we selected 
stars from three sources: members of young moving groups, 
stars with debris disks, and other B and A stars from 
the \textit{Hipparcos} catalog.  We describe 
target selection for the first two categories in more detail in 
\citet{moving_groups} and \citet{debris}, but in general we selected all B and 
A stars in nearby young moving groups as well as all of those that 
host debris disks for our input target list.

In addition, we used two 
other methods to flag young B and A stars from the Hipparcos catalog: 
main-sequence lifetime and position on the color-magnitude diagram.  
(1)~We chose stars 
with an early spectral type (earlier than $\sim$A5) which have main-sequence 
lifetimes of only hundreds of Myrs, providing a constraint on their ages 
even in the absence of any other age information.  
(2)~We added stars that have relatively faint absolute magnitudes 
on the color-magnitude diagram, which suggested they were potentially 
young.  (We discuss this 
technique in more detail in the next section.)  For the 
\textit{Hipparcos}-selected 
stars, we chose stars within 75 pc with B or A spectral types and removed 
any giants.  There was 
no distance cut placed on debris disk hosts or moving group targets in our 
sample, which can be found up to 172 pc from the Sun, though most stars are 
within 100 pc.  Stars 
with close binaries and those not observable from Gemini-South 
($\delta>$+20$^\circ$) were also removed from the sample.  Two stars, the 
debris disk host HD~85672 and the nearby HIP~54872, were also kept on the 
target list despite having declinations between +20$^\circ$ and +30$^\circ$.

To produce an observing list from this input catalog, we used a variation of 
the Monte Carlo 
simulations described in Section~\ref{planetfraction_section} to rank targets 
by the likelihood of imaging a young hot-start planet around them with NICI, 
given their age, distance, spectral type, and apparent magnitude (see 
\citealt{liunici}).  An ensemble of simulated planets was 
placed around each target star, following an extrapolation of the 
distribution of giant planets found by the RV method, and these 
were then compared to the expected 
NICI contrast curve.  Stars around which NICI could detect a 
larger fraction of these simulated planets were given higher priority compared 
to stars with a lower fraction of detectable planets.  This 
approach maximizes the probability of detecting a planet, and also produces 
a sample that offers the most stringent statistical constraints on the 
underlying planet population properties.  Our final observing list contains 
4 B and A stars that were added as members of moving groups, 24 that 
were added as debris disk hosts, and a further 4 stars that belong to both 
categories.  15 stars were added to the 
observing list with spectral types earlier than A5, and 33 were added that 
had faint $V$ magnitudes on the color-magnitude diagram.  In this way we 
obtained our final observing list of 70 B and A stars for the Gemini NICI 
Planet-Finding Campaign.

\section{Age Determination for Field B and A Stars}\label{ages_sec}

\subsection{Previous Methods}\label{previous_age}

Because planets and brown dwarfs cool and become fainter throughout 
their lifetime, determining the age of a substellar companion's host star is 
essential to determining the physical properties of the 
companion.  Therefore, direct imaging surveys 
must be able to reliably age-date their target stars to understand both 
detected companions and overall survey sensitivities.  
Age-dating 
methods are typically most successful for solar-type (FGK) stars, where a 
variety of techniques are available, including lithium absorption, calcium 
emission, and gyrochronology (e.g. \citealt{mamajek09}).  For stars with 
higher mass the most robust method is to identify stars in 
young moving groups which share a common age and determine a consistent 
age for all the stars in the group (e.g. \citealt{zs04}).  Alternatively, 
if an A star is in a binary system with a solar-type star, age-dating the 
solar-type secondary can yield the age of the higher-mass primary.  In 
cases where a high-mass star is single and does not belong to a kinematic 
group, however, we require an age-dating method that only utilizes the 
properties of the star itself.

Previous work has typically addressed the need to age-date 
B and A stars by turning to the color-magnitude diagram (CMD).  
In order to estimate the mass of the brown dwarf 
companion HR~7329~B, \citet{jura95} and \citet{lowrance00} placed its 
primary star HR 7329 A on an [$M_V$, $B-V$] color-magnitude diagram, 
along with A stars from nearby clusters and in the field 
(see Figure 3 of \citealt{lowrance00}).  They noted that young 
($\lesssim$100 Myr) clusters 
such as the Pleiades, Alpha Per, and IC 2391 have A stars near the bottom 
of this color-magnitude diagram (i.e., at fainter absolute $V$ magnitudes) 
compared to brighter A stars of the older ($\sim$600 Myr) 
Hyades and Praesepe clusters.  
They also noted that the famous young A stars $\beta$ Pic, HD 141569, 
and HR~4796 all lie at the 
very bottom of this diagram along with HR 7329 A, thus arguing for a young 
($\lesssim$ 100 Myr) age for HR 7329 system.  \citet{moor06} and 
\citet{rhee07} expanded this 
method to other high mass stars with debris disks, noting that many of these 
stars also lay 
near the bottom of the color-magnitude diagram.  \citet{vigan12} use a similar 
approach by defining the 
dereddened single-star sequence of high mass stars in the Pleiades on the 
same [$M_V$, $B-V$] 
color-magnitude diagram and then assigning the age of the Pleiades 
(125 Myr, \citealt{stauffer98}) 
to stars with similar color-magnitude positions as the 
Pleiades A stars.  However, this common approach to flagging young A stars is 
likely too optimistic in many cases, due to two effects: (1)~the degeneracy 
between the effects of age and metallicity, and (2)~the increase of 
main-sequence lifetime with decreasing stellar mass.

\begin{figure}
\epsscale{0.9}
\plotone{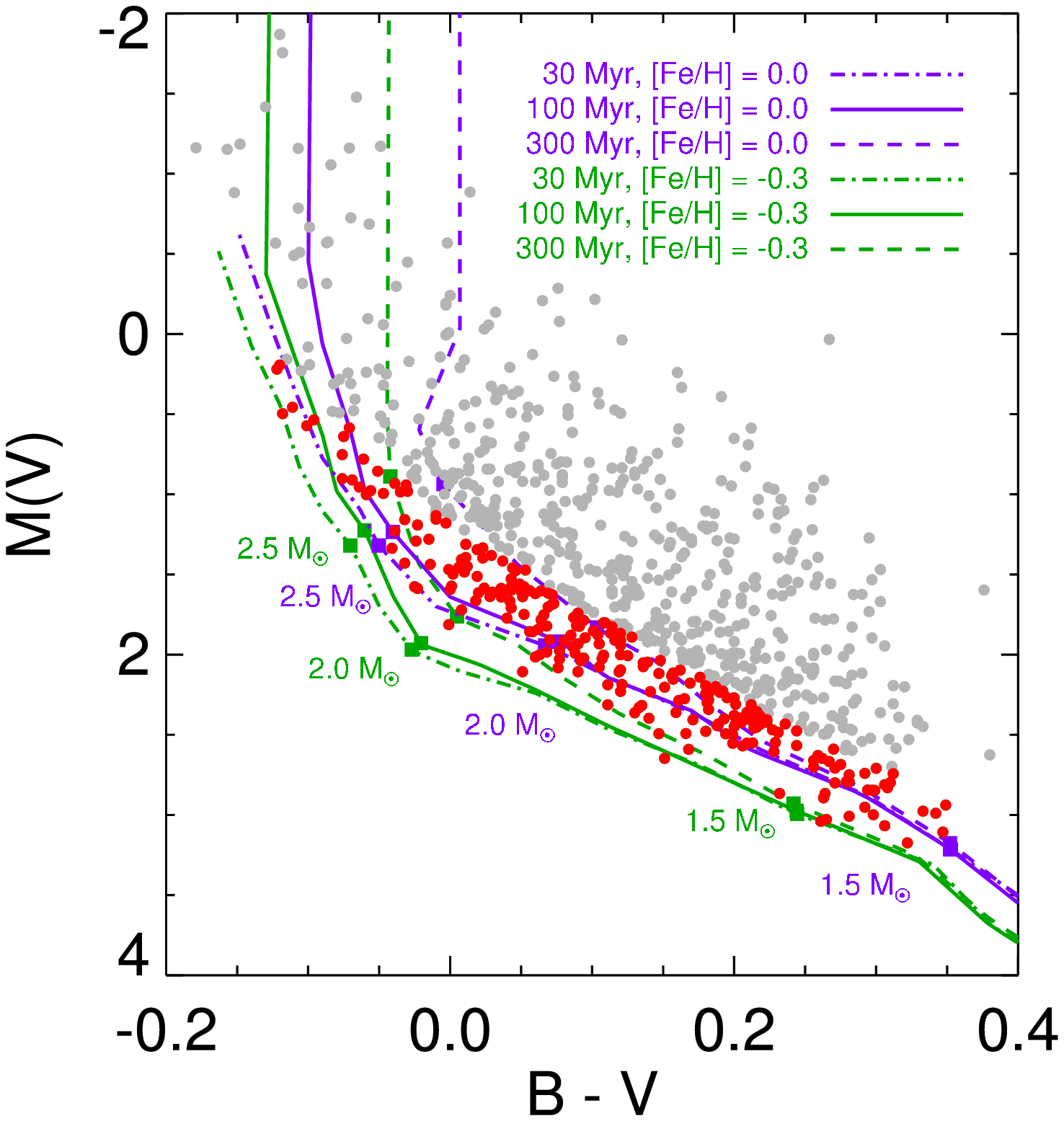}
\caption{Color-magnitude diagram for a volume-limited ($<$100 pc) 
sample of Hipparcos 
B and A stars, with isochrones from \citet{siess00} overlaid.  
Squares mark the locations on the isochrones of 
2.5, 2.0, and 1.5 M$_{\sun}$.  We select 
candidate young stars (red points) as those that lie below 
the fit to the young cluster A stars from \citet{lowrance00}.  These 
stars appear young when compared to the solar metallicity isochrones; 
however, the half-solar 
abundance isochrones ([Fe/H] = $-0.3$) with older ages ($\ge$ 100 Myr) 
can fit these ``low CMD'' stars equally well.
\label{hipcolmag1_fig}}
\end{figure}

In order to demonstrate the limitations of this approach, we construct a 
volume-limited sample of Hipparcos stars within 100 pc and 
with parallax uncertainties below 5\%, spectral types earlier than F0, 
luminosity class IV or V, $-0.2 < B-V < 0.4$, 
and purged of known binaries, resulting in 776 stars.  All stellar data 
are taken from the extended Hipparcos compilation of \citet{xhip}.  
Figure~\ref{hipcolmag1_fig} illustrates the difficulty in using 
color-magnitude diagram position as a youth indicator for this 
volume-limited sample of B and A stars.  
The \citet{siess00} isochrones show that the ``low CMD'' stars (defined 
here as lying below the young cluster fit of \citealt{lowrance00}, with all 
other B and A stars referred to as ``high CMD'' stars) can 
be equally explained as young solar-metallicity stars or older sub-solar 
metallicity stars.  Therefore a low 
position on the color-magnitude diagram is not a unique indicator of youth, 
but rather suggests either youth or low metallicity.

\begin{figure}
\epsscale{0.9}
\plotone{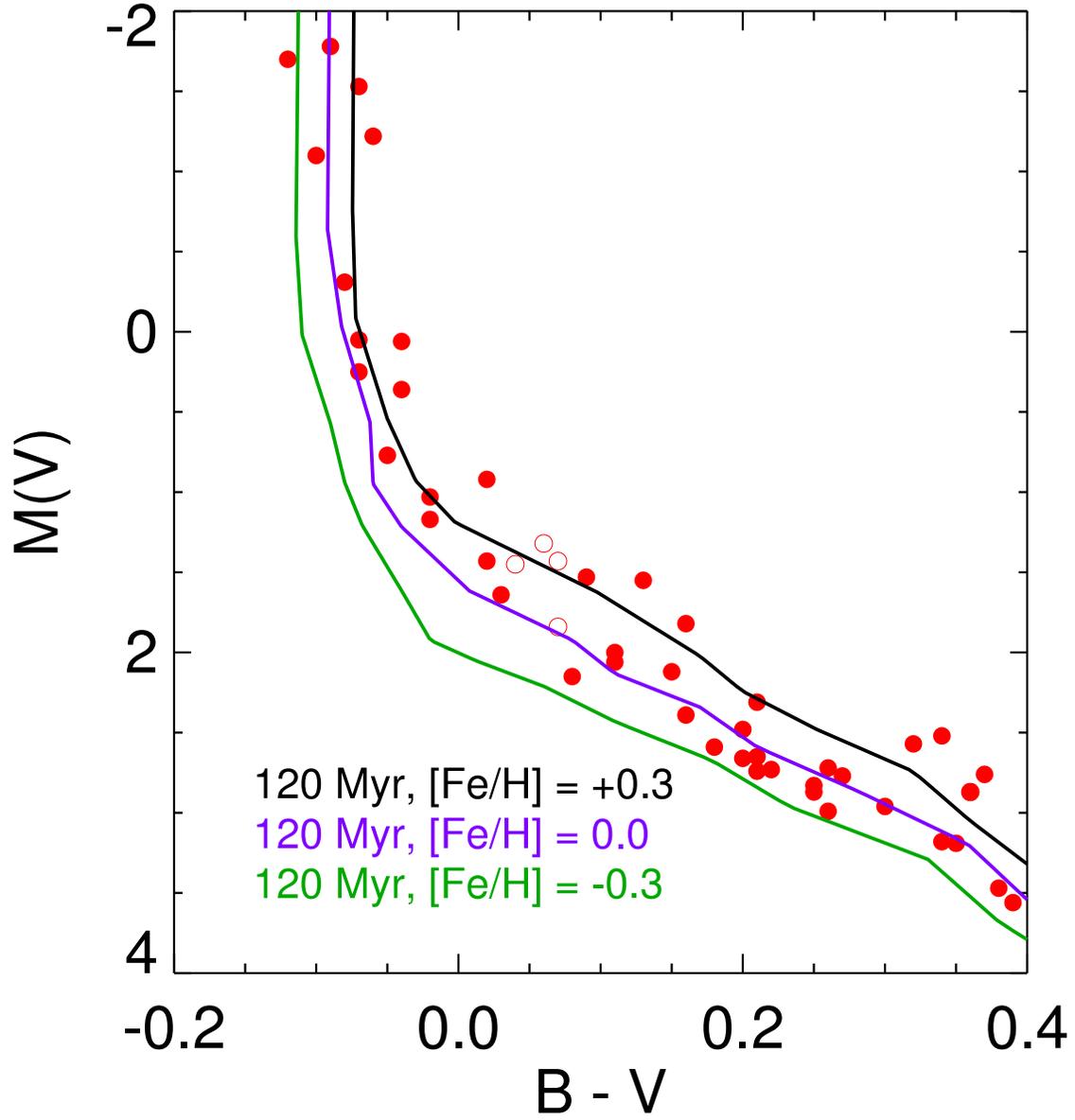}
\caption{High-mass Pleiades stars \citep{stauffer07}, with known binaries 
plotted as open circles, 
compared to 120 Myr 
tracks of three metallicities from \citet{siess00}.  The solar metallicity 
isochrone is consistent with the Pleiades single-star sequence.
\label{seiss_pleiades}}
\end{figure}

The \citet{siess00} isochrones themselves provide a good fit to B and A 
stars, as we demonstrate in Figure~\ref{seiss_pleiades} where we plot 
the \citet{siess00} 120 Myr isochrones 
for sub-solar, solar, and super-solar metallicities against Pleiades stars 
from \citet{stauffer07}, finding a good match between the solar track and the 
Pleiades single-star sequence.  A more detailed analysis of Pleiades stars 
and main-sequence binaries by \citet{isochrones_test} supports the validity 
of these models, showing that the \citet{siess00} 
isochrones are a good fit to both main-sequence and pre-main sequence 
stars.

\begin{figure}
\epsscale{0.9}
\plotone{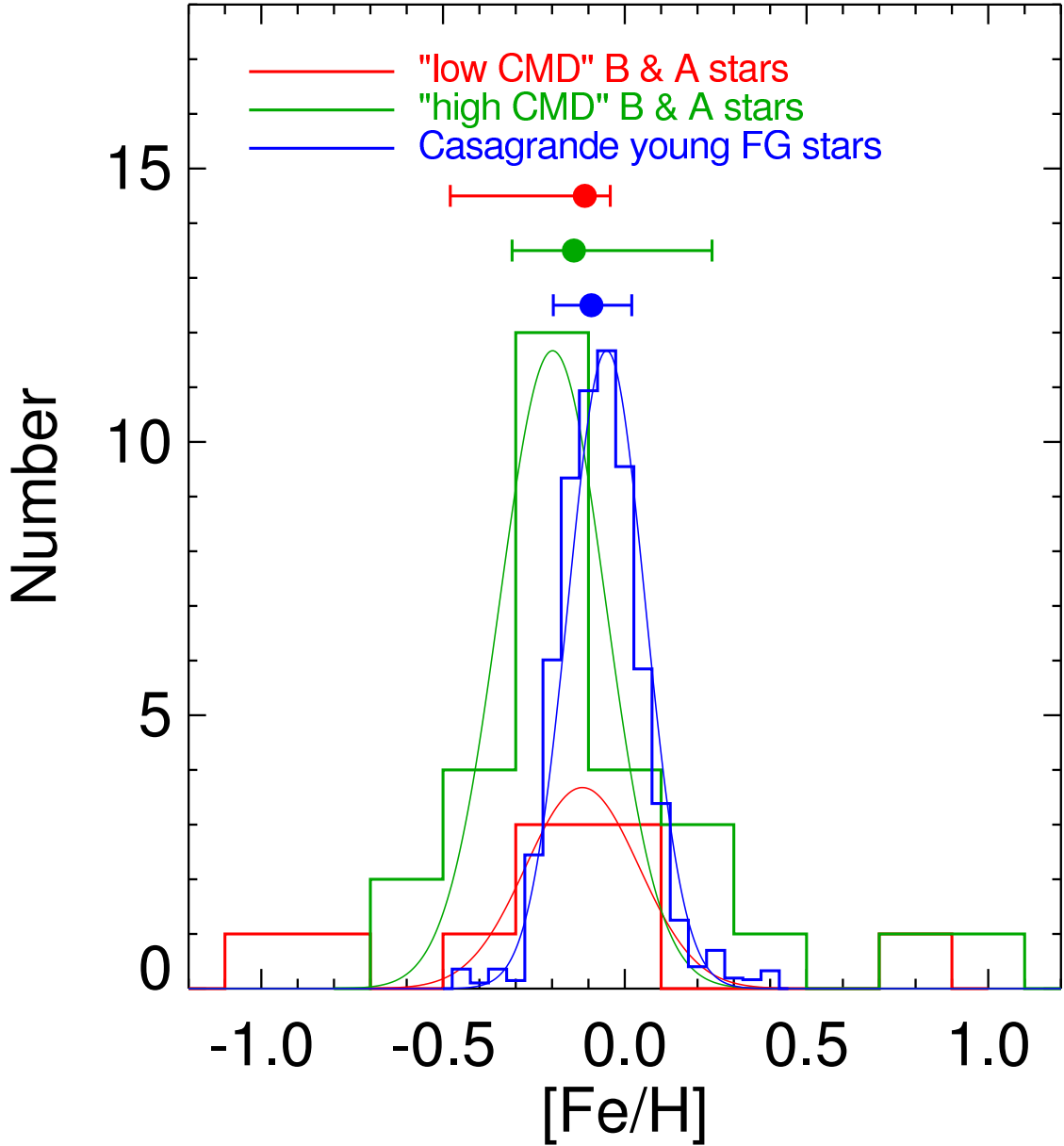}
\caption{Histogram of metallicities for Hipparcos B and A stars from 
literature measurements.  Both ``low CMD'' (red histogram) and 
``high CMD'' (green histogram) stars agree well with each other and also to 
the \citet{casagrande} measurement of young ($<$1~Gyr) solar 
neighborhood F and G stars (blue histogram).  
The \citet{casagrande} histogram has been extracted 
from their Figure 16 and normalized to match the peak of our fit 
for the ``high CMD'' stars.  Filled circles with error bars indicate the 
median and 68\% confidence intervals for each histogram.
\label{fehhist1_fig}}
\end{figure}

Metallicities between 0 and $-0.3$ dex are not particularly rare for B and A 
stars as illustrated by Figure~\ref{fehhist1_fig}, which shows B and A stars 
from the Hipparcos sample with literature metallicity measurements 
(again from the \citealt{xhip} compilation), divided into ``low CMD'' 
and ``high CMD'' samples.  No appreciable difference is seen between 
the two samples of stars, nor between these B and A stars and the 
\citet{casagrande} metallicity measurements of young ($<$1 Gyr) F and G 
dwarfs.  The lack 
of a trend between color-magnitude diagram position and metallicity 
means that while ``low CMD'' stars are not uniformly young, neither are they 
uniformly low metallicity.  They instead seem to have a similar metallicity 
distribution to the solar neighborhood.  Therefore there are young solar or 
supersolar metallicity stars in the sample of ``low CMD'' stars, mixed in with 
lower metallicity, older stars.

\begin{figure}
\epsscale{0.9}
\plotone{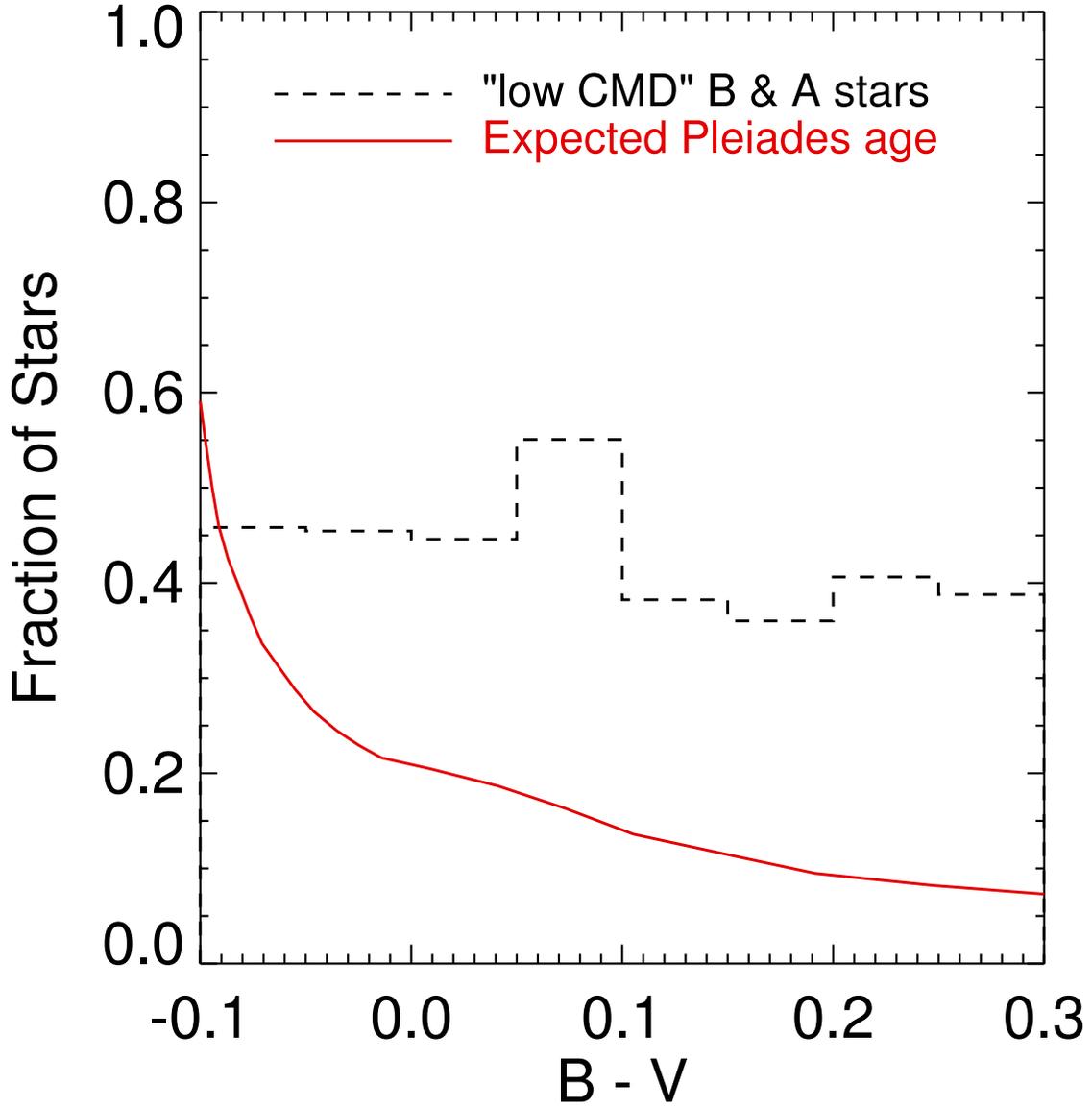}
\caption{The fraction of stars lying low on the color-magnitude diagram 
(``low CMD'') from our Hipparcos volume-limited sample 
as a function of $B-V$ color (black histogram), compared to the expected 
number of stars the same age as the Pleiades or younger ($\le$125 Myr, red 
curve), given the mean main-sequence lifetime of solar metallicity 
stars as a function of 
zero-age main-sequence color \citep{siess00} and assuming a constant star 
formation rate.  Even ignoring the presence of low-metallicity stars, 
assigning Pleiades age to all ``low CMD'' stars overpredicts the fraction of 
young A stars in a sample, especially at later spectral types.
\label{pleiadesfrac_fig}}
\end{figure}

In addition to metallicity, another important consideration is that most of 
a main-sequence star's evolution across the color-magnitude diagram occurs 
during the 
final two-thirds of a star's main-sequence lifetime 
(e.g. \citealt{soderblom2010}).  The \citet{siess00} models indicate that this 
trend holds for stellar masses between 1.5 and 4.5 M$_\sun$ (appropriate to 
our NICI sample), where stars show little change in $B-V$ or $V$ from the 
ZAMS to $\sim$1/3 of the main sequence lifetime, and then a brightening in $V$ 
over the remainder of the star's main sequence lifetime.  
This effect can be seen in Figure~\ref{hipcolmag1_fig}, where at $B-V >$~0.1, 
the 30 Myr and 100 Myr isochrones for a given metallicity overlap, while all 
isochrones between 30 and 300 Myr overlap for $B-V >$~0.2.  Even if 
we were to (incorrectly) assume that all stars have solar metallicity, while 
all Pleiades-age stars would lie low on the color-magnitude diagram, not all 
``low CMD'' stars would be Pleiades-age.  In this uniform solar metallicity 
scenario, at best we would be constraining the ages of stars to be 
in the first third of their 
main-sequence lifetime, which is true for all Pleiades A stars.  But 
the longer lifetimes for later-type A stars would make the default assigned 
age of 
125 Myr increasingly inaccurate.  For A9V stars, with an expected lifetime of 
1.5 Gyr, ``low CMD'' stars would have ages up to $\sim$500~Myr, a factor of 
4 older than 
the age of the Pleiades.  Figure~\ref{pleiadesfrac_fig} demonstrates this 
by plotting the fraction of ``low CMD'' stars as a function of $B-V$ from 
our Hipparcos volume-limited sample.  For comparison, we also plot the 
expected fraction of stars younger than the Pleiades, which is the 
quotient of 125 Myr and the main-sequence lifetime for solar-metallicity 
stars from \citet{siess00}, assuming a constant star formation rate in the 
solar neighborhood.  While the fraction of ``Pleiades-like'' stars derived 
from the color-magnitude diagram appears to be relatively constant for all 
values of 
$B-V$, the expected fraction of Pleiades-aged stars drops by a factor of 8 
over this range.  
In other words, the ``low CMD'' stars are expected to have an increasingly 
large age spread at lower masses simply from stellar evolution.  This effect 
is missed by the common procedure of assigning Pleiades ages to all ``low 
CMD'' stars.

\subsection{A Bayesian Approach to Estimating Ages from Isochrones}\label{bayes}

For a given location on a color-magnitude diagram, 
there are multiple combinations of age, 
mass, and metallicity that can reproduce the color and magnitude of a given 
star.  We turn to Bayesian inference in order to determine the relative 
likelihoods of these combinations and to incorporate prior knowledge 
about the 
distributions of these stellar parameters.  Such a technique has been 
previously used for solar-type stars by \citet{takeda07}, who determined the 
ages of target stars for RV planet searches 
through a Bayesian analysis of their 
spectroscopically-derived properties (T$_{eff}$, [Fe/H], and log(g)) combined 
with their $V$ magnitude.

To determine the ages for B and A stars using photometry, we begin with Bayes' 
theorem:

\begin{equation}
P(model|data) \propto P(data|model) P(model) .
\label{bayes1}
\end{equation}

\noindent That is, the probability of a model given a set of 
data ($P(model|data)$) is 
proportional to the product of the probability of the data given the model 
($P(data|model)$) and the probability of the model itself ($P(model)$).  
From left to right, the three terms above are referred to as 
the posterior probability density function (PDF), the likelihood, 
and the prior.

Our model is the set of three stellar 
parameters, age ($\tau$), mass ($M_*$), and metallicity ([Fe/H]), and 
our data are the measured absolute magnitude $M(V)$ and $B-V$ color 
for a given star.  At any combination of age, mass, and 
metallicity \citet{siess00} predict the $M(V)$ and $B-V$.  It is then 
straightforward to compute the chi-squared statistic for this model:

\begin{equation}
\chi^2(\tau,M_*,[Fe/H]) = \sum \frac{(O - E)^2}{\sigma^2} = 
\frac{((B-V)_O - (B-V)_E)^2}{\sigma_{B-V}^2} + 
\frac{(M(V)_O - M(V)_E)^2}{\sigma_{M(V)}^2}
\label{bayes2}
\end{equation}

\noindent where $(B-V)_O$ and $M(V)_O$ are the observed ($O$) color and absolute 
magnitude, with observational errors $\sigma_{B-V}$ and $\sigma_{M(V)}$, 
respectively.  $(B-V)_E$ 
and $M(V)_E$ are the expected ($E$) color and magnitude, predicted by 
\citet{siess00} for a given age, mass, 
and metallicity.  The relative likelihood of a specific 
combination of three stellar parameters is then

\begin{equation}
P(data|model) = P(B-V,M(V)|\tau,M_*,[Fe/H]) \propto e^{- \frac{1}{2} \chi^2(\tau,M_*,[Fe/H])}
\label{bayes3}
\end{equation}

\noindent This formulation assumes Gaussian statistics, which is a reasonable 
assumption given that our errors are from photometry and parallax measurements.

\citet{siess00} provide pre-main sequence and main-sequence 
evolutionary tracks, which we interpolate to a regular 3-dimensional grid 
in age, mass, and metallicity.  Age is logarithmically gridded between 1 Myr 
and 10 Gyr, with the predicted photometry becoming undefined when the star 
leaves the main sequence.  
Mass is uniformly gridded between 1 and 5 M$_\sun$.  Metallicity is 
uniformly gridded between $-0.3$ $<$ [Fe/H] $<$ $+0.3$, since a linear grid 
in [Fe/H] corresponds to a logarithmic grid in absolute metal abundance.

We adopt priors to incorporate knowledge of the solar 
neighborhood when deriving probability distributions of stellar parameters.  

\begin{enumerate}
\item For metallicity, we turn to the \citet{casagrande} metallicities for 
a magnitude-limited sample ($V <$ 8.3 mag) of 16000 F and G dwarfs in 
the solar neighborhood.  In 
particular, we use their Figure 16 to provide a metallicity 
distribution for stars younger than 1 Gyr, an appropriate age 
range for field B and A stars.  We model the \citet{casagrande} 
metallicity histogram with a normal distribution 
with mean of $-$0.05 and $\sigma$ of 0.11 dex and 
use this as our metallicity prior.  We note that the \citet{casagrande} 
metallicity distribution of young F and G stars in 
the solar neighborhood is consistent with the \citet{nieva12} metallicity 
results for 20 nearby ($<500$ pc) early-B stars 
([Fe/H] = 0.0~$\pm$~0.1 dex).  This good match to the young F and G dwarf 
measurements occurs 
despite the fact that the lifetimes of early-B stars are over an 
order of magnitude younger than 1 Gyr.  Further support for the near solar 
metallicity of young stars is given by \citet{biazzo12}, who find 
[Fe/H]~=~0.10~$\pm$~0.03~dex for five solar-type stars in the AB Dor moving group 
(100 Myr).  So all the evidence to date suggests that solar neighborhood 
B and A stars have a metallicity distribution similar to young F and G stars.

\item For our age prior, since the star formation rate is basically constant in the 
solar neighborhood over at least the last 2 Gyr \citep{localsfr}, we adopt 
a prior that is uniform in linear age for B and A stars.  Our 
gridding of the \citet{siess00} models is logarithmic in age, and so we 
weight each grid point such that equal linear age intervals 
have equal probability.  The relative probability of age bin $i$ is given by 
$\delta \tau_i/T$, where $\delta \tau_i$ is the age range 
encompassed by the $i$-th bin, and $T$ is the total age range for all bins.  
This prior has the effect of pushing the posterior PDF 
toward older ages, since older logarithmic age bins encompass 
a greater span of time than younger ones.

\item Finally, we apply a Salpeter IMF of the form 
$dN/dM \propto M^{-2.35}$ \citep{salpeter} as our mass prior.  Each mass 
bin is weighted by this IMF, favoring lower masses over higher ones.  We 
note that the effect of this prior is minimal, since each star has a narrow, 
well-defined mass posterior PDF.  As a check, we computed the shift in the 
median of the 
mass and age posterior PDFs for all our target stars with and without 
the IMF prior.  When the IMF prior is applied, the median shift is 
0.2\% toward lower masses and 1.4\% to older ages.  The standard deviation of 
the shift is 0.15\% in mass and 1.4\% in age.
\end{enumerate}

\begin{figure}
\epsscale{0.9}
\plotone{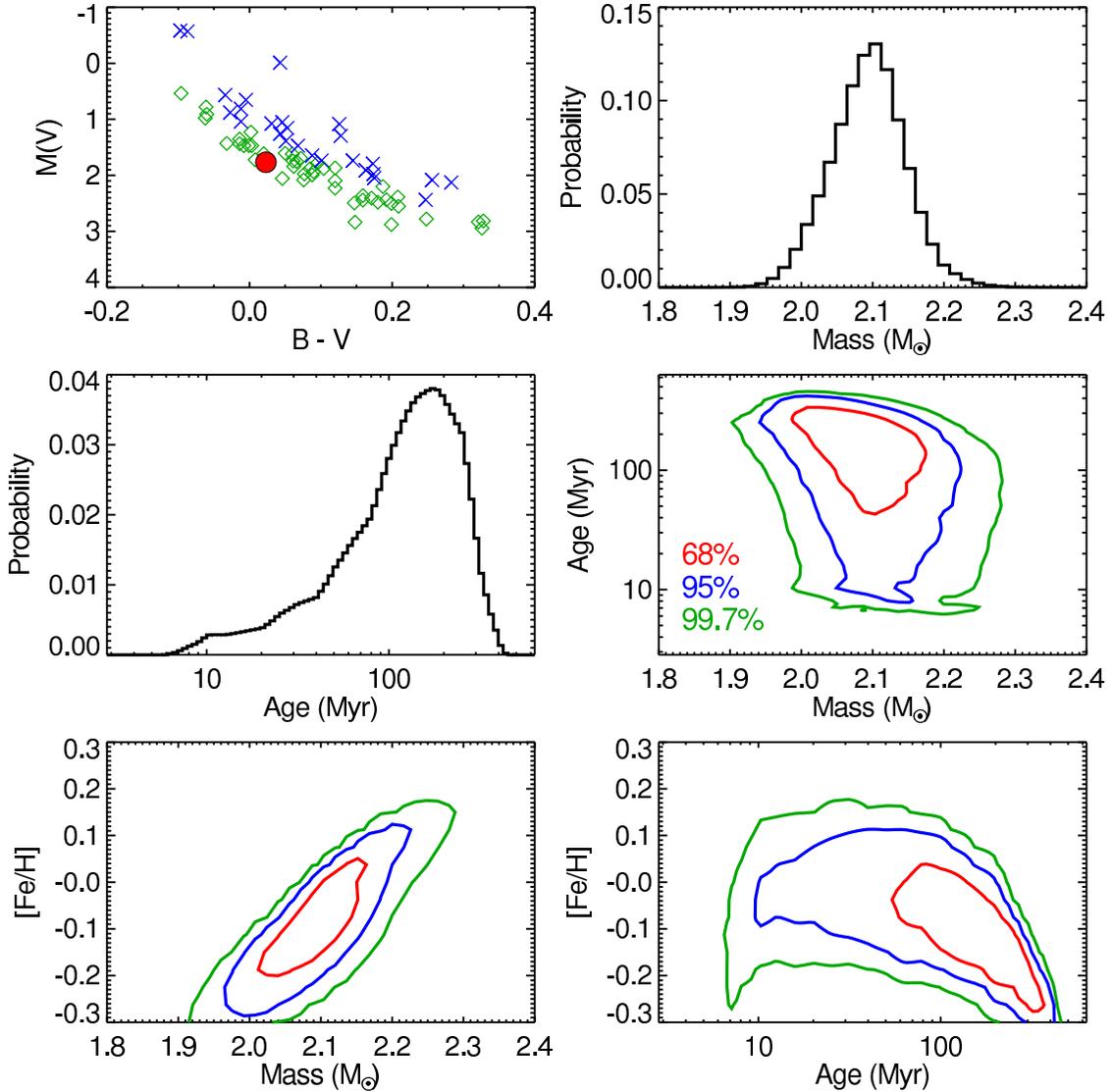}
\caption{An example of our Bayesian method for determining ages, applied to 
the A0 star HD~24966.  Top left: the position of 
HD 24966 (filled red circle) on the color-magnitude diagram compared to 
the other NICI Campaign B and A stars; the star lies among the ``low CMD'' 
(green diamonds) stars.  Blue crosses denote the  ``high CMD'' NICI B and 
A target stars.  Top right and middle left panels: the 
posterior PDFs for mass and age from our Bayesian analysis, 
showing a well-defined mass and a relatively wide age 
distribution.  The median age is 128 Myr with 68\% confidence limit 
from 51--227 Myr.  Middle right, bottom left, and bottom right panels: 
2D contour plots indicating the covariance of the three stellar 
parameters.  In general, older ages correspond to 
lower metallicities for this star.
\label{bayes_fig}}
\end{figure}

Our final age probability distribution for each star is then given by 
marginalizing over all metallicities and masses.  This is done after 
determining all likelihoods for each model (given by Equations~\ref{bayes2} 
and \ref{bayes3}), weighted by the metallicity and age priors 
(Equation~\ref{bayes1}).  Figure~\ref{bayes_fig} shows an example of our 
Bayesian age analysis for the A0V star HD 24966.  Since the main-sequence 
lifetime for a 2.1 $M_\sun$ star is 600 Myr, a median age for HD 24966 of 
128~Myr (and 68\% confidence level between 51--227 Myr) suggests that this 
star is in the first third of its main-sequence lifetime.

\begin{figure}
\epsscale{0.9}
\plotone{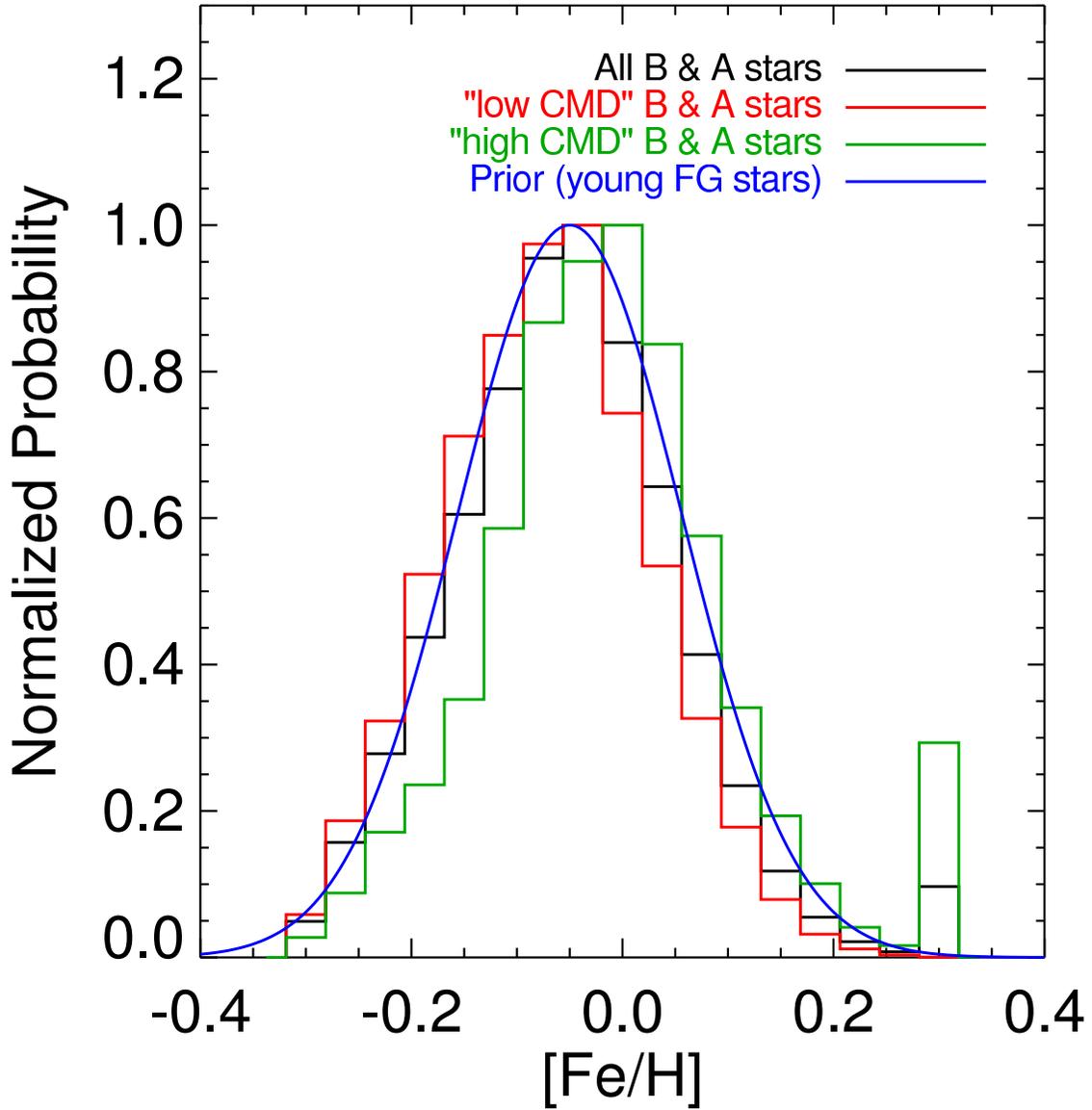}
\caption{Combined metallicity posterior PDFs for all B and A stars 
in our NICI sample (black histogram) compared to our adopted 
metallicity prior, the 
\citet{casagrande} measurements for F and G stars younger than 1 Gyr (blue 
curve).  In general, our Bayesian method finds lower metallicities 
(consistent with the prior) for ``low CMD'' B and A stars 
(red histogram) and 
slightly higher metallicities for ``high CMD'' stars (green histogram).
\label{metallicityout_fig}}
\end{figure}

Figure~\ref{metallicityout_fig} shows the combined [Fe/H] 
posterior PDFs marginalized over age and mass 
from all 70 target stars compared to our adopted 
\citet{casagrande} 
prior.  The posterior and prior closely match each other for the ``low CMD'' 
stars, while there is a small 0.05~dex offset toward higher metallicities 
for the ``high CMD'' stars.  This is as expected, since ``high CMD'' stars 
are best fit by higher metallicity isochrones.

\begin{figure}
\epsscale{0.9}
\plotone{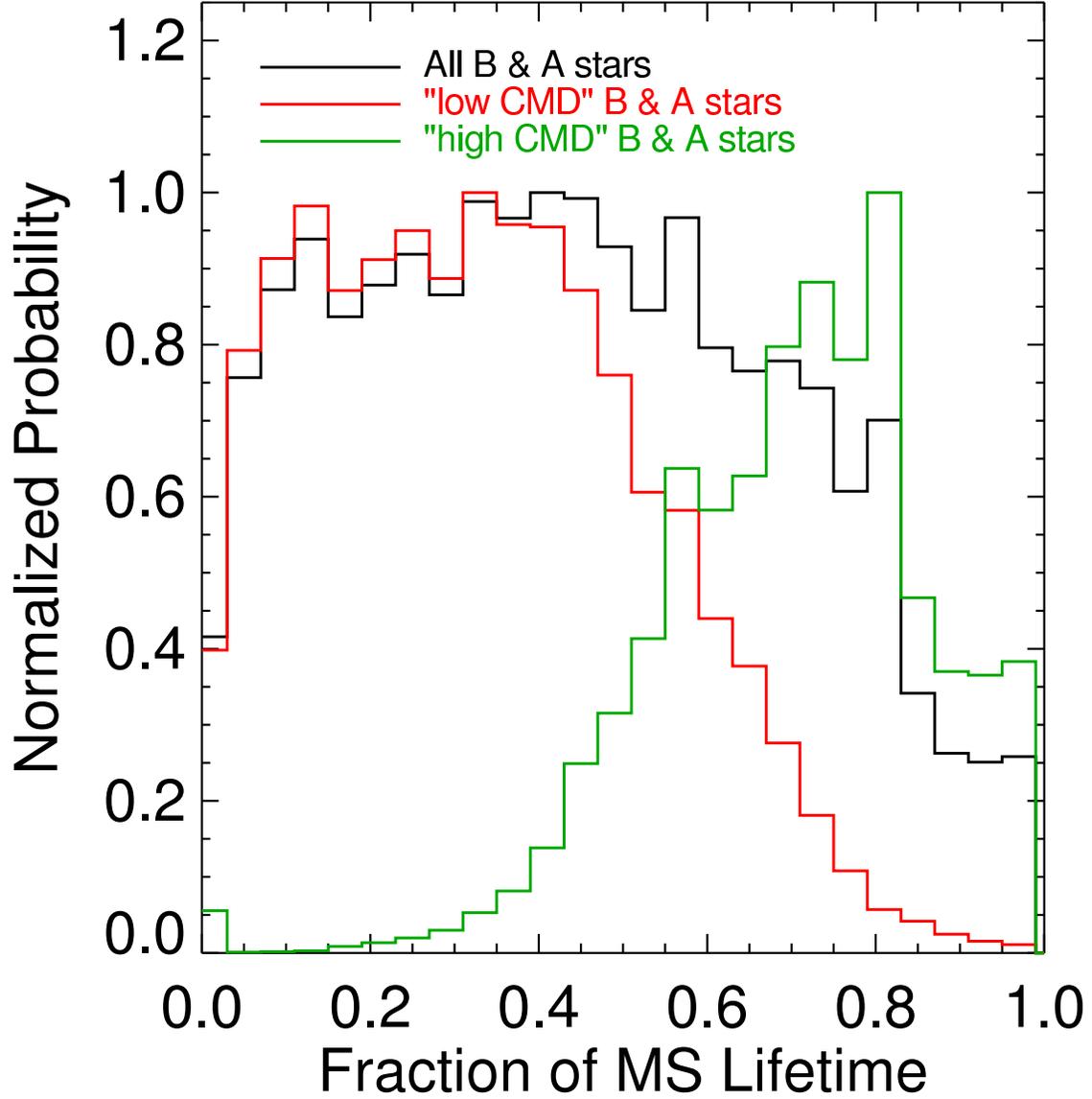}
\caption{Combined age posterior PDFs for our NICI sample of B and A stars 
normalized to the inferred main-sequence lifetime (black histogram).  As 
expected, the ``low CMD'' stars (red histogram) are systematically closer to 
the beginning of 
the main sequence compared to ``high CMD'' stars (green histogram), 
even when metallicity is taken into account.
\label{ageout2_fig}}
\end{figure}

In addition to age, mass, and metallicity, we 
can also produce a posterior PDF of main-sequence lifetime for each star 
from the same likelihood and priors that produce posterior PDFs for age, 
mass and metallicity (though main-sequence lifetime is a function only of 
mass and metallicity and independent of age).  We express the combined age 
PDFs for all stars in Figure~\ref{ageout2_fig} as a fraction of their 
main-sequence 
lifetime, using the lifetime PDF for each star.  Despite allowing 
metallicity to vary (though following the \citealt{casagrande} prior), 
our Bayesian technique finds ``low CMD'' stars to be systematically younger 
than ``high CMD'' stars as expected.

\begin{figure}
\epsscale{0.9}
\plotone{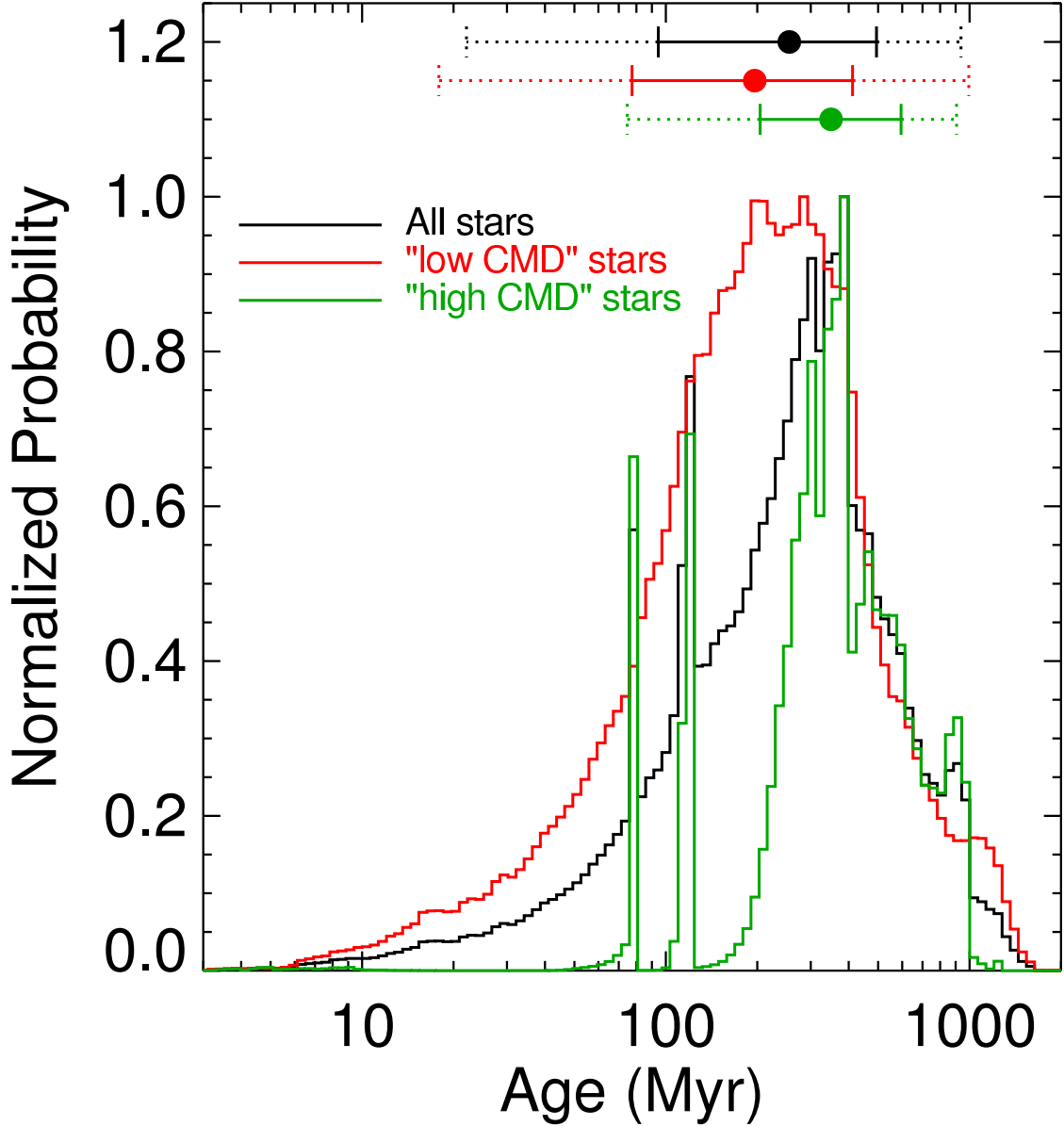}
\caption{As in Figure~\ref{ageout2_fig}, the combined age PDFs for our 
targets but 
now with absolute ages.  At the top of the plot we note the location of the 
median (filled circle), 68\% confidence level (inner brackets) and 
95\% confidence level (outer brackets) for each combined PDF.  
The overall sample is peaked at 300 Myr, 
falling quickly to $\sim$1 Gyr, with a young age tail extending to $\sim$10 
Myr.  Spikes in the distribution for ``high CMD'' stars at 78, 115, and 118 
Myr are due to the stars HIP 74785, HIP 49669, and HIP 36188, which have 
spectral types B8, B7, and B8 respectively.  The posterior age PDFs of 
these stars are essentially delta functions at the oldest allowable ages 
in the \citet{siess00} isochrones at the $B-V$ of these stars, since their 
$V$ magnitudes are significantly brighter than the isochrone values at the 
end of the main sequence.
\label{ageout1_fig}}
\end{figure}

\begin{figure}
\epsscale{0.9}
\plotone{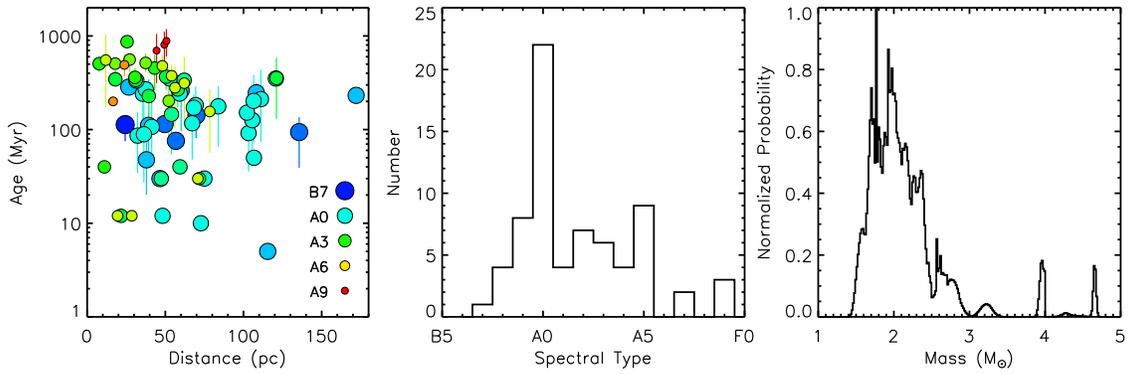}
\caption{\small{\
Properties of the 70 B and A stars observed by the Gemini NICI 
Planet-Finding Campaign.  (\textit{left}) Age and distance of our target stars, with 
symbol size and color corresponding to spectral type.  Stars in a moving 
group are presented at a single age, while stars where we derive the age from 
our Bayesian method are plotted at the median age, with error bars indicating 
the 68\% confidence interval.  (\textit{middle}) A histogram showing the spectral types 
of our target stars.  (\textit{right}) The mass distribution of our sample, 
produced by combining the mass posterior PDFs for each of our target stars 
from our Bayesian analysis.  The median of the mass distribution is 
2~M$_{\sun}$, 60 of our 70 stars have a median mass between 
1.5--2.5~M$_{\sun}$, and 66 have a median mass less than 3~M$_{\sun}$.
}
\label{targets_fig}}
\end{figure}

In Figure~\ref{ageout1_fig}, we display the 
same combined age PDF as a function of absolute age.  The most common ages 
for our target stars are between 50 and 500 Myr, with low-probability 
tails extending below 10 Myr and above 1 Gyr.  The median age, 68\% 
confidence limit, and 95\% confidence limit for each of our 70 
target stars are given in Table~\ref{table1}.  
We also show the properties of our target stars in Figure~\ref{targets_fig}.  
About a fifth (14 out of 70) of 
our stars have independent age measurements from moving group 
membership, as indicated in Table~\ref{table1}, which we adopt as the final 
ages for these stars.  Also, we adopt the age of 5 Myr for HD 141569 from 
\citet{weinberger00}, based on their study of its two M-star companions.  
These independent ages are often significantly younger than the ages derived 
by our Bayesian analysis, which at best can only place a star within the 
first third of its main sequence lifetime.

\subsection{Comparison with Previous Ages}

Stars with debris disks are generally well-studied and have literature 
age measurements that are widely cited (e.g. \citealt{moor06}, 
\citealt{rhee07}).  Nevertheless, for most of the debris disk stars in 
our B and A star sample, we do not use these literature ages and instead 
use our Bayesian ages.  Of the 28 stars in our sample with debris disks, 
10 belong to well-known moving groups, and we adopt that moving group age 
for these stars.  Also, we use the 5~Myr age for HD~141569 from 
\citet{weinberger00}.  Of the remaining 17 stars, 
3 (HIP~85340, $\gamma$~Oph, and Fomalhaut) 
have ages taken from isochrones, while 
14 (HD~10939, HD~17848, HD~24966, HD~31295, HD~32297, HD~54341, HD~71155, 
HD~85672, HD~110411, HD~131835, HD~138965, HD~176638, HD~182681, and 
HD~196544) are flagged as young for their low position on the color-magnitude 
diagram.  For these our Bayesian analysis provides more accurate 
representations of the ages than isochrones or color-magnitude 
diagram position alone, so we override the literature ages for these 17 stars 
with our Bayesian ages.

\begin{figure}
\epsscale{0.9}
\plotone{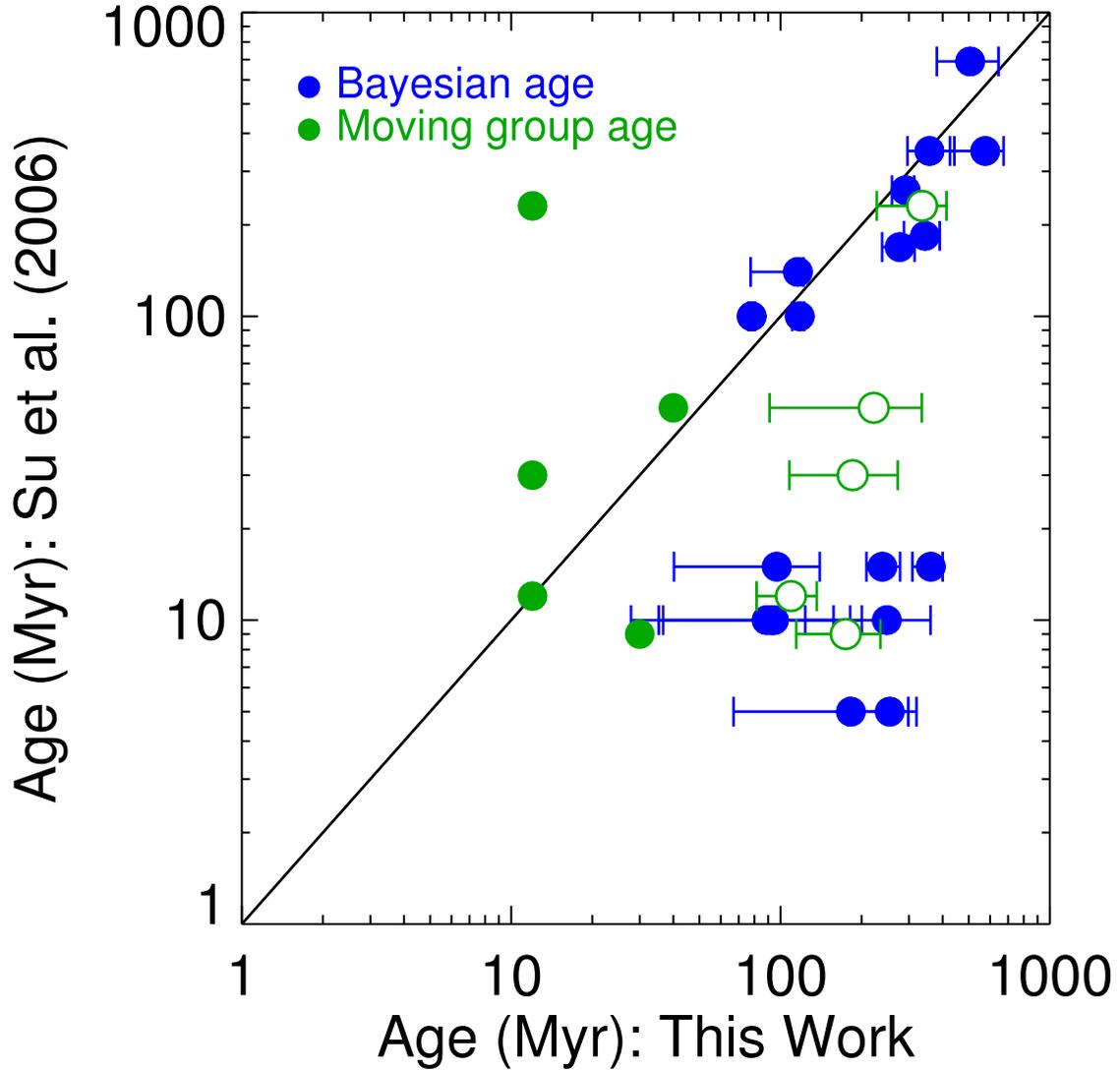}
\caption{\small{
Comparison of the ages derived using our Bayesian analysis to 
ages from \citet{su06} based on CMD analysis.  For stars that 
do not belong to known moving groups 
we give the Bayesian age as the median and 68\% confidence interval (filled 
blue circle and error bars).  Stars with an independent age measurement from 
membership in a moving group (as 
detailed in Table~\ref{table1}) are plotted twice, once as an empty green 
circle with error bars indicating the results of the Bayesian analysis and 
again as a filled green circle indicating the age of the moving group, which 
we adopt 
as the final age for that star.  There is good agreement between 
our work and \citet{su06} for stars that \citet{su06} flag as older than 
100 Myr, but below that age our Bayesian analysis finds systematically older 
ages as expected.  For the moving group stars, since the Bayesian analysis of 
photometry can only 
place a star in the first third of its main-sequence lifetime, 
additional information is needed to age-date 
young stars with greater accuracy.}
\label{agesu_fig}}
\end{figure}

Figure~\ref{agesu_fig} compares our ages to stars that we have in common 
with \citet{su06}.  For the ages derived by our 
Bayesian analysis, we find good agreement for stars which \citet{su06} 
determine to be older than 100 Myr.  Stars that \citet{su06} find to be 
younger than 100 Myr show a systematic disagreement, with the median of 
our Bayesian age distribution significantly older than the single age found 
by \citet{su06}.  This is as we would expect, since most movement across 
the color-magnitude diagram happens in the final two-thirds of a star's 
main-sequence lifetime, so older stars can be age-dated more definitively 
while younger stars can be anywhere in the first third of their 
main-sequence lifetimes.  Our Bayesian method takes this uncertainty into 
account, rather than assigning very young ages ($\lesssim$100 Myr) for 
``low CMD'' stars.  Also, stars that belong to a well-known moving group 
(as well as HD 141569) have adopted ages that are uniformly younger than the 
Bayesian distributions.  This is not surprising, as nearby moving groups are 
significantly younger than the field population.

\begin{figure}
\epsscale{0.9}
\plotone{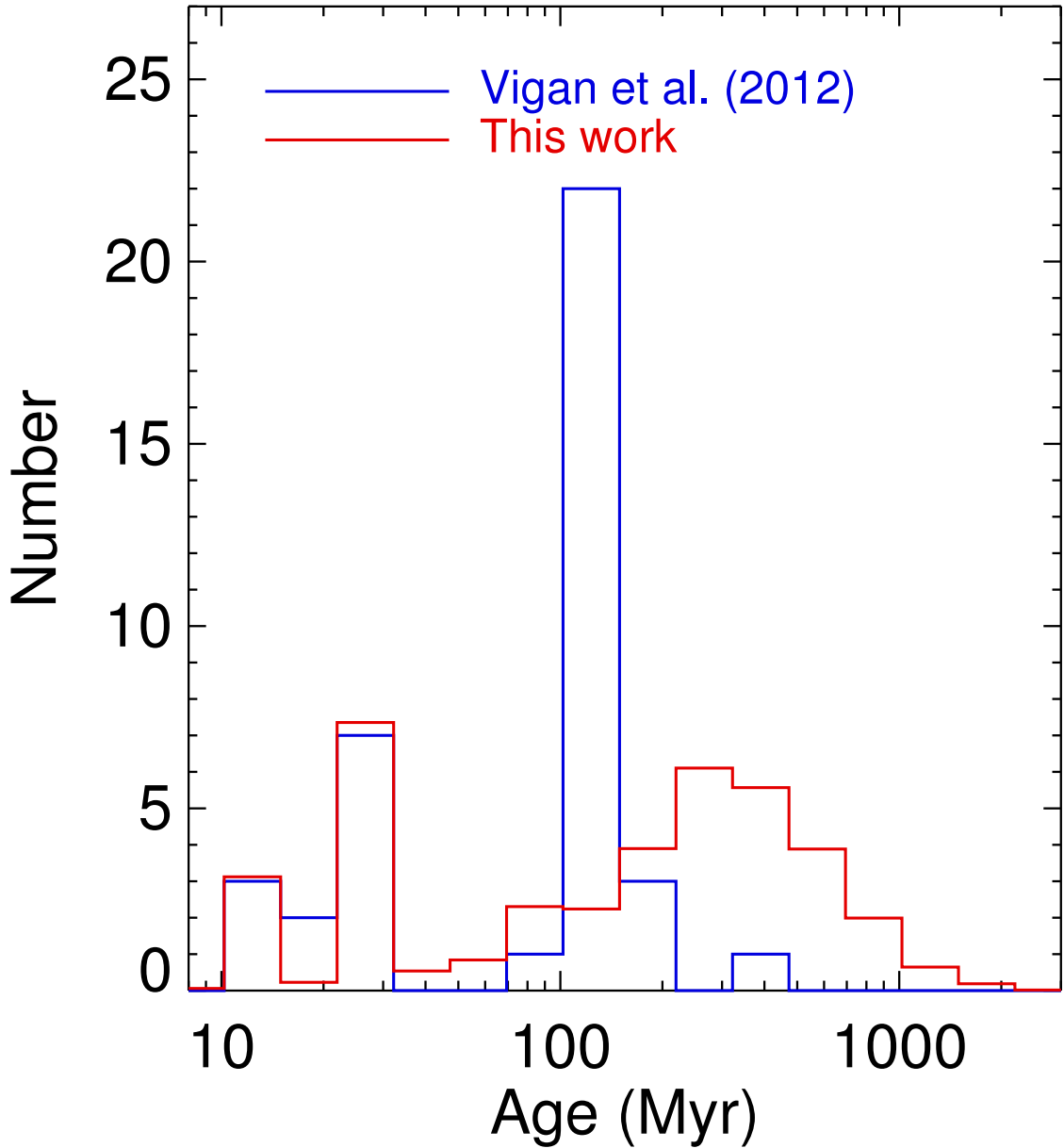}
\caption{Ages for stars in the \citet{vigan12} sample, using their ages 
(blue) and 
ages from our Bayesian analysis (red).  Ages for stars in moving 
groups (at 12, 30 and 70 Myr) 
are the same in both histograms, while for the remaining 
stars the posterior PDFs from our Bayesian analysis are used to populate 
each age bin.  \citet{vigan12} have a significant spike at 125 Myr, where 17 
of their 39 stars are assumed to have the age of the Pleiades.  Our 
Bayesian analysis suggests much older ages for these stars, with a peak at 
300 Myr and a tail extending beyond 1 Gyr.
\label{vigan_fig}}
\end{figure}

Finally, we consider the ages derived by \citet{vigan12} for their sample of 
young A stars.  We apply our Bayesian analysis to their sample and 
compare the histogram of our derived ages to theirs in 
Figure~\ref{vigan_fig}.  Out of the 39 stars from their sample, 11 
belong to moving groups, and we adopt the same moving group ages for these 
stars.  For the remaining 28 stars, we derive ages from our Bayesian 
analysis and populate each bin of our histogram with the appropriate 
contribution from the posterior age PDF.  As expected, the Bayesian 
approach results in a significant tail at large ages ($>$200 Myr).  
By contrast, almost half of the \citet{vigan12} ages (17/39) are incorrectly 
assumed to be 125 Myr, the age of the Pleiades.  For stars not in young moving 
groups, we compute the offset from the \citet{vigan12} ages to the median 
age from our Bayesian analysis.  The median of this offset for these 28 stars 
is 135 Myr, an increase in age of 55\%, which significantly increases the 
minimum detectable mass around these stars.  The typical contrast achieved 
at 2'' in \citet{vigan12} is 14.2 mags, which for 26 of these 28 stars 
corresponds to planetary masses using their ages and 
the \citet{cond} evolutionary models for companion 
luminosity.  When we instead adopt the median of our Bayesian age 
distributions for these stars, only 18 stars reach planetary masses at 2'', 
and the median increase in detectable mass at 2'' for these 18 stars is 
22\%.  The magnitude of this increase in detectable mass ranges from 0 
to 9 M$_{Jup}$, with a median of 2.2 $M_{Jup}$.  Therefore, if our older 
ages were to be used, the constraints on exoplanets from the \citet{vigan12} 
dataset would become weaker.

\begin{figure}
\epsscale{0.9}
\plotone{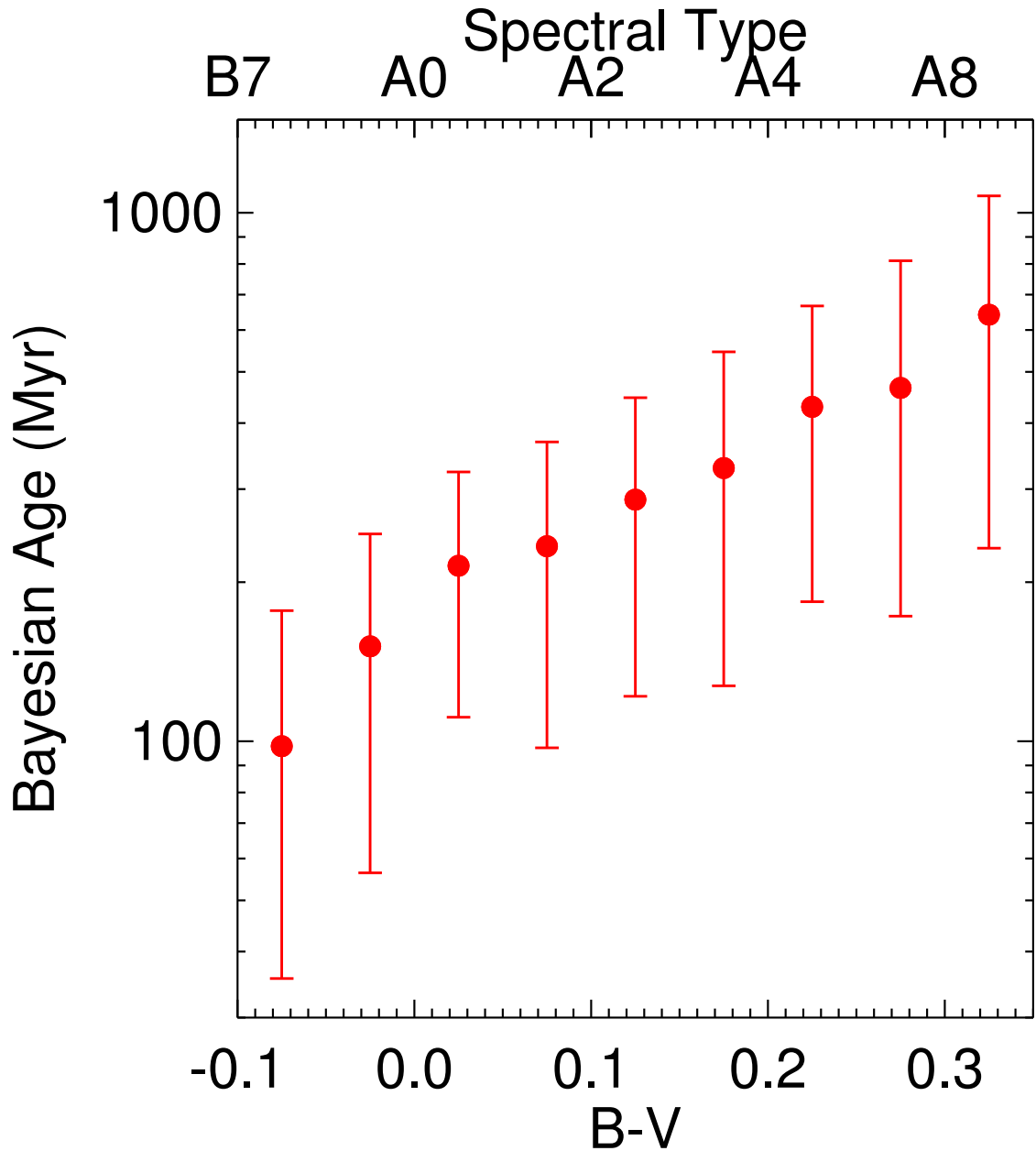}
\caption{Bayesian ages for ``low CMD'' stars in our 
volume-limited sample.  
The stars are divided into bins by B-V color, and the age posterior PDFs are 
combined within each bin.  Approximate spectral type as a function of 
$B-V$ color is given on the top axis.  The filled circle 
and error bars then give the median and 68\% confidence interval for that 
bin, with most bins containing about 25 stars.  Assigning Pleiades ages 
(125 Myr) to these stars is reasonable only at the bluest colors; 
later-type A stars that are faint on the color-magnitude diagram 
are likely to be significantly older.
\label{bayesspread}}
\end{figure}

In Figure~\ref{bayesspread} we examine how this age 
uncertainty for ``low CMD'' stars depends on $B-V$ color.  We divide 
``low CMD'' stars in our volume-limited sample of \textit{Hipparcos} 
stars into 9 color bins, and combine 
the marginalized PDFs for age for the stars in that bin.  As expected, the 
median age in each color bin increases with redder color.  Assuming a 
Pleiades age for ``low CMD'' stars is a fair representation of stars 
bluer than $B-V$ of $-0.05$, but for redder stars such an assumption 
significantly underestimates the age.

To summarize the results of our age analysis, the standard practice of 
assigning an age of 125 Myr to 
stars in the Pleiades locus on the color-magnitude diagram produces 
systematically younger ages than we find with our Bayesian method.  
Figure~\ref{ageout1_fig} shows that these Pleiades-like stars in the NICI 
Campaign sample have a median age of 200 Myr with a 
68\% confidence level between 160 and 840 Myr.  This is expected since 
identifying a star as similar to a Pleiades 
A star only places it within the first third of its main-sequence 
lifetime, which for late-type A stars can imply ages much 
larger than 125~Myr.

\section{Observations}

Observations for this work were carried out as part of the Gemini NICI 
Planet-Finding Campaign between 2008 and 2012, using the NICI instrument at 
Gemini-South.  We describe the performance of NICI and the observing process 
in greater detail in \citet{pipeline}.  Briefly, we utilize two observing 
modes to optimize our sensitivity to all types of planetary companions 
over a wide range of angular separations: Angular Differential Imaging (ADI) 
and Angular and Spectral Differential Imaging (ASDI).  In ADI mode, we 
observe in the 
broadband $H$ filter through a partially transparent focal plane mask with 
the telescope rotator off.  This provides the highest sensitivity at large 
separations ($\gtrsim$1.5$''$), where sensitivity to faint companions is 
limited by background noise.  Large-separation companions also 
move the most as the field rotates.

In ASDI mode, we again observe with the focal plane mask in place and the 
rotator off, but a 50/50 beamsplitter 
sends light to two cameras simultaneously, which contain methane-on 
($CH_4L$, central wavelength 1.652 $\mu$m) and methane-off ($CH_4S$, central 
wavelength 1.578 $\mu$m) 4\% narrow-band filters in the $H$ band.  
These two simultaneous images are scaled and then subtracted to remove 
speckle noise and the stellar halo and leave behind the flux from a 
real companion.  
ASDI is most sensitive to methanated companions (which will have 
greater flux in the methane-off filter than the methane-on filter) but 
can detect companions of all spectral types, and 
produces higher contrasts $\lesssim$1.5$''$ than ADI alone.  
Typically we observe targets with both observing modes.  We note that 
observations of 
planets around HR 8799 suggest that the methane absorption in the spectra of
these planets is significantly weaker than predicted by theoretical models 
(e.g. \citealt{bowler_8799,barman11,skemer12}).  If this is a general feature 
of young giant planets it could indicate that our expected sensitivity to 
these planets in the ASDI mode is overly optimistic.  It will not, 
however, impact our expected sensitivity in the ADI mode, or at larger 
separations in ASDI mode where self-subtraction is not an issue (see 
Section~\ref{planetfraction_section}).

For each target star, 
Table~\ref{tab:observing} gives the observing 
epoch, observing mode, number of images, total exposure time, and amount of 
sky rotation.  The typical exposure time for ADI observations was 1200 s, and 
2700 s for ASDI.  In addition, we obtained $\sim$2--3 times deeper 
exposures of Fomalhaut and $\beta$ Pic, given the existence of their planetary 
companions.

\section{Results}

\subsection{Candidates and Follow-up}

Given NICI's ability to achieve very high contrasts (median of $\sim$15 
magnitudes---a flux ratio of 10$^6$---at 1$''$) 
and its 18$''$x18$''$ field of view, it is expected that our images will 
contain a large number of background objects in addition to any common proper 
motion (CPM) companions.  The Campaign data 
were reduced and checked for candidate companions by both an automated 
procedure and visual inspection of the images as described in 
\citet{pipeline}.  Stars with candidate companions were 
flagged, and archival high contrast imaging data (from the VLT, HST, 
Gemini-North, 
or Keck) were retrieved and analyzed to see if the candidates could be 
recovered.  If archival data were insufficient to determine whether the 
candidates were background or CPM, an additional NICI observation was 
typically obtained to measure the 
astrometry of the companion(s) at a second epoch.  In a few cases multiple 
additional epochs were required to definitively classify each candidate.  
We followed up all candidates within 400 AU except 
for most stars with large numbers ($\gtrsim$10) of candidates due to 
these stars being in the foreground of the galactic bulge or at galactic 
latitude close to 0$^\circ$.

A candidate companion observed at multiple epochs with sufficient 
astrometric precision will either be in the same location as in the first 
epoch (CPM, with the possibility of a small amount of orbital motion) or 
displaced from the first epoch position, following the expected motion of a 
background object.  A distant ($\gtrsim$500 pc) background object will have 
negligible proper motion ($\lesssim$2~mas/yr) and thus appear to move with 
respect to the target star in the opposite direction 
indicated by the target star's (much larger) proper and parallactic 
motion.  To quantitatively classify stars as background (bg) or CPM, we 
compute the chi-square statistic for the two scenarios:

\begin{equation} \chi^2_{bg} = \sum_{i} \left ( \frac{(\rho_{obs,i} -
\rho_{bg,i})^2}{\sigma_{\rho,i}^2} + \frac{(PA_{obs,i} -
PA_{bg,i})^2}{\sigma_{PA,i}^2} \right )
\end{equation}

\begin{equation} \chi^2_{CPM} = \sum_{i} \left ( \frac{(\rho_{obs,i} -
\rho_{0})^2}{\sigma_{\rho,i}^2} + \frac{(PA_{obs,i} -
PA_{0})^2}{\sigma_{PA,i}^2} \right )
\end{equation}

\noindent where the summation is over all epochs except the reference epoch; 
$\rho_{obs,i}$ and $PA_{obs,i}$ are the observed separation and position angle 
at the $i$-th epoch; $\rho_{bg,i}$ and $PA_{bg,i}$ are the separation 
and position angle predicted for a background object given the 
position at 
the reference epoch; $\rho_{0}$ and $PA_{0}$ are the separation and 
position angle at the reference epoch; and $\sigma_{\rho,i}$ and 
$\sigma_{PA,i}$ are the uncertainties in separation and position angle at 
the $i$-th epoch.  We convert chi-square to reduced chi-square ($\chi^2_\nu$) 
using 
the number of degrees of freedom (dof), which is $(N_{epochs} - 1) \times 2$, 
where $N_{epochs}$ is the total number of epochs.  The number of epochs in the 
dof calculation is reduced by one, 
since the background and CPM predictions are tied to the reference epoch, 
and then multiplied by 2, because there is a separation and PA 
measurement at each epoch.  The result for each of our candidates is 
unambiguous, with $\chi^2_\nu$ for background close to one and $\chi^2_\nu$ 
for CPM much larger than one, or vice-versa.

\begin{figure}
\centerline{
\includegraphics[width=2.0in]{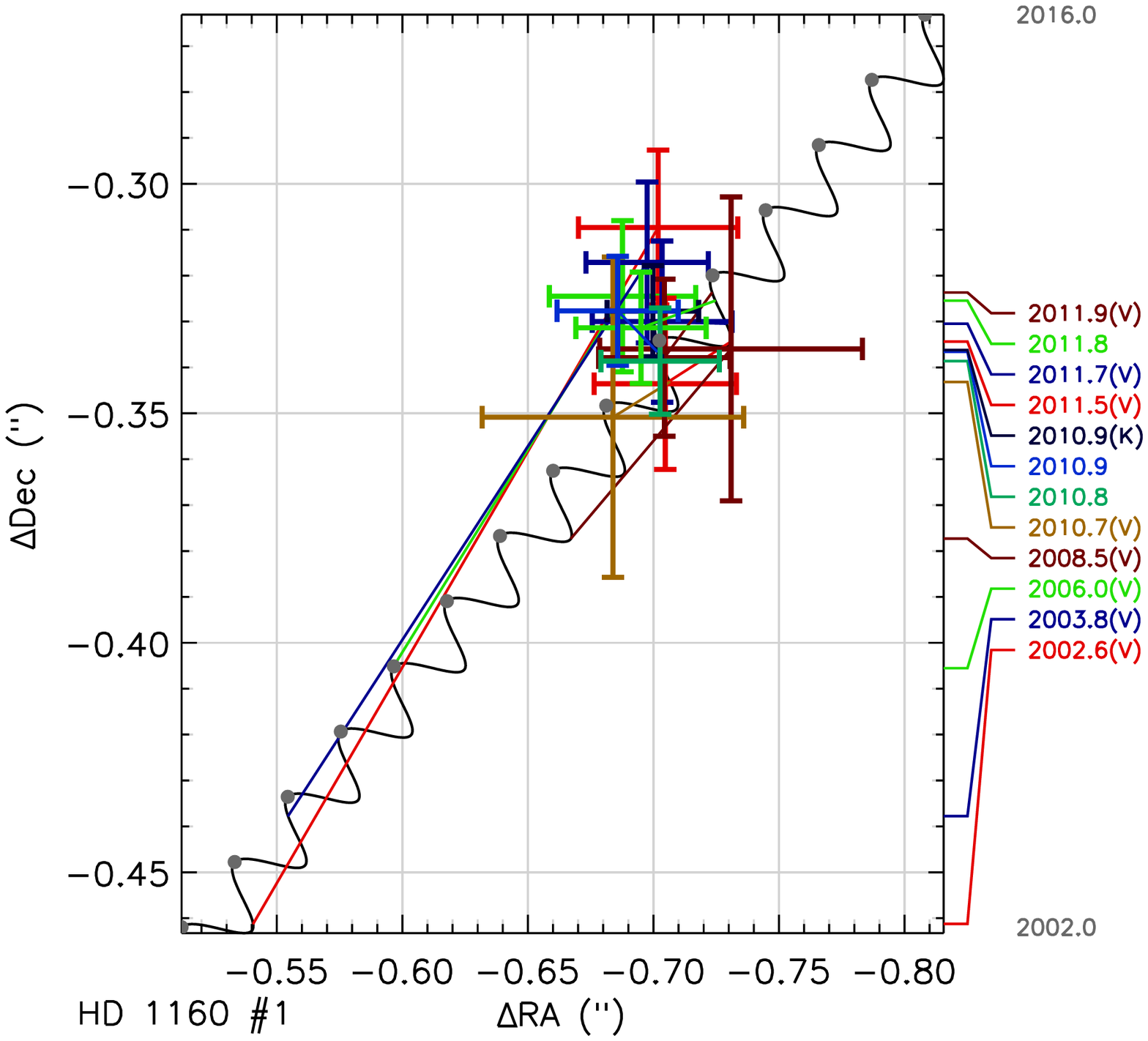}
\hskip -0.3in
\includegraphics[width=2.0in]{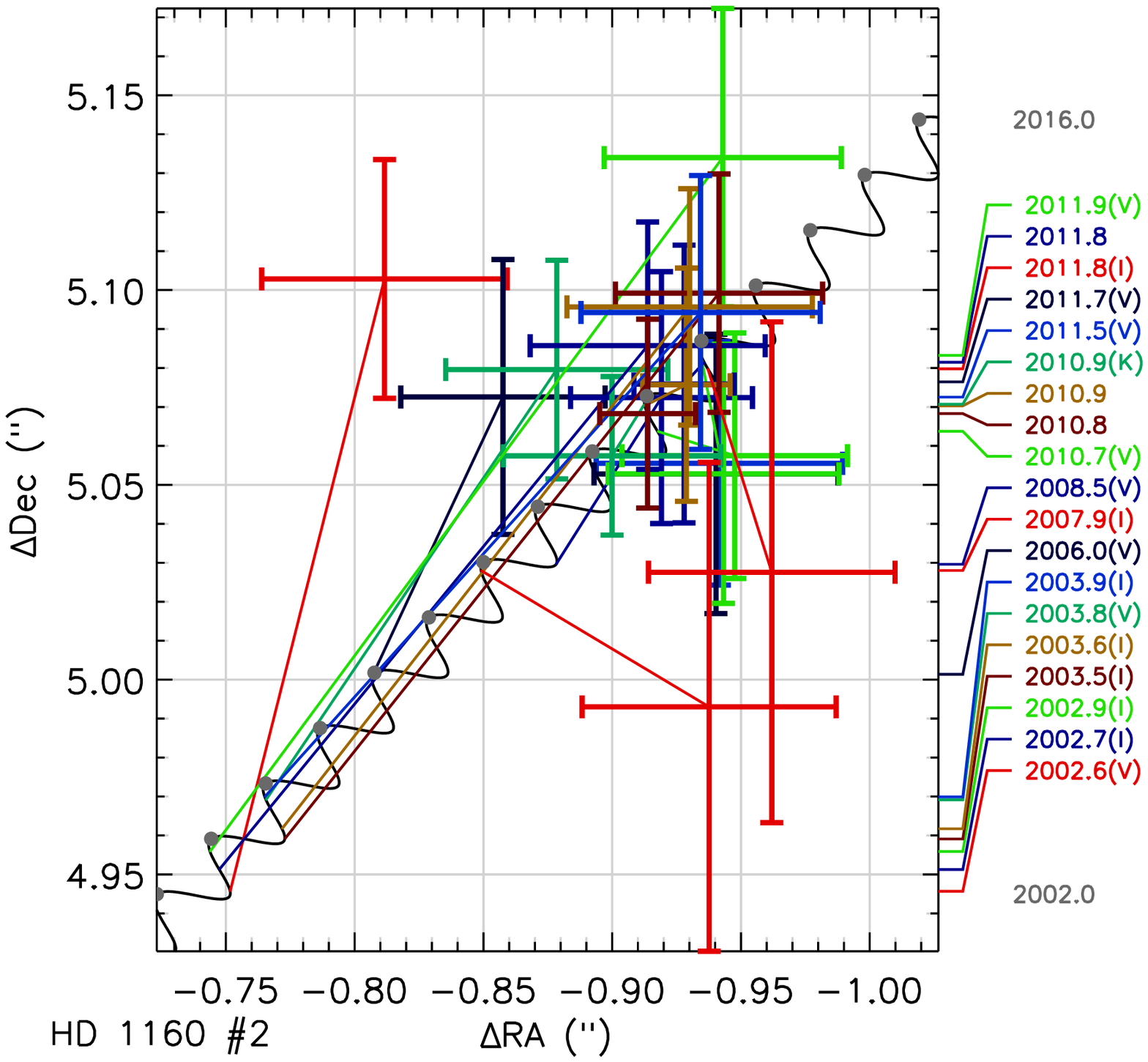}
\hskip -0.3in
\includegraphics[width=2.0in]{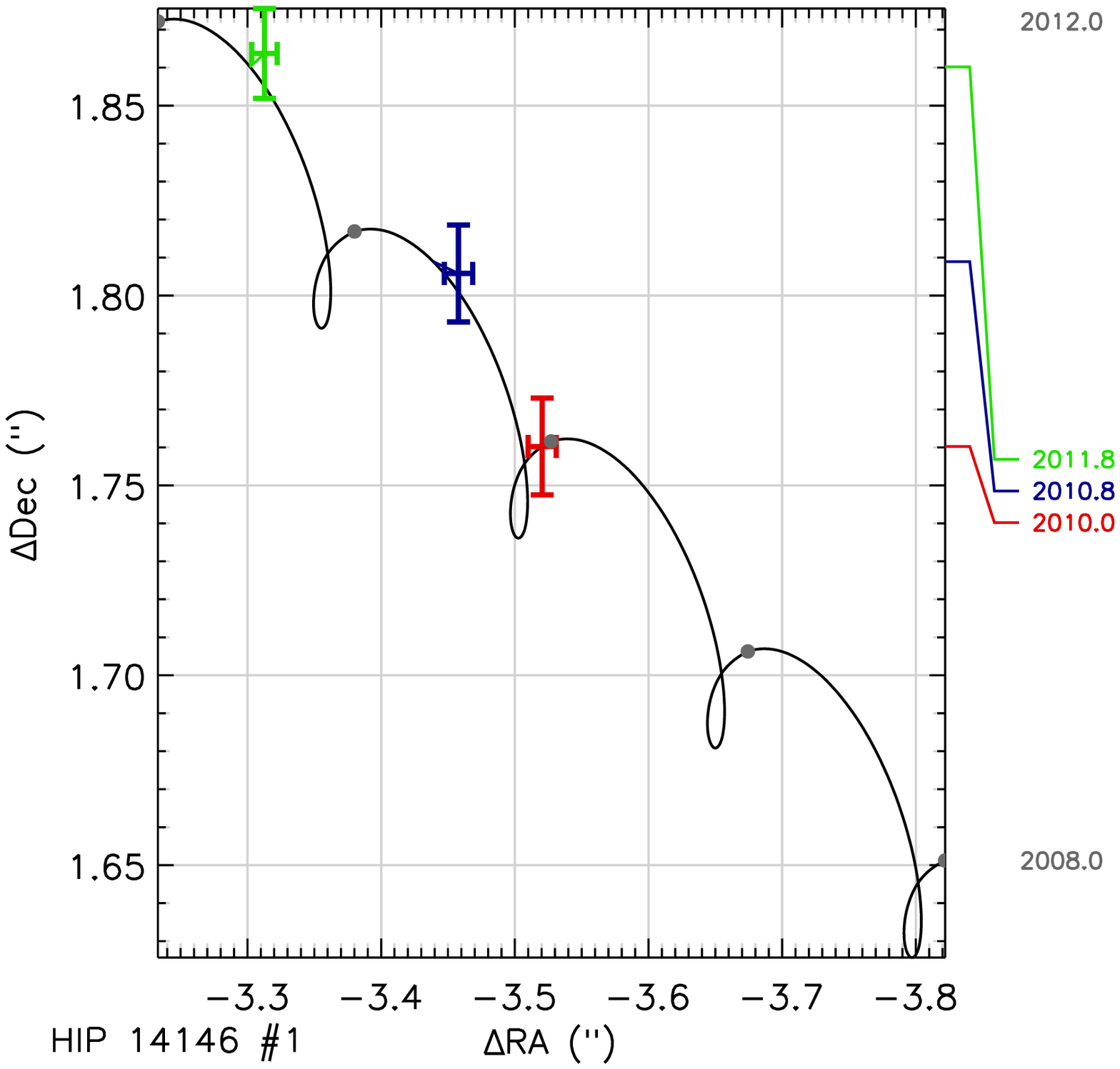}
\hskip -0.3in
\includegraphics[width=2.0in]{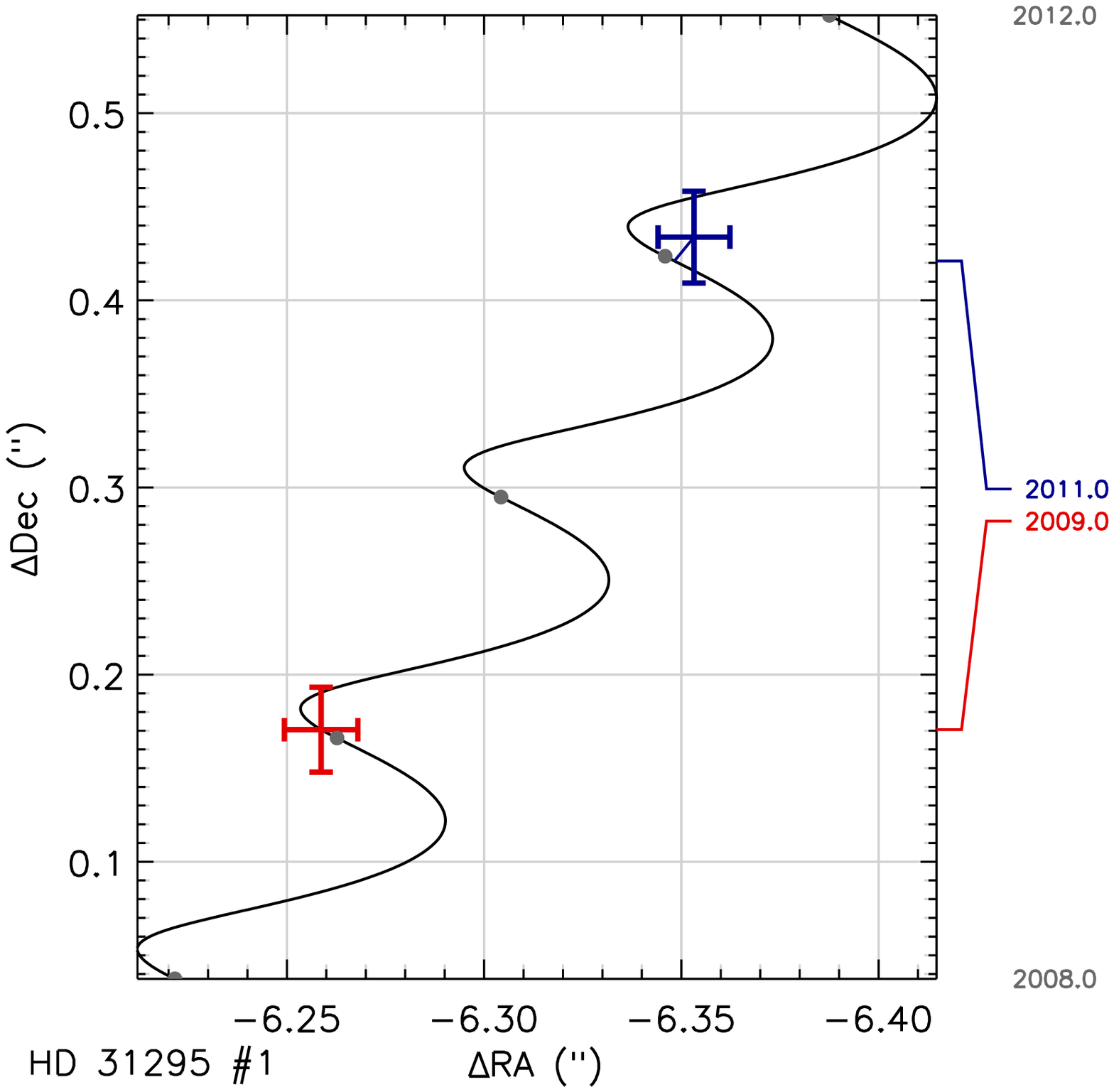}
}
\vskip -0.2in
\centerline{
\includegraphics[width=2.0in]{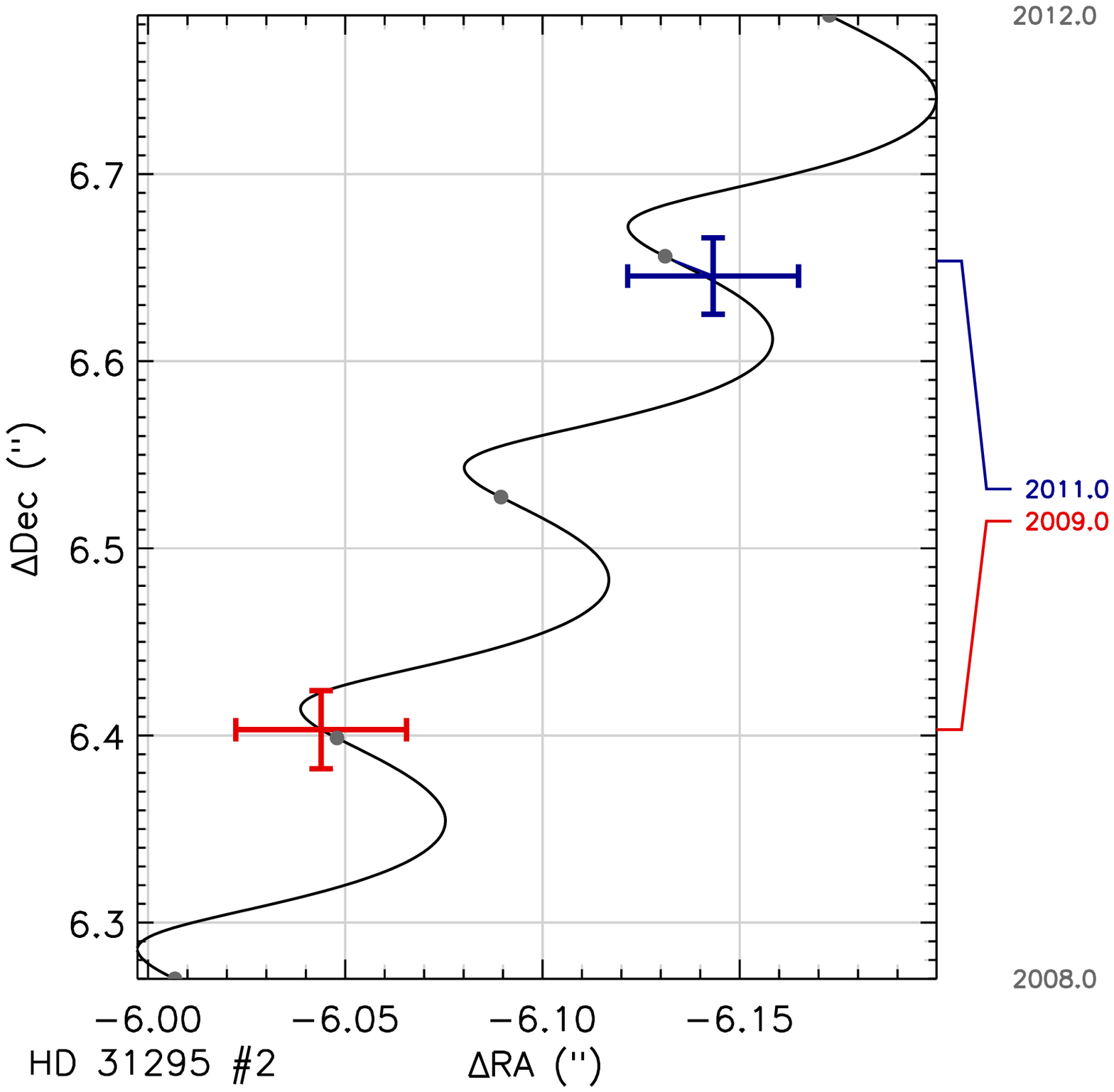}
\hskip -0.3in
\includegraphics[width=2.0in]{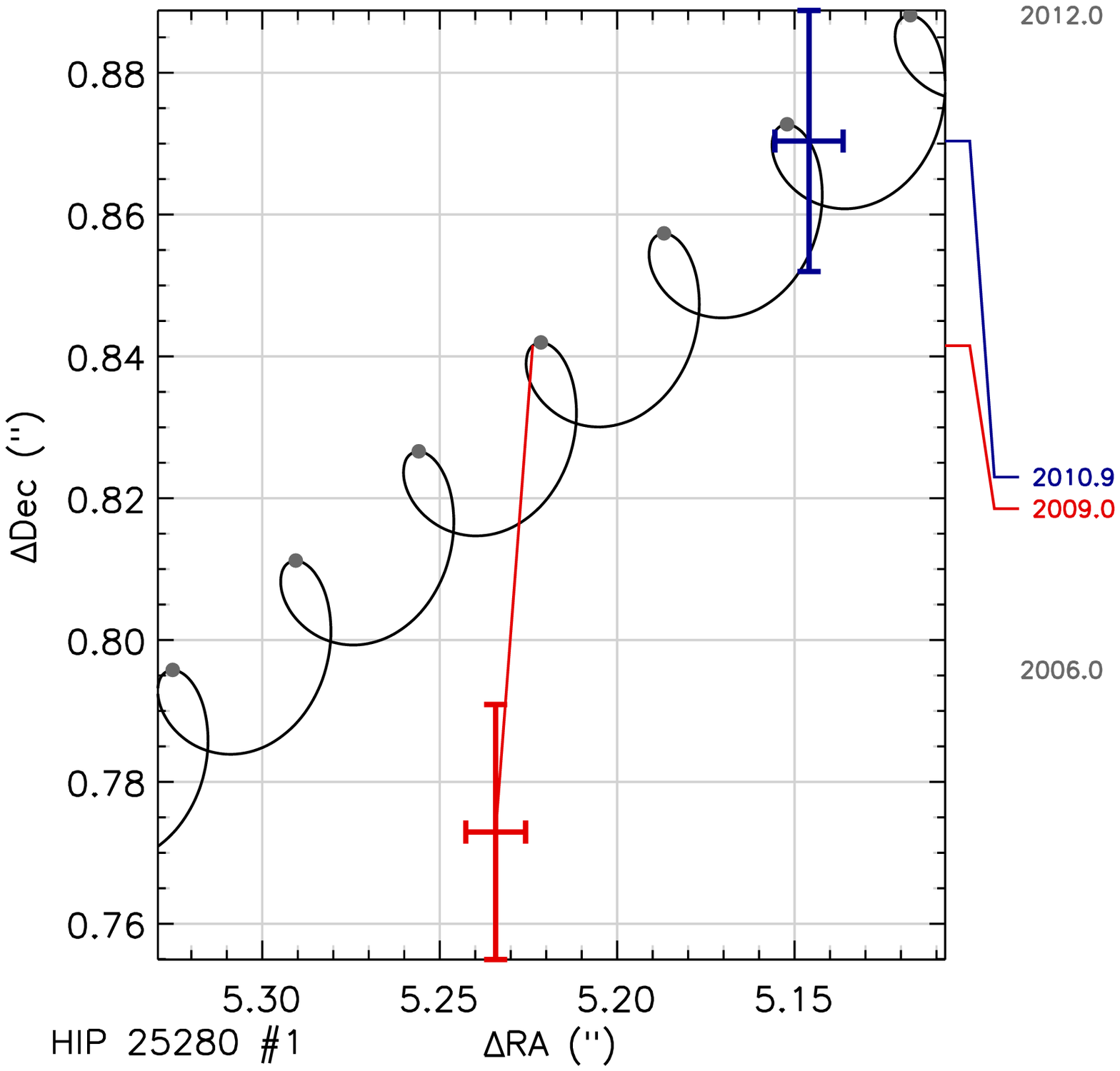}
\hskip -0.3in
\includegraphics[width=2.0in]{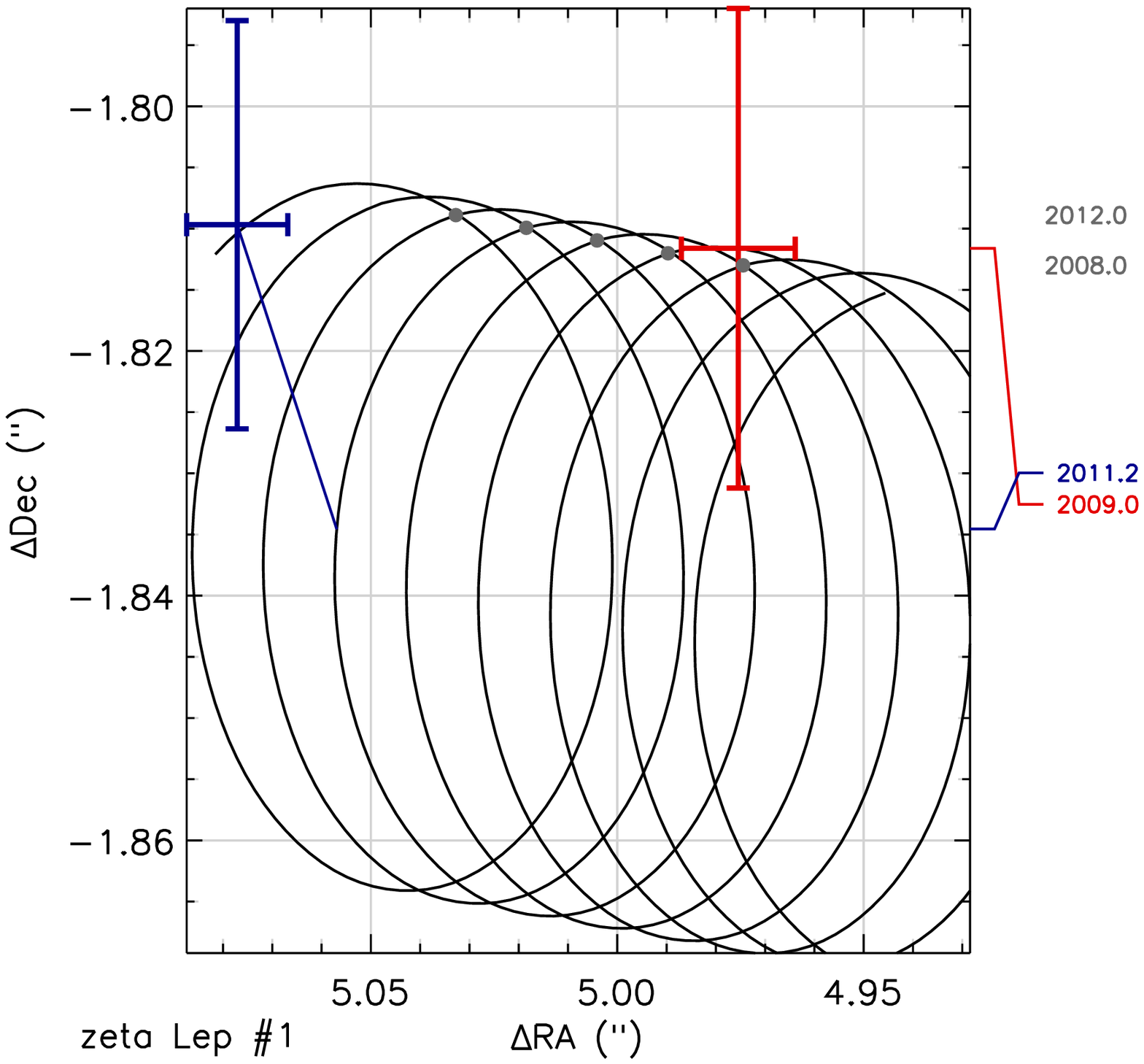}
\hskip -0.3in
\includegraphics[width=2.0in]{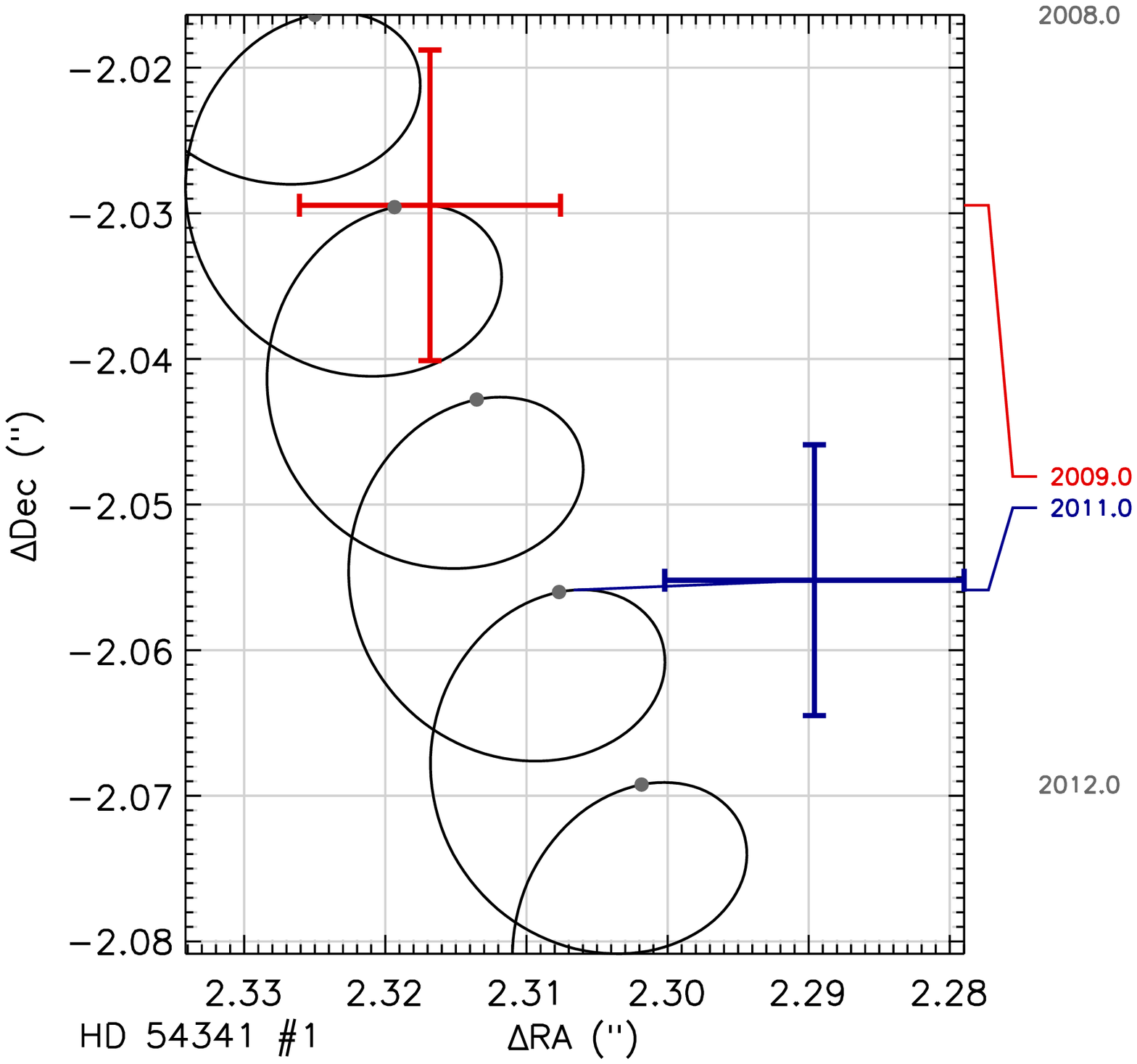}
}
\vskip -0.2in
\centerline{
\includegraphics[width=2.0in]{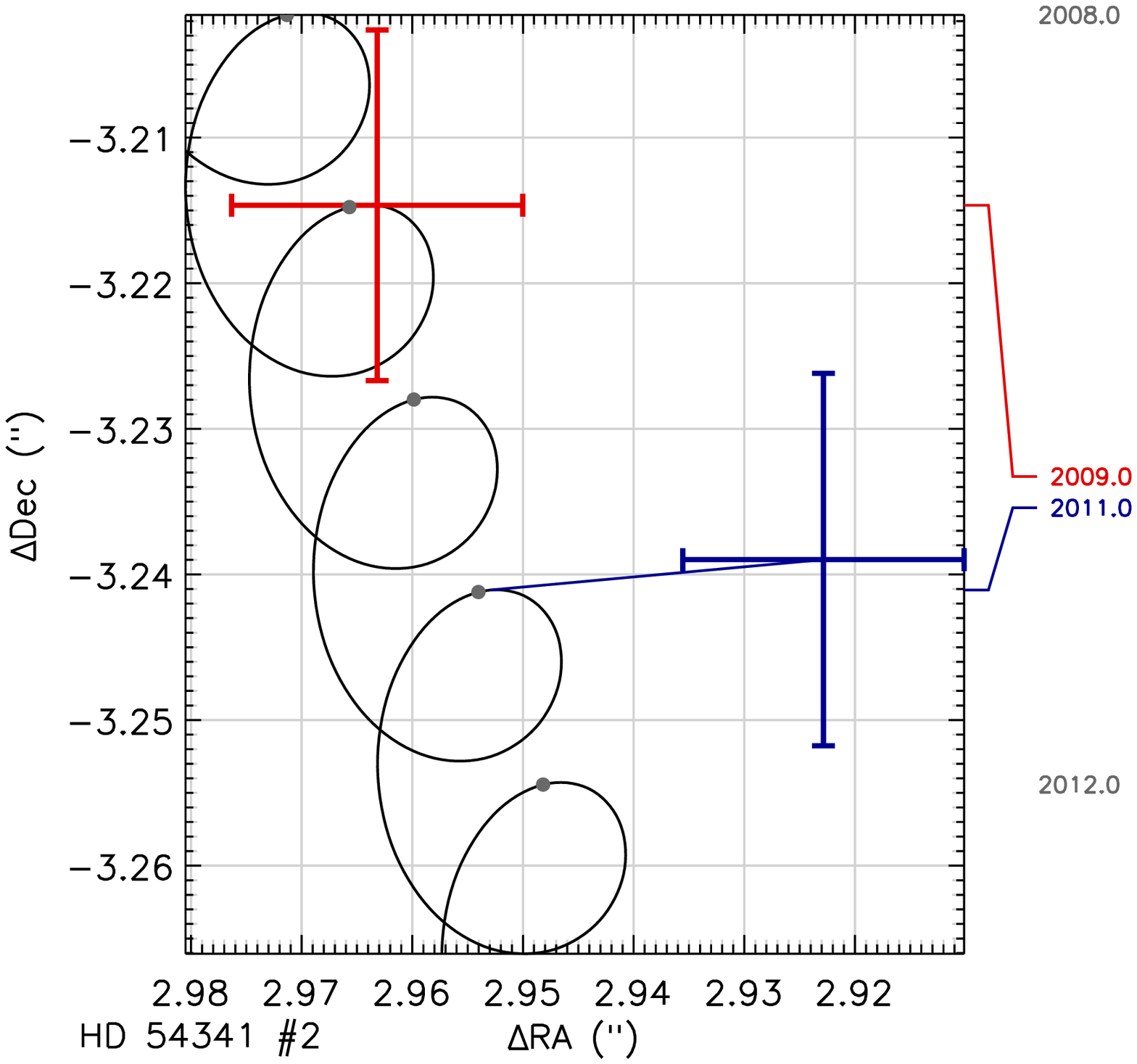}
\hskip -0.3in
\includegraphics[width=2.0in]{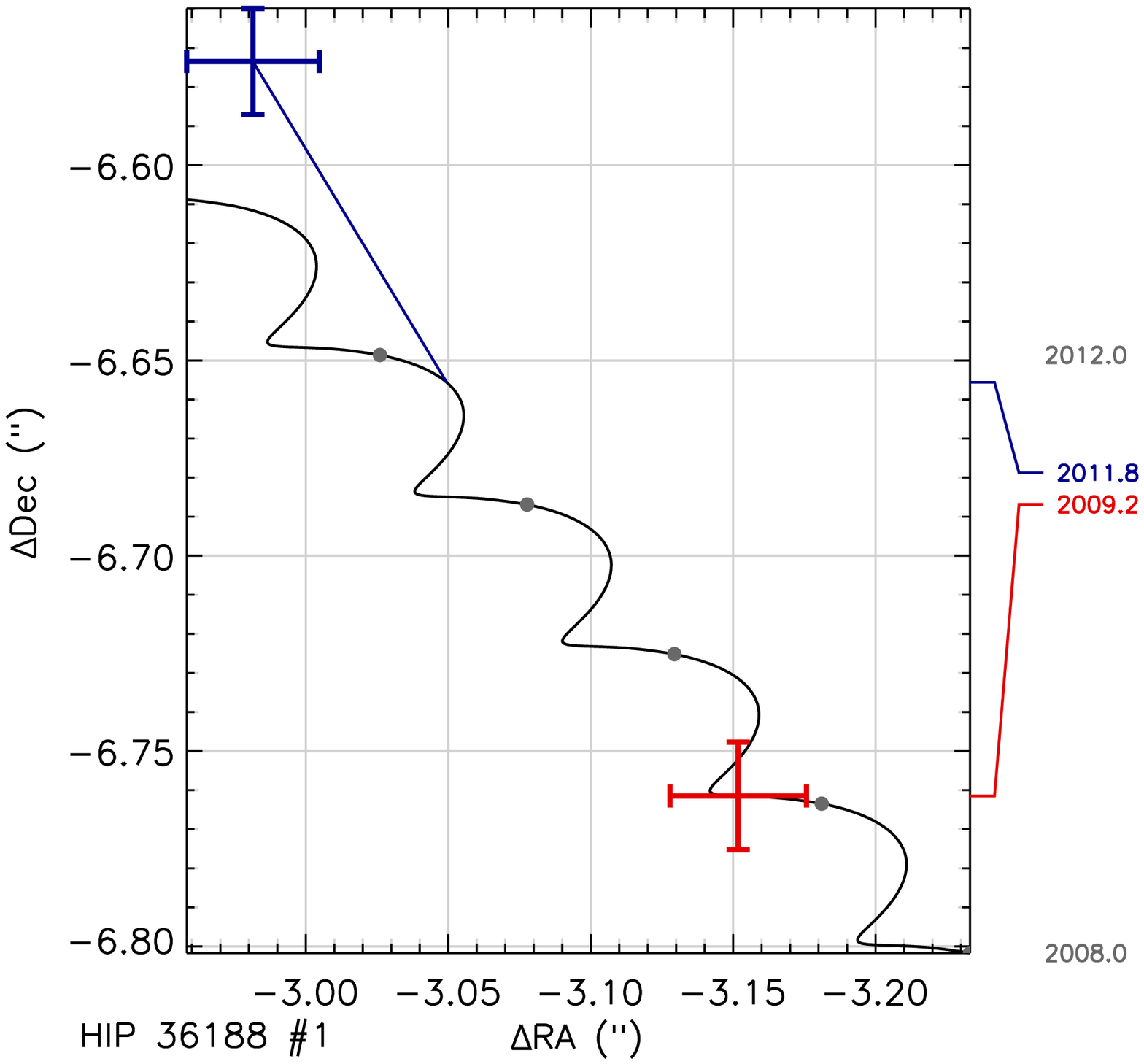}
\hskip -0.3in
\includegraphics[width=2.0in]{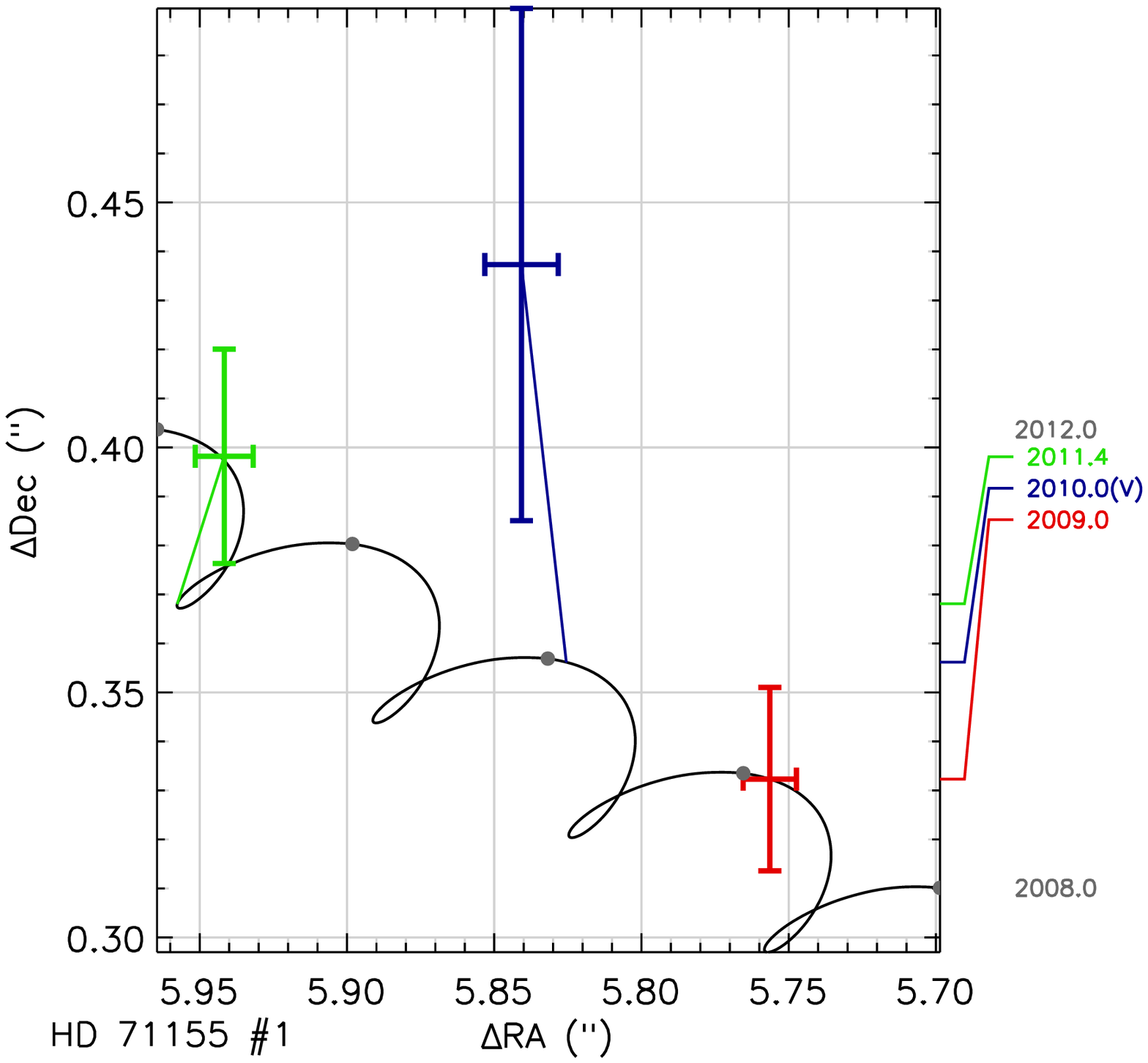}
\hskip -0.3in
\includegraphics[width=2.0in]{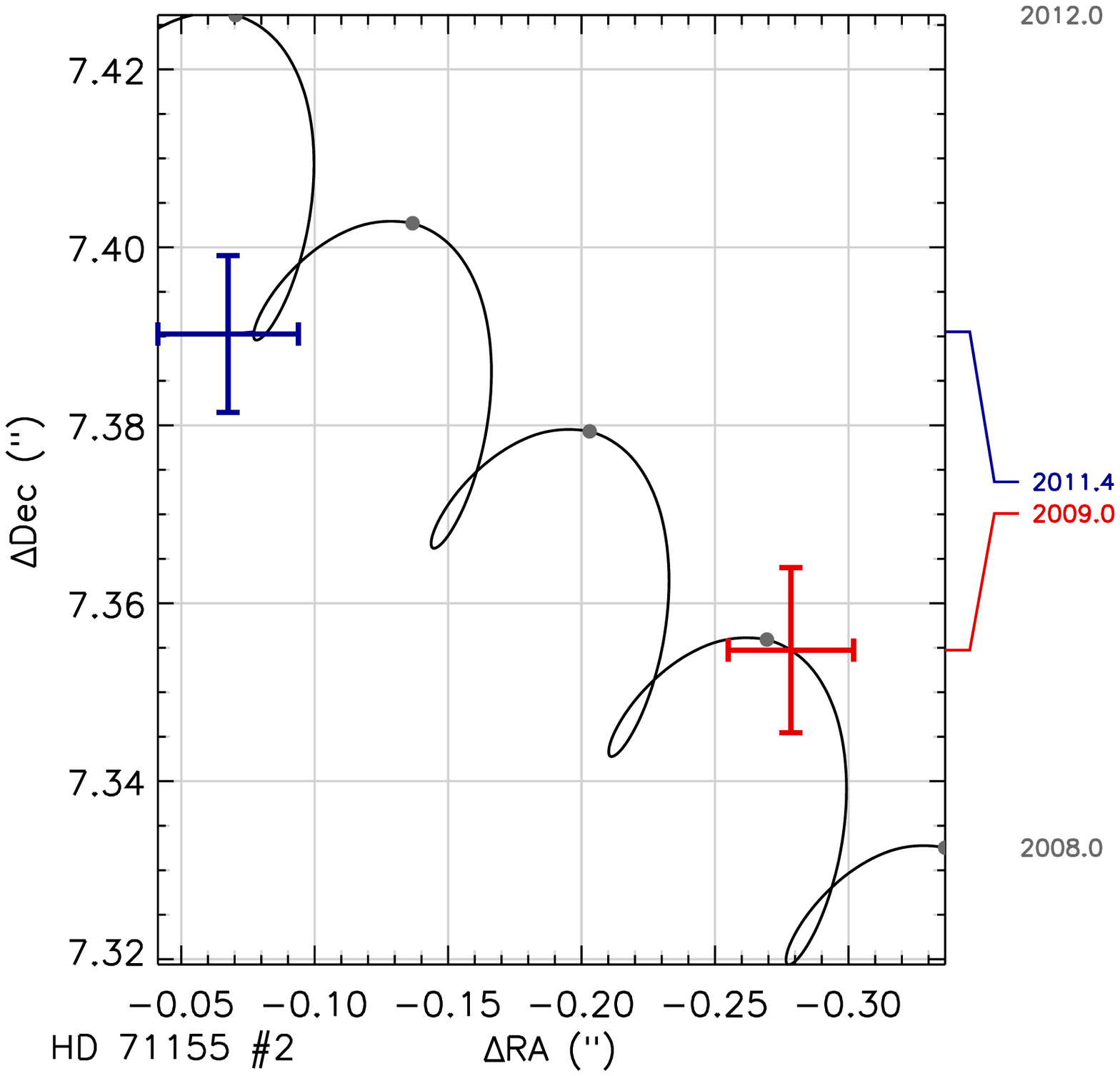}
}
\vskip -0.2in
\centerline{
\includegraphics[width=2.0in]{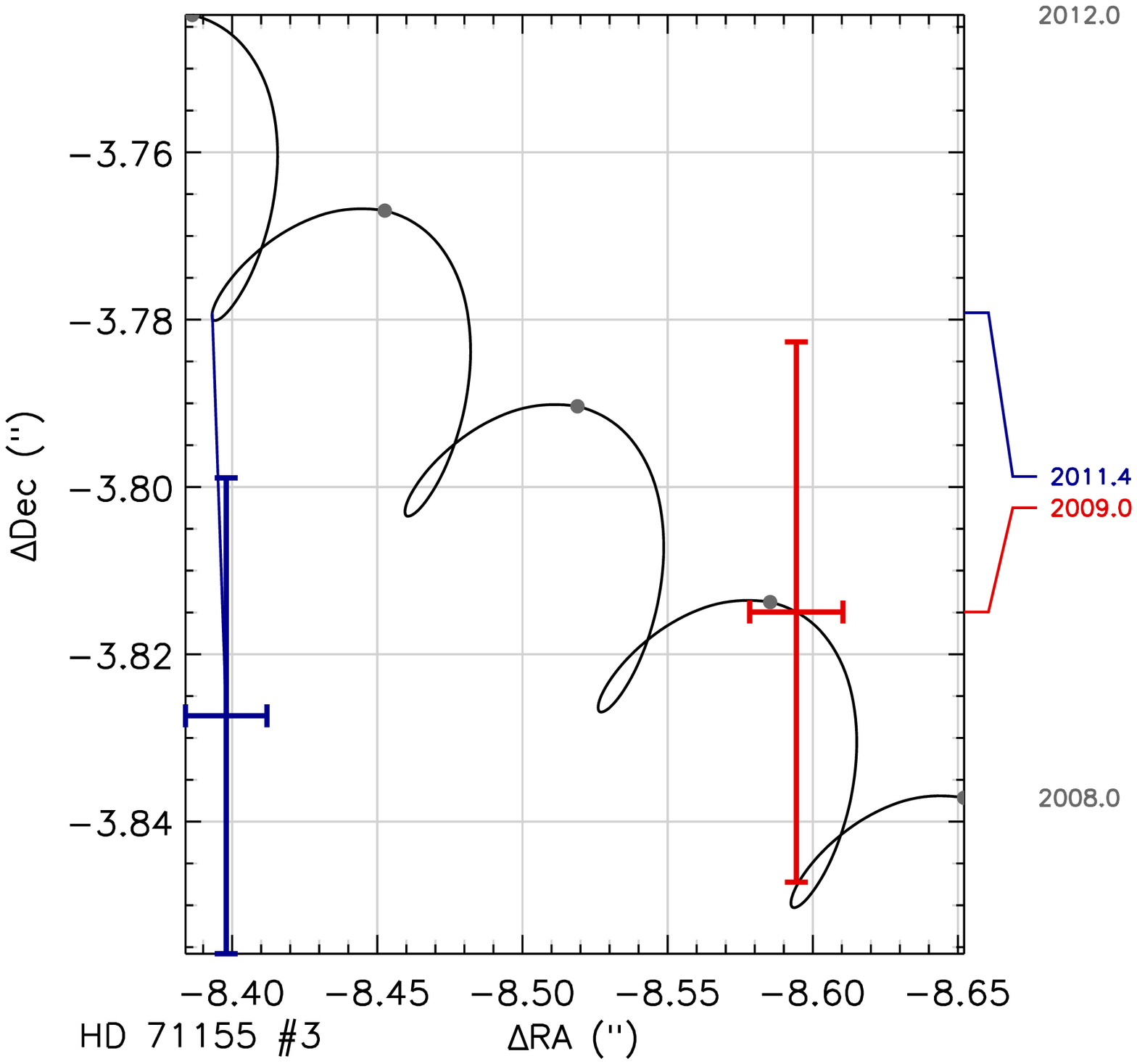}
\hskip -0.3in
\includegraphics[width=2.0in]{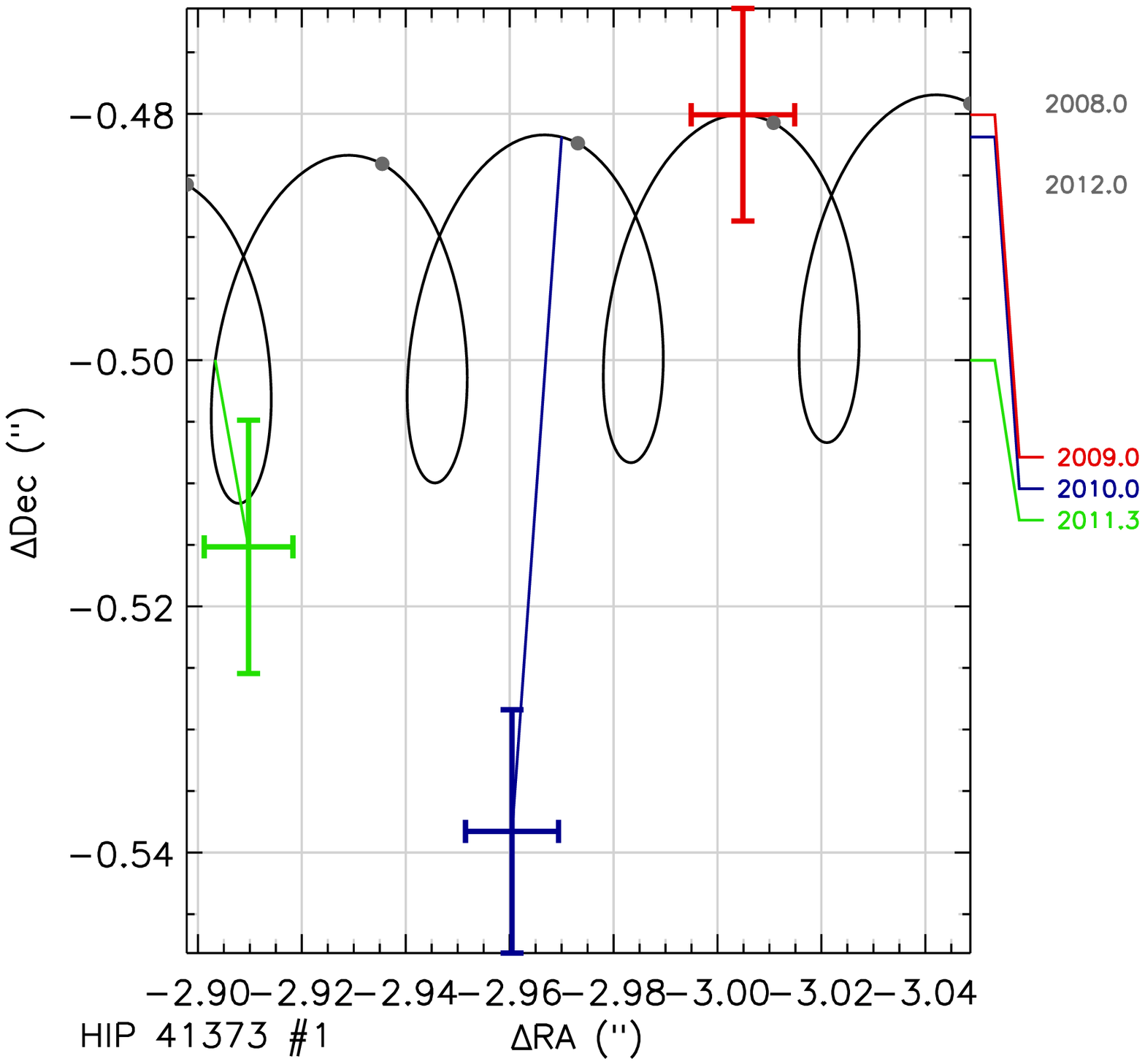}
\hskip -0.3in
\includegraphics[width=2.0in]{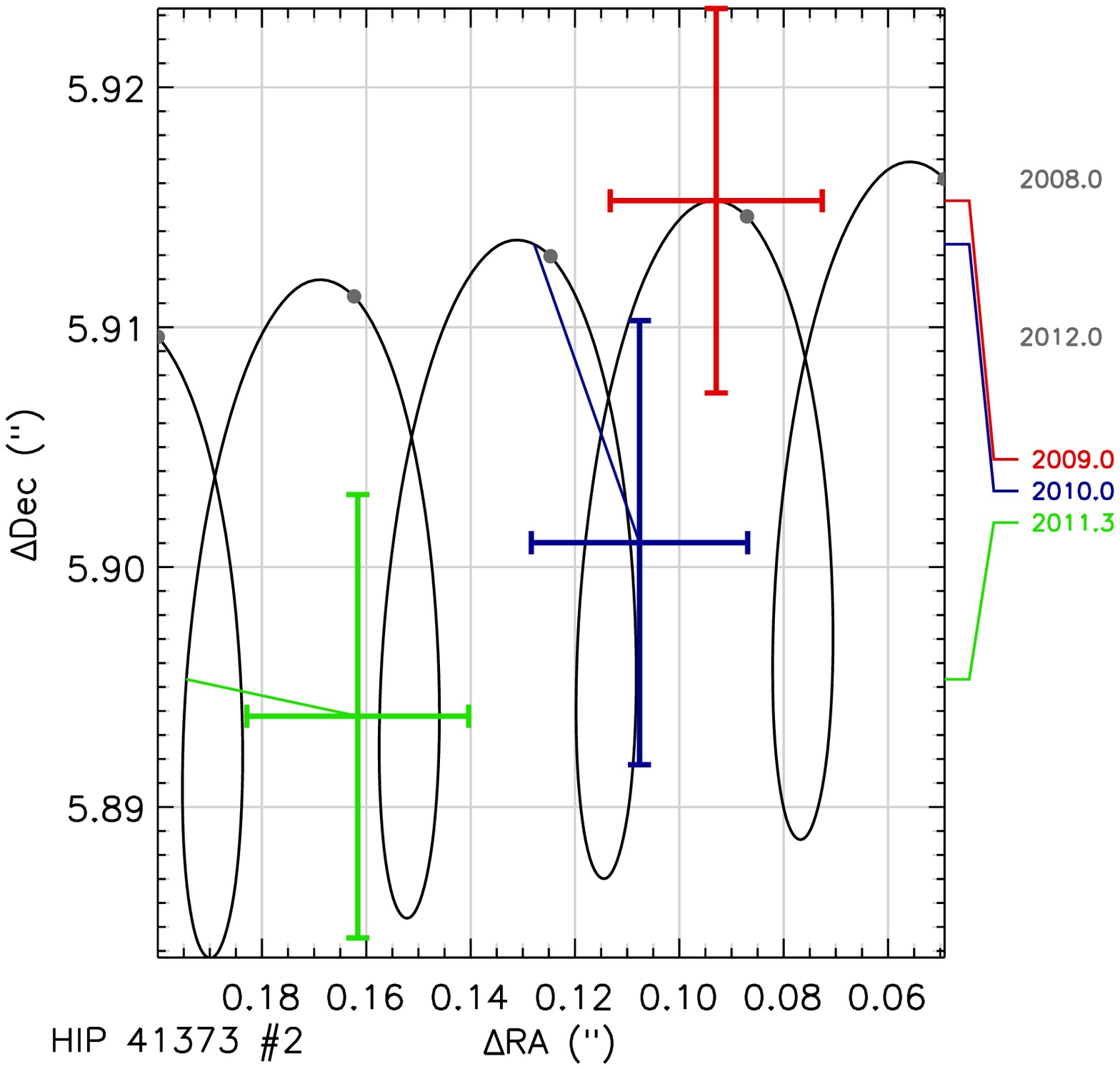}
\hskip -0.3in
\includegraphics[width=2.0in]{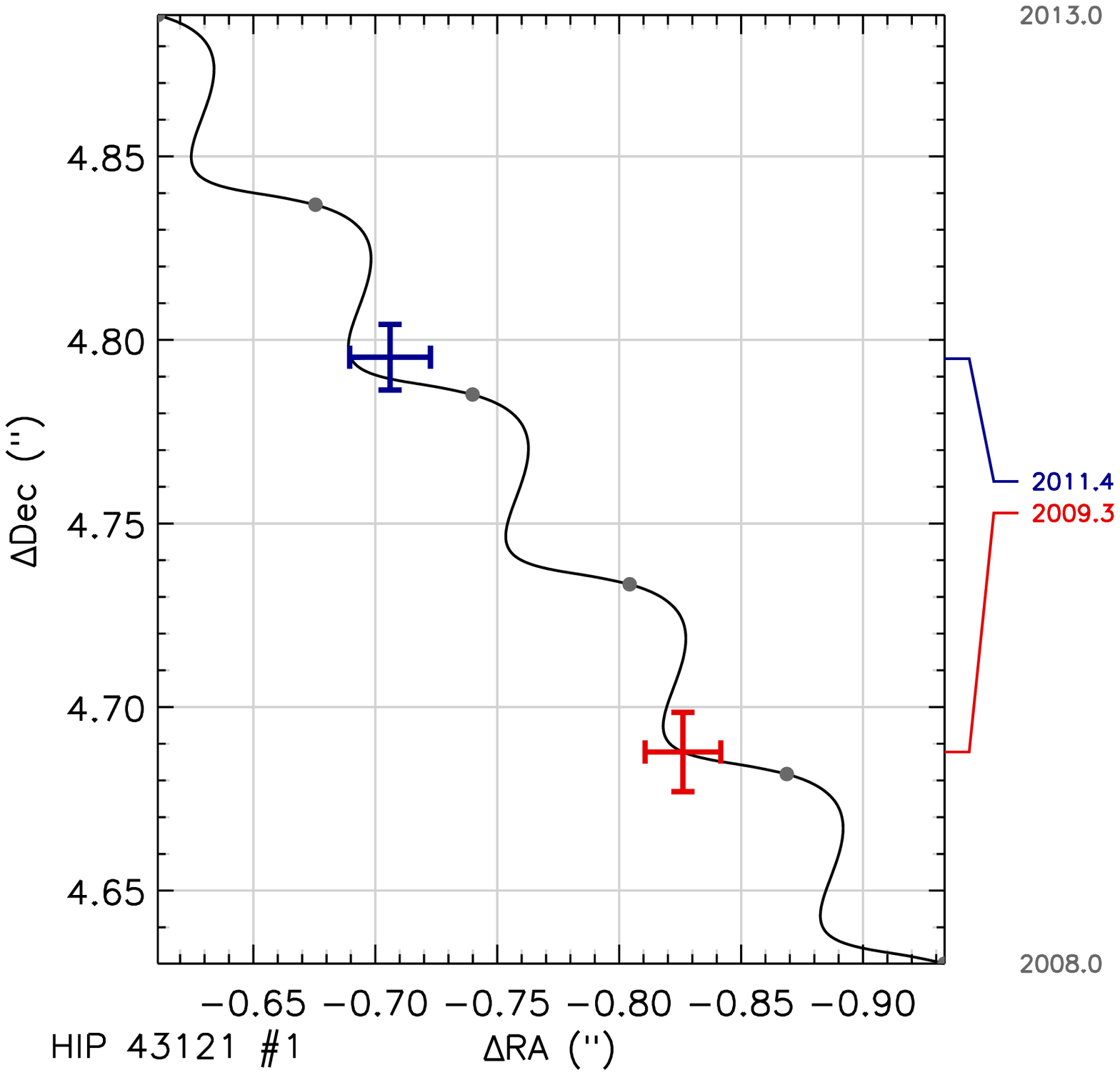}
}
\vskip -0.2in
\caption{
\footnotesize{On-sky motion of candidate companions listed in 
Table~\ref{table_comp2a}.  For each candidate, the background track (black 
curve) is calculated from the proper motion and parallax of the star 
and position of the candidate at the initial 
reference epoch.  Astrometry at the reference epoch and 
additional epochs are shown as points with error bars, and a colored line 
connects the position at additional epochs to the expected position on the 
background track.  The labels at the right of each plot give the epochs of 
each astrometric data point, at the vertical position corresponding to the 
location on the background track for that epoch.  When the epoch is given 
alone, the observation was conducted with the NICI instrument.  Otherwise 
observational data are taken from VLT NACO (V), Keck NIRC2 (K), HST NICMOS 
(H), VLT ISAAC 
(I), ESO 3.6m COME-ON-PLUS/ADONIS (E), and Gemini-North NIRI (G).}
}\label{tiled_fig1}
\end{figure}

\begin{figure}
\centerline{
\includegraphics[width=2.0in]{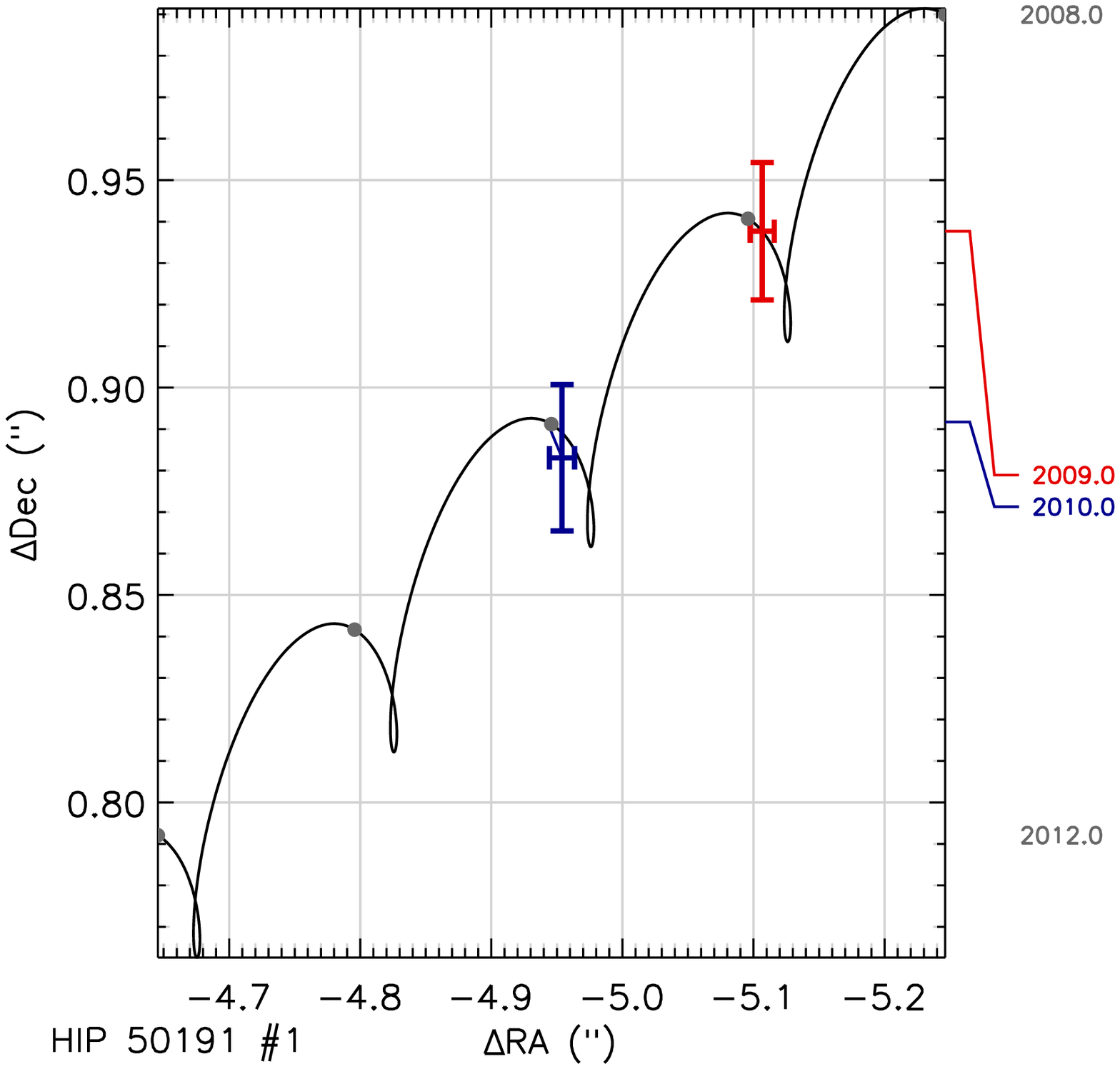}
\hskip -0.3in
\includegraphics[width=2.0in]{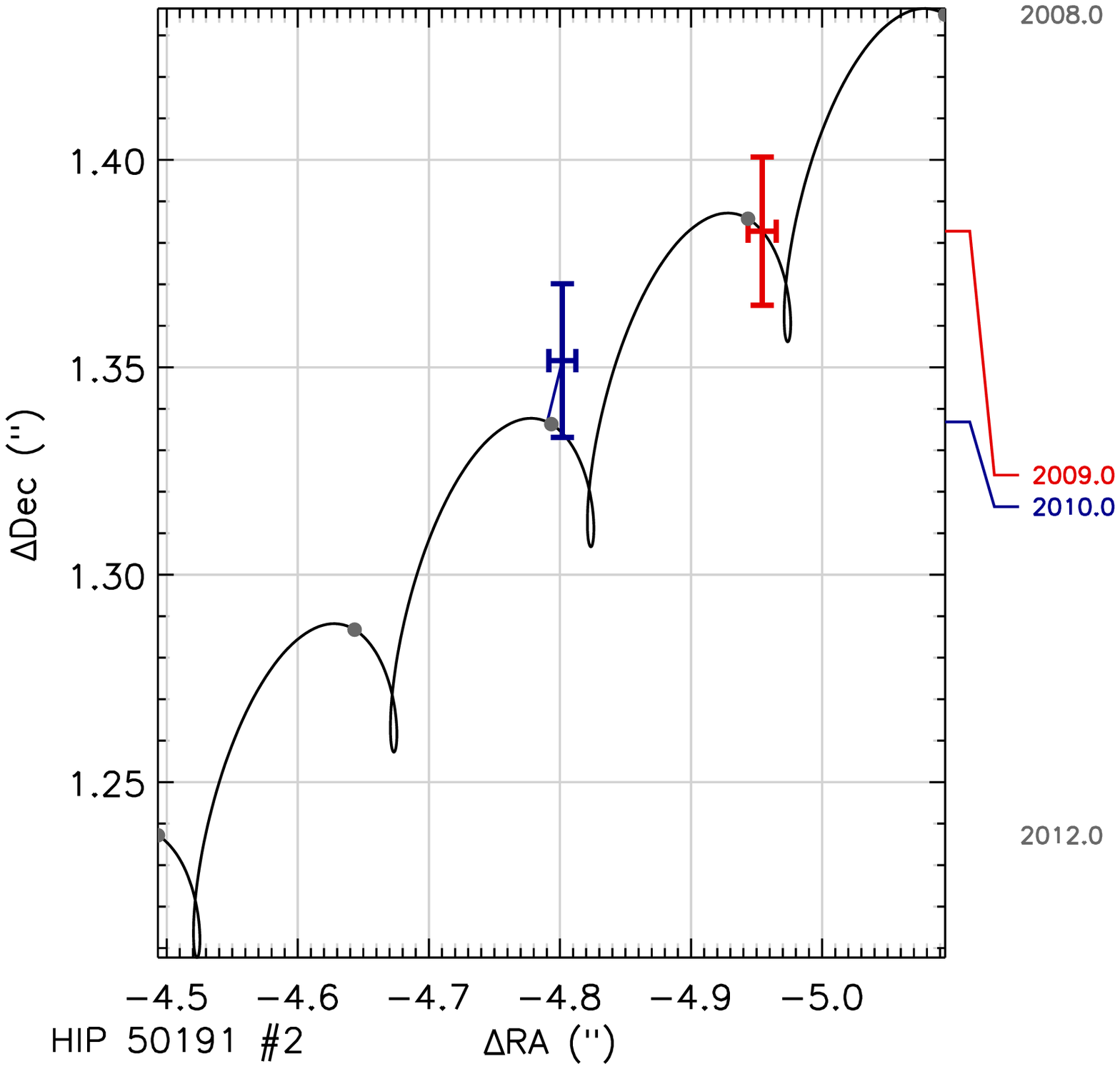}
\hskip -0.3in
\includegraphics[width=2.0in]{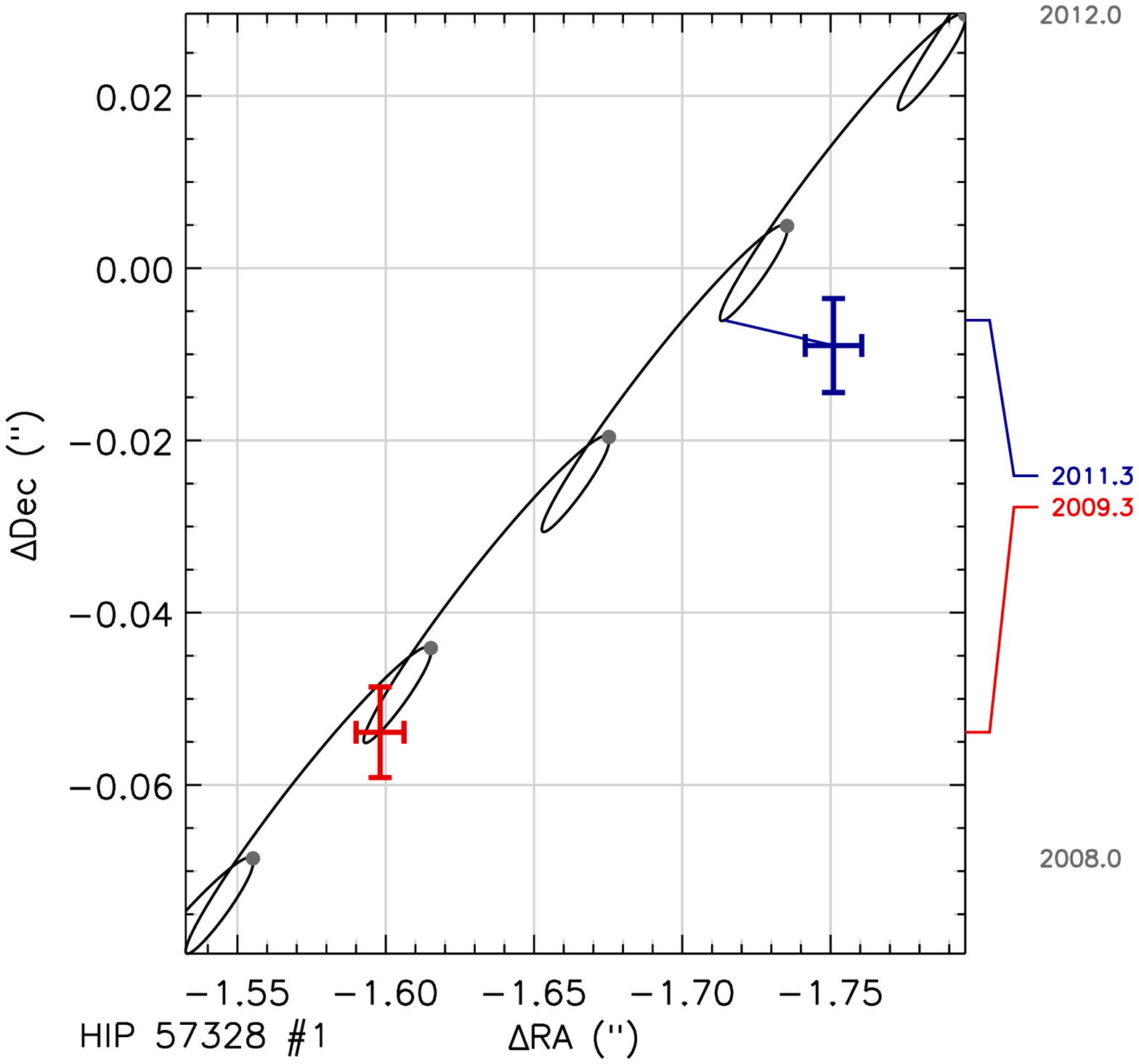}
\hskip -0.3in
\includegraphics[width=2.0in]{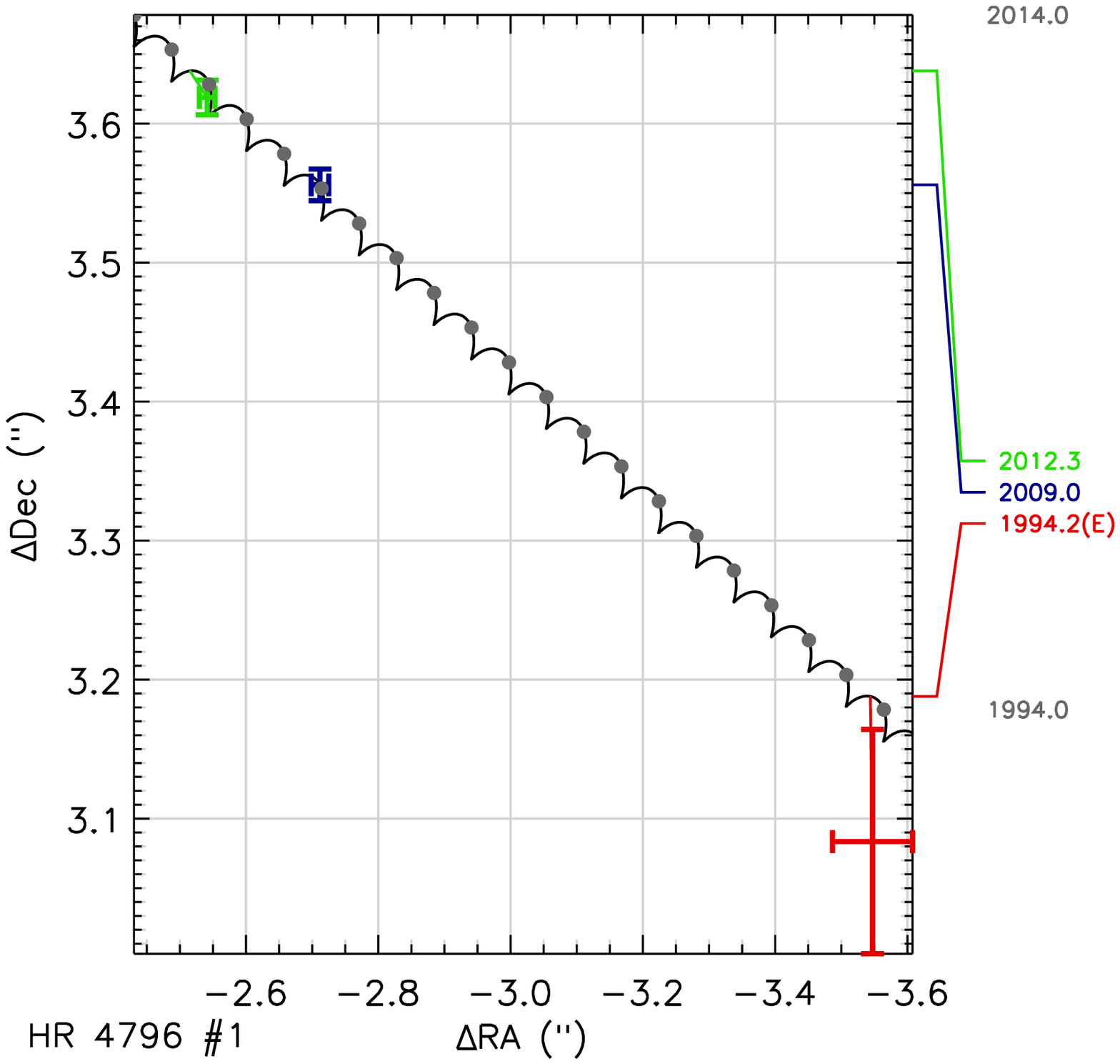}
}
\vskip -0.2in
\centerline{
\includegraphics[width=2.0in]{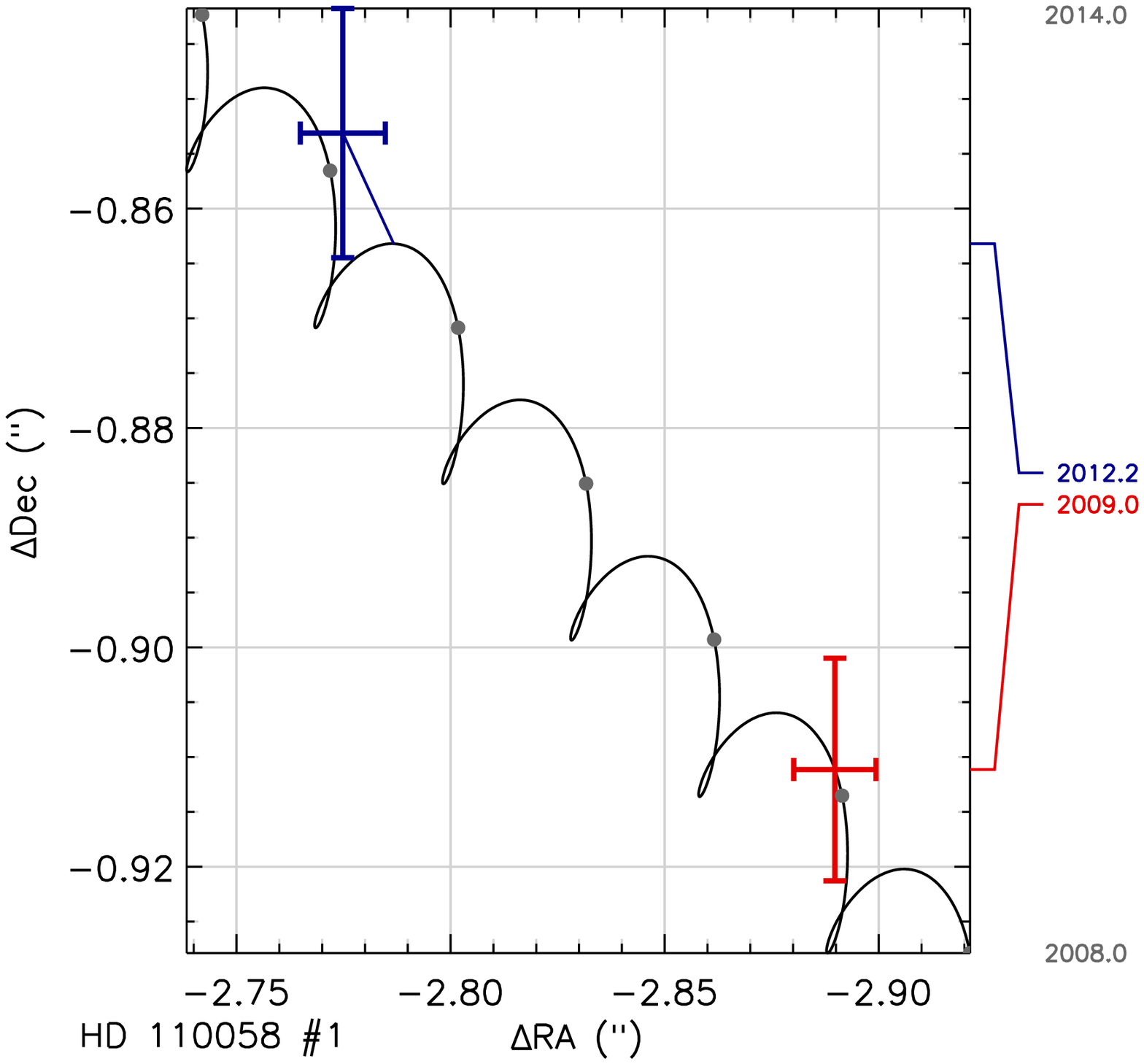}
\hskip -0.3in
\includegraphics[width=2.0in]{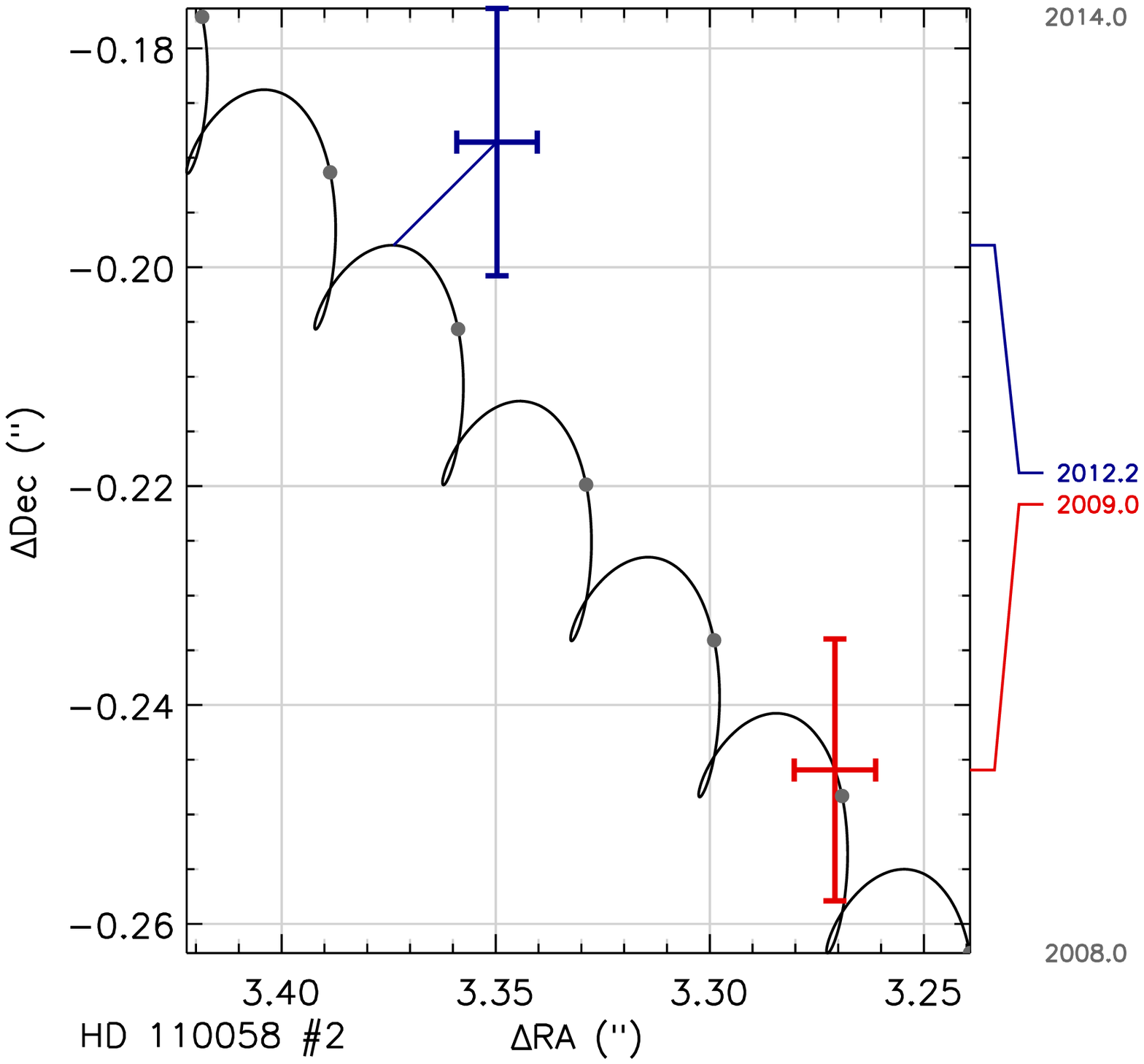}
\hskip -0.3in
\includegraphics[width=2.0in]{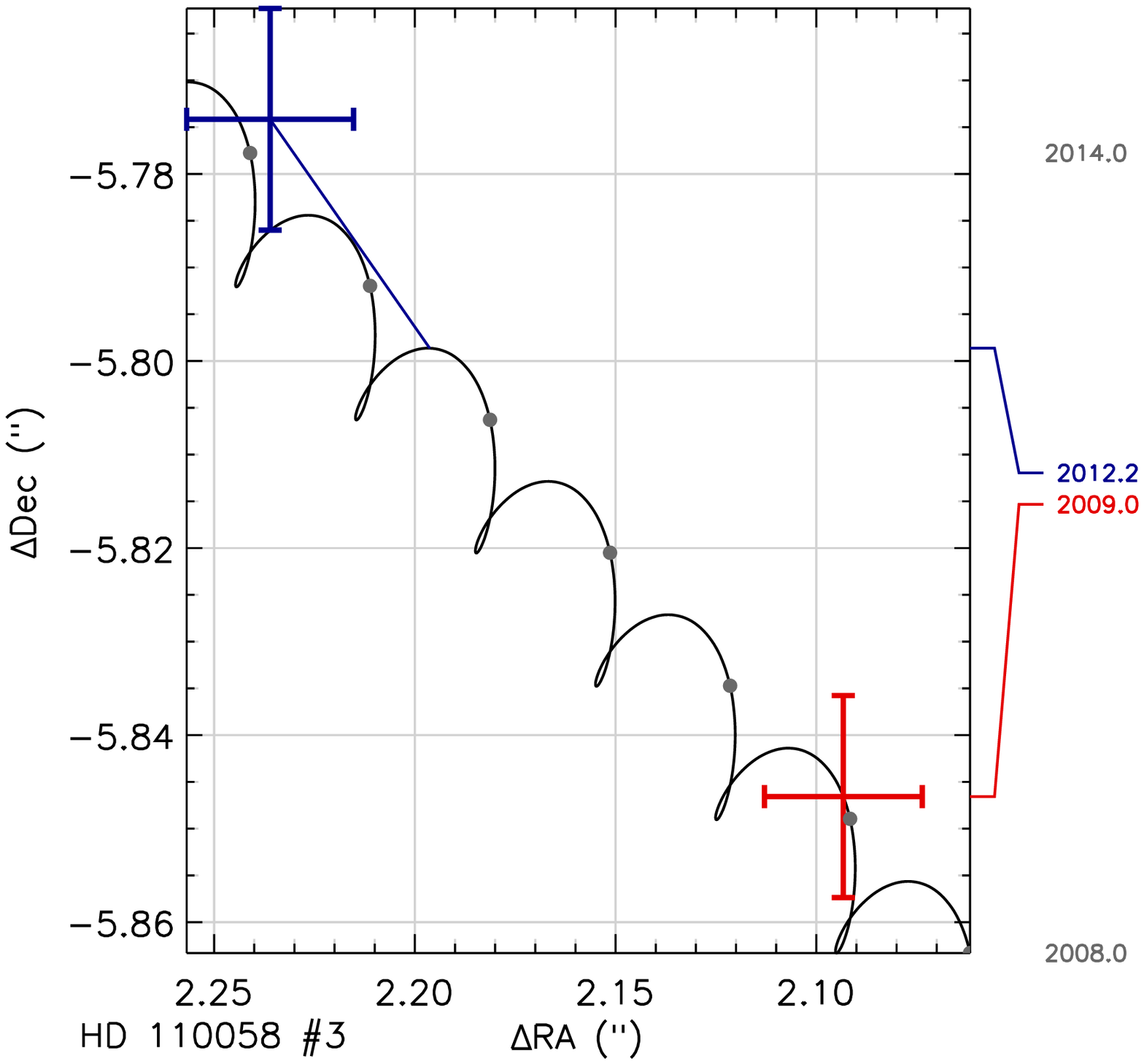}
\hskip -0.3in
\includegraphics[width=2.0in]{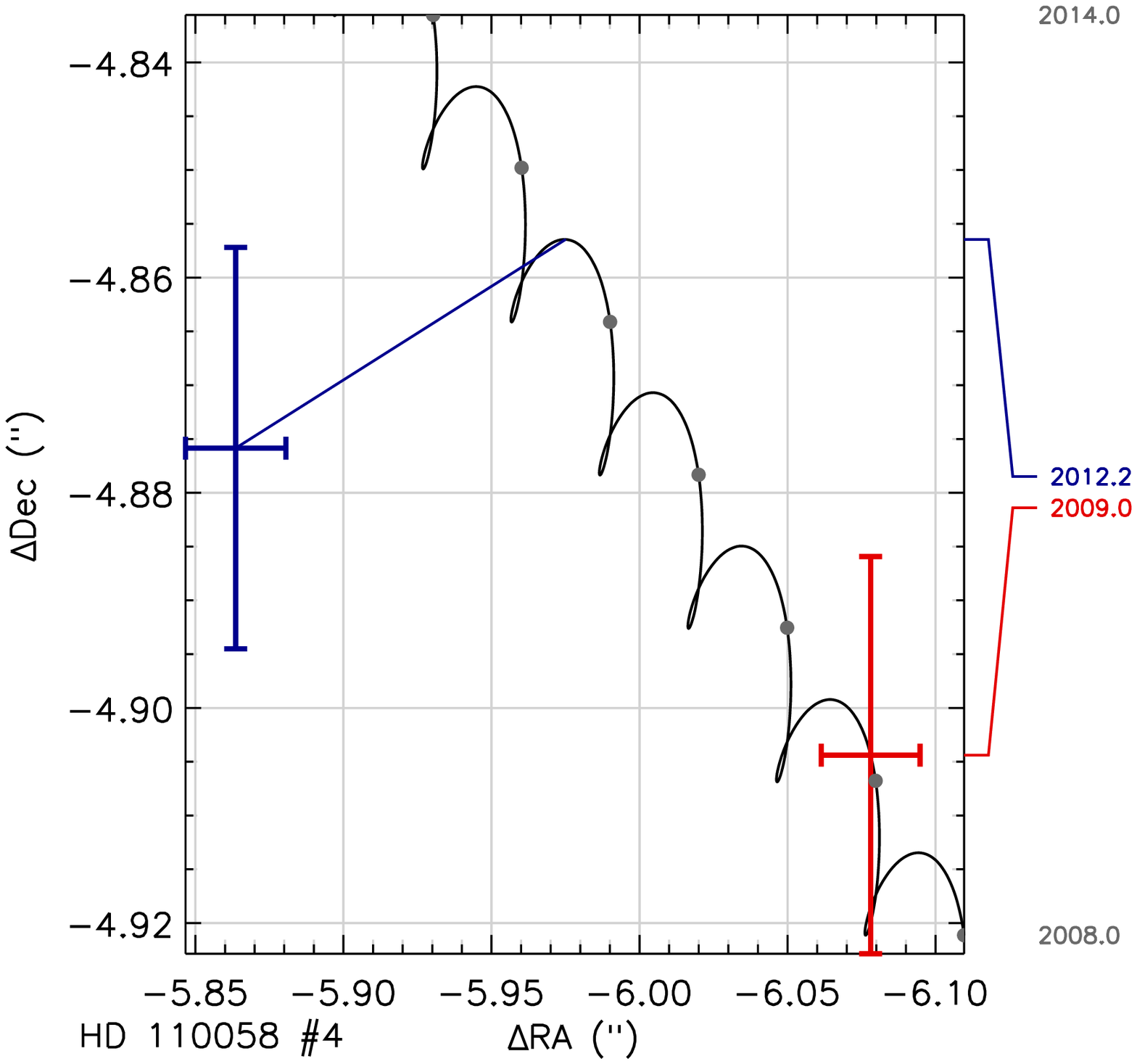}
}
\vskip -0.2in
\centerline{
\includegraphics[width=2.0in]{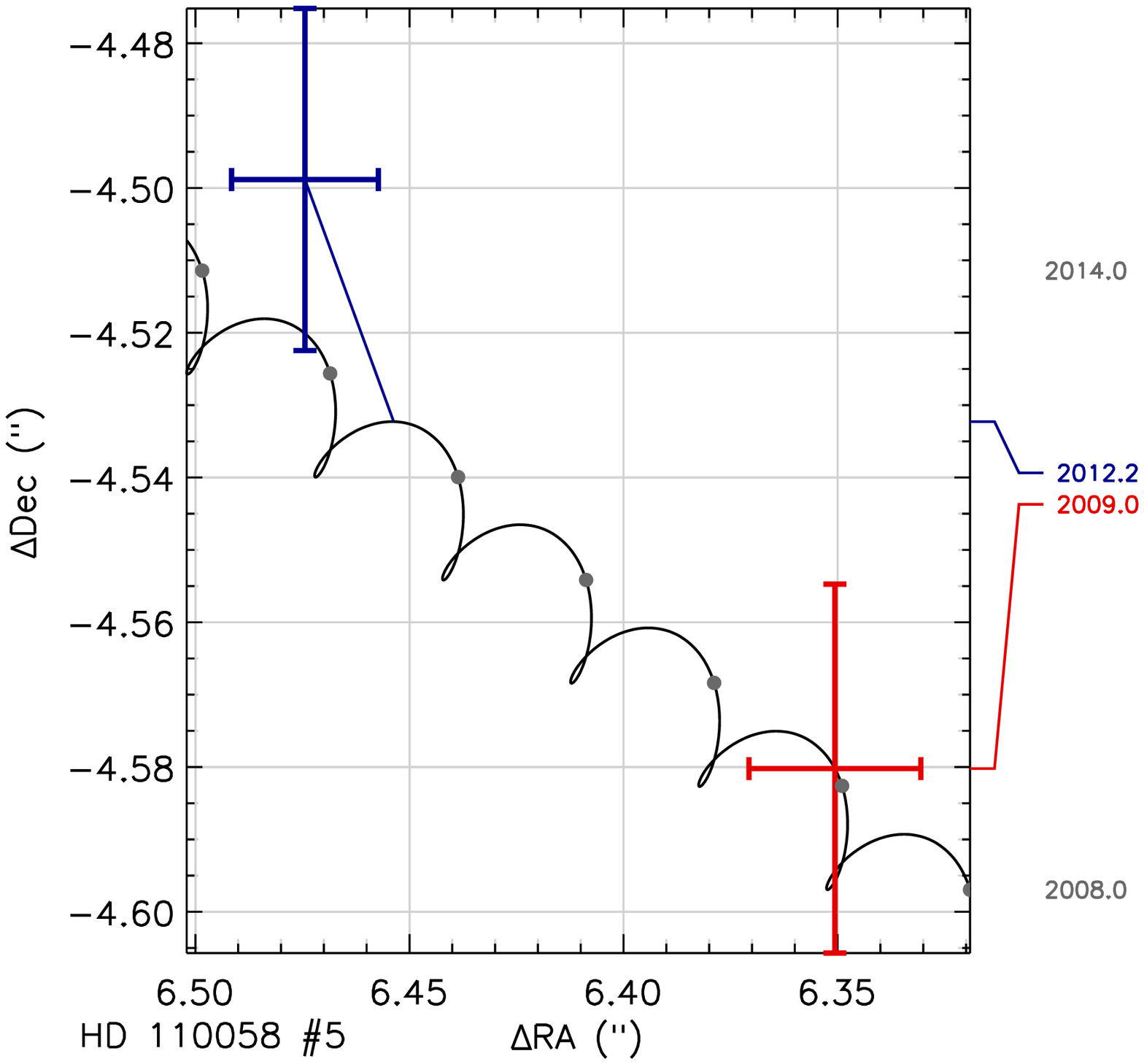}
\hskip -0.3in
\includegraphics[width=2.0in]{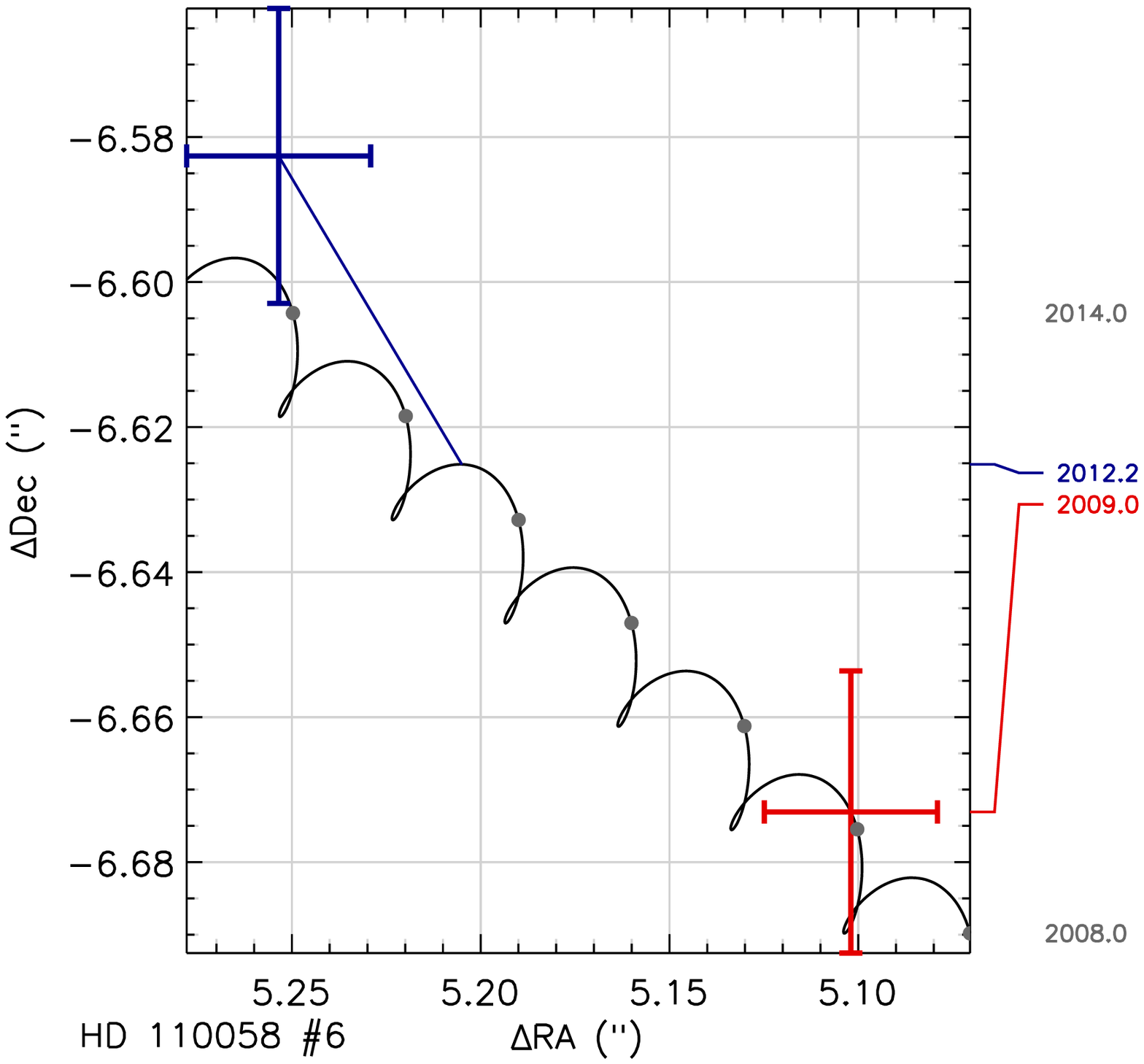}
\hskip -0.3in
\includegraphics[width=2.0in]{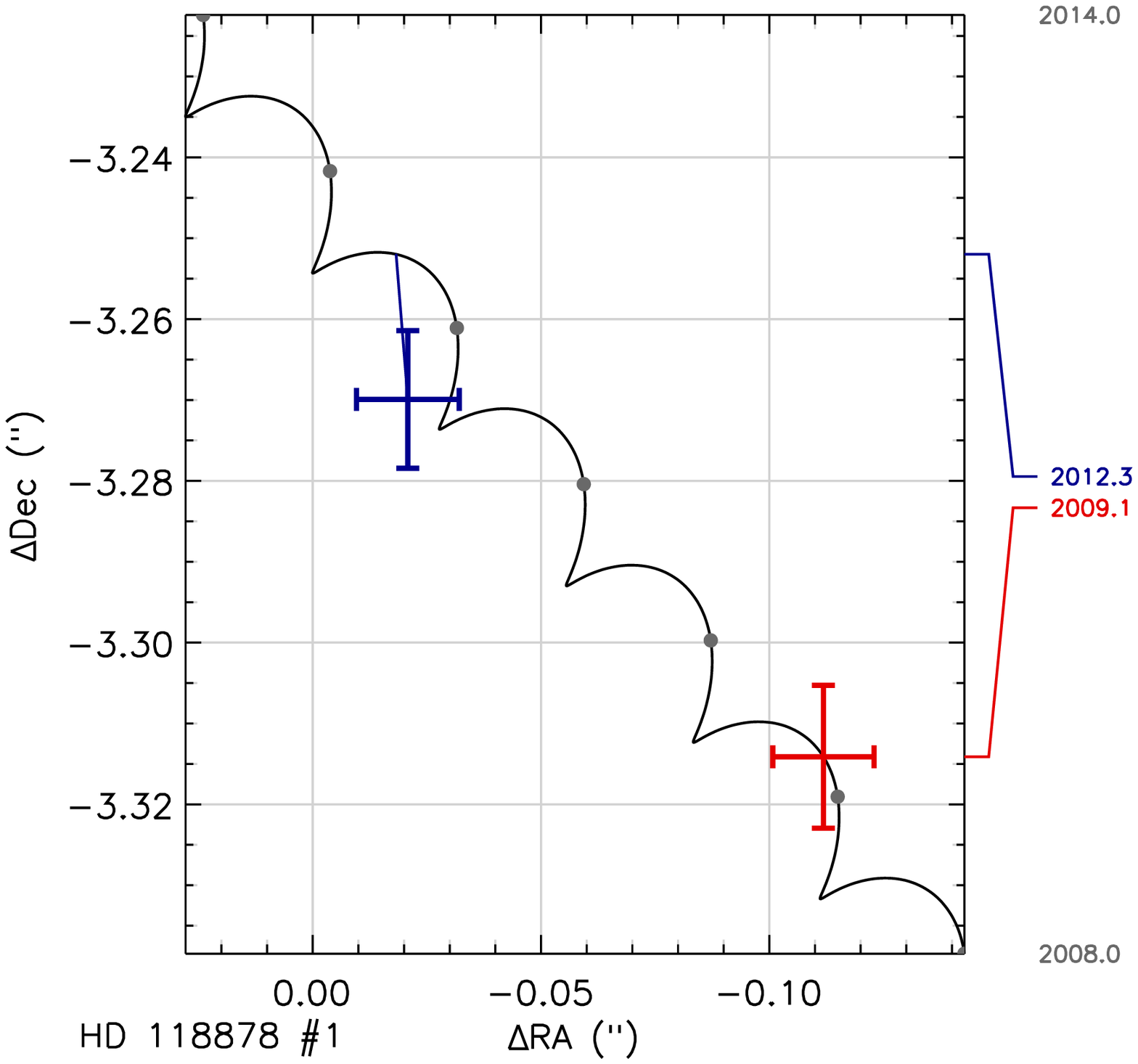}
\hskip -0.3in
\includegraphics[width=2.0in]{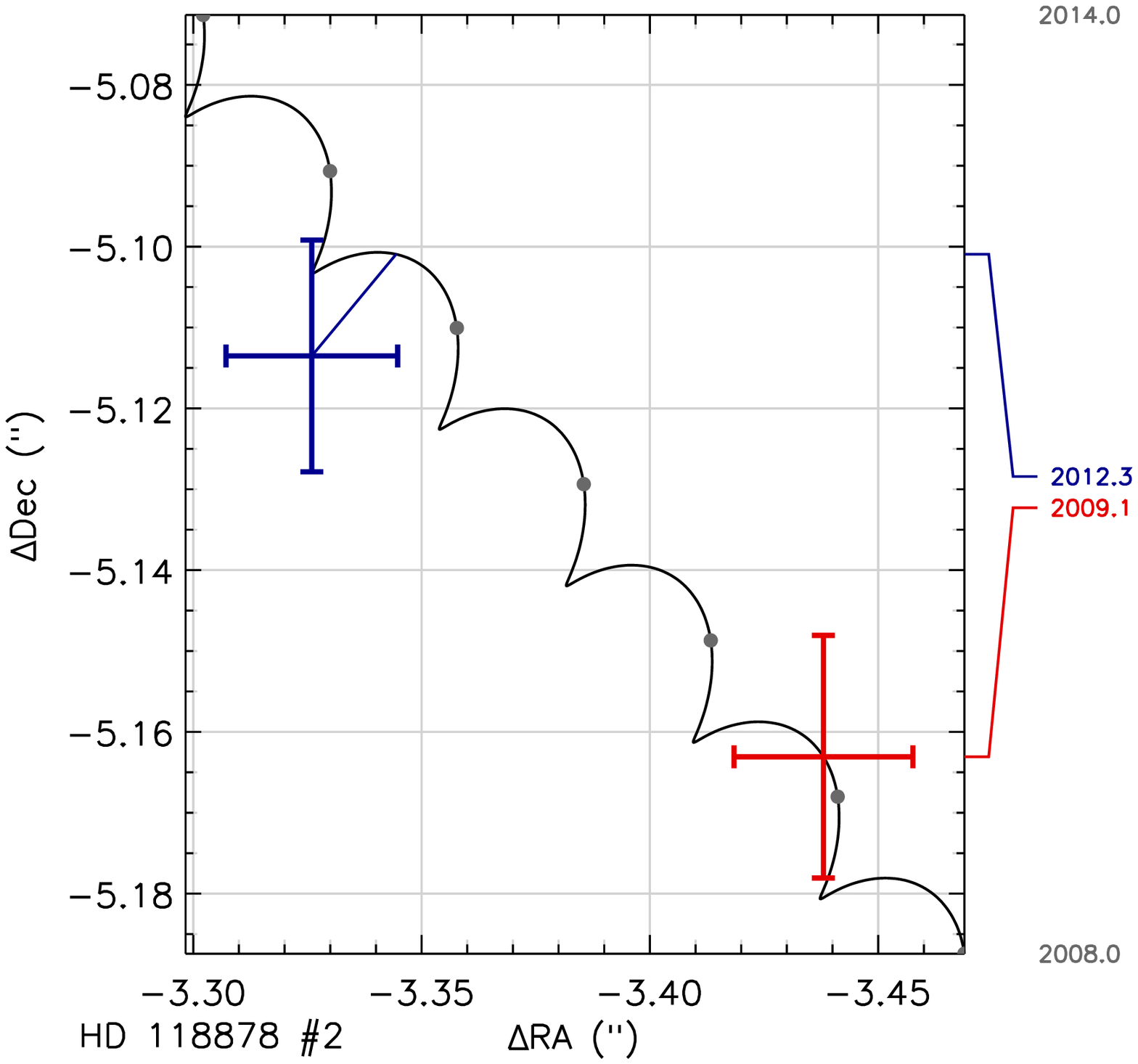}
}
\vskip -0.2in
\centerline{
\includegraphics[width=2.0in]{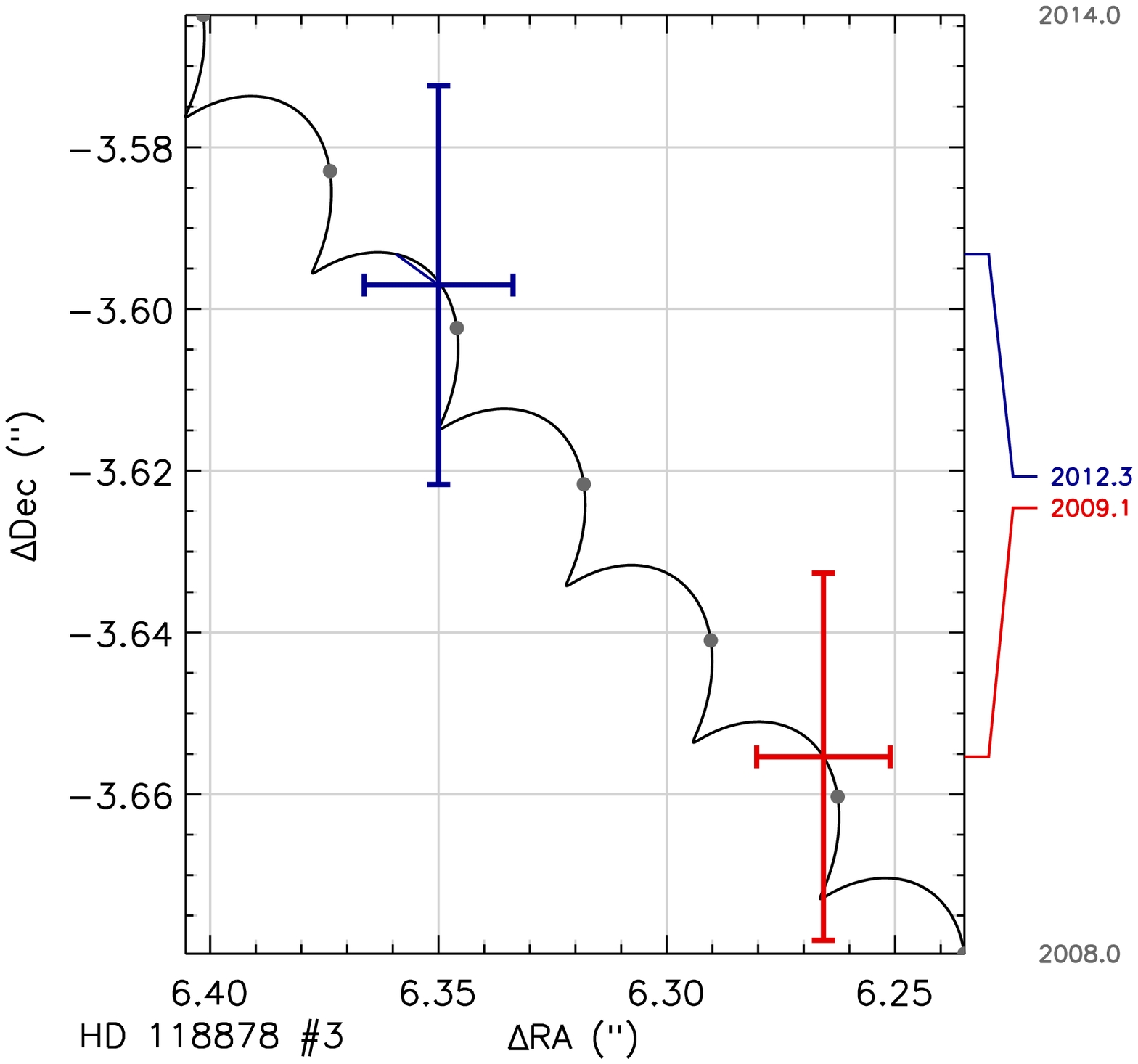}
\hskip -0.3in
\includegraphics[width=2.0in]{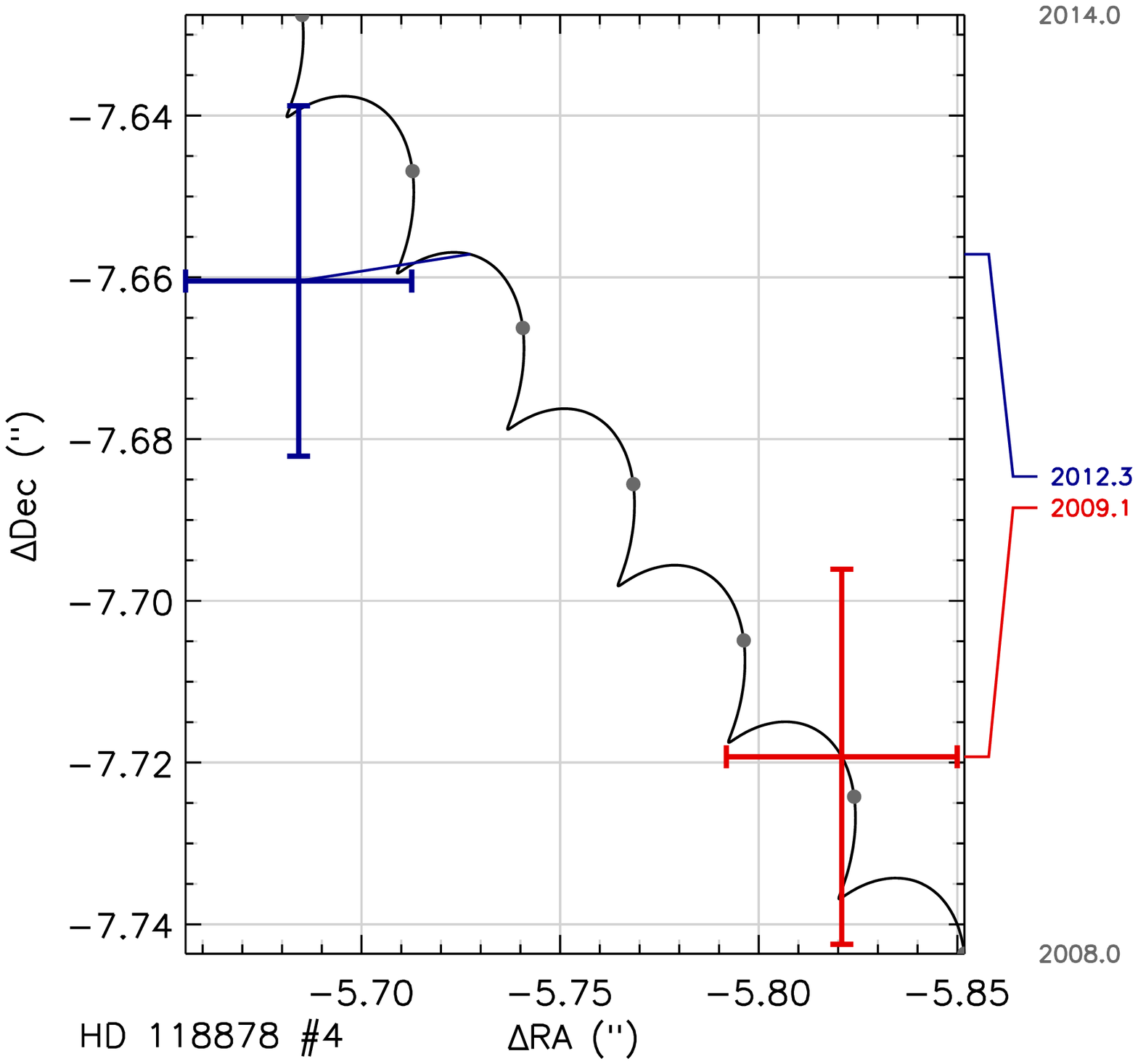}
\hskip -0.3in
\includegraphics[width=2.0in]{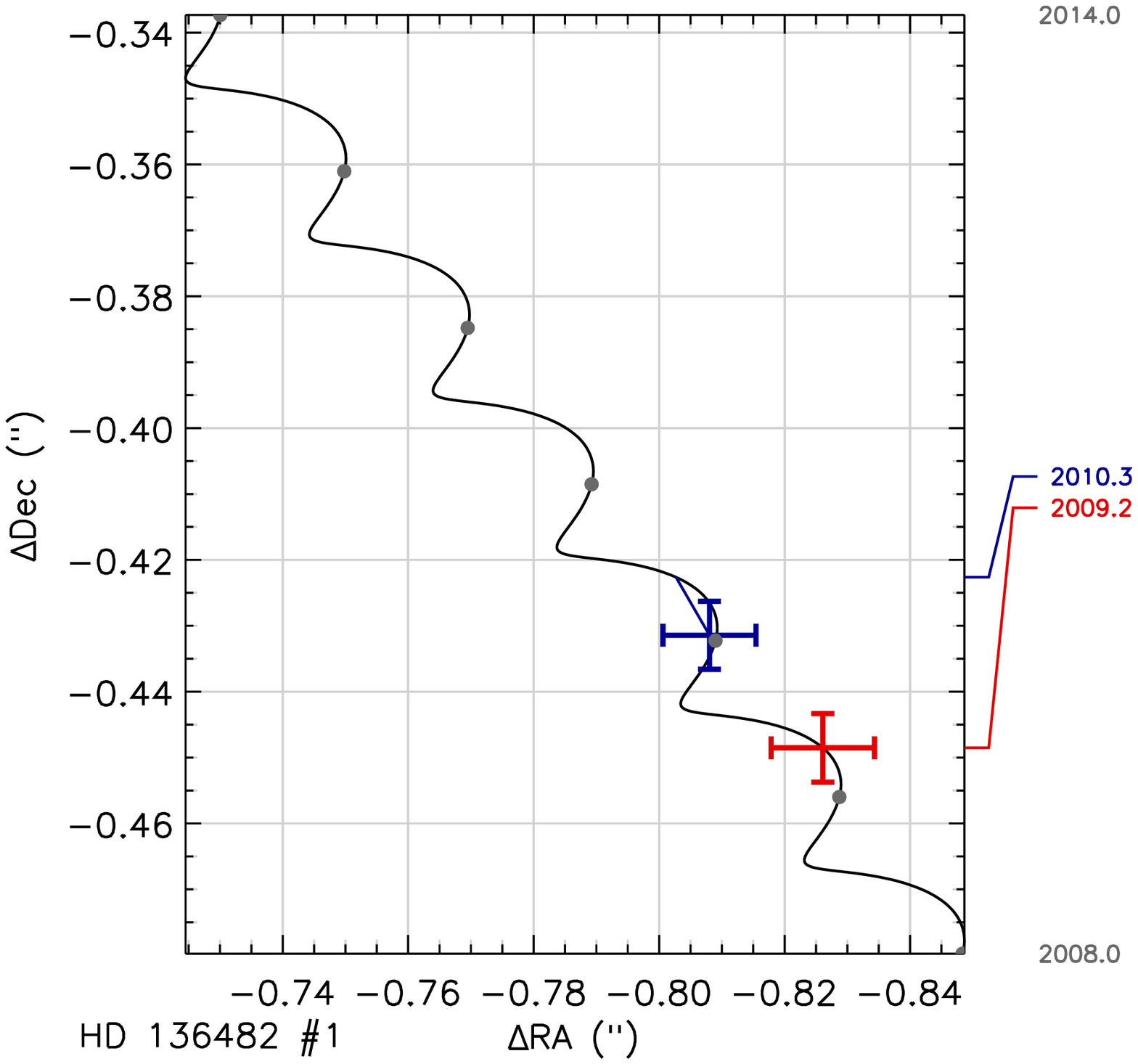}
\hskip -0.3in
\includegraphics[width=2.0in]{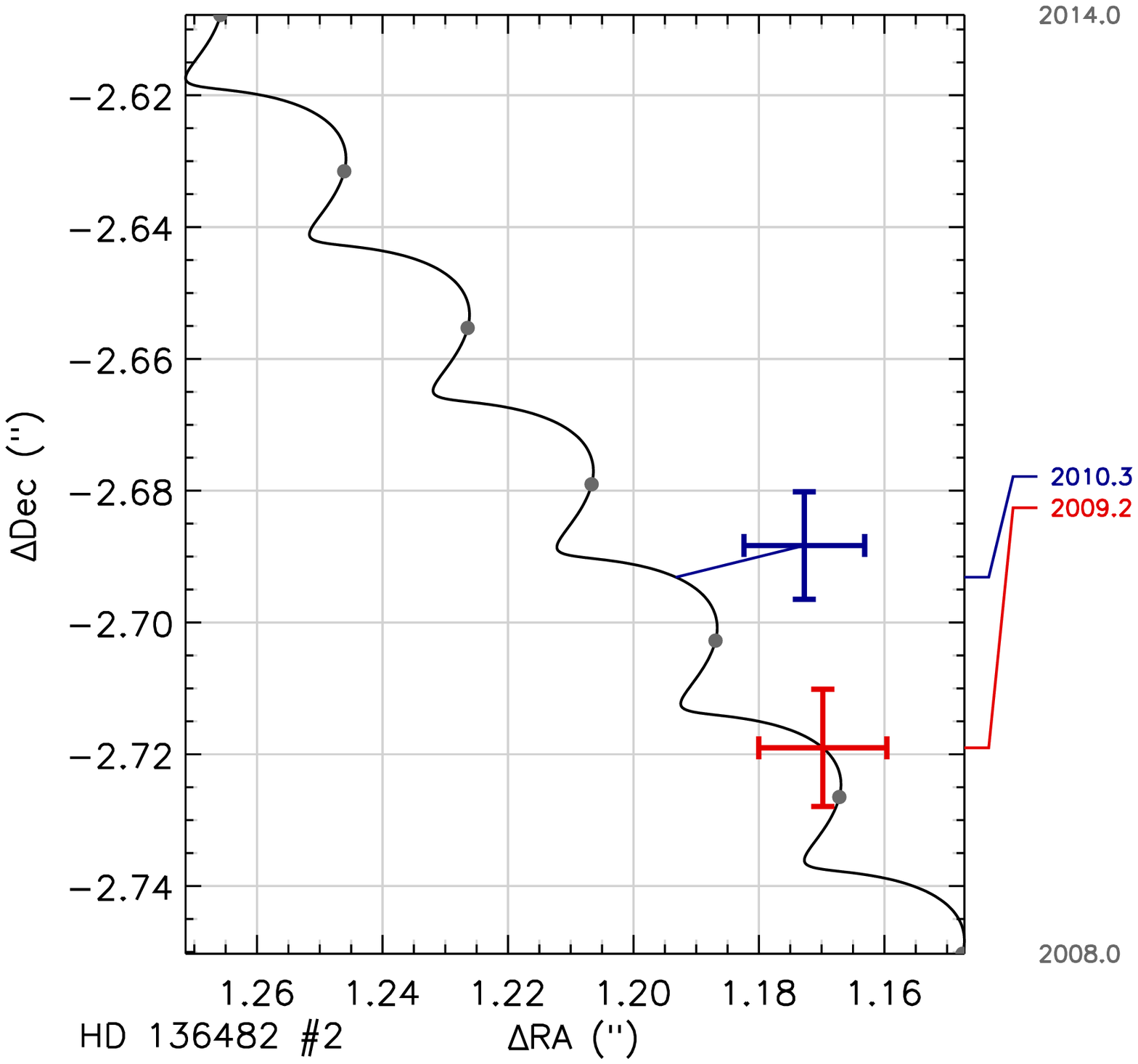}
}
\vskip -0.2in
\caption{Candidate companion on-sky motion, continued from 
Figure~\ref{tiled_fig1}.}\label{tiled_fig2}
\end{figure}

\begin{figure}
\centerline{
\includegraphics[width=2.0in]{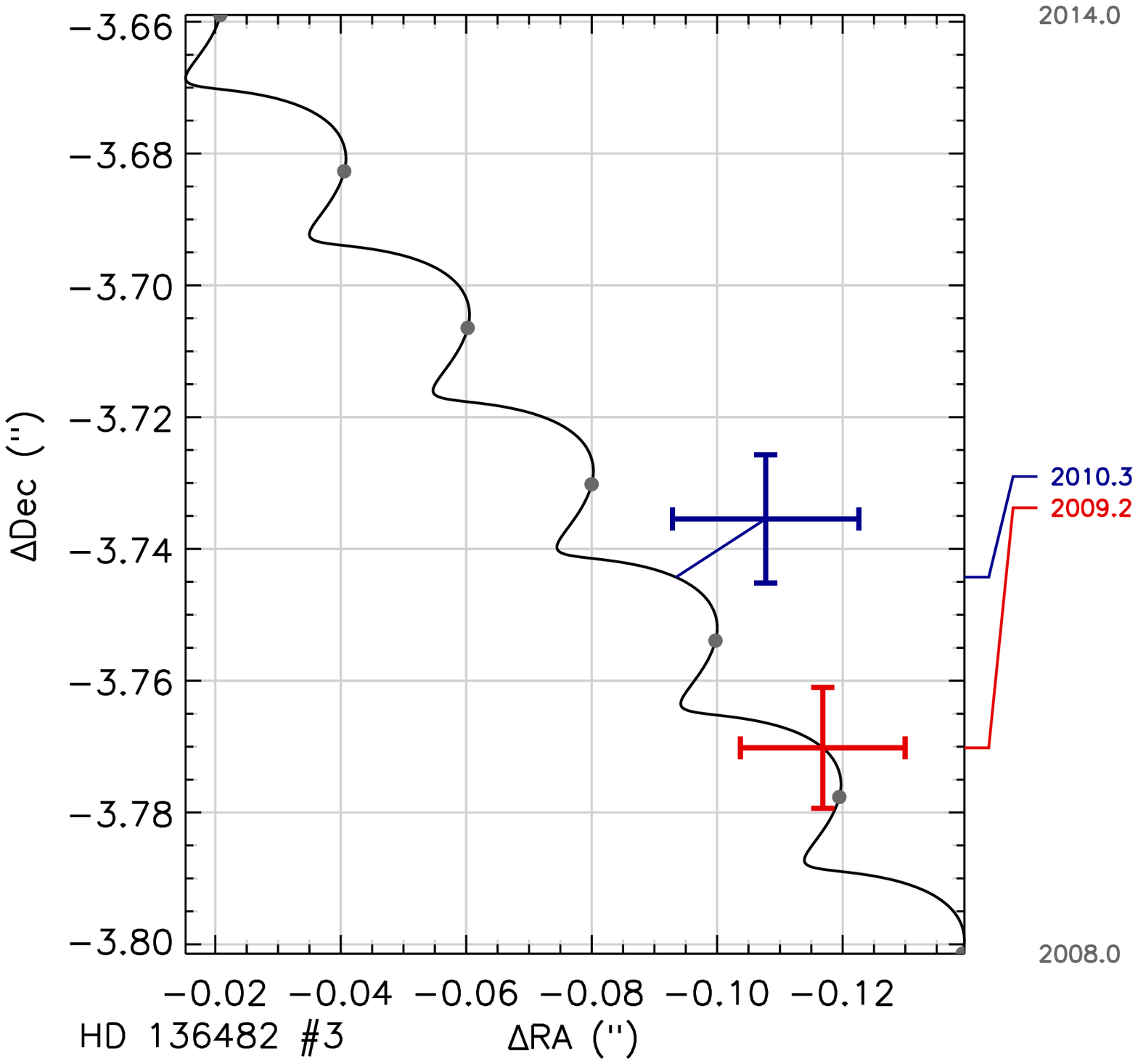}
\hskip -0.3in
\includegraphics[width=2.0in]{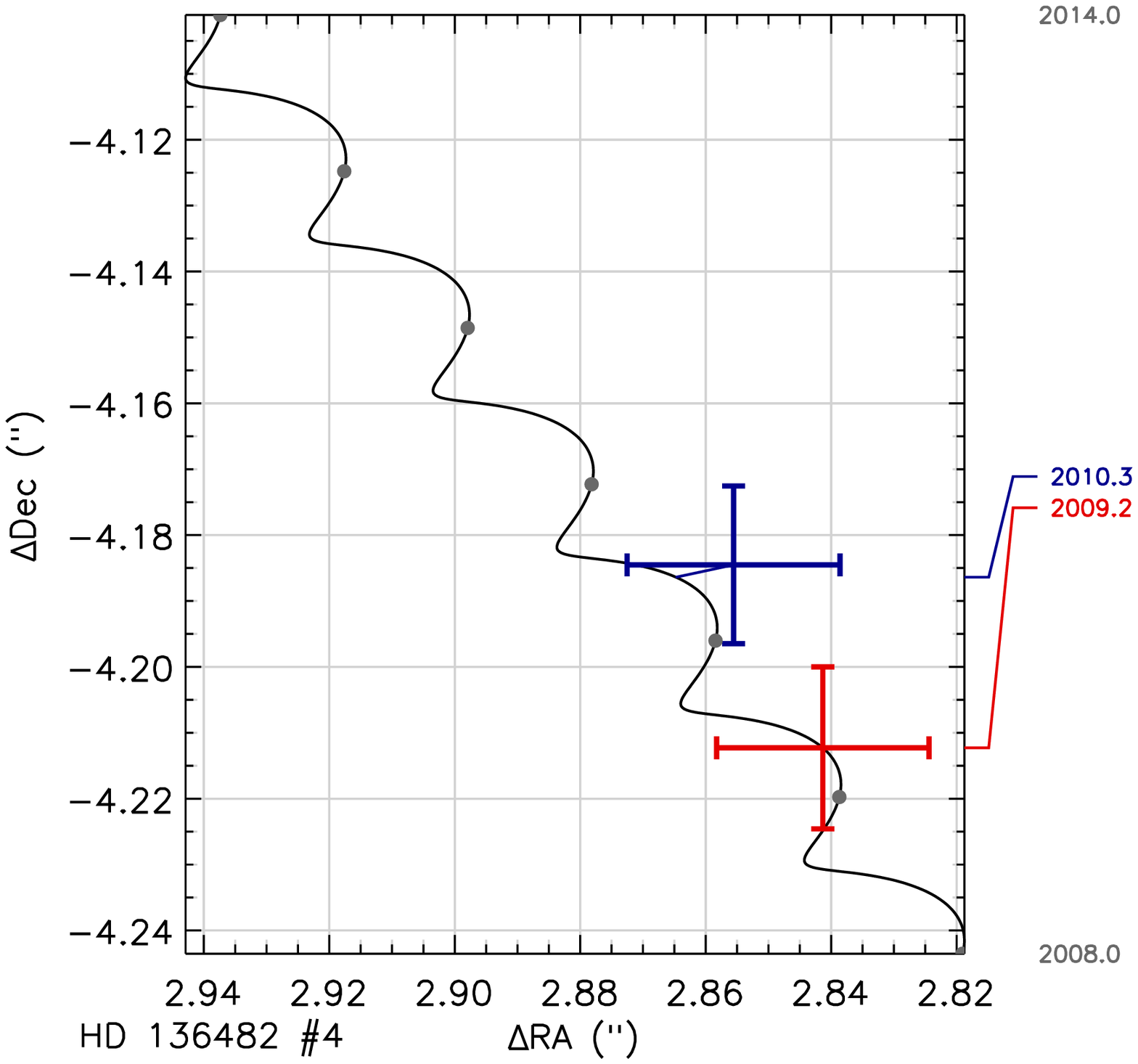}
\hskip -0.3in
\includegraphics[width=2.0in]{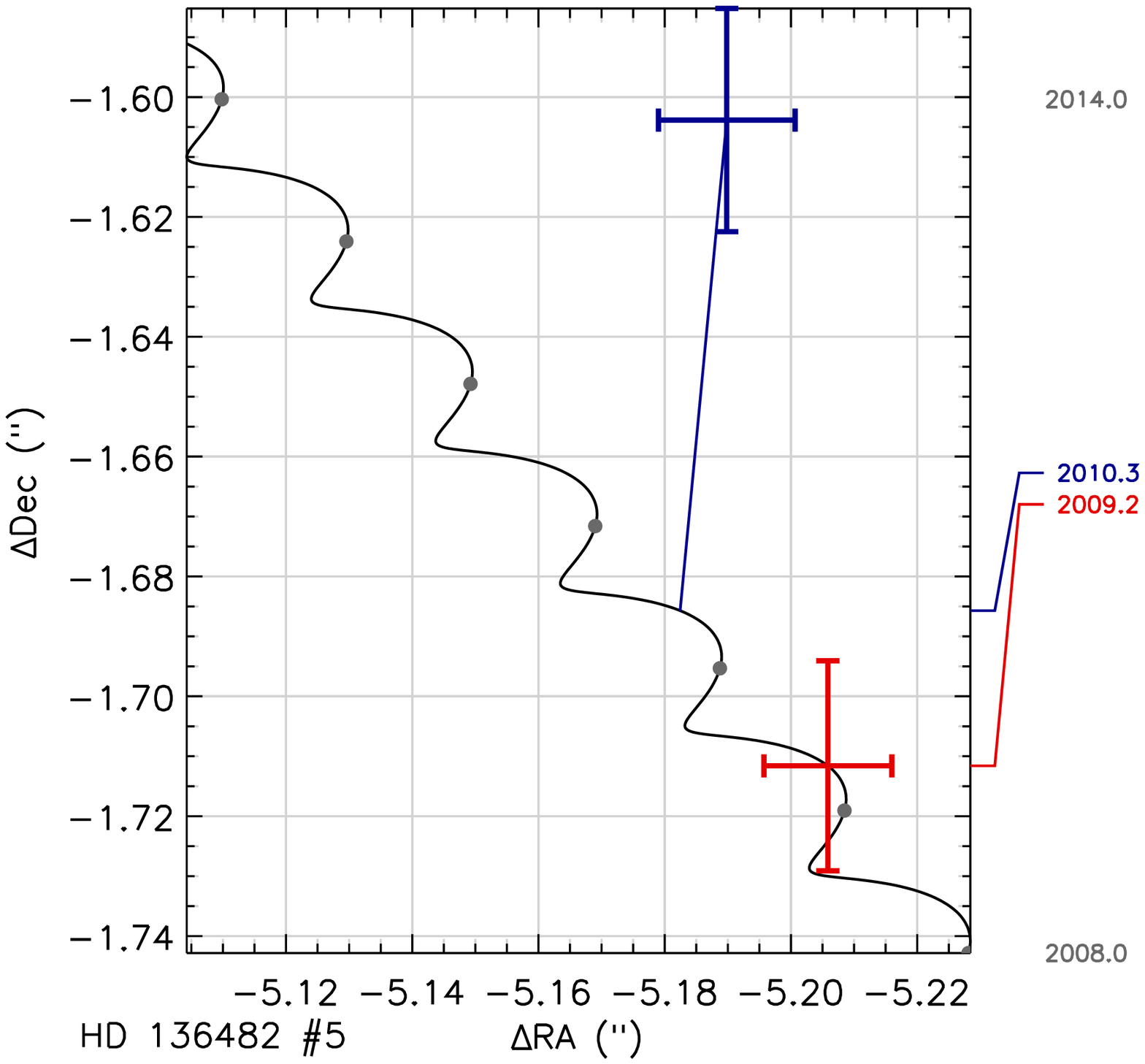}
\hskip -0.3in
\includegraphics[width=2.0in]{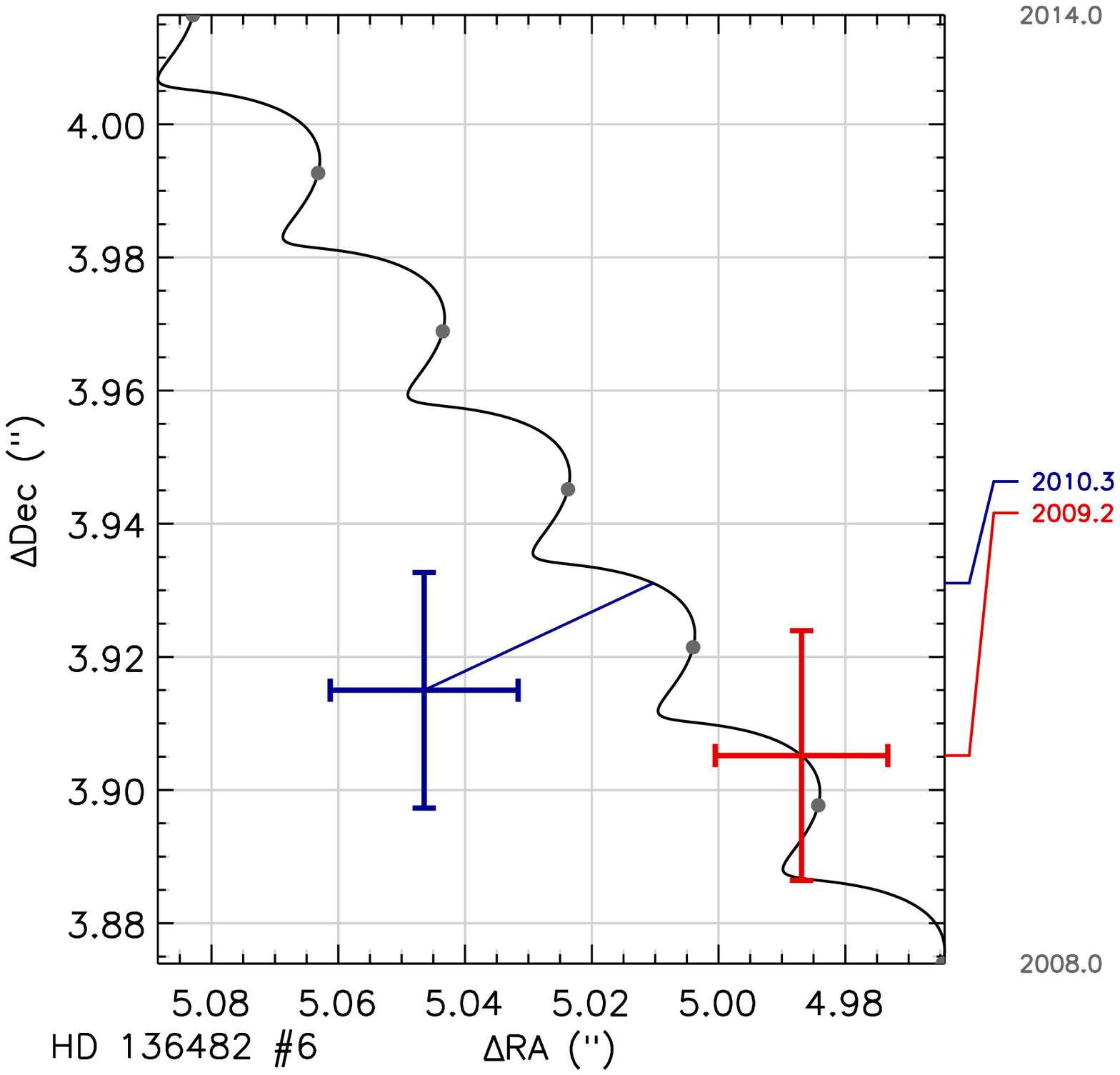}
}
\vskip -0.2in
\centerline{
\includegraphics[width=2.0in]{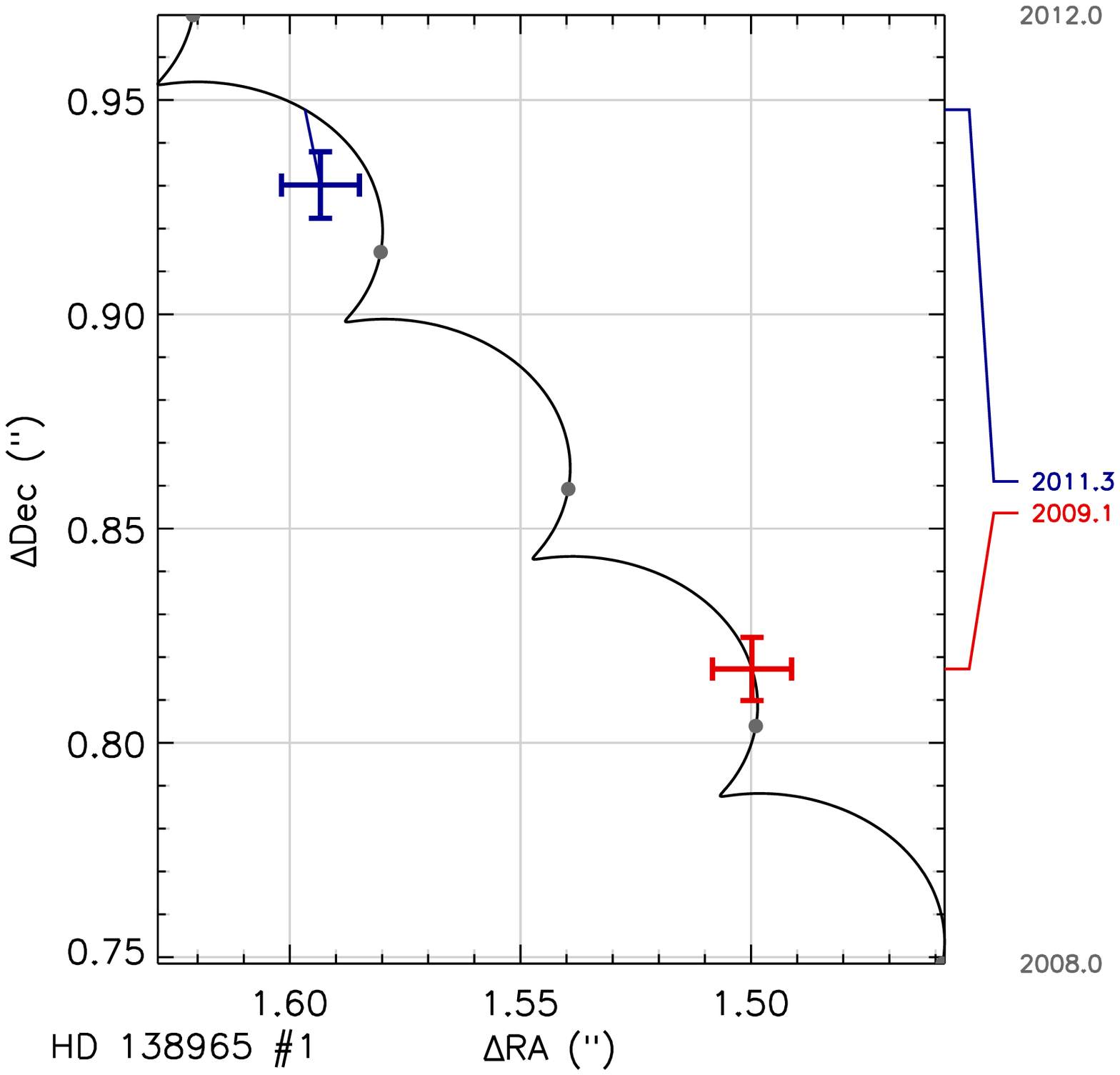}
\hskip -0.3in
\includegraphics[width=2.0in]{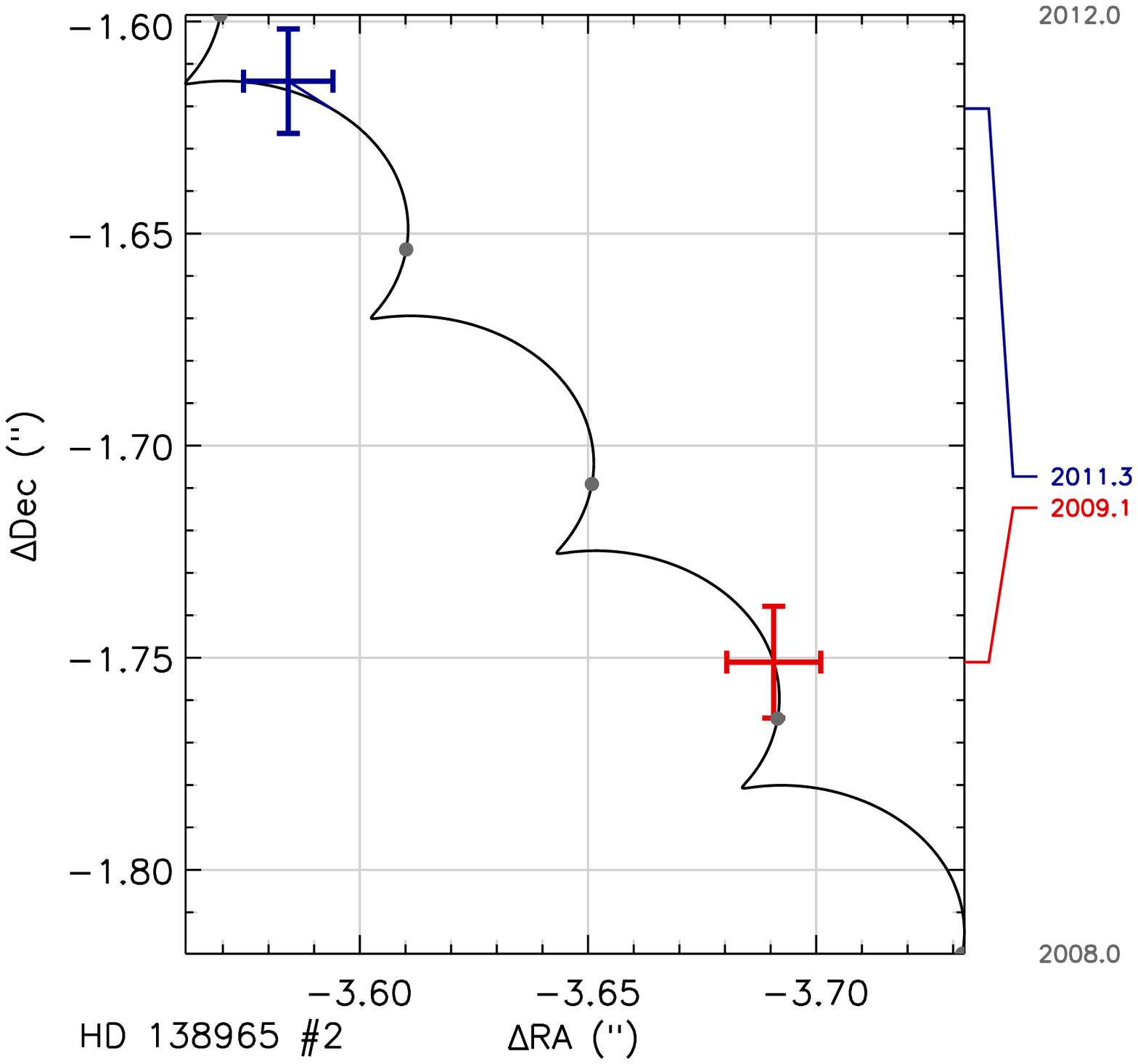}
\hskip -0.3in
\includegraphics[width=2.0in]{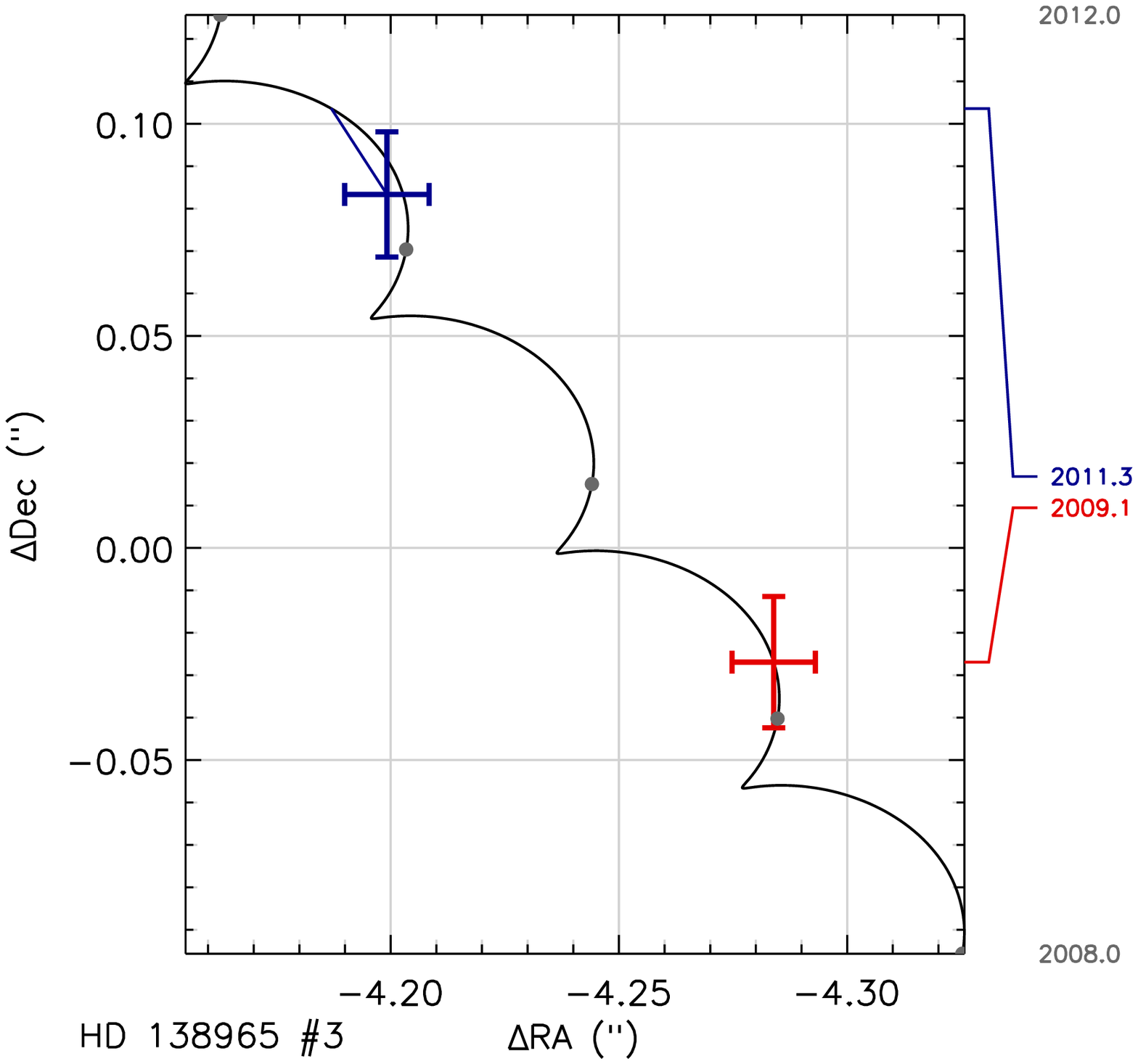}
\hskip -0.3in
\includegraphics[width=2.0in]{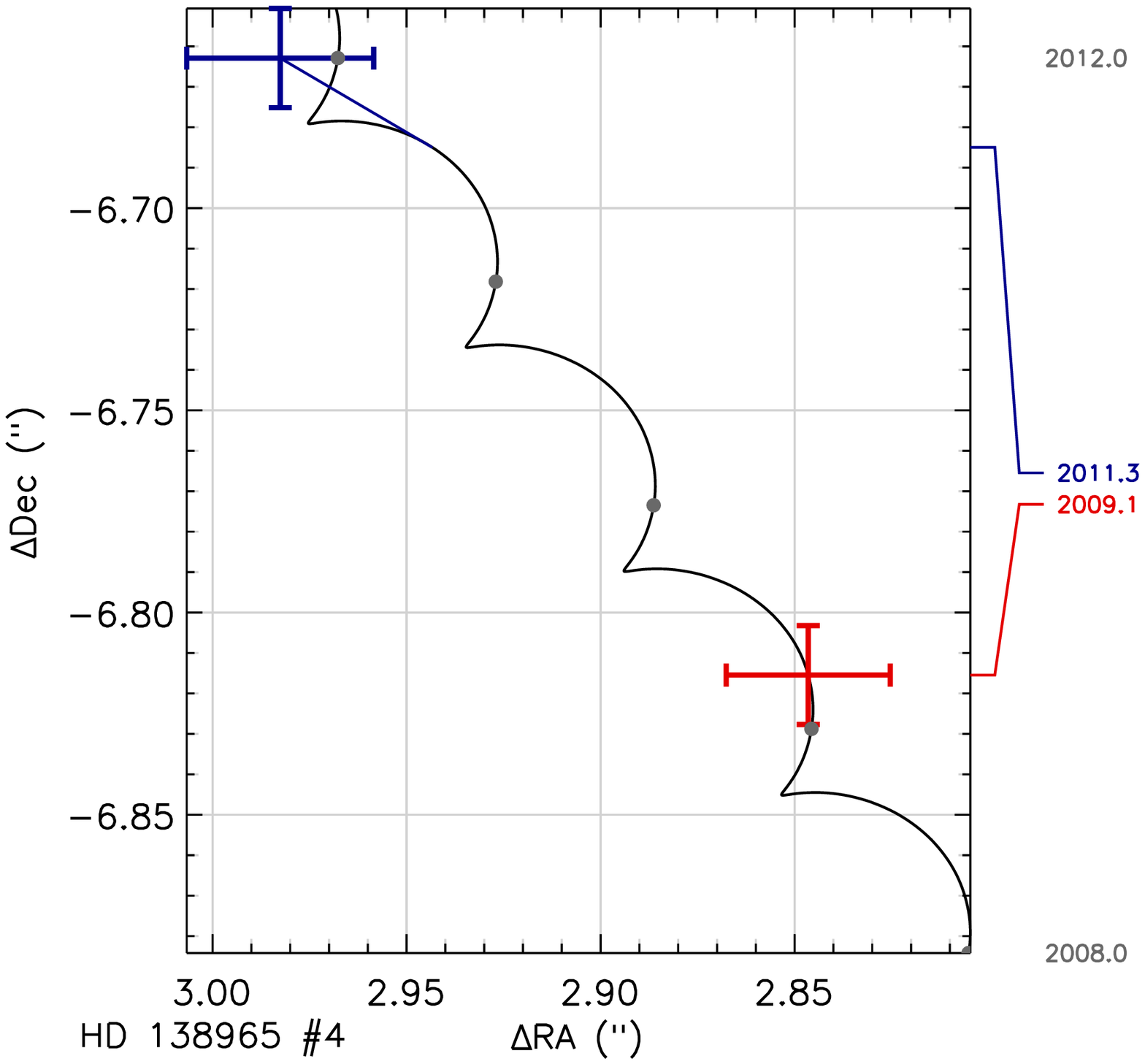}
}
\vskip -0.2in
\centerline{
\includegraphics[width=2.0in]{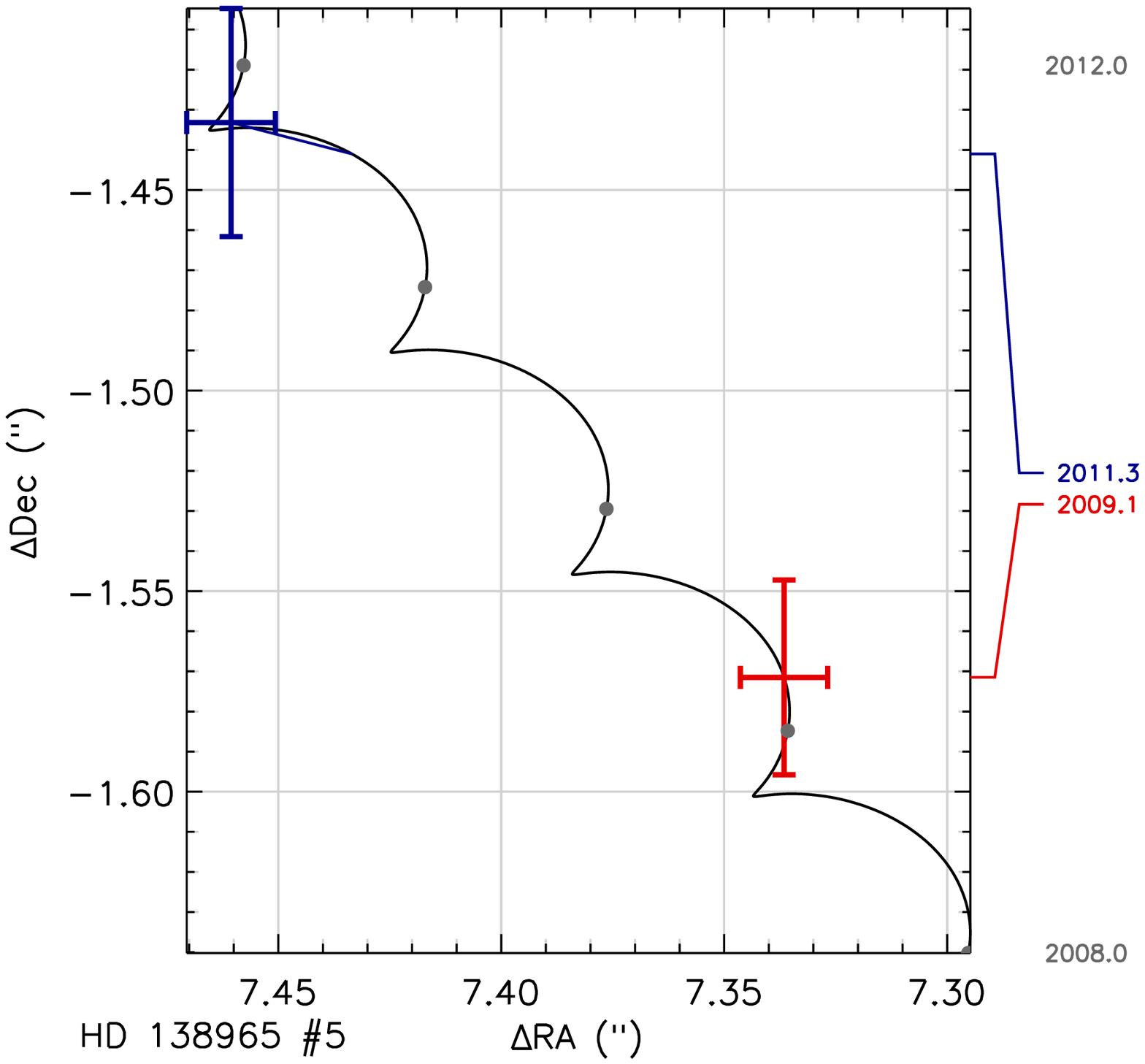}
\hskip -0.3in
\includegraphics[width=2.0in]{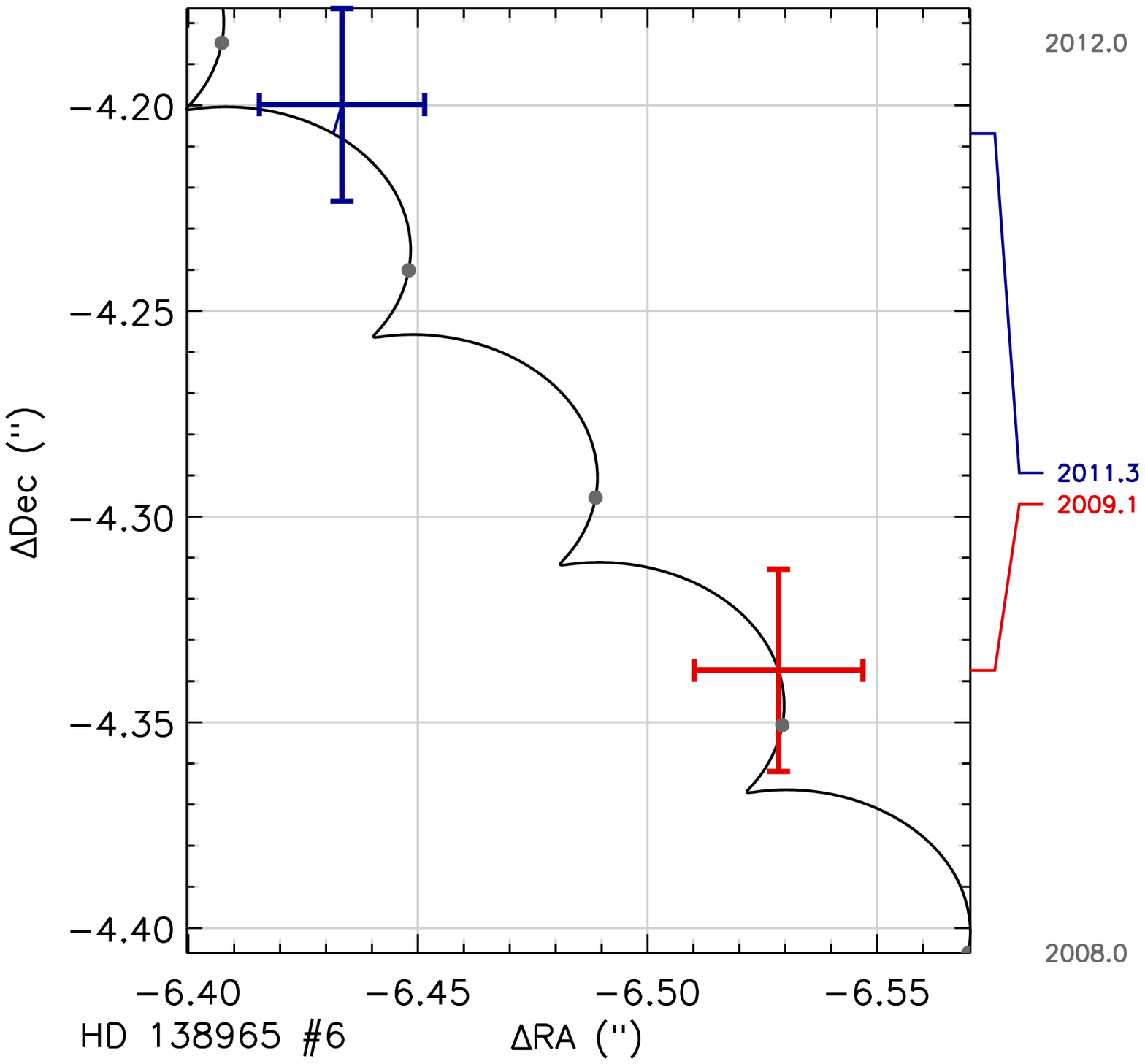}
\hskip -0.3in
\includegraphics[width=2.0in]{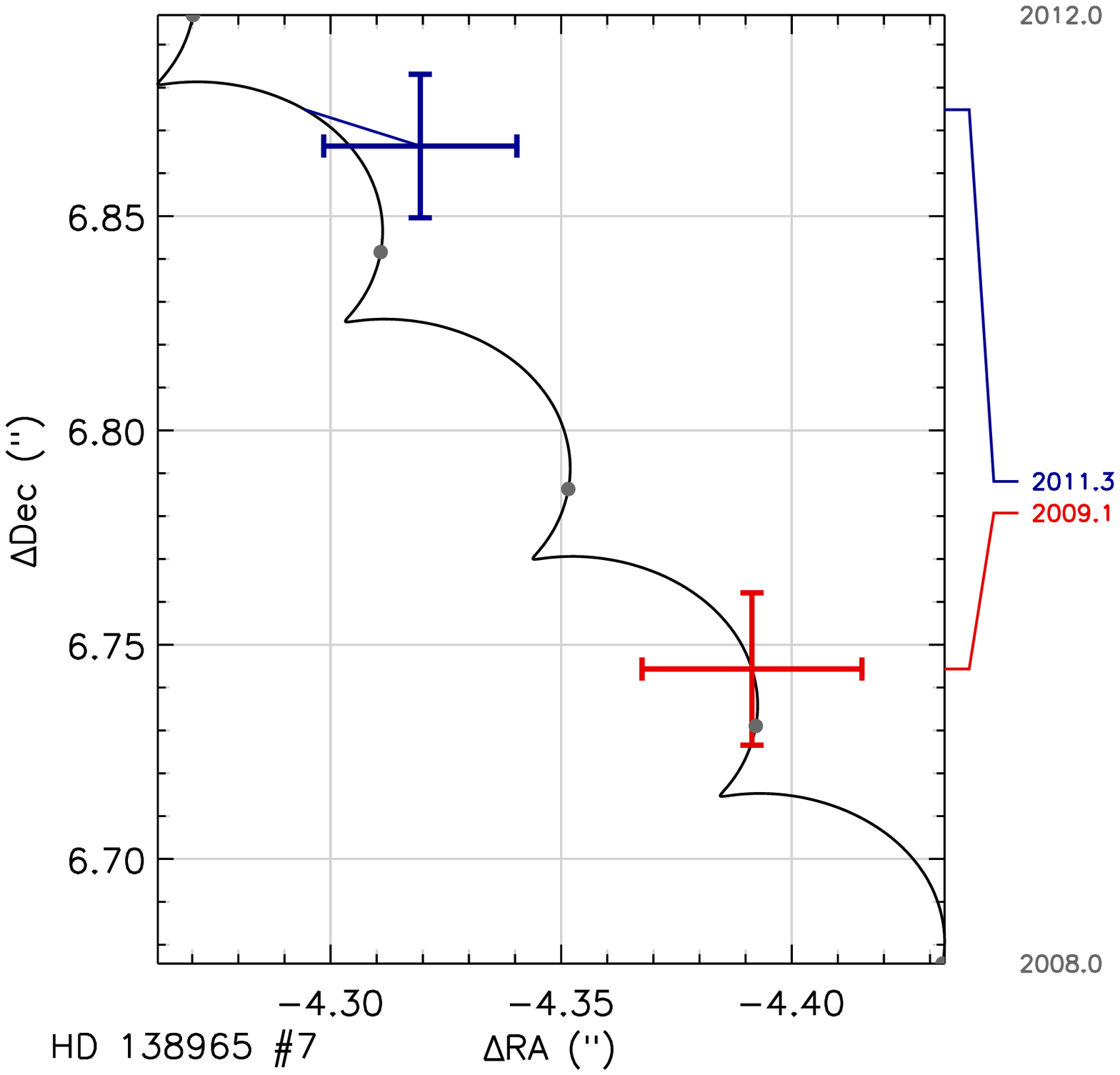}
\hskip -0.3in
\includegraphics[width=2.0in]{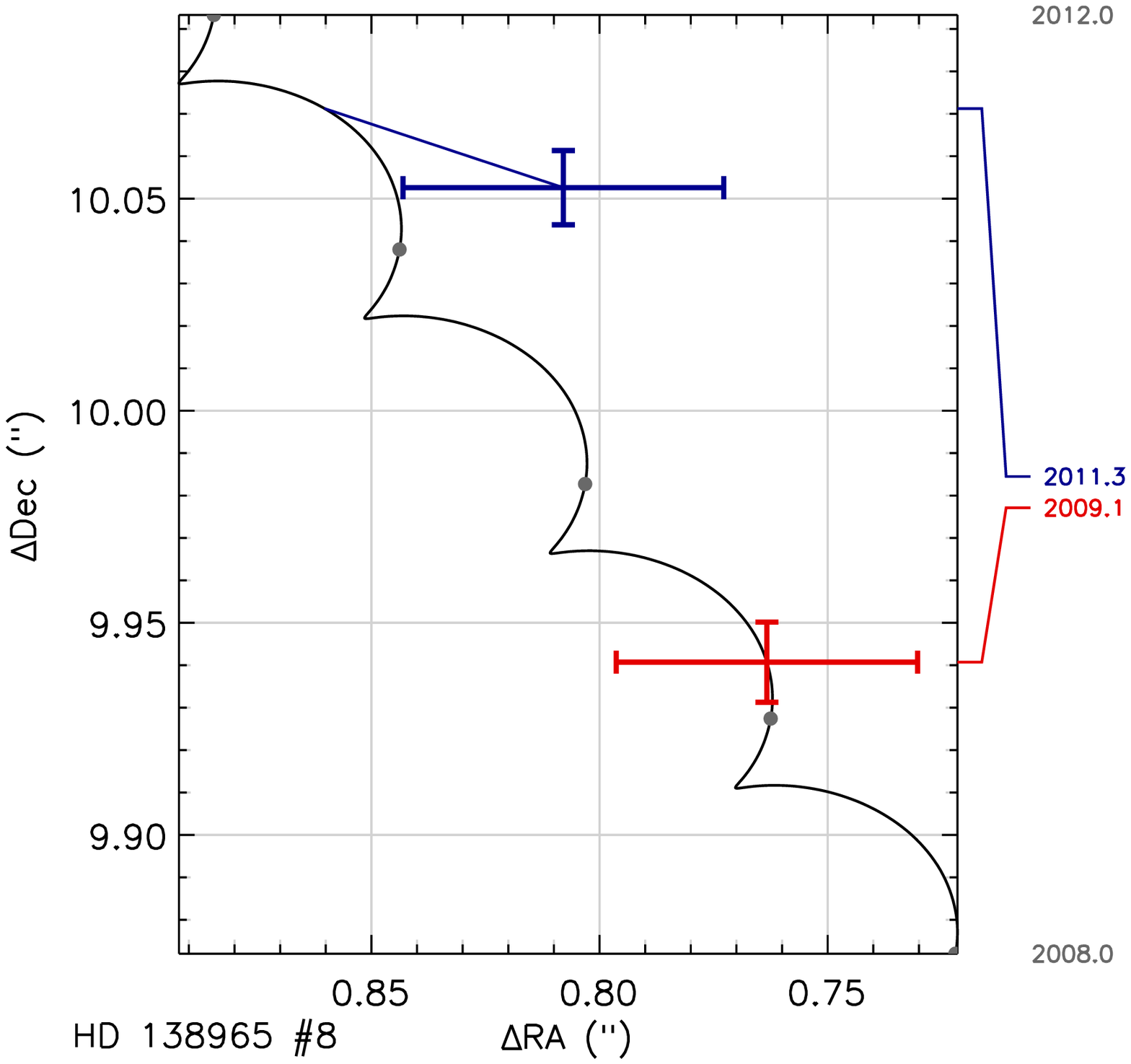}
}
\vskip -0.2in
\centerline{
\includegraphics[width=2.0in]{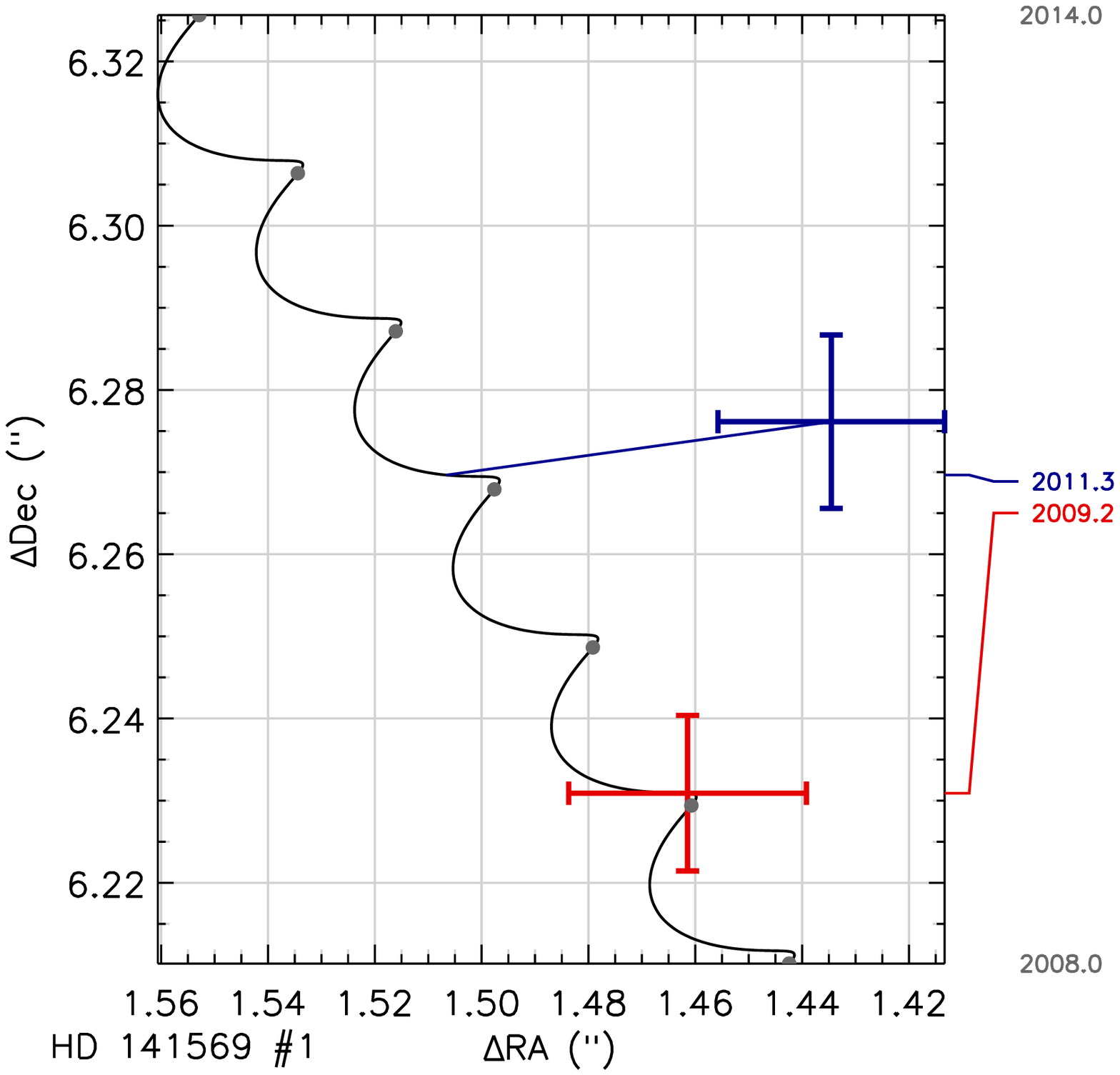}
\hskip -0.3in
\includegraphics[width=2.0in]{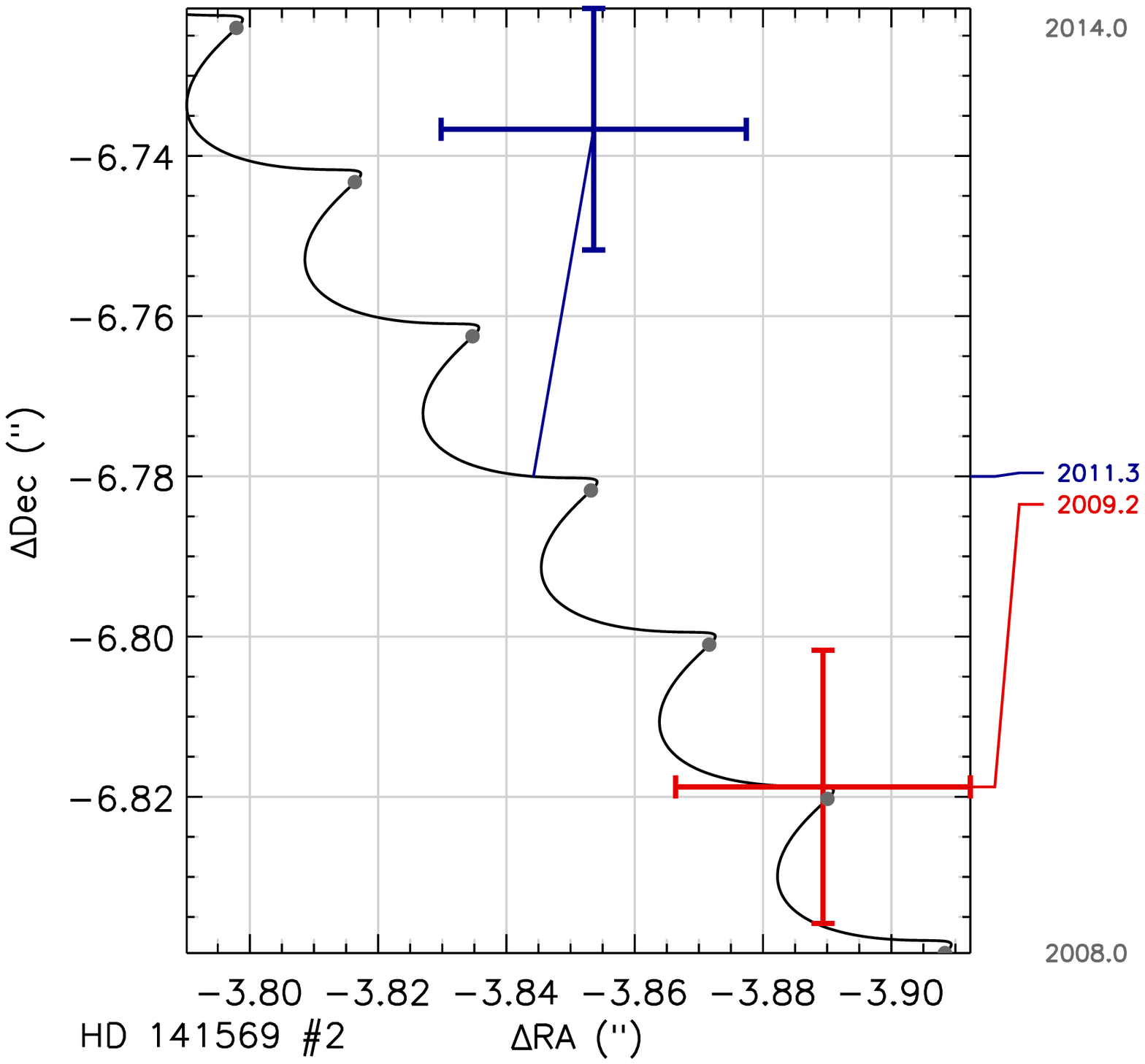}
\hskip -0.3in
\includegraphics[width=2.0in]{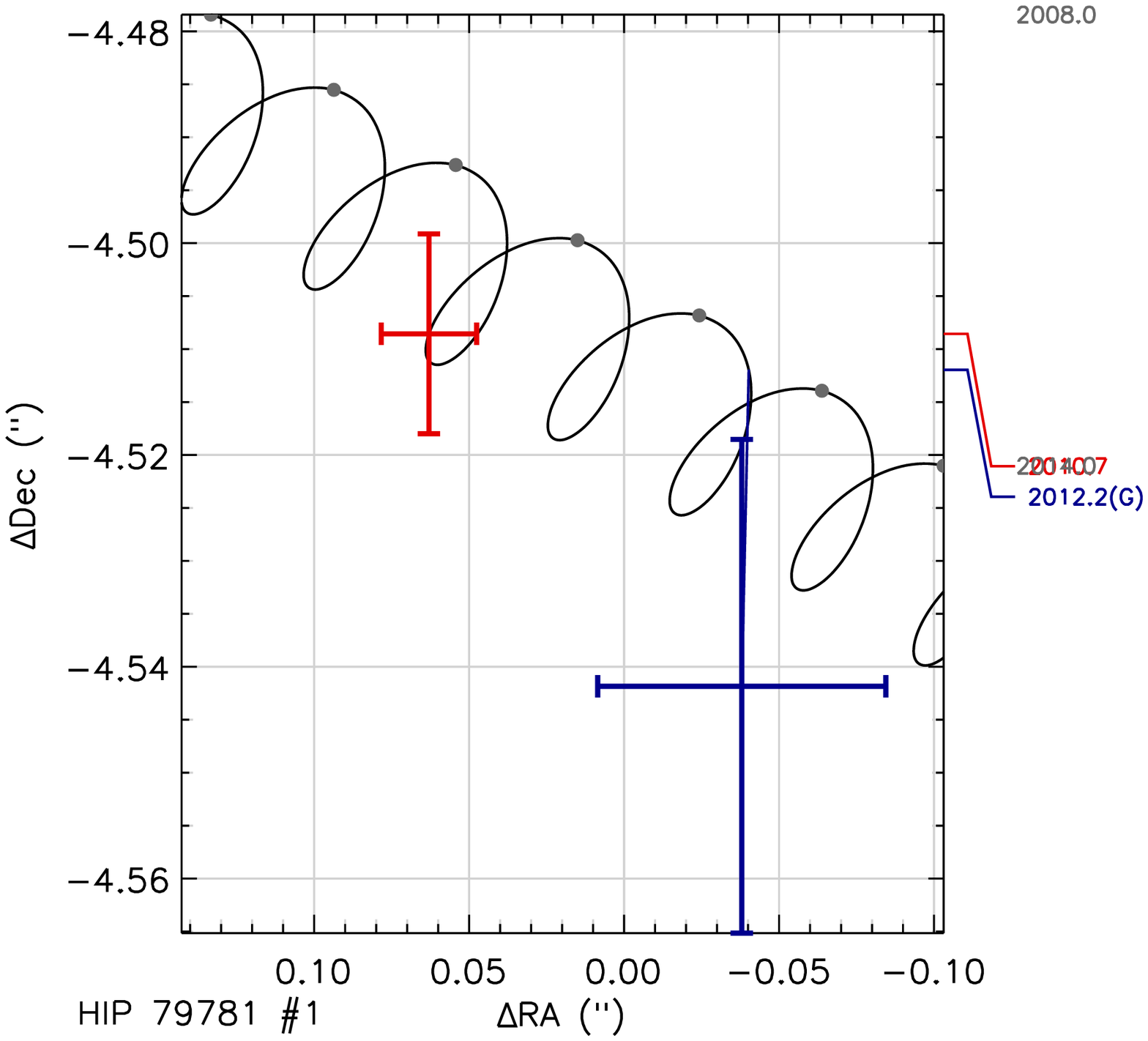}
\hskip -0.3in
\includegraphics[width=2.0in]{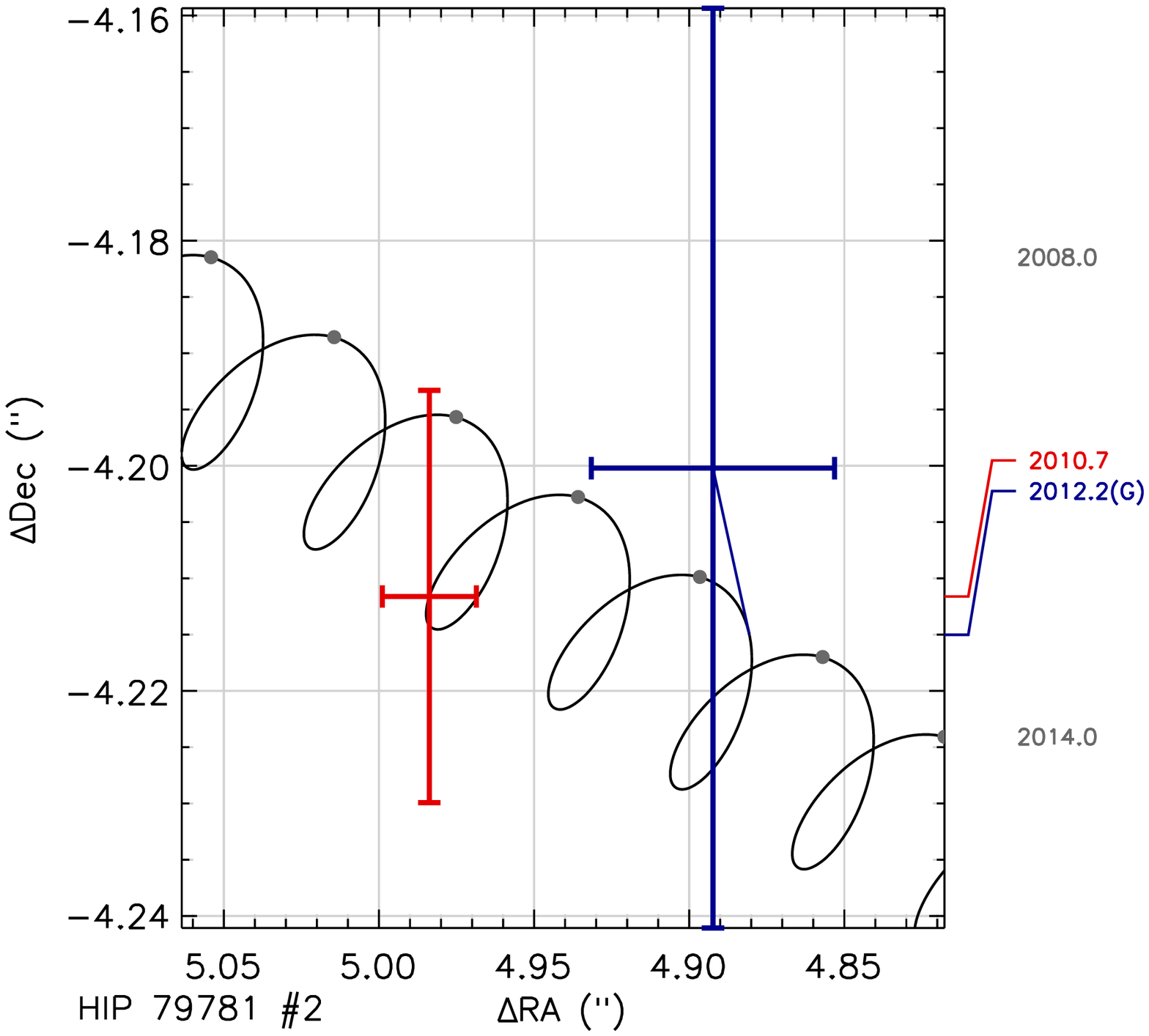}
}
\vskip -0.2in
\caption{Candidate companion on-sky motion, continued from
Figures~\ref{tiled_fig1} and~\ref{tiled_fig2}.}\label{tiled_fig3}
\end{figure}

\begin{figure}
\centerline{
\includegraphics[width=2.0in]{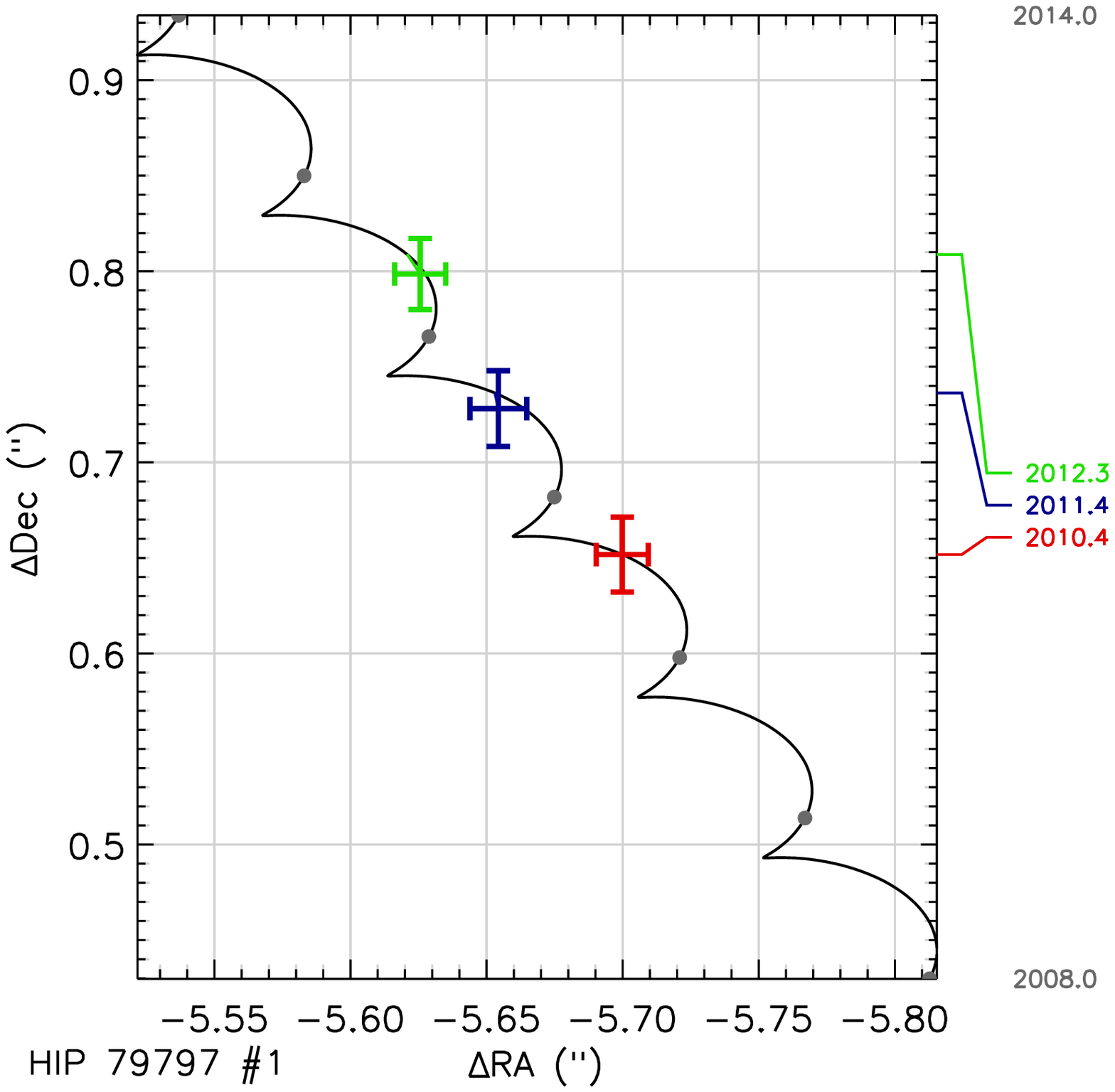}
\hskip -0.3in
\includegraphics[width=2.0in]{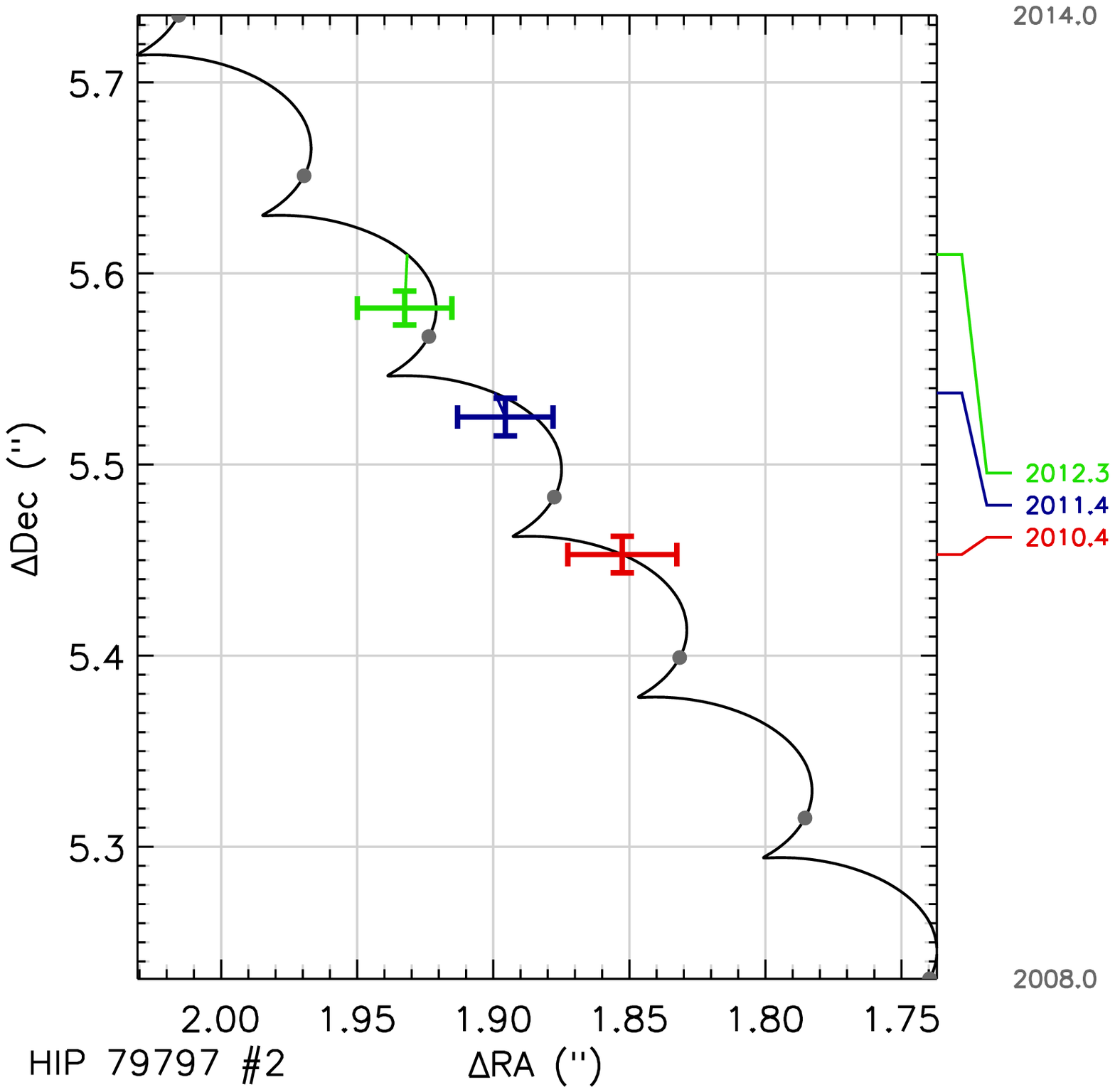}
\hskip -0.3in
\includegraphics[width=2.0in]{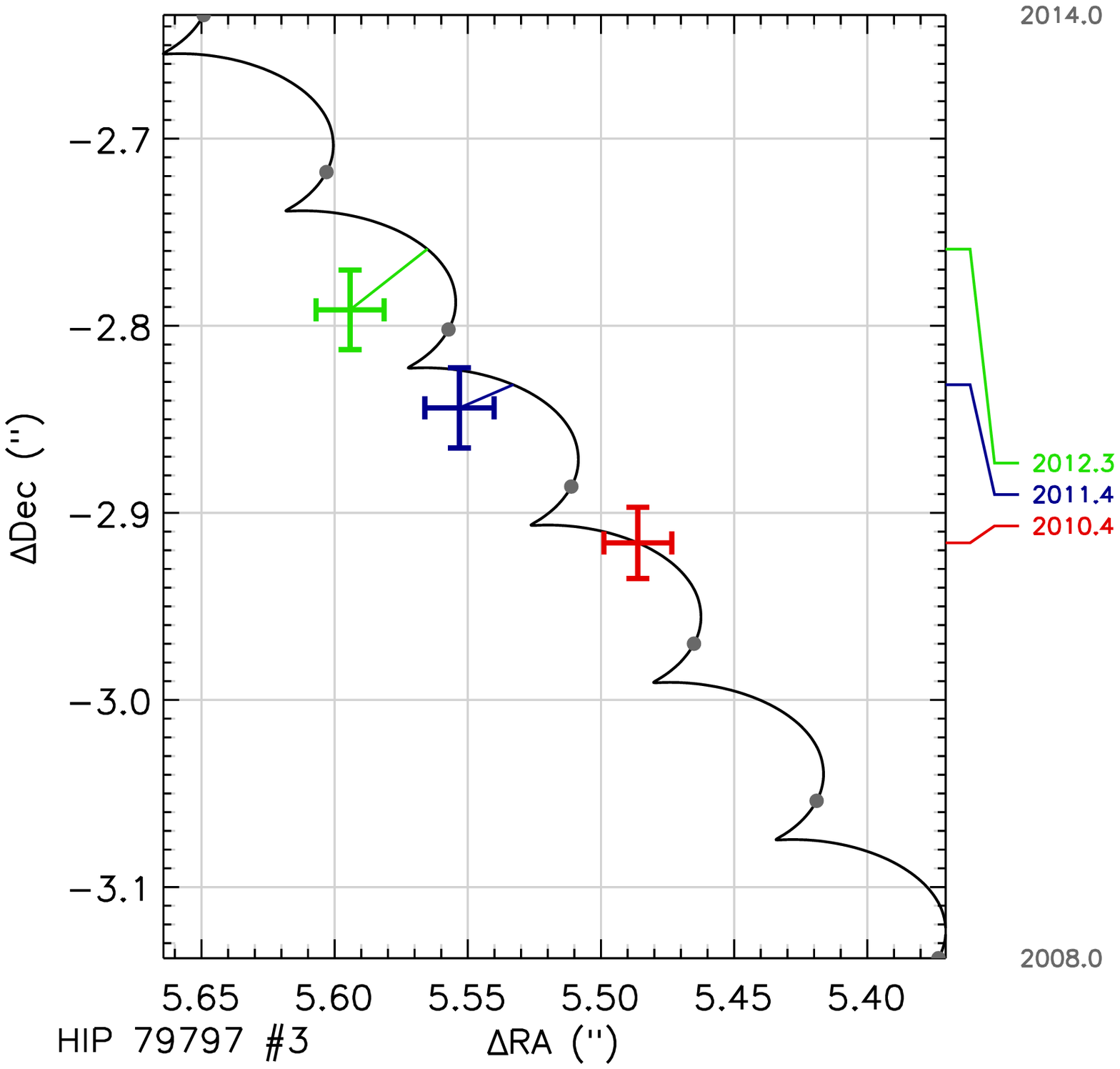}
\hskip -0.3in
\includegraphics[width=2.0in]{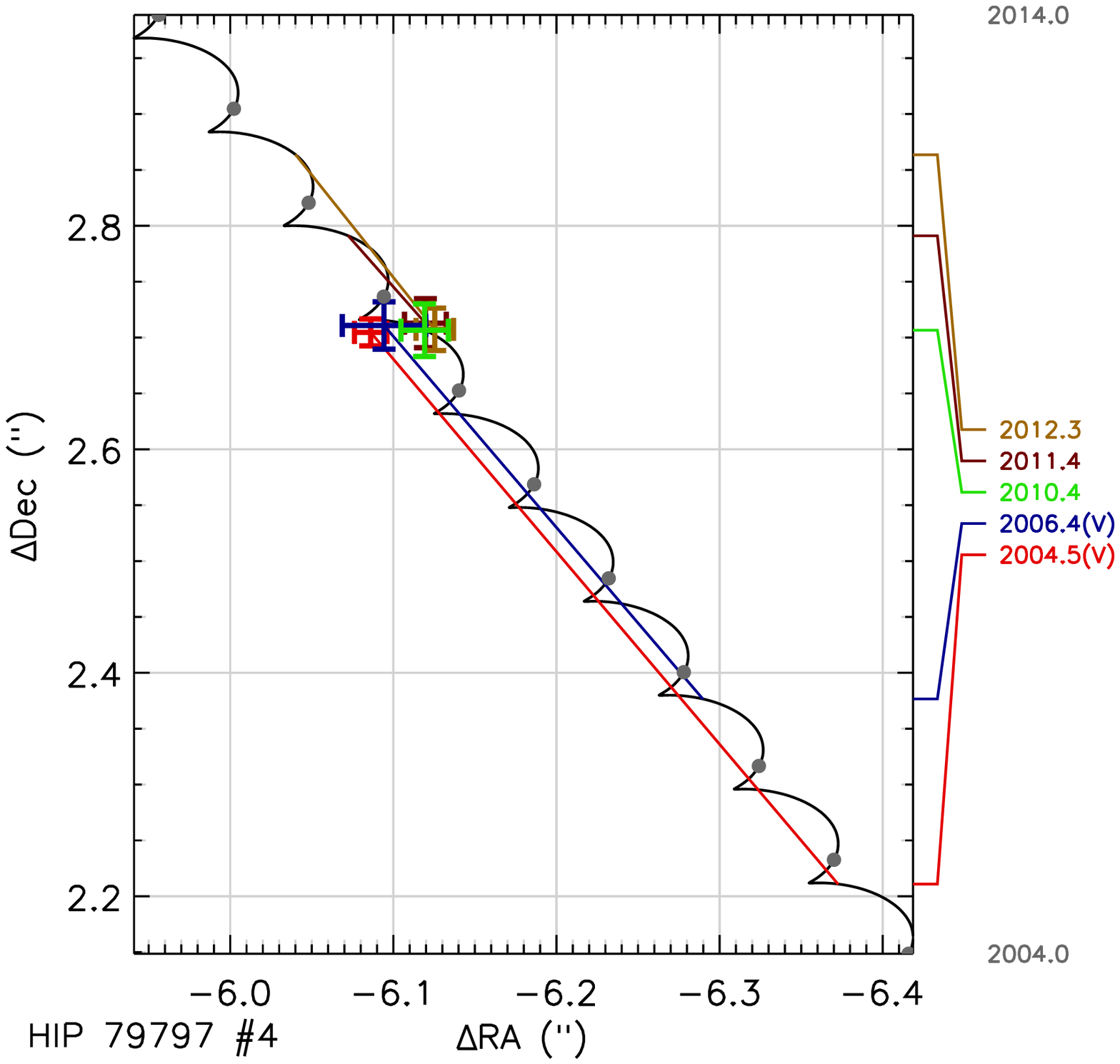}
}
\vskip -0.2in
\centerline{
\includegraphics[width=2.0in]{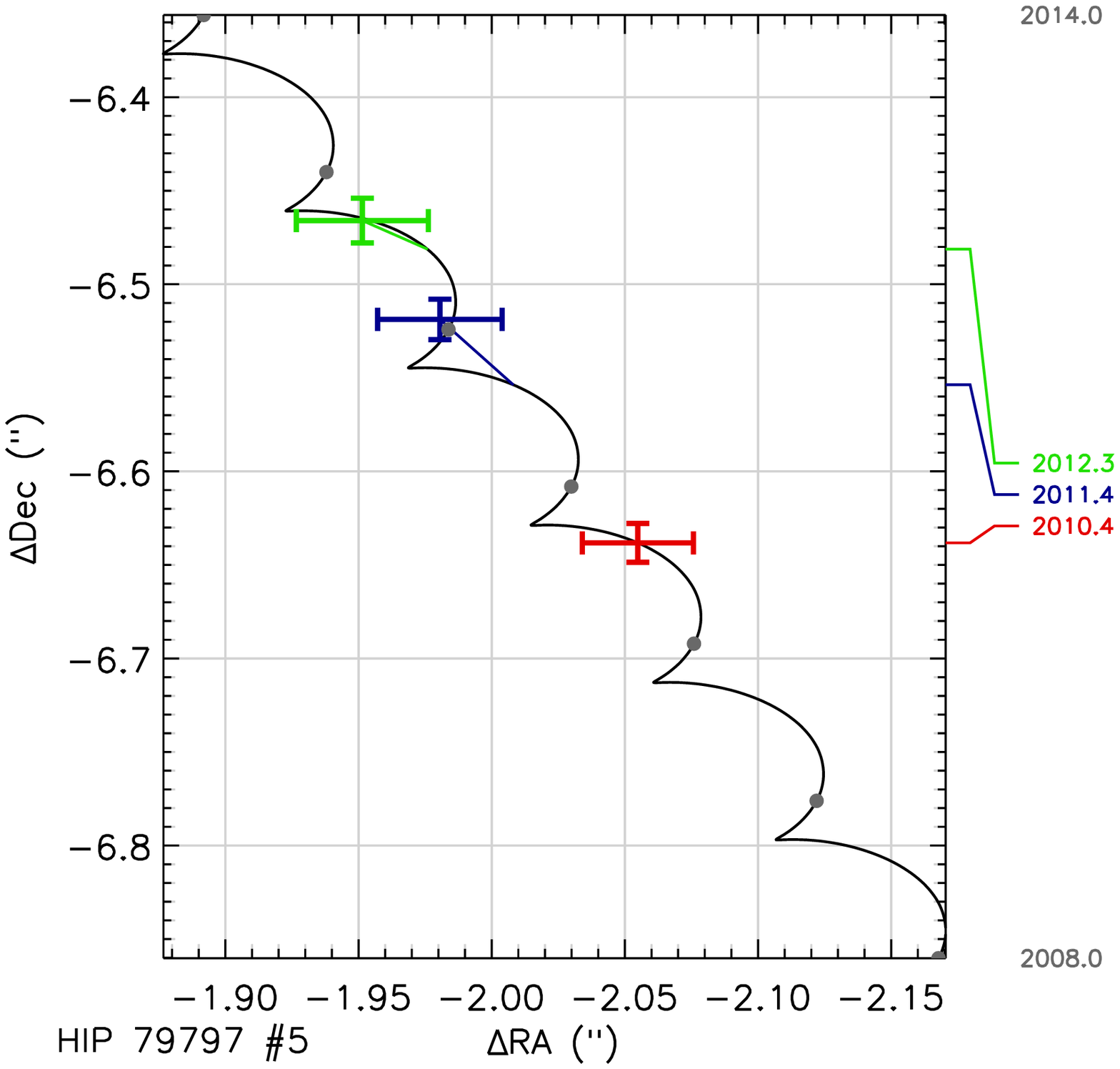}
\hskip -0.3in
\includegraphics[width=2.0in]{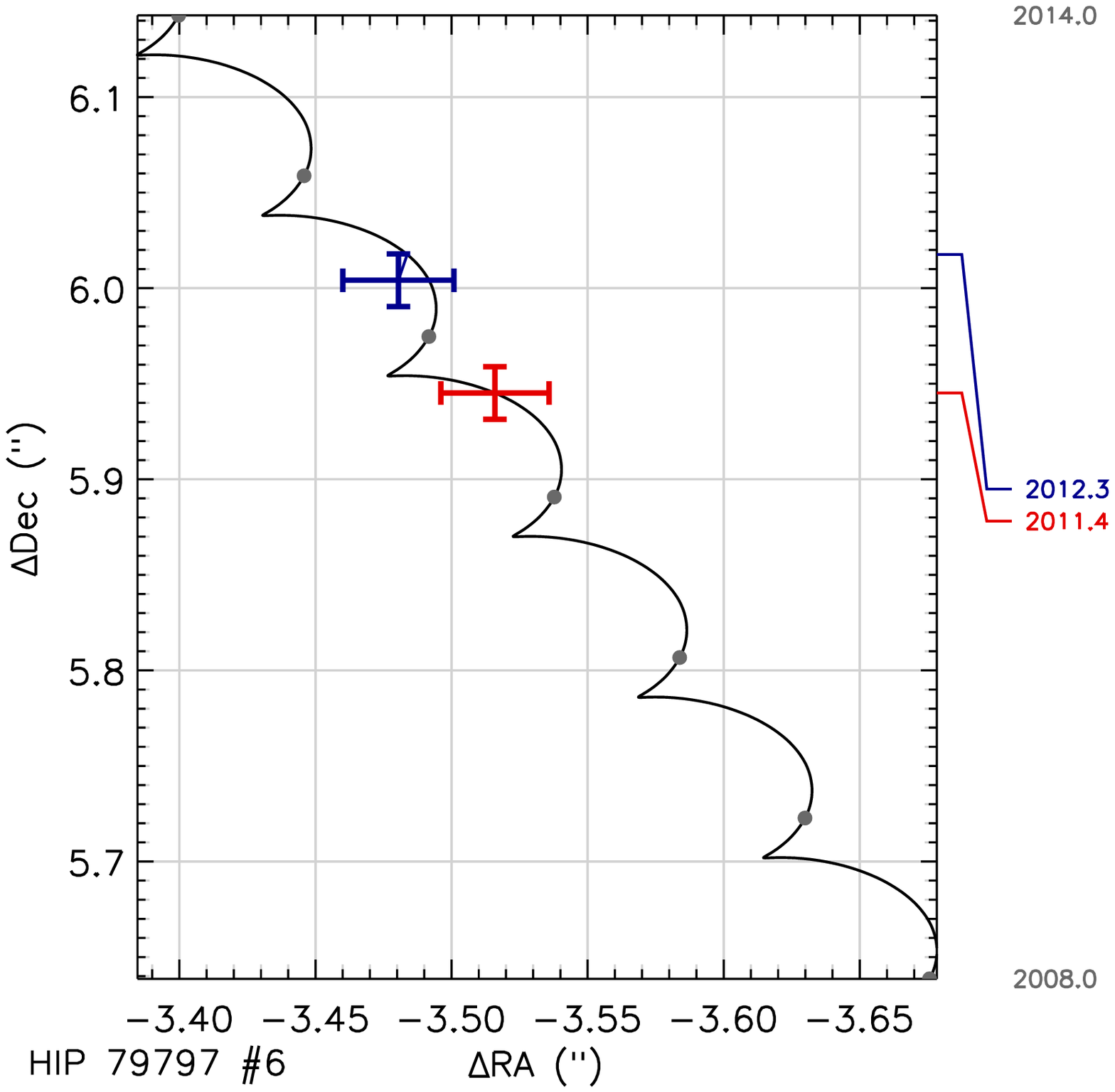}
\hskip -0.3in
\includegraphics[width=2.0in]{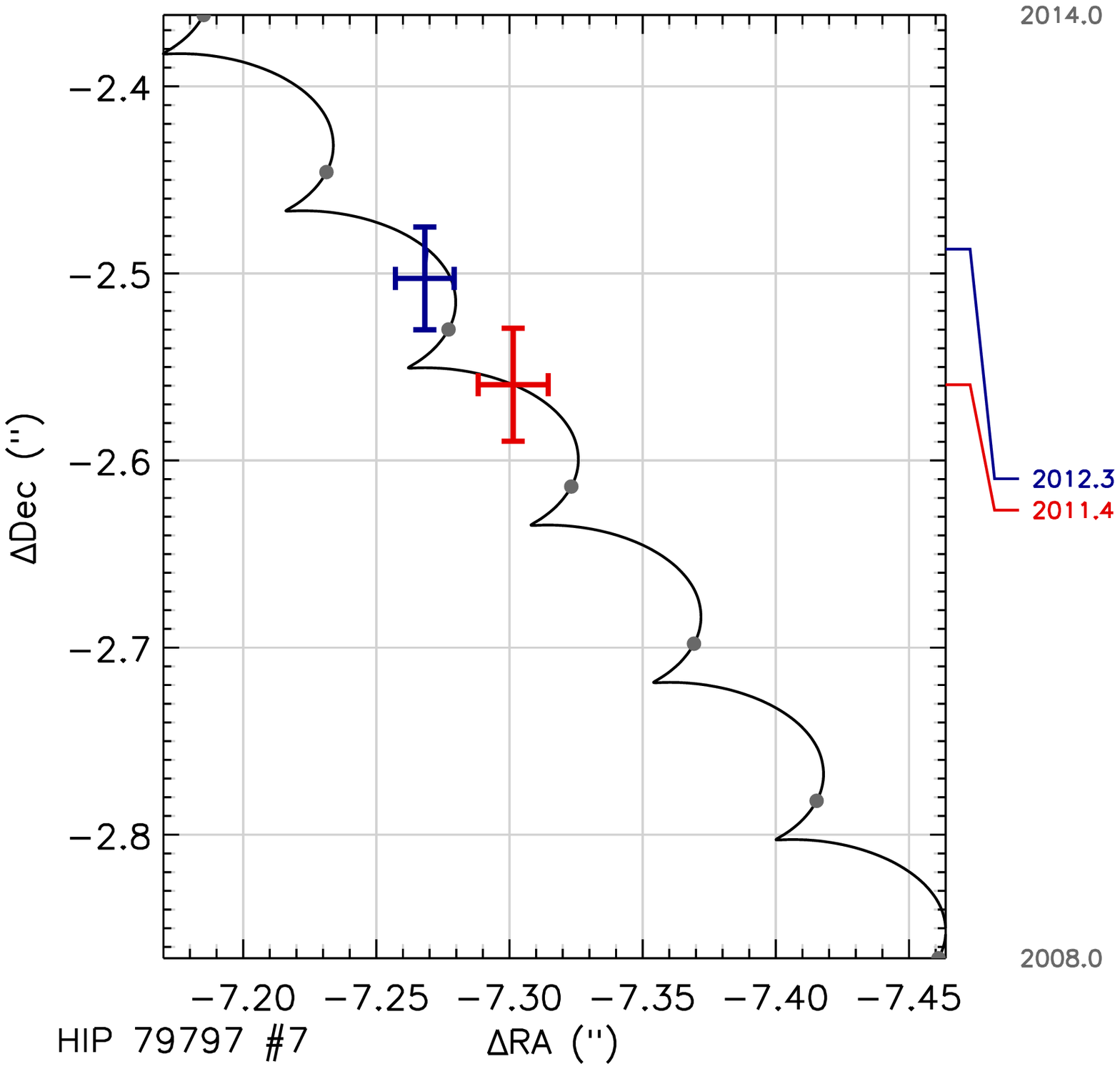}
\hskip -0.3in
\includegraphics[width=2.0in]{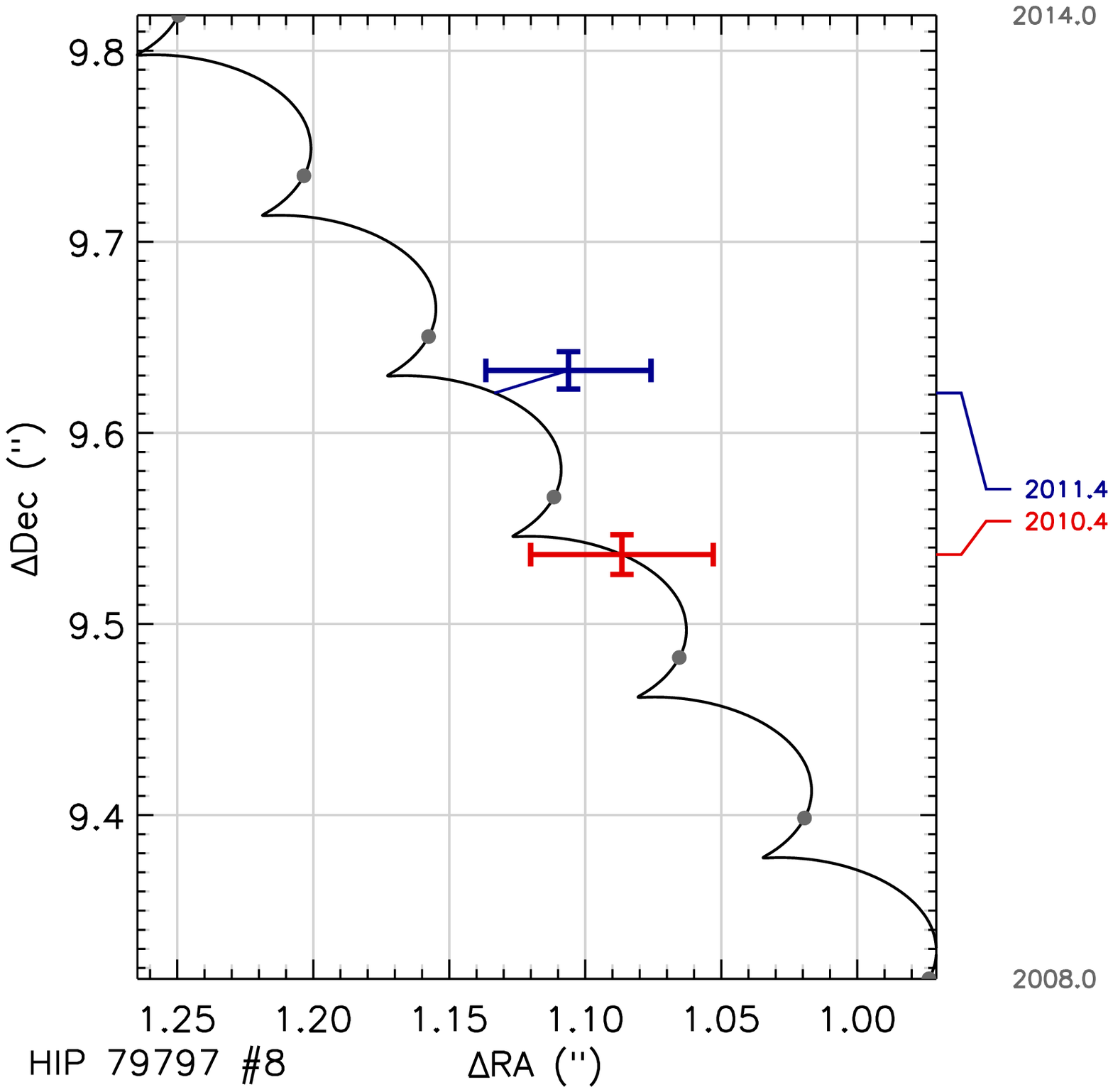}
}
\vskip -0.2in
\centerline{
\includegraphics[width=2.0in]{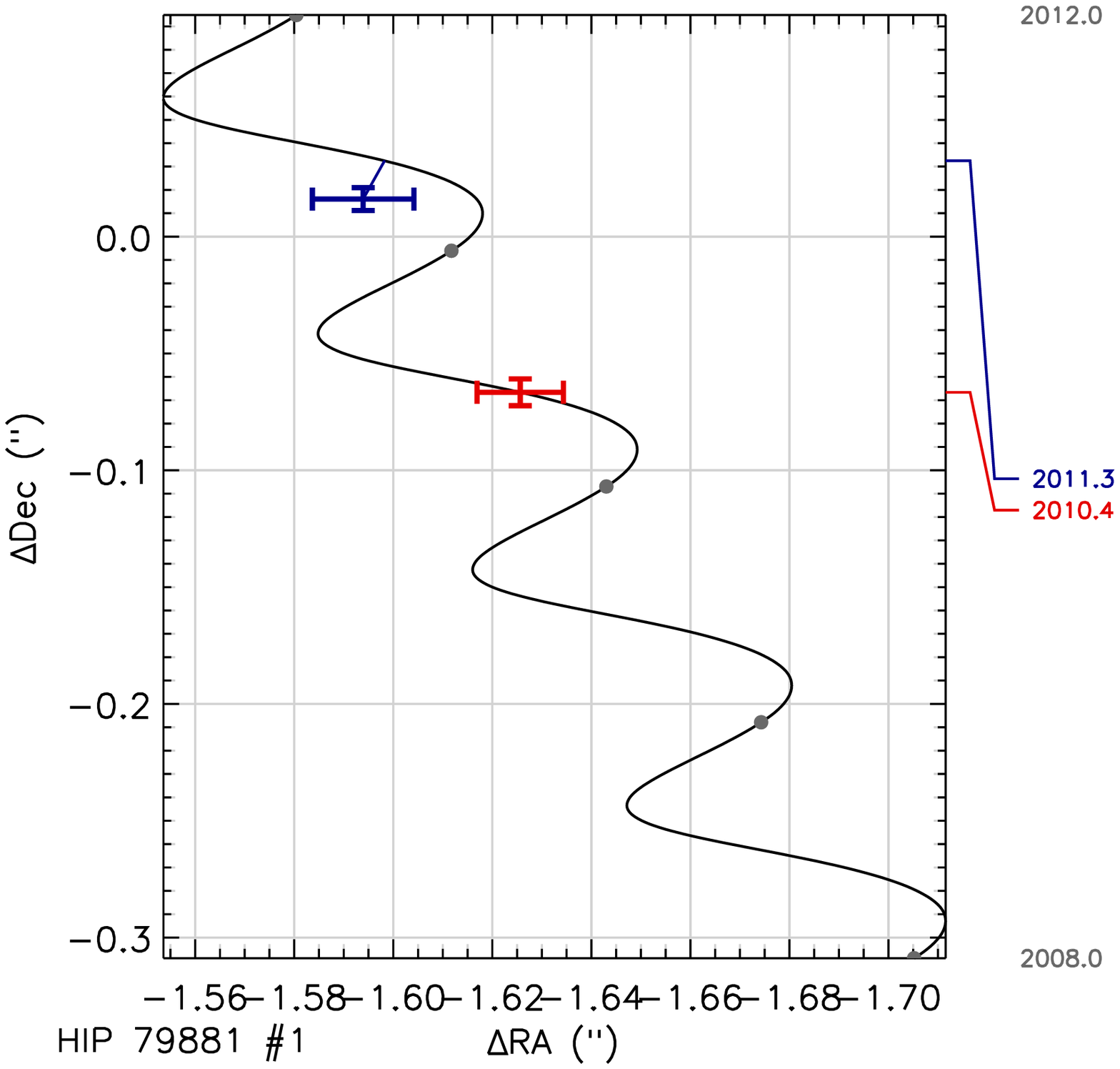}
\hskip -0.3in
\includegraphics[width=2.0in]{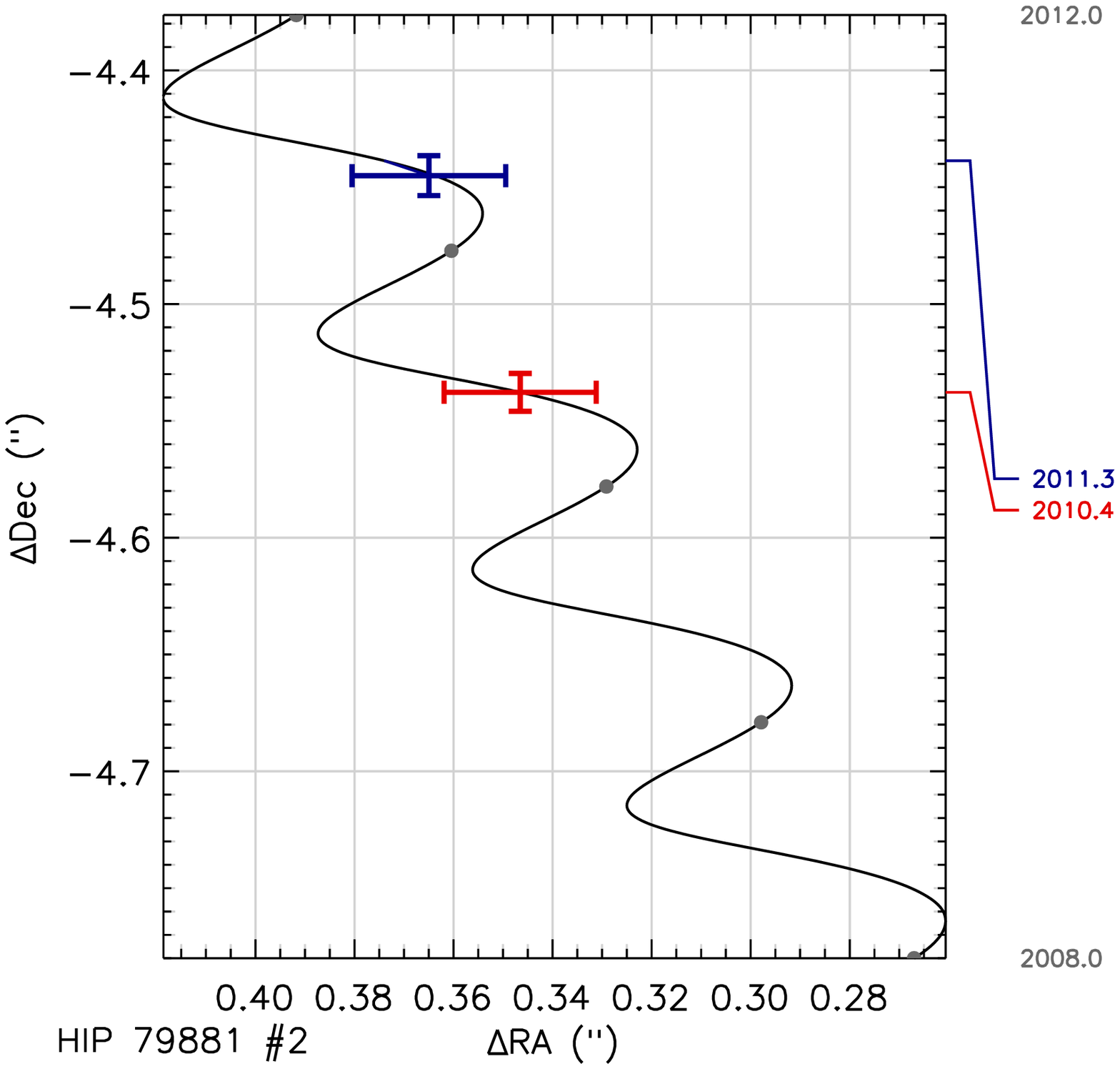}
\hskip -0.3in
\includegraphics[width=2.0in]{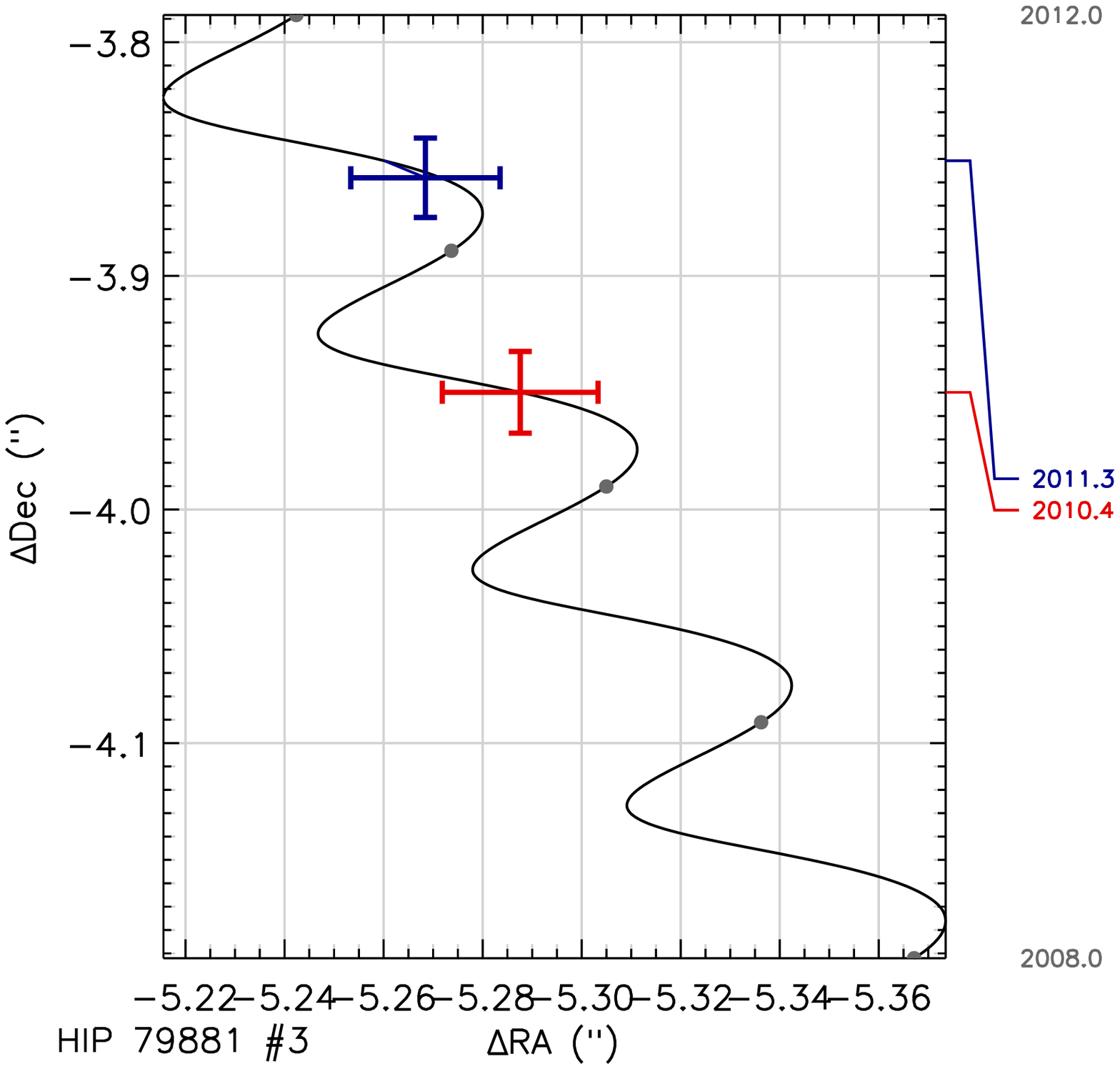}
\hskip -0.3in
\includegraphics[width=2.0in]{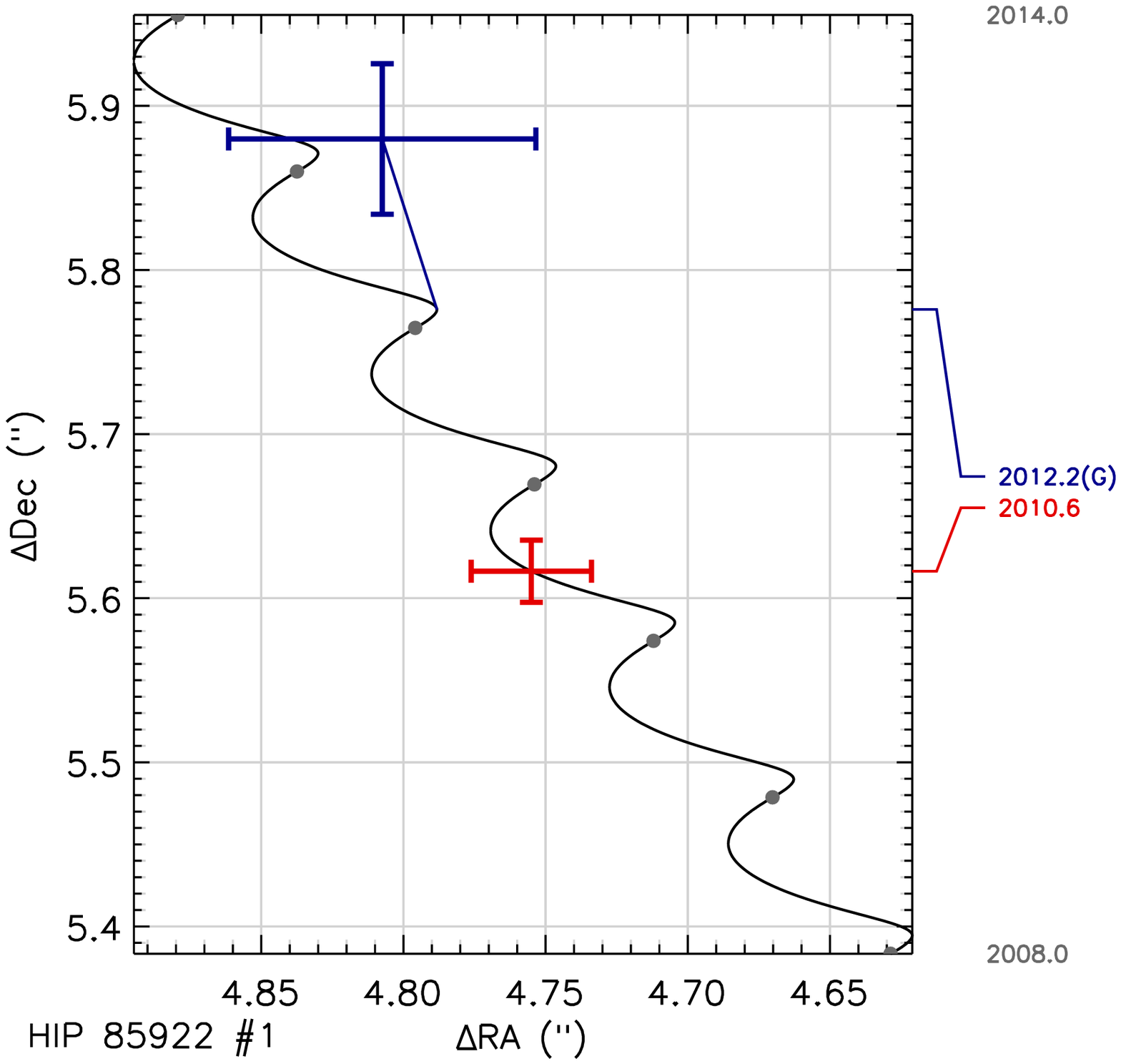}
}
\vskip -0.2in
\centerline{
\includegraphics[width=2.0in]{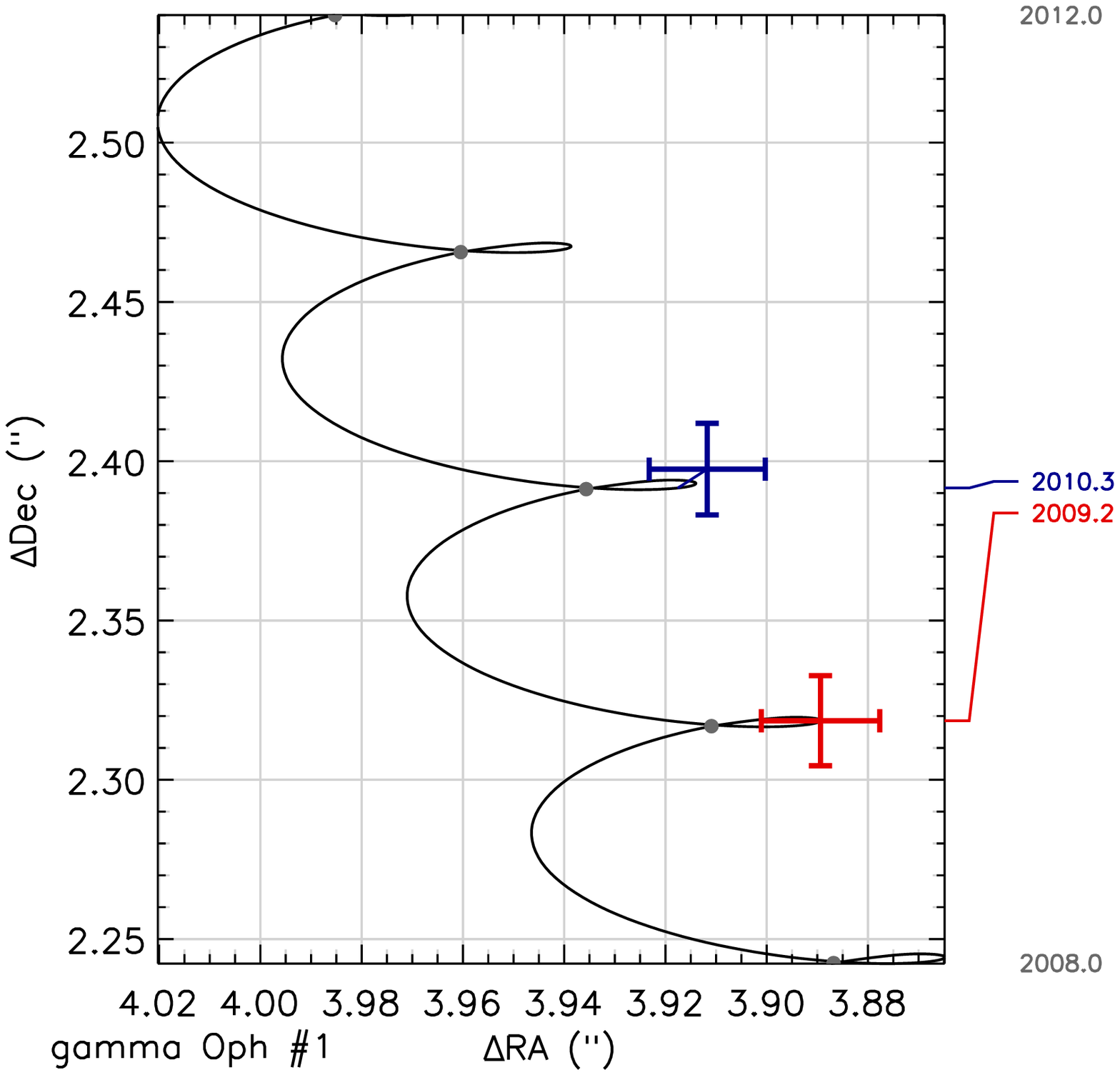}
\hskip -0.3in
\includegraphics[width=2.0in]{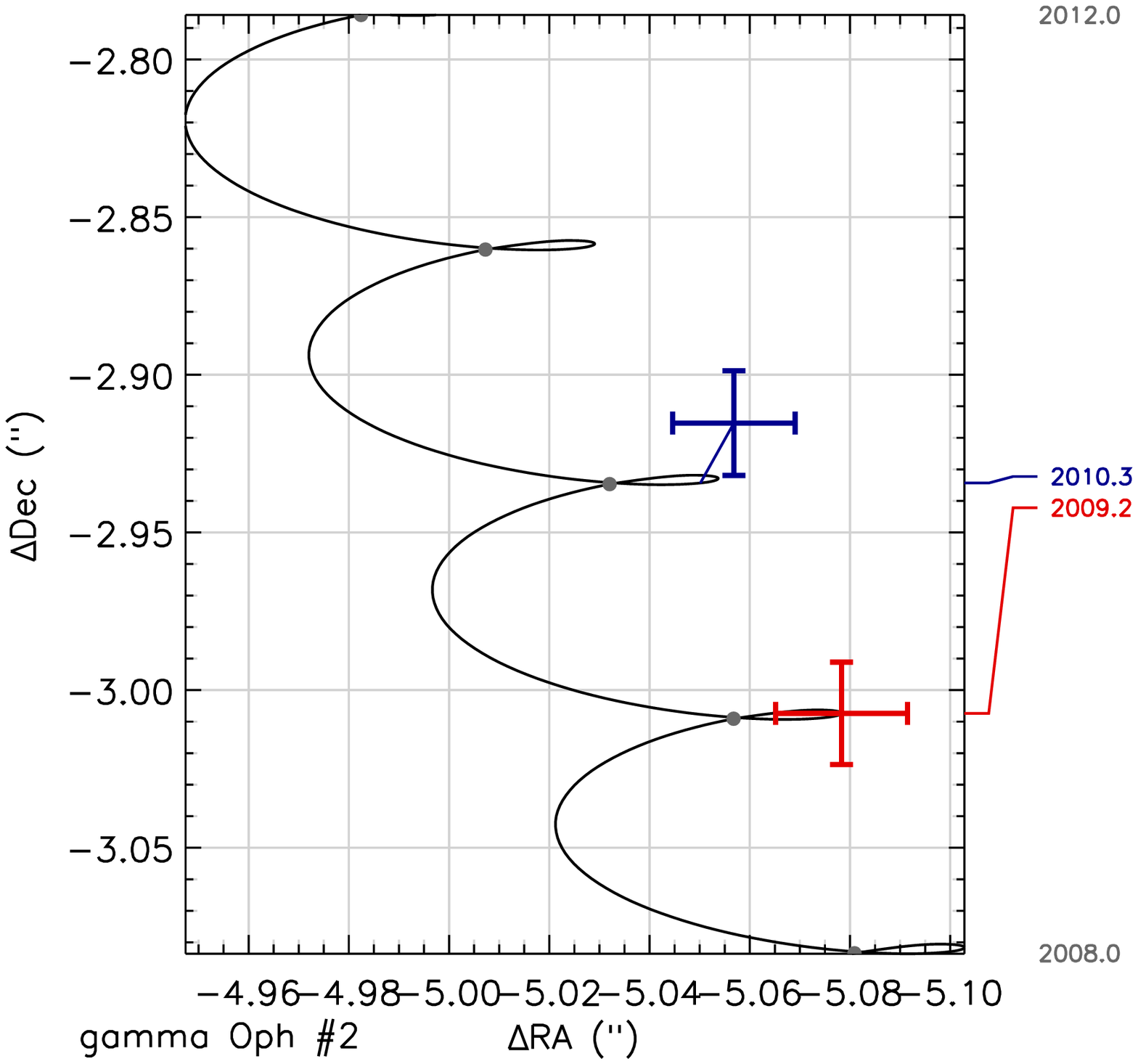}
\hskip -0.3in
\includegraphics[width=2.0in]{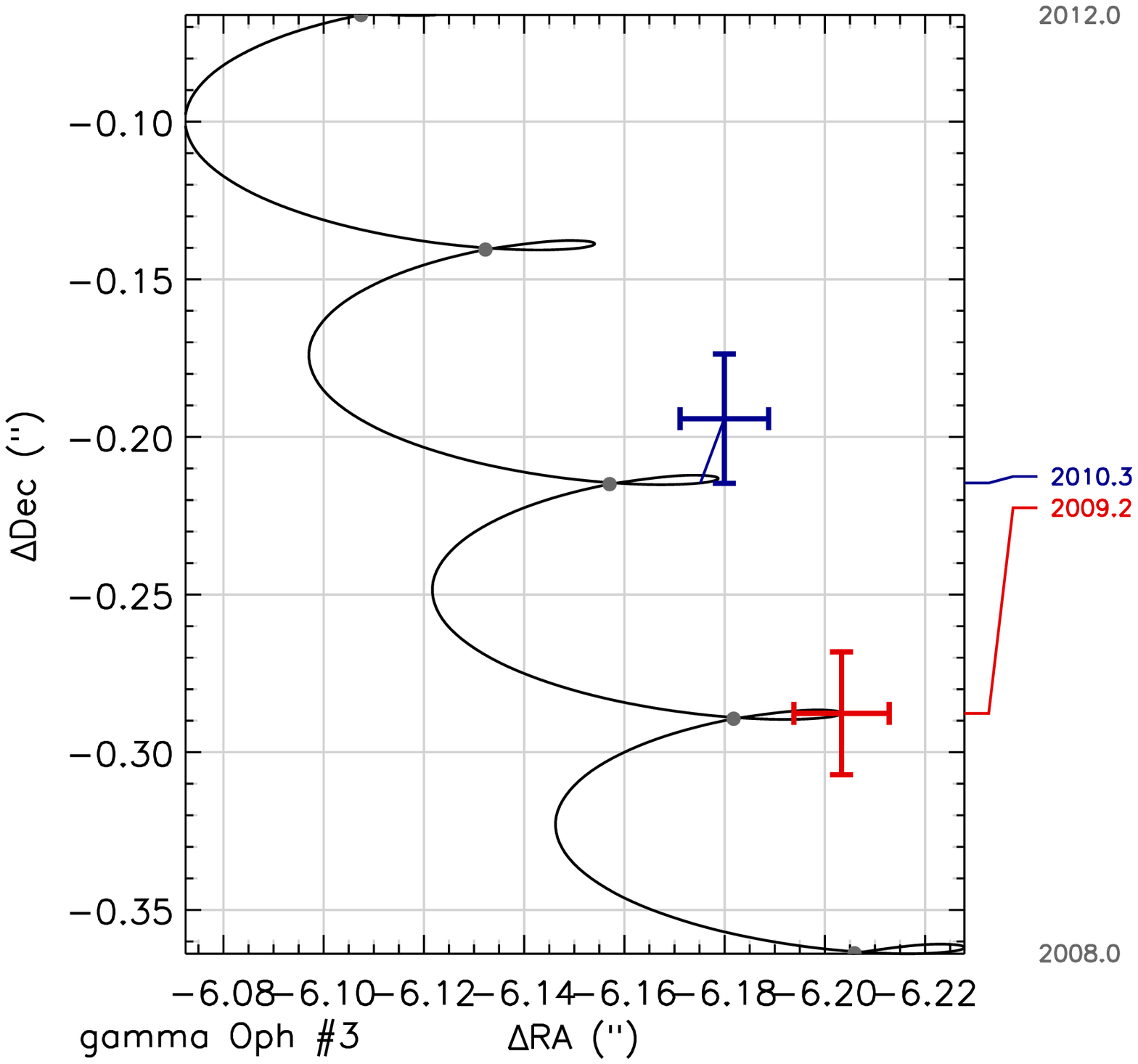}
\hskip -0.3in
\includegraphics[width=2.0in]{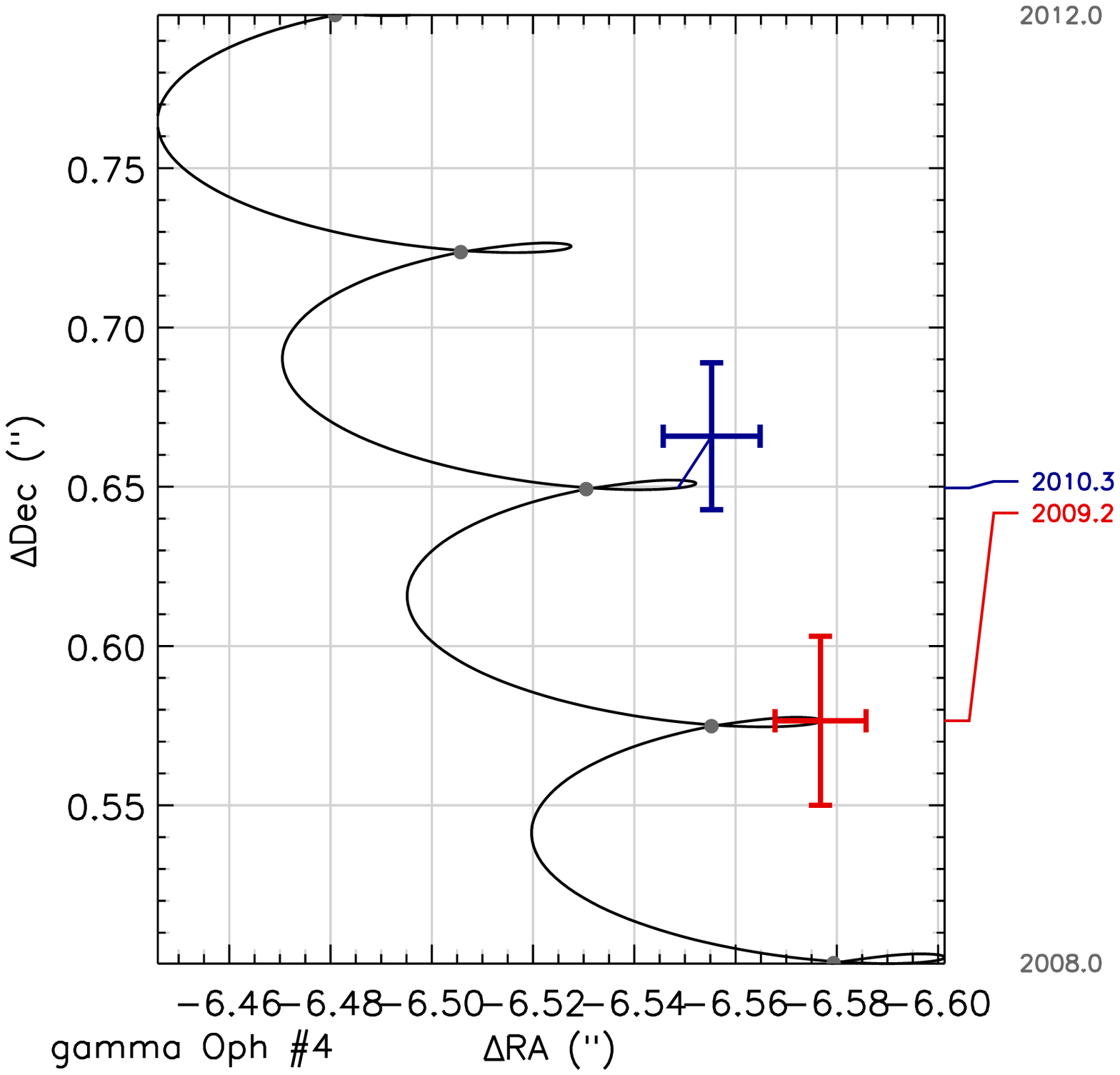}
}
\vskip -0.2in
\caption{Candidate companion on-sky motion, continued from
Figures~\ref{tiled_fig1}--\ref{tiled_fig3}.}\label{tiled_fig4}
\end{figure}

\begin{figure}
\centerline{
\includegraphics[width=2.0in]{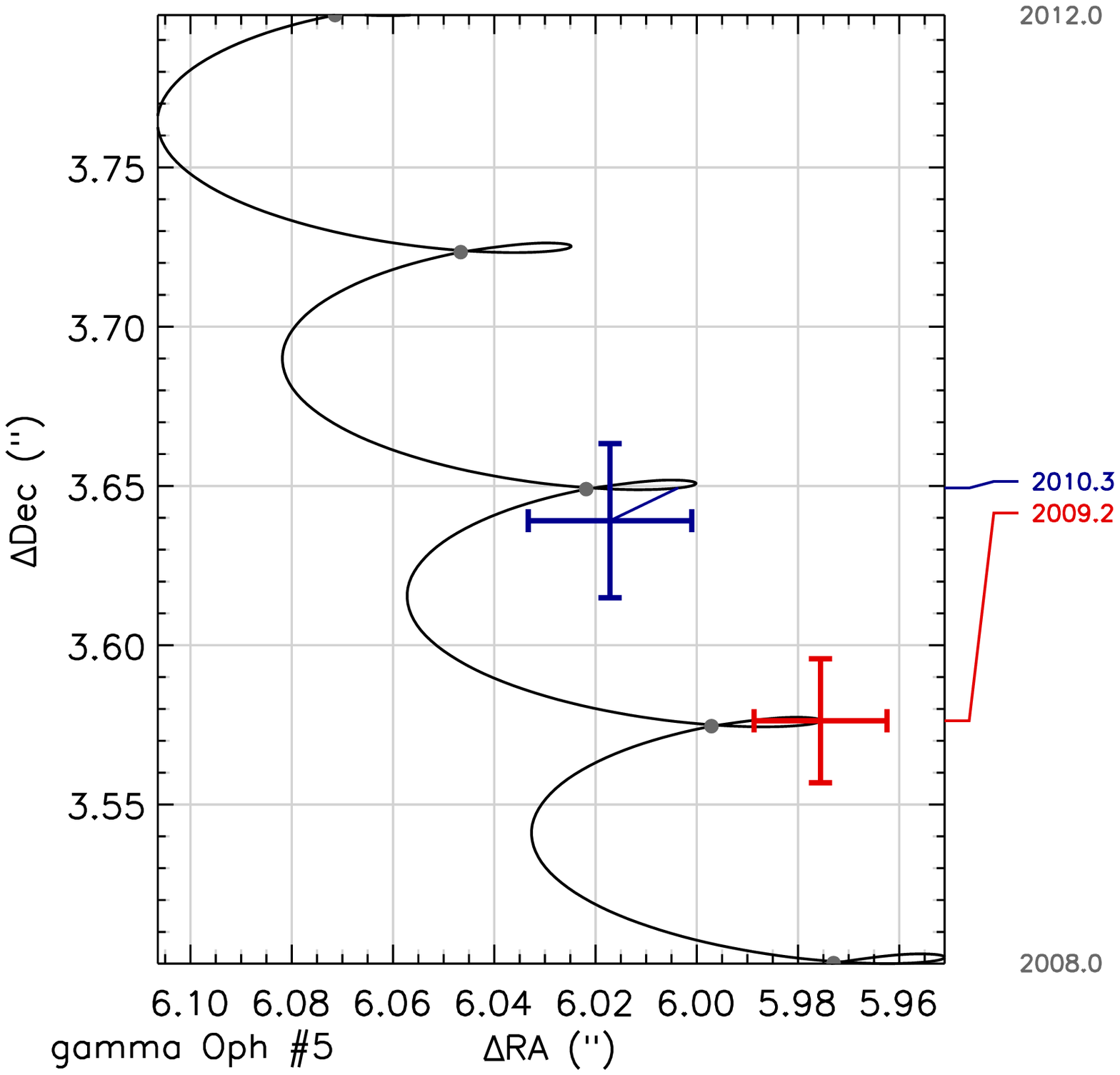}
\hskip -0.3in
\includegraphics[width=2.0in]{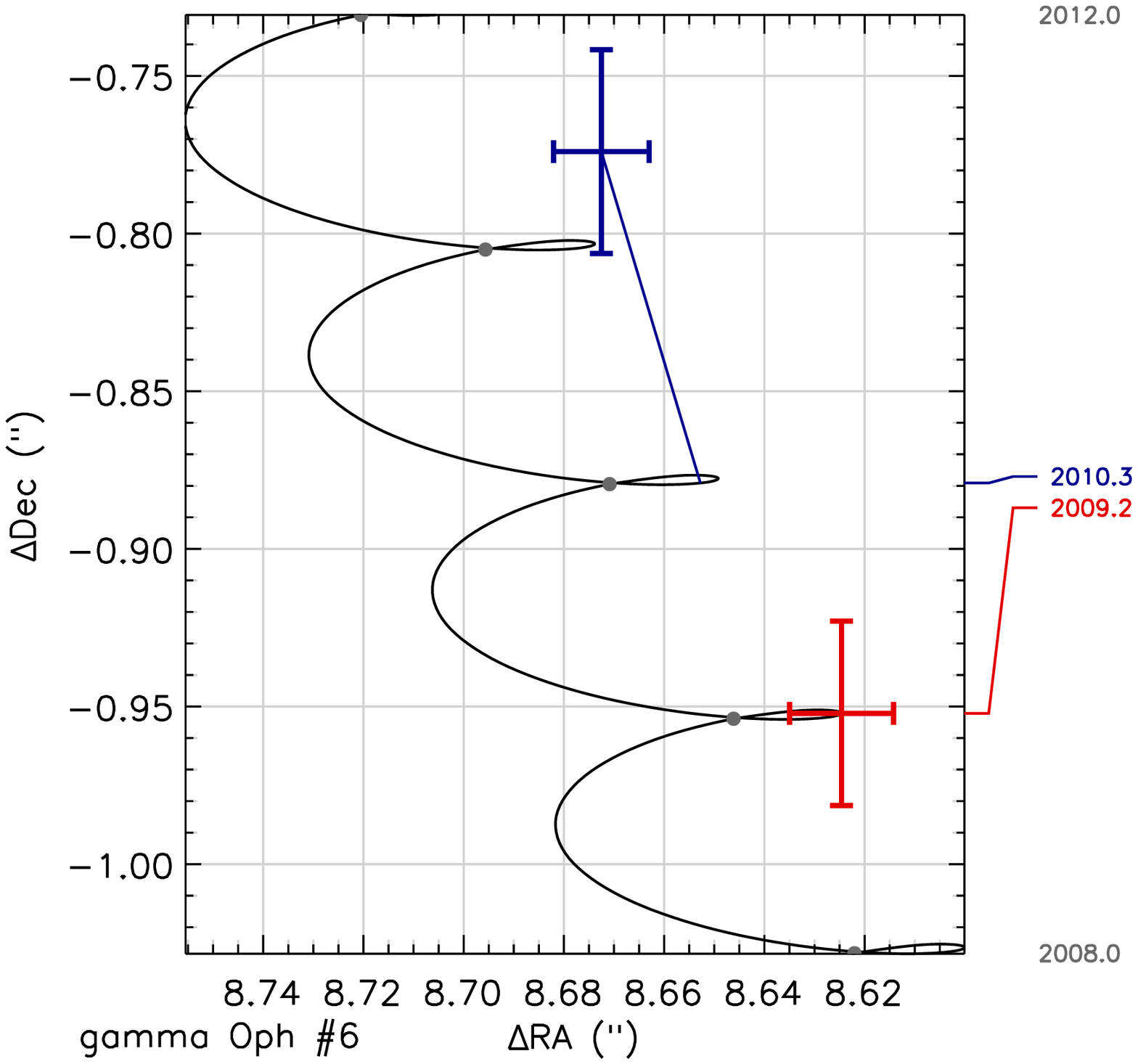}
\hskip -0.3in
\includegraphics[width=2.0in]{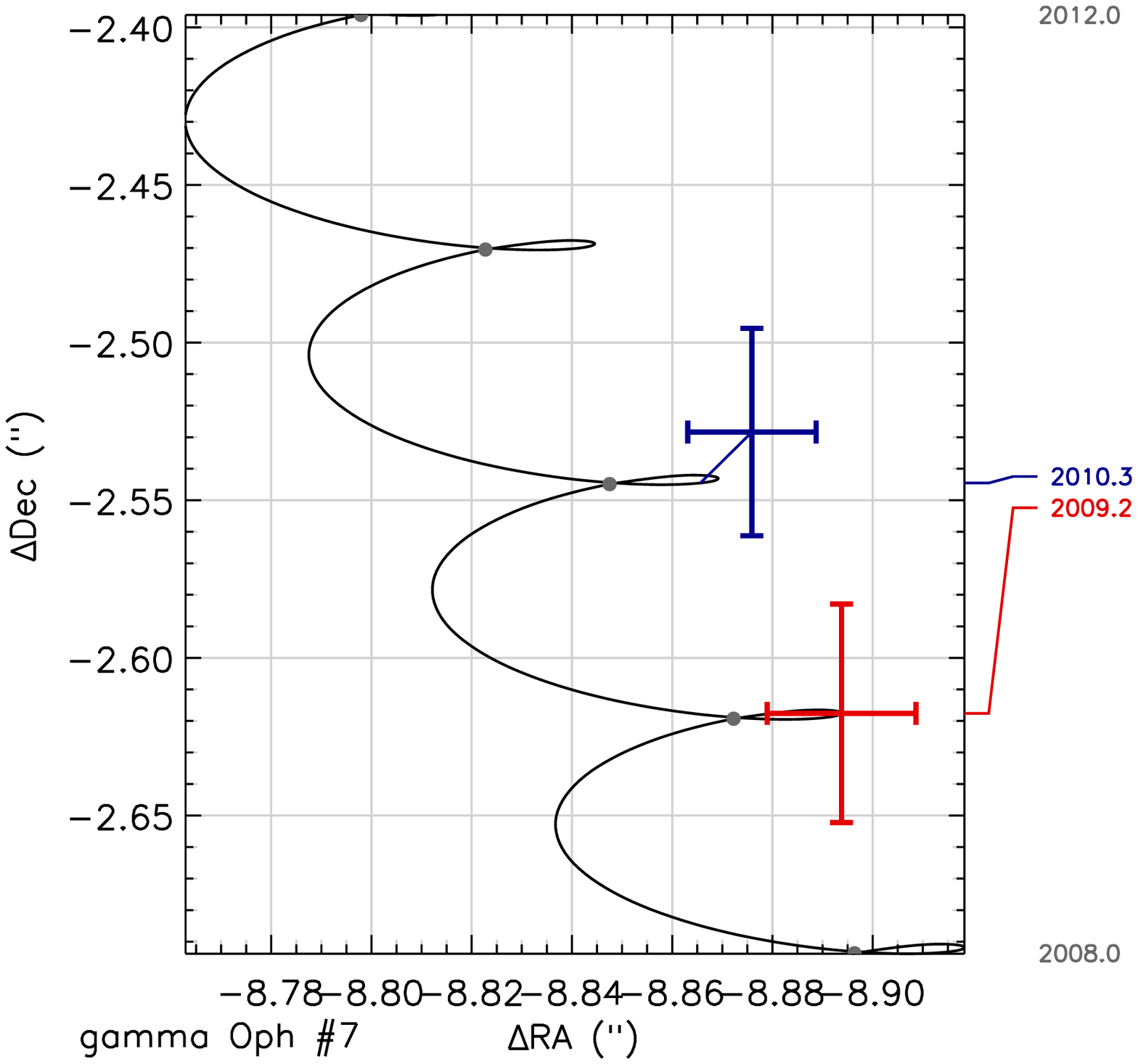}
\hskip -0.3in
\includegraphics[width=2.0in]{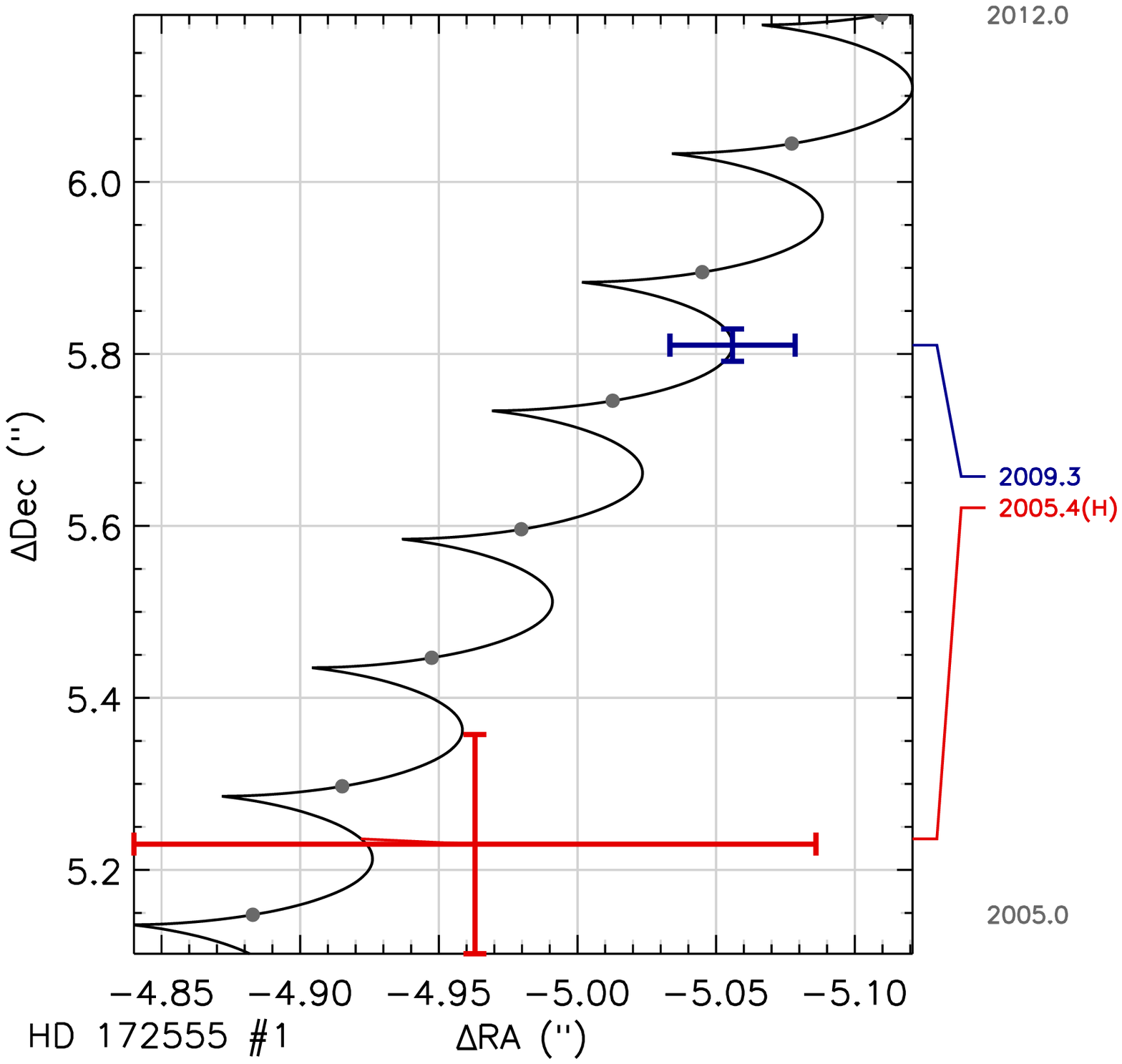}
}
\vskip -0.2in
\centerline{
\includegraphics[width=2.0in]{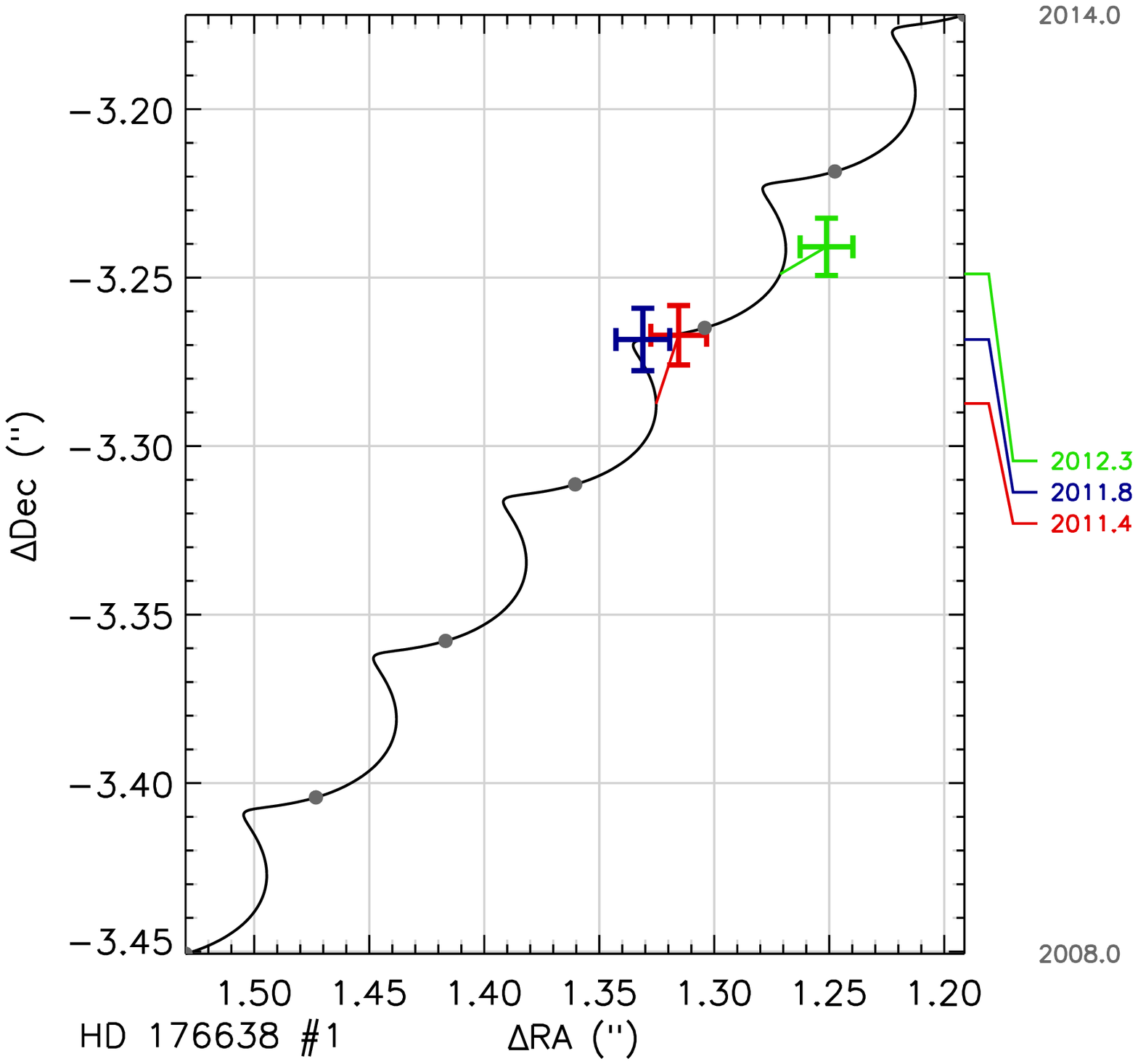}
\hskip -0.3in
\includegraphics[width=2.0in]{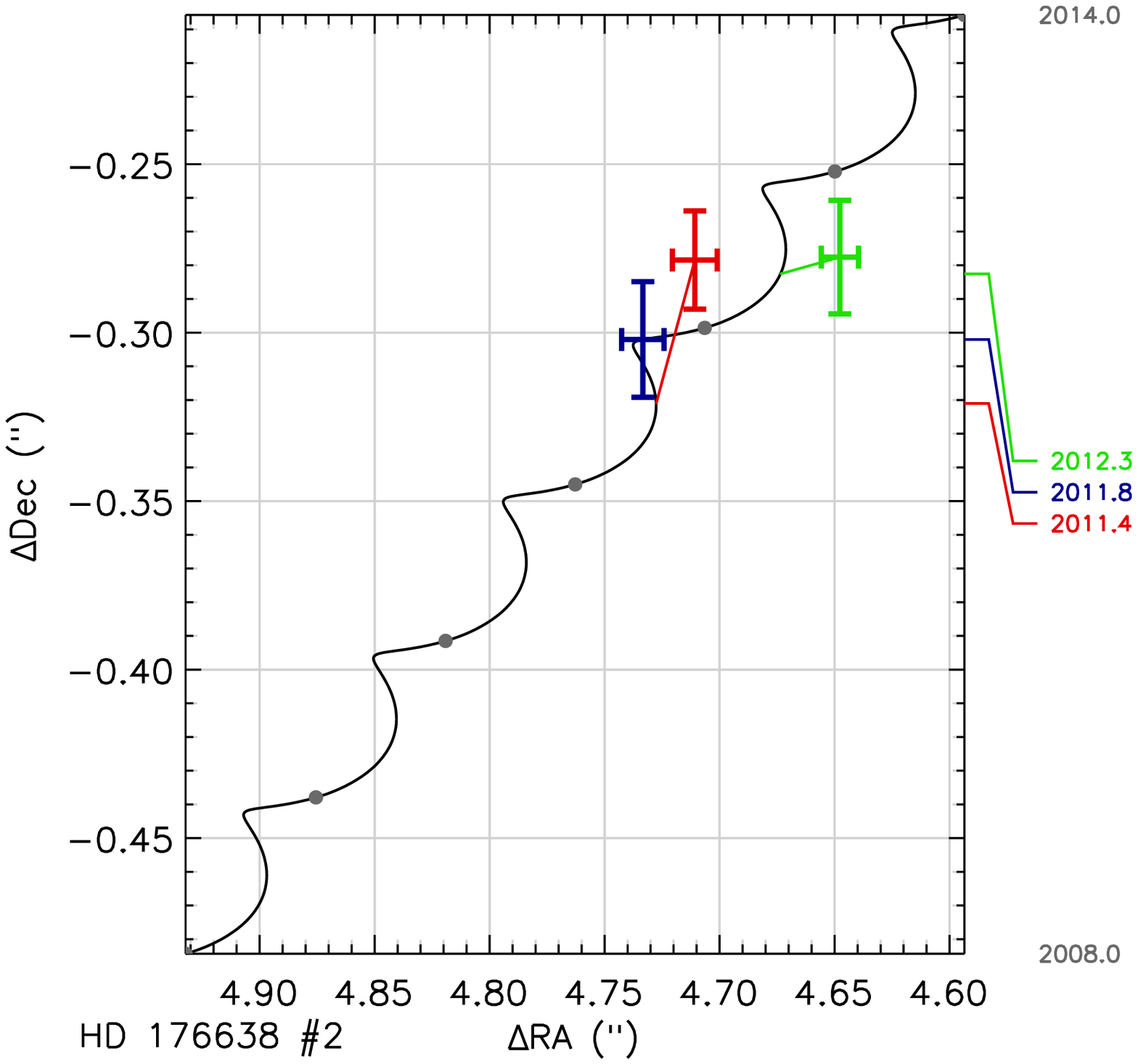}
\hskip -0.3in
\includegraphics[width=2.0in]{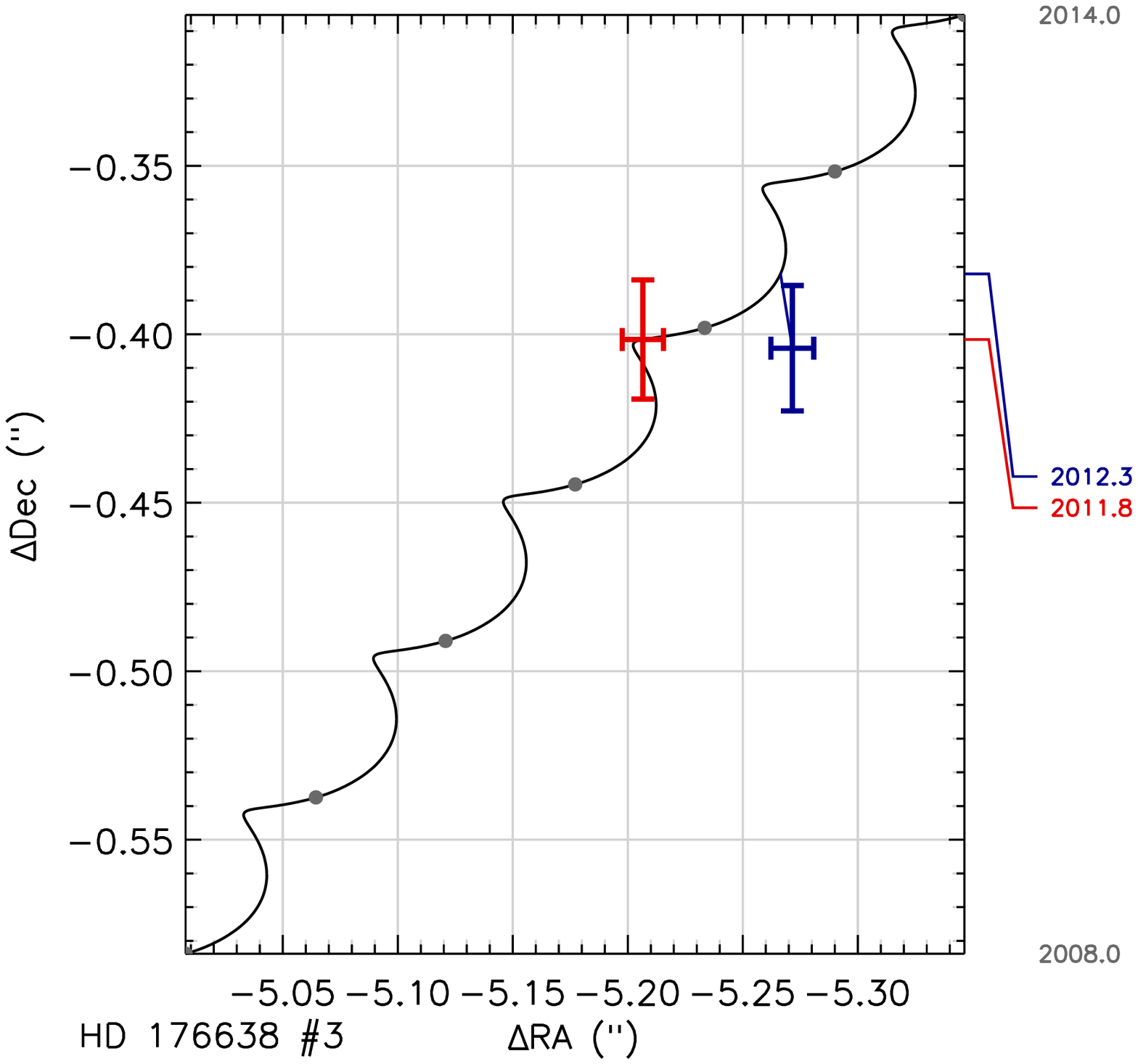}
\hskip -0.3in
\includegraphics[width=2.0in]{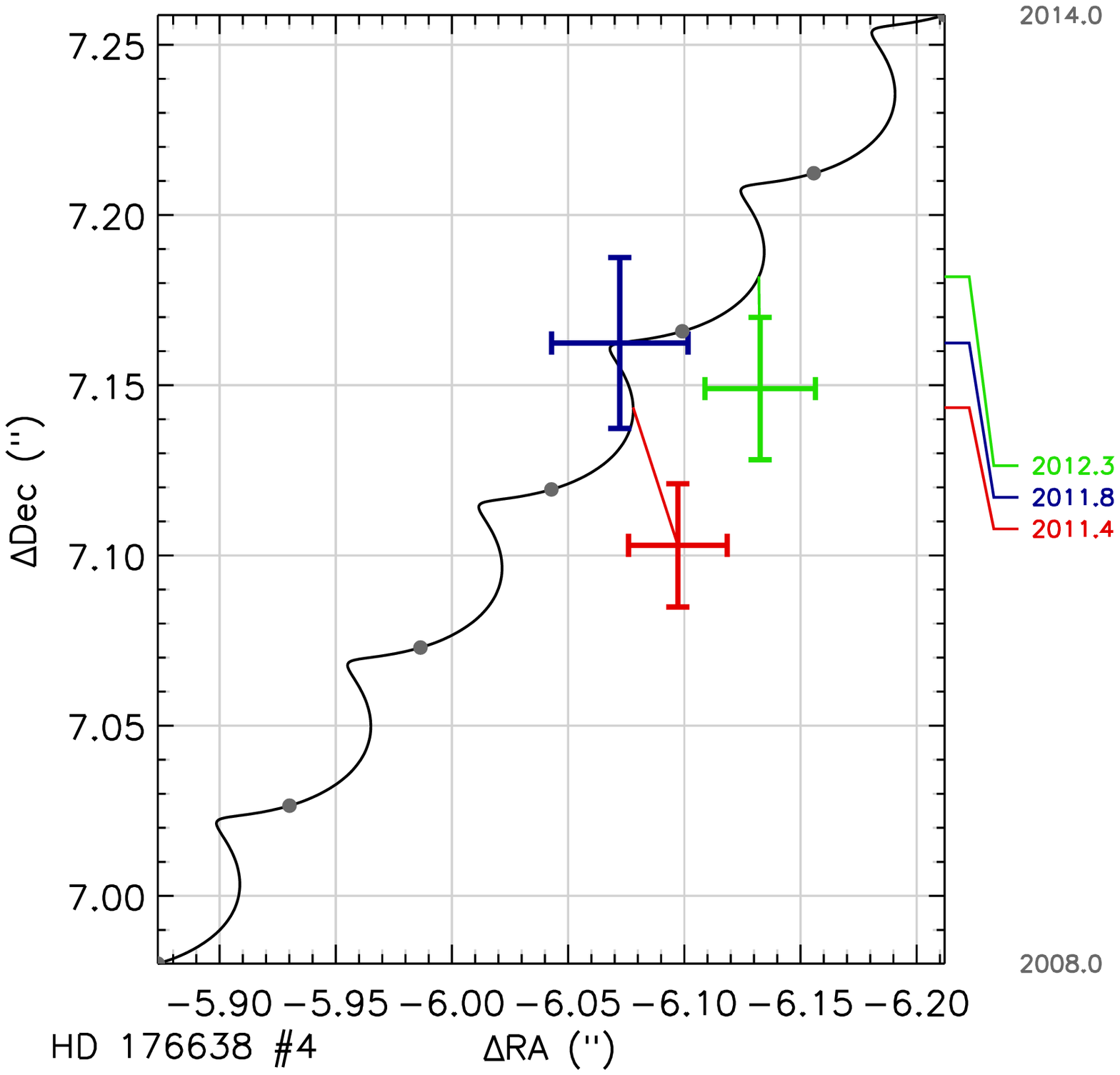}
}
\vskip -0.2in
\centerline{
\includegraphics[width=2.0in]{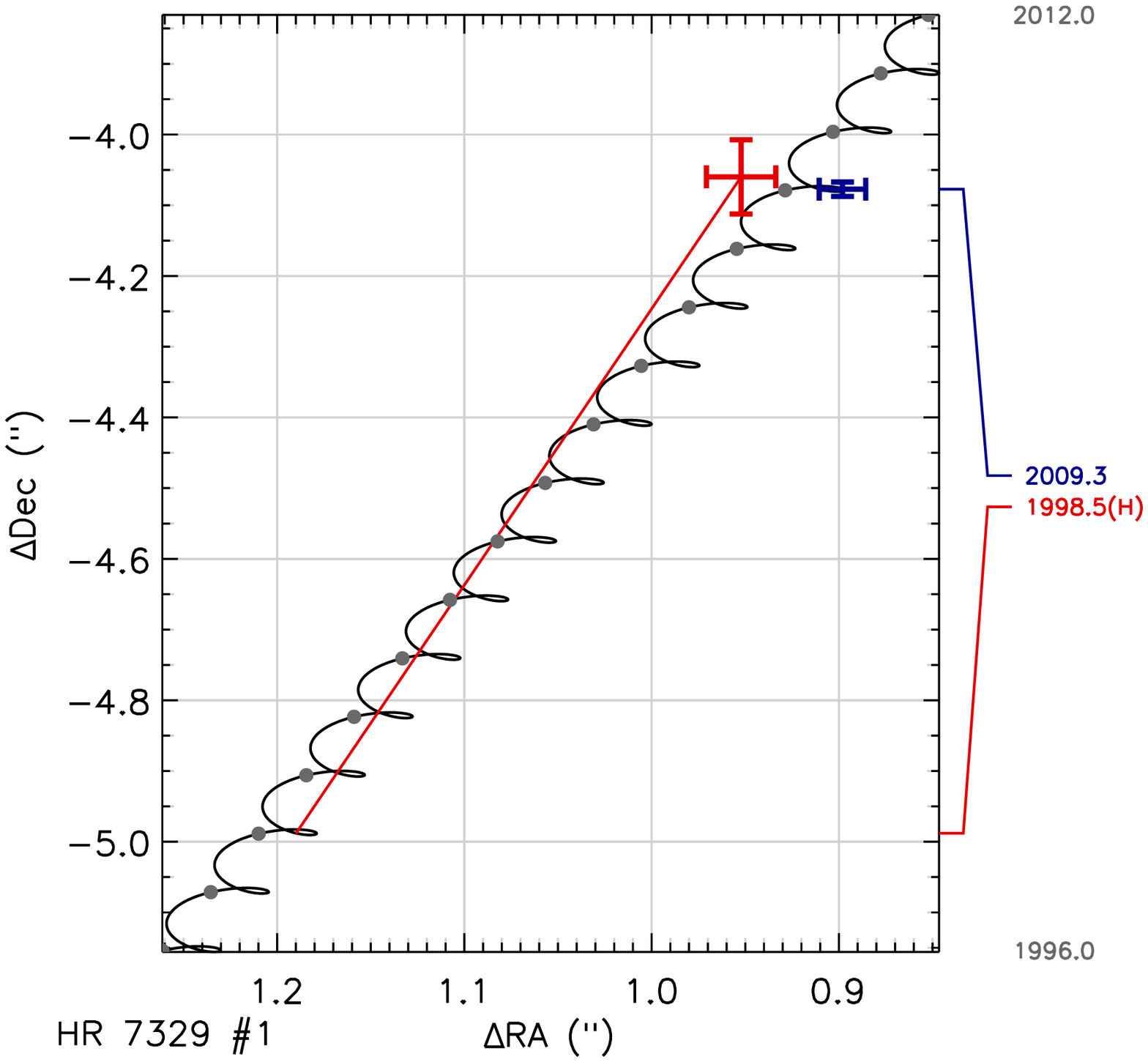}
\hskip -0.3in
\includegraphics[width=2.0in]{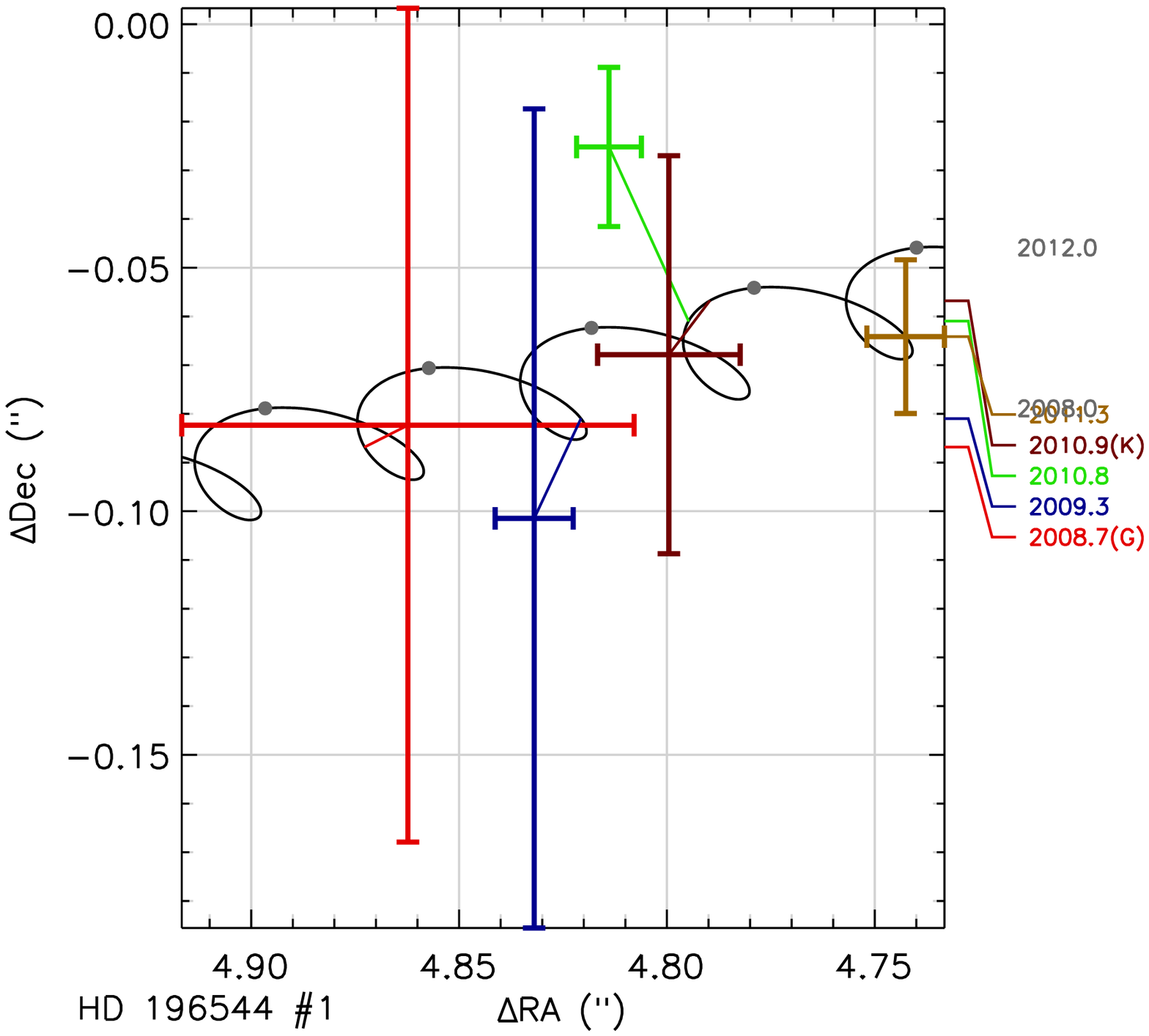}
\hskip -0.3in
\includegraphics[width=2.0in]{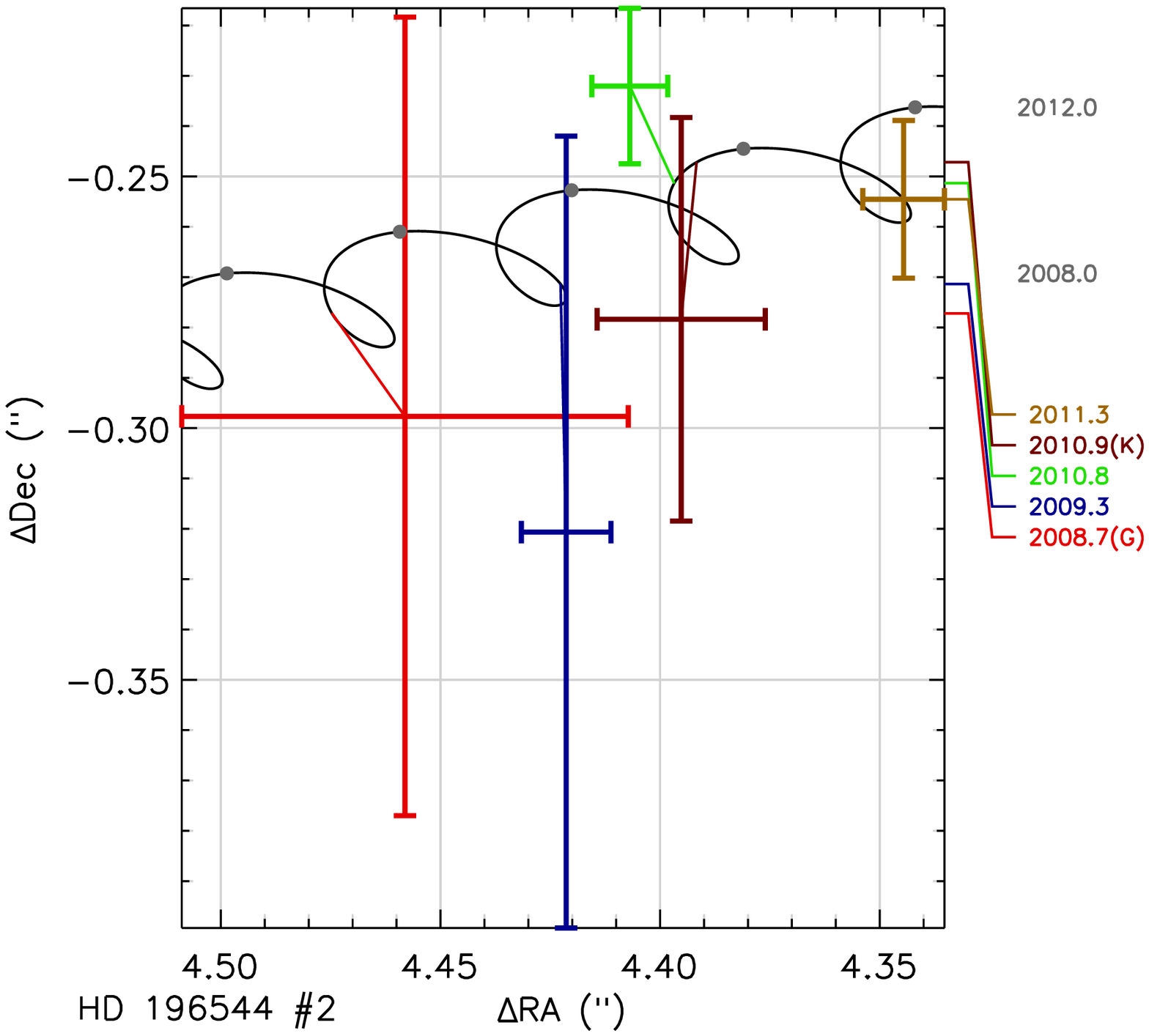}
\hskip -0.3in
\includegraphics[width=2.0in]{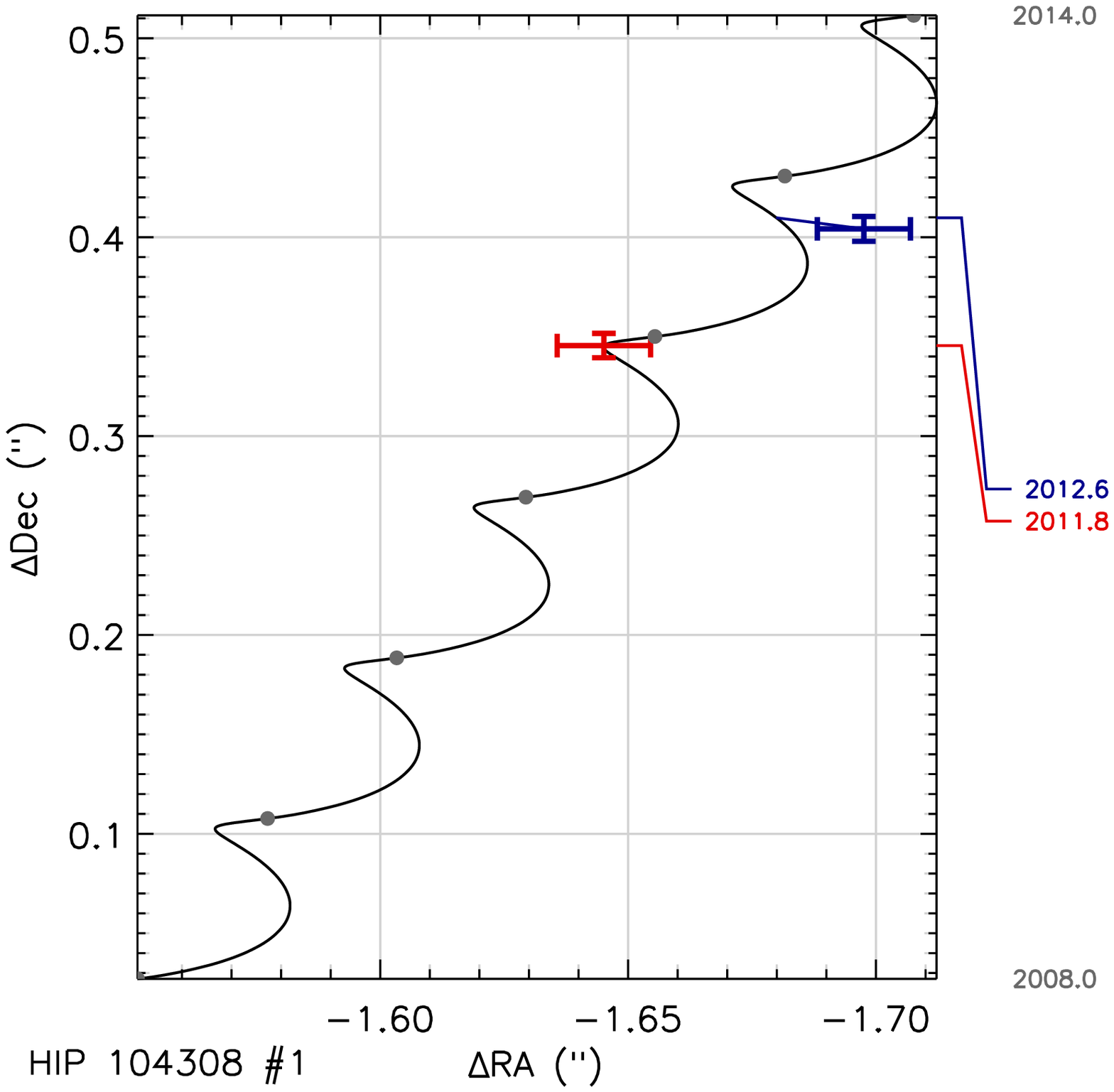}
}
\vskip -0.2in
\centerline{
\includegraphics[width=2.0in]{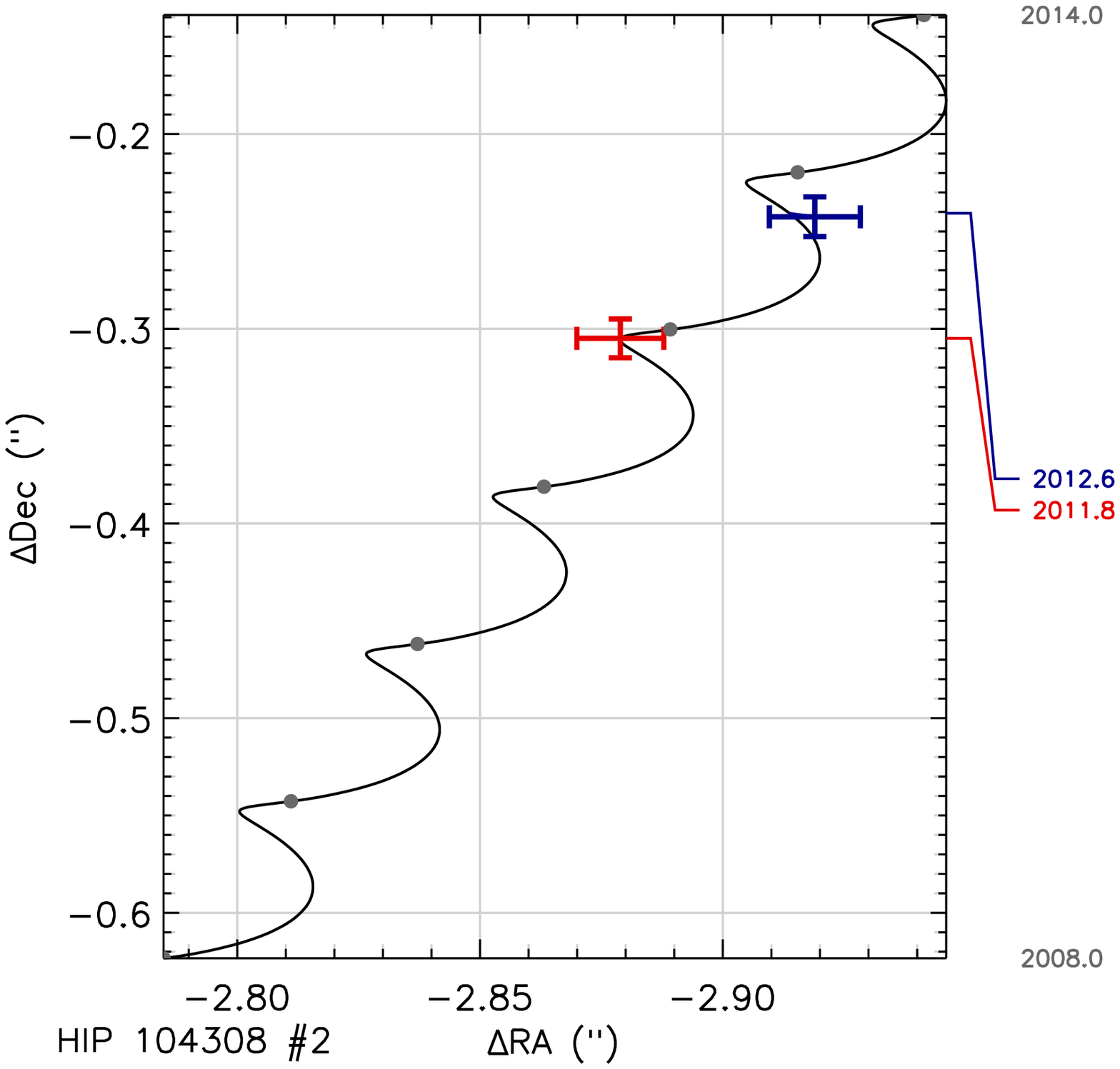}
\hskip -0.3in
\includegraphics[width=2.0in]{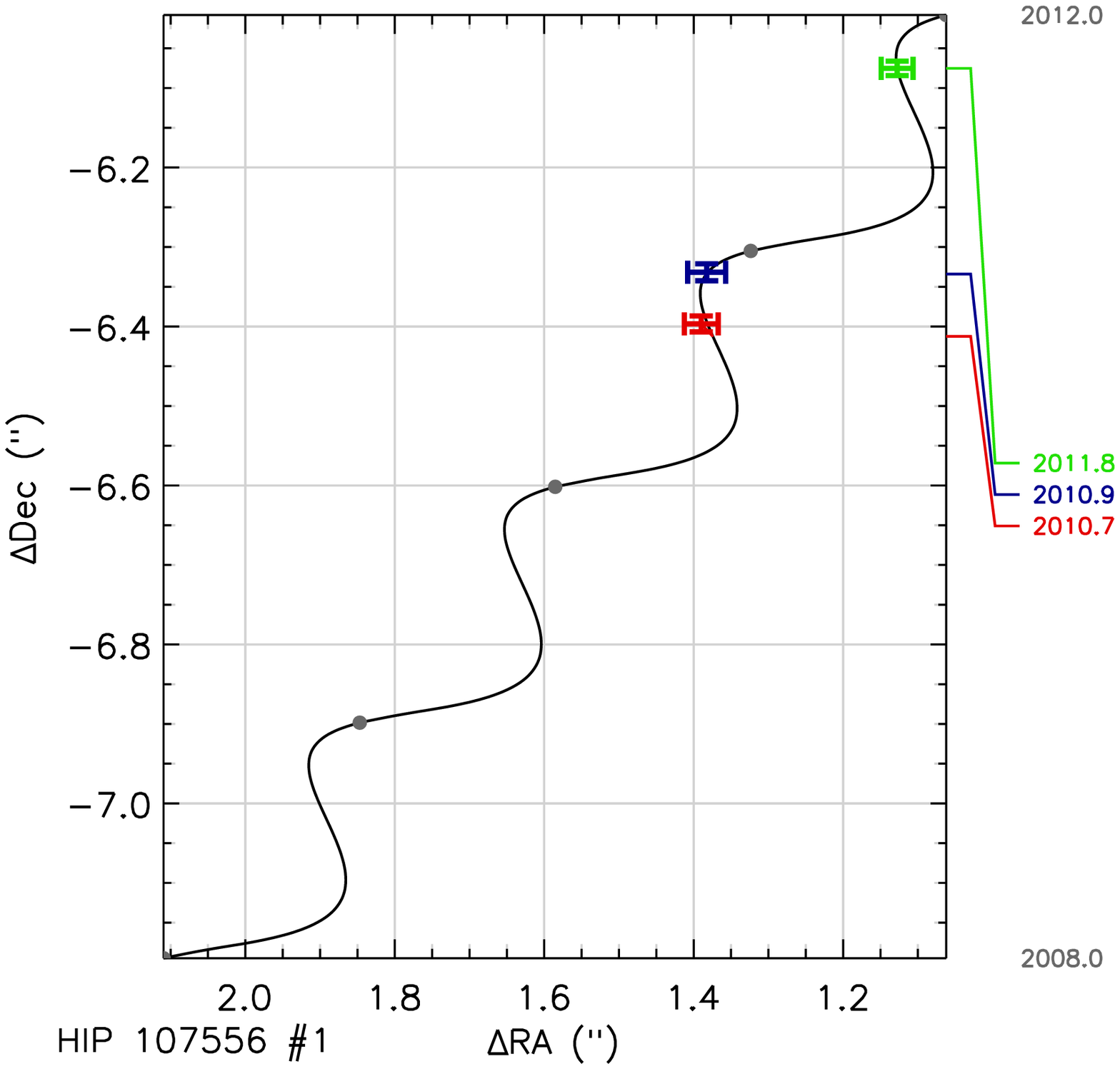}
\hskip -0.3in
}
\caption{Candidate companion on-sky motion, continued from
Figures~\ref{tiled_fig1}--\ref{tiled_fig4}.}\label{tiled_fig5}
\end{figure}

Table~\ref{table_comp2a} lists the properties of each candidate companion 
for which we have more than one epoch.  For each of these companions we 
then give the astrometry at each epoch in Table~\ref{table_comp2b}.  
Errors in the background track are computed using a Monte Carlo method, 
combining astrometric errors at the reference epoch for the candidate 
and errors in the proper motion and parallax of the target star.  
All proper motion and parallax measurements for our targets are from 
\citet{newhip}.  We typically choose the reference epoch to be 
the first NICI epoch for that candidate.  We also 
show the motion of each companion on the sky with respect to the 
background track in Figures~\ref{tiled_fig1}--\ref{tiled_fig5}.

\subsection{Detected Companions: HD 1160 BC, HR 7329 B, and HIP 79797 Bab}

Of the candidate companions imaged around our target stars, there are five  
that are CPM companions, found around HD 1160, HR 7329, and HIP 79797.  
HD 1160 B is a brown dwarf with mass 33$^{+12}_{-9}$M$_{Jup}$, and 
HD 1160 C is an M3.5 star with mass 0.22$^{+0.03}_{-0.04}$M$_{\sun}$ 
\citep{hd1160}.  We also detect the known substellar companion to 
HR 7329 B \citep{lowrance00}, which is a brown dwarf with mass between 20 
and 50 M$_{Jup}$ \citep{hr7329}.  The companion to HIP 79797 (also known 
as HR 6037) was independently discovered as a single object 
by \citet{hip79797} who determined a mass of 62$\pm$20 M$_{Jup}$ for the 
6.7$''$ companion.

\begin{figure}
\epsscale{0.9}
\plotone{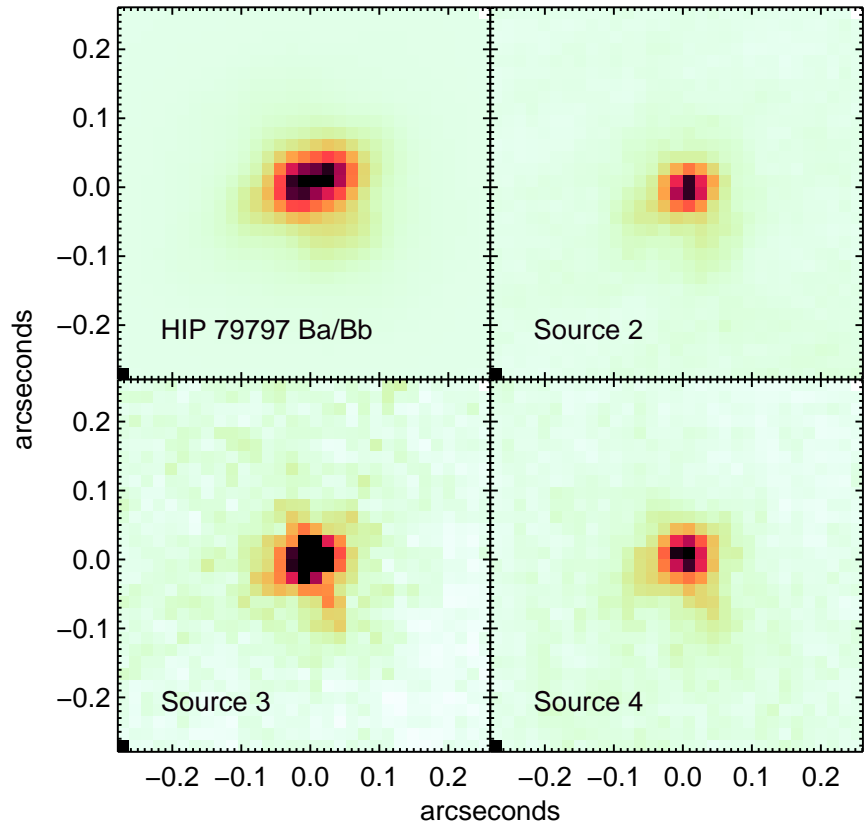}
\caption{
\footnotesize{
(upper left panel) The binary HIP 79797 
Ba/Bb in the $H$-band as imaged by NICI on UT 2012-04-06.  Ba is down 
and to the left, Bb is up and to the right.  The binary is clearly 
resolved, and the two components have near equal flux.  The other 
three panels show three background objects in the same image.  
All the background objects appear circular, while only HIP 79797 Ba/Bb is 
extended.  The separation of the 
components is 60 $\pm$ 6 mas.  The images are oriented with north 
up and east to the left.
}
\label{hip79797}}
\end{figure}

NICI observations of HIP 79797 not only further confirm that HIP 79797 B is a 
CPM companion but reveal that the companion is in fact a binary, with a 
separation of 0.06$''$ and a near-unity flux ratio at 
$JHK_S$.  
We have retrieved the archival 2004 and 2006 VLT data used by 
\citet{hip79797} but are unable to split the binary in these images, given 
the low S/N of the data.  In five NICI epochs 
(2010-2012), however, we are able to resolve the binary 
(Figure~\ref{hip79797}) and determine relative astrometry for the two 
components (Table~\ref{hip79797_table}).  We summarize 
the properties of the HIP~79797 system in Table~\ref{hip79797_properties}.

We estimate the masses of the now-resolved brown dwarf 
binary companions using our 
measurement of the difference in the $K_S$ magnitudes of the combined two 
components with respect to the primary (7.31~$\pm$~0.17~mags), 
the $K_S$-band flux ratio of the two components (0.20 $\pm$ 0.06 mags), 
and our Bayesian age distribution for HIP 79797 (median 203 Myr, 
68\% confidence interval between 71--372~Myr).  We also assume the 
spectral type of M9 measured by \citet{hip79797} applies to both 
objects (which is reasonable given the nearly identical colors of the 
two components), and use a $K$-band bolometric correction from 
\citet{sptyperef2} to determine the luminosity of each component, which 
we then convert to mass using the Lyon/DUSTY evolutionary models 
\citep{chabrier00}.  We determine uncertainties through a Monte Carlo method, 
assuming Gaussian photometric and astrometric errors and using our posterior 
PDF for age.  We find similar masses for the two 
components (as expected given their small flux ratio) of 58$^{+21}_{-20}$ 
M$_{Jup}$ and 55$^{+20}_{-19}$ M$_{Jup}$, with a mass ratio of 
0.93$\pm$0.03.  These are similar values to the mass estimated by 
\citet{hip79797} based on their $K_S$-band photometry assuming the system 
was only a single object, meaning our mass estimates are discrepant with 
theirs.  While our median age for the system of 203 Myr is slightly 
younger than their age of 
300 Myr, we also find a brighter $K_S$-band flux for the combined Bab system 
of 13.0 $\pm$ 0.17 mag, compared 
to the 13.9~$\pm$~0.1 mag and 14.4 $\pm$ 0.2 mag 
measured by \citet{hip79797} in 
their two VLT/NACO datasets.  We note that the NACO 
data were taken with a neutral density filter to avoid saturation, 
and so HIP 79797 Ba/Bb is detected at low signal-to-noise in 
these images.

To assess the orbital properties, 
we have performed a preliminary Markov Chain Monte Carlo (MCMC) analysis 
on our 2 years of astrometric data for 
HIP 79797 Ba and Bb.  Any orbital motion over our 
short time baseline is less than half a 
NICI pixel, so we do not attempt to measure a dynamical mass for 
HIP~79797~Ba~and~Bb.  Instead, we fix the combined mass of the brown dwarfs 
to 113 M$_{Jup}$ and attempt to constrain the 
possible orbital parameters.  Our MCMC fit favors 
an edge-on orbit ($i$~=~87$\pm$5$^{\circ}$) with semi-major axis of 
0.48$^{+1.87}_{-0.37}$$''$ (25$^{+97}_{-19}$ AU).  This 
preference for inclined orbits is understandable given the change in the 
separation from the first to the last epoch (17 mas) is only slightly larger 
than the mean error on a single epoch (12 mas), while the PA of the binary 
does not appear to change (within errors).  That 
is, our data suggest, at low confidence, that the projected 
sky motion of 
HIP 79797 Bb is directly away from HIP 79797 Ba, as would 
be expected in an edge-on orbit.  Similarly, the semi-major axis must be 
large enough for there to be such a small amount of motion over two years.  
The corresponding orbital 
period is then 380$^{+3700}_{-330}$ years.  We stress that 
these are preliminary results, based on a short time baseline with 
very little on-sky motion, and so the 68\% confidence interval for 
orbital periods spans almost 
two orders of magnitude.  If the orbital period is on the 
shorter side of the estimated period 
range, future astrometric monitoring could plausibly 
yield a precise orbit and 
dynamical masses, as $\sim$30\% coverage of an astrometric orbit is 
sufficient to yield these results (e.g. \citealt{2mass1534,dupuy11}).  
In particular, a dynamical mass would 
be of immense value since the only other brown dwarf binary with a 
dynamical mass and an independent age estimate is the L4+L4 binary 
HD~130948 \citep{hd130948}.  Such benchmark brown 
dwarf binaries are key to testing models of brown dwarf atmospheres and 
evolution.

\subsection{Single Epoch Candidates}\label{oneepochsec}

For 20 of our 70 B and A stars, there are candidate 
companions that only appear in one epoch of NICI data and have no detections 
in archival observations.  These 
candidates fall into three categories.  (1) The most common case was that 
follow-up of these companions was given low priority, as we focused our 
observing time on candidates with the smallest projected physical 
separations ($<$400 AU).
(2) In the second case follow-up was attempted but 
the second epoch observation was unable to recover the candidate, 
because the candidate was especially faint or close to the target star.  In 
some cases with multiple candidates around a single star, we were able to 
confirm the nature of some candidates but not others.  (3) Finally, since 
the target star is not centered on the NICI detector, the orientation of 
the detector on the 
sky depends on the parallactic angle of the observation.  For wide-separation 
candidates ($\gtrsim$6.3'') this can mean a candidate is on the detector in 
one epoch and off the detector in another.  While we typically tried to 
schedule our second epoch observations to detect these wide candidates, 
it was not always possible to do this so some remain unvalidated.

For the candidates around these 20 
stars, we are unable to confirm whether these are background sources or 
co-moving companions.  Given that only 4 of the 78 candidate companions 
in Table~\ref{table_comp2a} are not background objects, it is likely that few, 
if any, of these single-epoch candidates are co-moving companions, especially 
given that most of them are at wider separations.  Nevertheless, we present 
the astrometry for these candidates in Table~\ref{table_oneepoch} as a 
reference for future observations of these stars.

\subsection{Notes on Individual Stars}

\subsubsection{$\beta$ Pic}

We assign the planet-host $\beta$ Pic an age of 12 
Myr, given its membership in the $\beta$ Pic moving group \citep{zsbw01}.  
Our Bayesian 
age analysis would suggest a significantly older age with a 
median of 109 Myr and 68\% confidence between 82 and 134 Myr.  While $\beta$ Pic 
does lie ``low'' on the CMD (lower than any other star at that $B-V$ in 
Figure~\ref{hipcolmag1_fig}), its spectral type of A5 and $B-V$ color of 
0.17 correspond to a main sequence lifetime about 10 times longer than 
12 Myr.  From CMD position alone it would only be possible to place $\beta$ 
Pic near the beginning of its main sequence lifetime.  The additional 
information of its membership in a moving group provides 
a more accurate age.  Indeed, \citet{barrado99} note 
the spread in isochronal ages found by different authors 
for $\beta$ Pic, with previous ages ranging 
from the zero-age main sequence to 100 Myr to 300 Myr, thus providing 
motivation to instead determine the age of other (lower-mass) 
stars in the same moving group.

In our initial NICI epoch (UT 
2008-11-22) using our standard campaign observing strategy 
we did not detect the faint planet $\beta$ Pic b.  Subsequent NICI 
observations beginning in 2009 did detect the planet, as a result of longer 
exposure times and the planet's orbit leading to a larger angular 
separation from its star.  Since these special observations were conducted 
separate from the NICI Campaign, we do not consider them here when deriving 
upper limits on the fraction of B~and~A stars that harbor planets.  
For this work, we only 
consider the 2008-11-22 contrast curve for $\beta$ Pic and treat our survey 
as having detected zero planets, since in this 
observation $\beta$ Pic b was not detectable.

\subsubsection{Fomalhaut}

From our Bayesian analysis, the planet-host Fomalhaut 
is assigned a median of 515 Myr and 68\% confidence 
interval between 418 and 623 Myr.  Fomalhaut has a similar $B-V$ color to 
$\beta$ Pic but is 0.6 magnitudes brighter on the CMD, placing it 
significantly above the Pleiades sequence.  Being beyond the first third 
of its main sequence lifetime allows the age of Fomalhaut to be more 
accurately estimated from its CMD position compared to younger stars.  Indeed, 
an analysis by \citet{fomalhaut_mamajek} notes that Fomalhaut has a 
very wide (57000 AU) 
K-dwarf companion, TW~PsA, with an estimated age of 440$\pm$40 
Myr, consistent with the 68\% confidence interval from our Bayesian age 
analysis.  We note that this age is twice as old as the 200$\pm$100 Myr age 
estimated for the Fomalhaut/TW PsA system by \citet{fomalhaut_old}.

\subsubsection{$\gamma$ Oph}

$\gamma$ Oph is an A0 star with a resolved debris disk that extends out to 
260 and 520 AU at 24 and 70 $\mu$m, respectively \citep{su08}.  
\citet{song01} assign an age of 184 Myr to $\gamma$ Oph with a range 
between 50 and 277 Myr, based on Str\"omgren photometry and isochrones.  This 
is significantly younger than our Bayesian age, which has median 342 Myr and 
a 68\% confidence interval between 285 and 388 Myr.  As with other stars 
with literature ages based on isochrones or CMD position we adopt our Bayesian 
age distribution for this star.  $\gamma$ Oph is high
on the color-magnitude diagram compared to Pleiades stars of the same 
color.  We detect 8 candidates companions to $\gamma$ Oph, the 7 innermost 
of which are determined to be background objects from two epochs of NICI 
data.  The final candidate, at a separation of 9.2'', is only visible in 
one epoch of NICI data and so we are unable to definitively determine if 
it is background or CPM.  Given its large separation and the large 
number of background objects in the field, this eighth candidate 
is most likely also a background object.

\subsubsection{49 Cet}

49 Cet is an A1 star hosting a disk with significant amounts 
of molecular gas \citep{zs12}.  For its color 49 Cet is relatively 
low on the CMD, comparable to Pleiades A stars; as a result, we derive a 
Bayesian age of 225 Myr with 68\% confidence interval between 91 and 377 
Myr.  This wide range is again a consequence of 49 Cet's location in the CMD
being consistent with it being in the first third of its main sequence 
lifetime.  \citet{zs12} determine 49 Cet to be a member of the Argus 
association, allowing them to derive a much younger age of 40 Myr.  We 
adopt this moving group age of 40 Myr.

\subsubsection{$\zeta$ Lep}

$\zeta$ Lep has a small debris disk that has only been 
resolved at 18.3 $\mu$m, with an extent of only 3 AU \citep{moerchen07}.  
Our Bayesian analysis finds an age of 330 Myr with 68\% confidence interval 
between 224 and 410 Myr, consistent with its position on the CMD in the 
midst of Pleiades A stars.  \citet{nakajima12} find $\zeta$ Lep to be a 
member of the $\beta$ Pic moving group, and so we assign it an age of 12 Myr.
We find a single candidate companion to $\zeta$ Lep at 5.3'' which we 
determine to be background based on two epochs of NICI data.

\subsubsection{HD 31295}\label{hd31295}

HD 31295 has been shown to have an IR excess with MIPS 
24 and 70 $\mu$m data by \citet{rhee07}, who estimate an age of 100 Myr 
based on its low position on the CMD.  As expected, our Bayesian analysis 
shows an older age, with a median of 241 Myr and 
a 68\% confidence interval that extends from 119 to 355 Myr.  We note that 
HD~31295 is listed in CCDM \citep{ccdm} as having a 40'' M-star companion 
(BD+09 683B).  We have 
examined the 1975 DSS image and the 2000 2MASS image of HD 31295, and found 
that BD+09 683B is in fact a background object ($\chi^2_{\nu , bg}$=0.21, 
$\chi^2_{\nu , CPM}$=2.94).  In addition, from our NICI images 
we have detected four candidate 
companions to this star, two of which we rule out as background objects.  
The remaining two (at 7.8'' and 8.7'') were only visible in one NICI epoch, 
leaving us unable to determine if they are background objects or CPM 
companions.

\subsubsection{HD 1160}

HD 1160 is an infrared photometric standard 
from \citet{elias82} and has been 
observed numerous times at the VLT to calibrate observations of nearby 
stars.  \citet{hd1160} combined these archival data with new NICI data 
to show that HD 1160 is a triple system consisting of an A0 star, an 
$\sim$L0 brown dwarf at 80 AU, and an M3.5 star at 530 AU.  Our Bayesian age 
for HD 1160 of 92 Myr, with 68\% confidence between 36 and 178 Myr, is 
consistent with the \citet{hd1160} age range of 50$^{+50}_{-40}$ Myr, derived 
by comparing the CMD positions of all three objects to young clusters of 
various ages.

\subsubsection{HR 7329}

HR 7329 is an A0 star that hosts a debris disk 
\citep{hr7329_1993}, which 
\citet{hr7329disk} find to be edge-on at 18.3 $\mu$m.  In addition, 
it hosts the brown dwarf companion HR 7329 B at 4.2'' \citep{lowrance00}, 
whose mass is 
estimated by \citet{hr7329} to be between 20 and 50 M$_{Jup}$.  We 
confirm that HR~7329~B is a CPM companion with our NICI observations, but 
find no other candidate companions in our NICI images.  Like other 
members of young moving groups our Bayesian method 
only places HR 7329 in the beginning of its main sequence lifetime: our 
median age for HR 7329 is 182 Myr, with 68\% confidence between 101 and 
268 Myr.  HR 7329 is a member of the $\beta$ Pic moving group 
\citep{zsbw01}, and so we adopt an age of 12 Myr for the star.

\subsubsection{HR 4796}

HR 4796 is an A0 star that hosts a resolved narrow ring 
of debris with a sharp inner and outer edge 
\citep{hr4796_1998_1, hr4796_1998_2, hr4796_2005, hr4796_2010}.  
We assign the star an 
age of 10 Myr as \citet{zs04} identify HR 4796 as a member of the TW Hydra 
Association (TWA).  We designed our NICI observations to place the known M2 
companion HR 4796 B \citep{jura95} off the detector so that its light 
would not limit the dynamic range we could reach.  We detect scattered 
light at the position angle HR 4796 B would appear, as it should have been 
less than 1'' from the edge of the detector.  We also detect a background 
star at 4.47'' which is referred to as HR 4796 D by \citet{hr4796_1997} 
who identify it as a background object based on its colors.  With 8 years of 
astrometry, including the initial \citet{hr4796_1997} detection, we 
confirm that this object is unambiguously a background star.  We detect 
no other objects around HR 4796.

\subsubsection{HD 141569}

HD 141569 is a B9 star that hosts a resolved disk 
\citep{hd141569_disk} containing both gas and dust \citep{merin04}.  Our 
Bayesian method find a median age of 318 Myr with a 68\% confidence 
interval between 163 and 440 Myr.  \citet{weinberger00} note that the 
two M-dwarf companions to HD 141569 show lithium absorption and X-ray flux 
consistent with youth, and find an age for the system of 5 Myr.  
\citet{merin04} examined high-resolution spectra of HD 141569 and confirm 
that it is a pre-main sequence star with low metallicity ([Fe/H] = $-$0.5).  
As with other very young B and A stars 
additional information beyond photometry (in this case the ages of the 
two M-dwarf companions) helps refine the age.  The two M-dwarf 
  companions are off the detector in our NICI images.  We detect two
additional candidate companions at 6.40'' and 7.85'' and find them both to be 
background objects.

\subsubsection{HD 71155}

HD 71155 is an A0 star with a debris disk detectable as an 
IR excess with MIPS 24 and 70 $\mu$m data \citep{rhee07}.  The disk 
of HD 71155 is barely resolvable at 10 $\mu$m, with a 
radius of $\sim$2 AU \citep{moerchen10}.  We find a median age for the star of 
273 Myr, with a 68\% confidence interval between 237 and 313 Myr.  We 
detected 3 candidate companions to HD 711155, all of which were determined to 
be background, as well as a one-epoch candidate companion at 9.93'' that we 
are unable to classify as CPM or background at this time.

\subsubsection{HD 172555}

HD 172555 is a member of the $\beta$ Pic moving group and 
so has an estimated age of 12 Myr \citep{zsbw01}.  It also has a wide K5 
companion, CD-64 1208, with a separation of 71.3'' (2040 AU, 
\citealt{feigelson06}).  The disk of HD 172555 contains 
both dust and gas \citep{riviere12}, possibly resulting from a recent 
impact event during planet formation \citep{hd172555_12}.  Our age for 
HD 172555 is 
276 Myr, with a 68\% confidence interval between 83 and 462 Myr.  
As with other later-type A stars 
in young moving groups we find an age range that corresponds to the 
first third of the main sequence lifetime, rather than the much 
younger age for the group itself.  We do not detect any closer 
components to this binary system, finding only a background 
object at 7.7'' separation.

\subsection{Upper Limits on Planet Fraction}\label{planetfraction_section}

Given the non-detection of planets from the NICI Campaign survey of B and 
A stars, 
we can set an upper limit on the fraction of stars with planets using 
the measured contrast curves for each star and Monte 
Carlo simulations of completeness to planets.

\subsubsection{Contrast Curves}\label{concurve_sec}

\begin{figure}
\epsscale{0.9}
\plotone{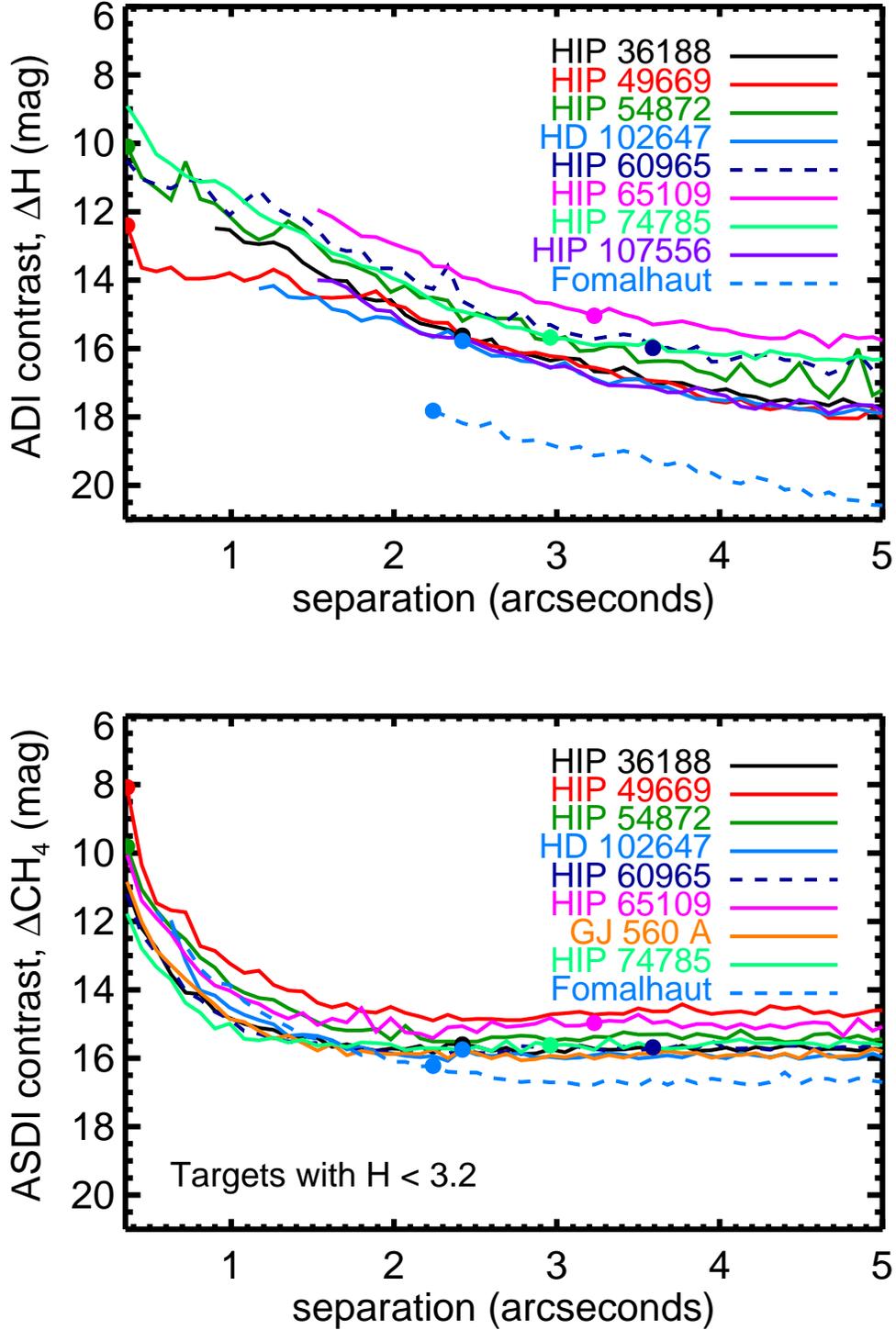}
\caption{
\footnotesize{
95\% completeness contrast plots for the brightest 10 stars, with $H <$ 3.2 
mag, given for ADI (top) and ASDI (bottom).  Some stars have 
observations only in one mode.  For stars with both ADI and ASDI data we 
mark with a filled circle the angular separation where the ADI curve reaches 
larger contrasts than the ASDI curve.  A contrast of 15 
magnitudes represents a flux ratio of 10$^6$.}
\label{con_plot0}}
\end{figure}

\begin{figure}
\epsscale{0.9}
\plotone{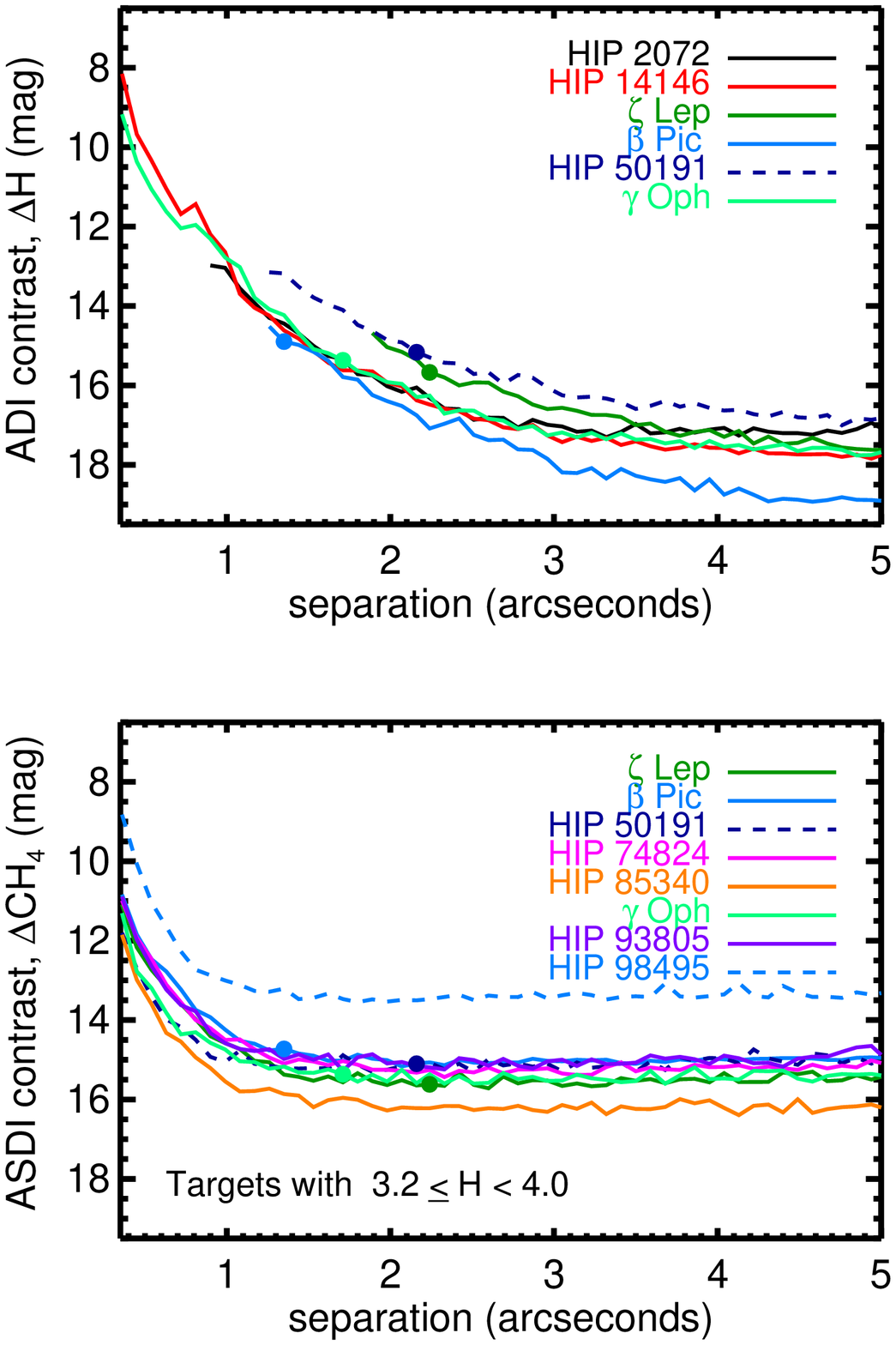}
\caption{Contrast plots for stars with 3.2 $\le H <$ 4.0 mag.  See 
Figure~\ref{con_plot0} caption for details.
\label{con_plot1}}
\end{figure}

\begin{figure}
\epsscale{0.9}
\plotone{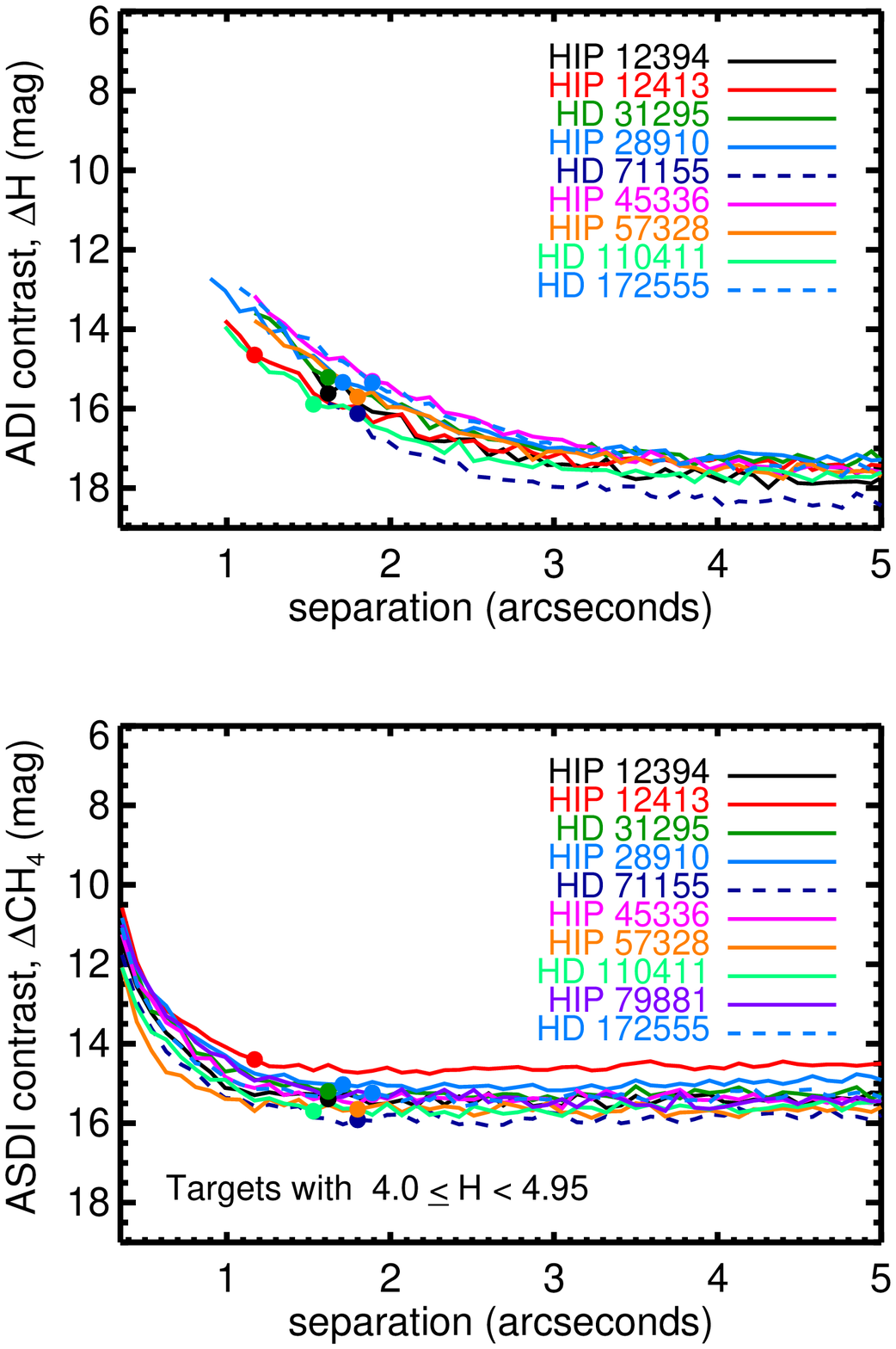}
\caption{Contrast plots for stars with 4.0 $\le H <$ 4.95 mag.  See 
Figure~\ref{con_plot0} caption for details.
\label{con_plot2}}
\end{figure}

\begin{figure}
\epsscale{0.9}
\plotone{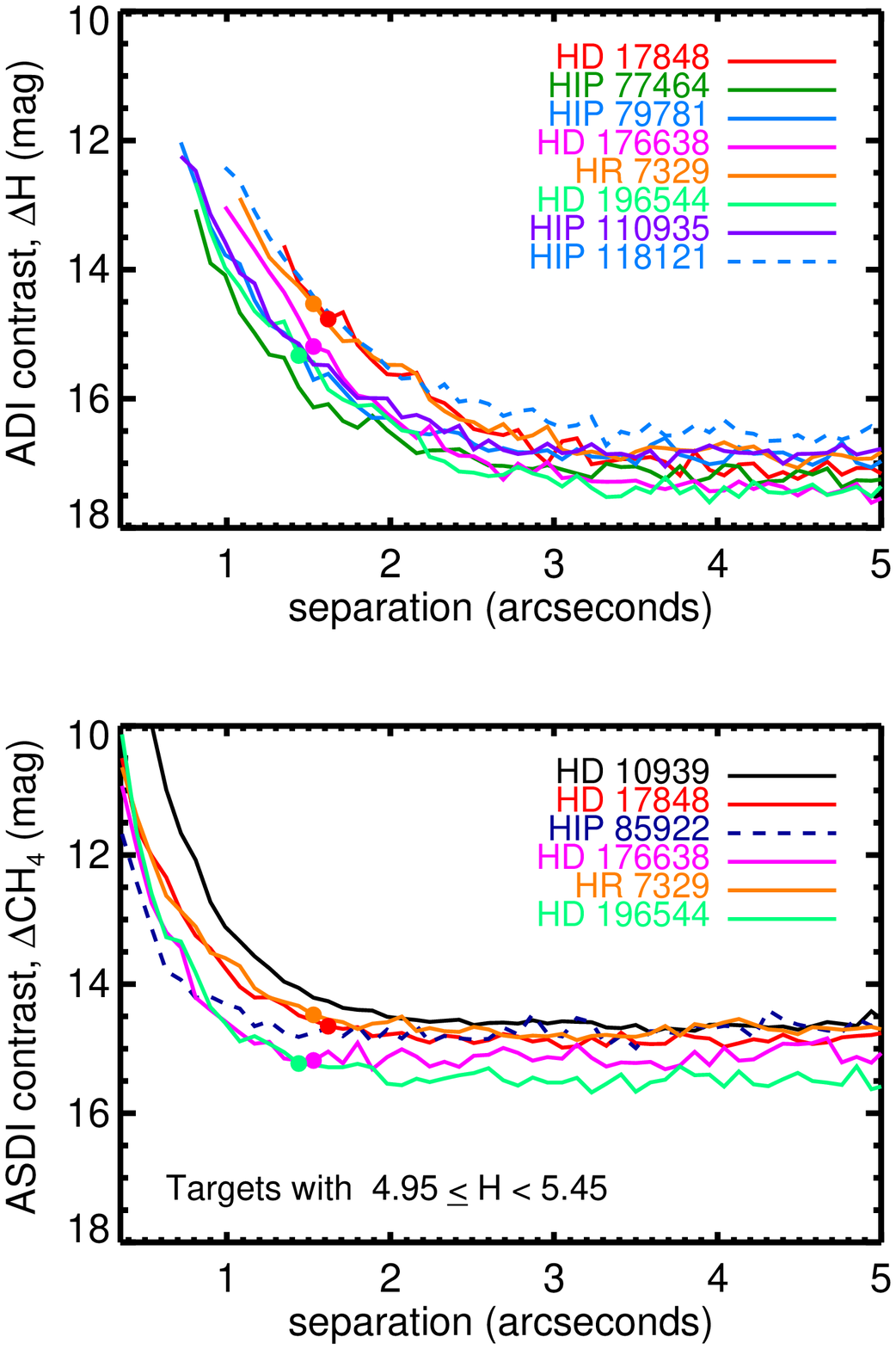}
\caption{Contrast plots for stars with 4.95 $\le H <$ 5.45 mag.  See 
Figure~\ref{con_plot0} caption for details.
\label{con_plot3}}
\end{figure}

\begin{figure}
\epsscale{0.9}
\plotone{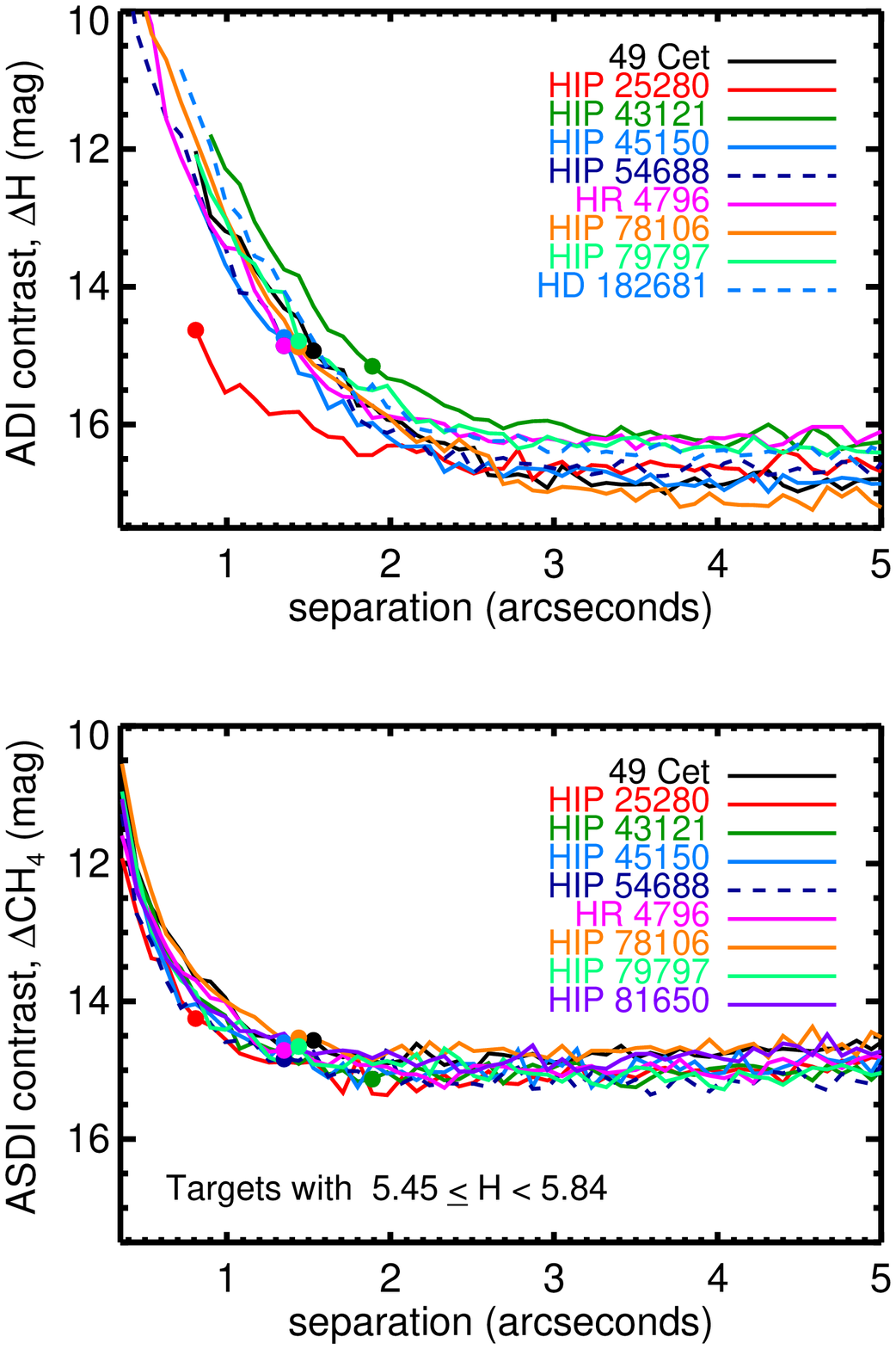}
\caption{Contrast plots for stars with 5.45 $\le H <$ 5.84 mag.  See 
Figure~\ref{con_plot0} caption for details.
\label{con_plot4}}
\end{figure}

\begin{figure}
\epsscale{0.9}
\plotone{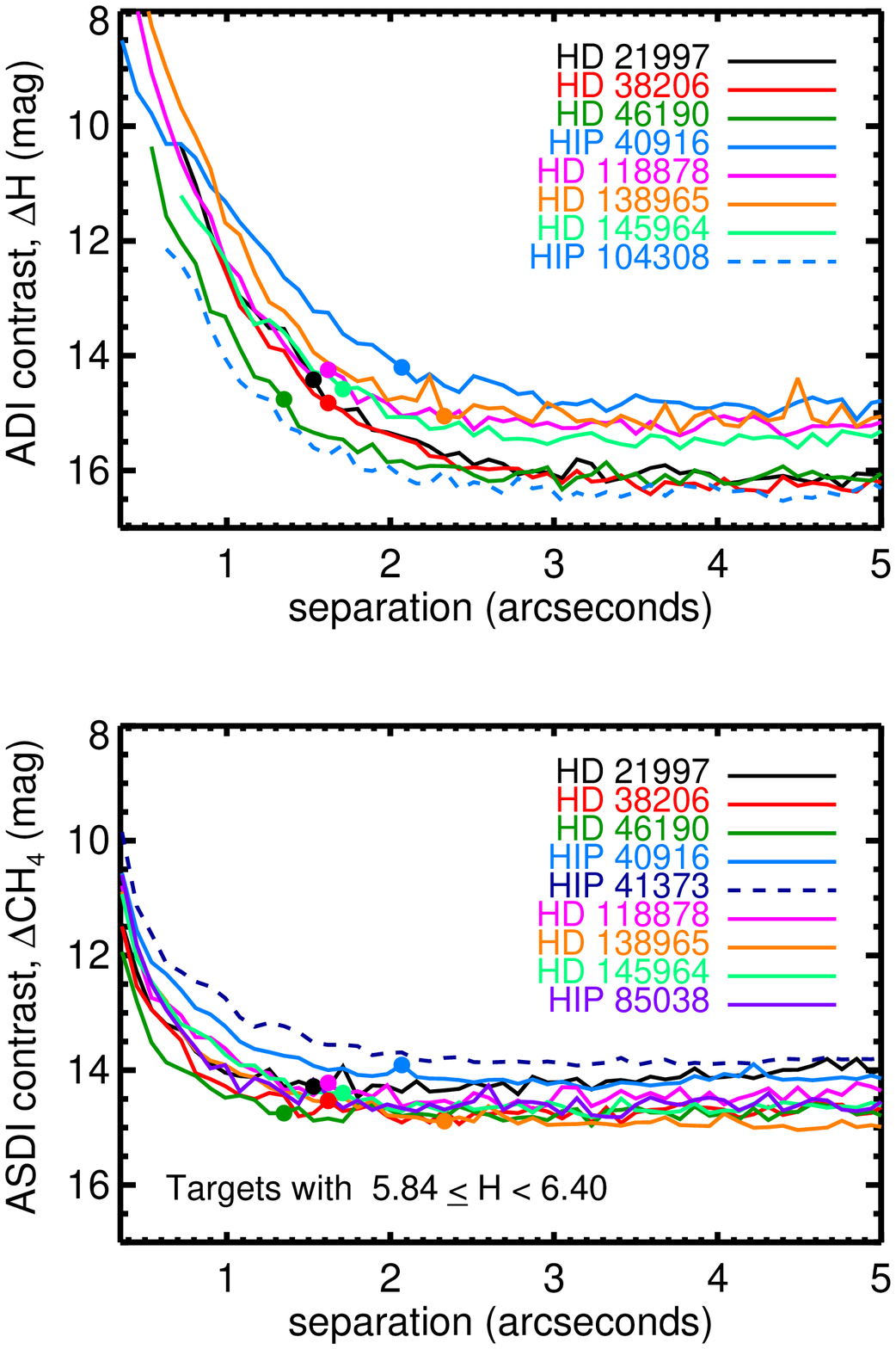}
\caption{Contrast plots for stars with 5.84 $\le H <$ 6.40 mag.  See 
Figure~\ref{con_plot0} caption for details.
\label{con_plot5}}
\end{figure}

\begin{figure}
\epsscale{0.9}
\plotone{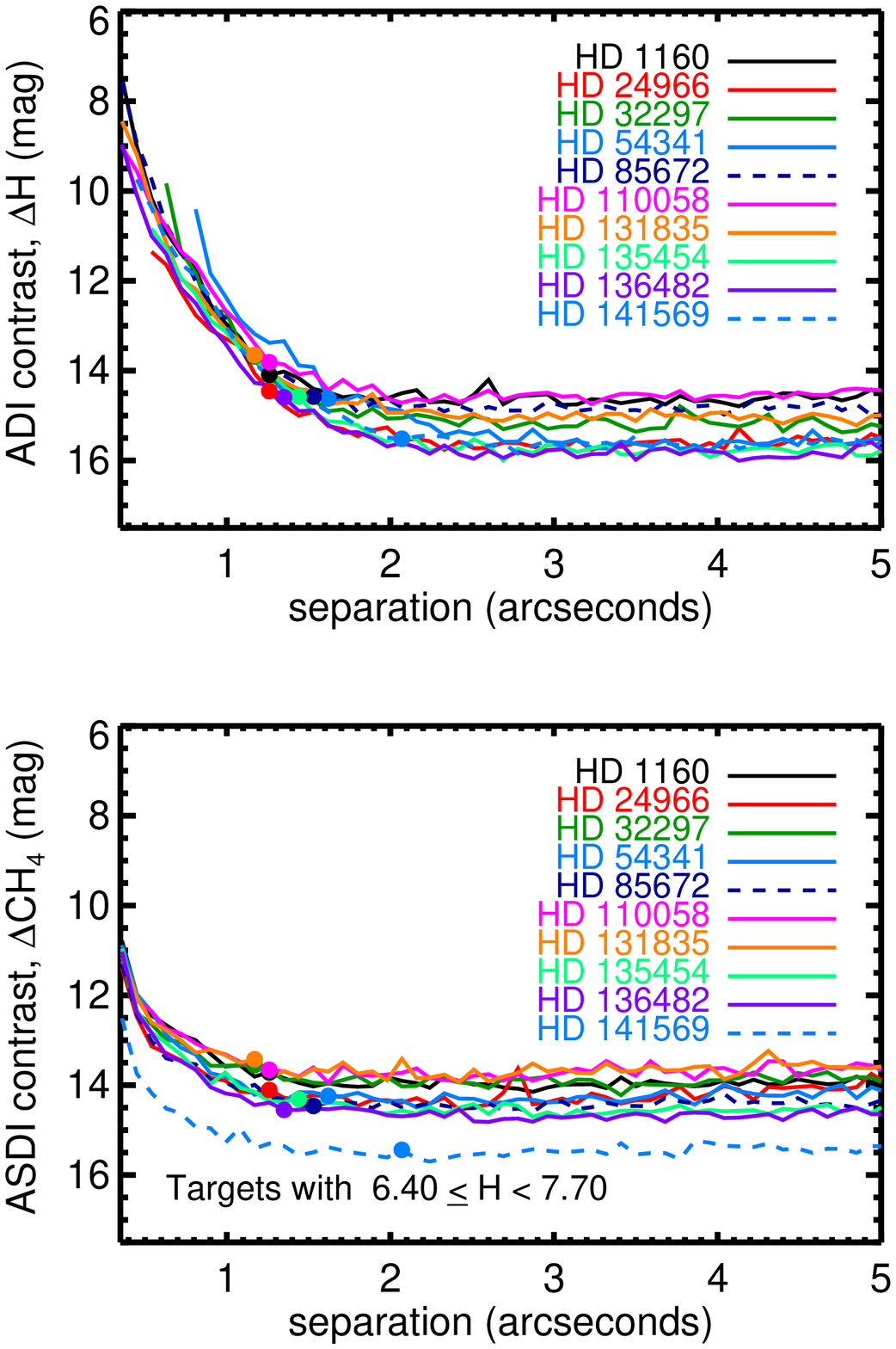}
\caption{Contrast plots for stars with 6.40 $\le H <$ 7.70 mag.  See 
Figure~\ref{con_plot0} caption for details.
\label{con_plot6}}
\end{figure}

For every observed star 
contrast curves are produced for each observing mode (ADI or ASDI) and 
give the 95\% completeness to companions as described in detail 
by \citet{pipeline}.  Briefly, simulated planets at a grid of flux ratios 
and separations are 
inserted into the raw data from each star, which are then reduced with the 
Campaign pipeline and the candidates are recovered with an automated 
detection method.  The resulting contrast curves give 
the flux ratio as a function of separation where 95\% of the 
simulated companions can be recovered.  
These 95\% completeness contrast curves are shown in 
Figures~\ref{con_plot0}--\ref{con_plot6}, grouped by apparent $H$ magnitude 
of the target star, and tabulated in Table~\ref{tab:contrasts}.  

As described in \citet{pipeline}, contrast curves for both ADI and ASDI 
datasets must be corrected for four effects to account for the true flux of 
the companions after emerging from the data reduction pipeline.  (1)~Light 
from a close companion passes through the 
partially transparent coronagraphic mask, and the companion's flux is 
diminished.  (2)~Over a 
single ADI exposure, as the sky rotates with respect to the detector, 
a companion can be smeared over multiple pixels, especially at larger 
angular 
separations.  (3)~ADI self-subtraction occurs for small-separation companions 
when the total amount of rotation for the companion is small, because some 
companion flux ends up in the combined PSF image and is then subtracted from 
the companion.  (4)~Finally, SDI reduction leads to self-subtraction of the 
companion flux, when subtracting the on-methane narrowband image from the 
off-methane image.

The first three effects (mask attenuation, smearing, and ADI self-subtraction) 
are functions of radius alone and are applied as corrections to the 
contrast curve itself (see \citealt{pipeline}).  The final effect, SDI 
self-subtraction, is a 
function of both radius and the flux ratio of the companion between the 
two narrowband filters, and so is dealt with in the Monte Carlo simulations, 
as described in the next section.

For stars with unconfirmed candidates due to having only one 
epoch of astrometric data (Section~\ref{oneepochsec} and 
Table~\ref{table_oneepoch}), we adjust the 
contrast curves to reflect the fact that we have not completely ruled out 
the presence of companions that are detectable by NICI.  
For each star with such candidates we make the following changes to the 
measured contrast curves: 

\noindent 
 (1) If there are 
unconfirmed candidate companions outside the radius with 100\% 
angular coverage (i.e., where some position angles at that radius are off 
the edge of the NICI detector, typically $>$6.3'') and brighter than 
the contrast curves at any of the observation epochs, 
then all contrast curves for that star are truncated at the angular 
separation of the innermost companion in the region with less than 100\% 
coverage.  For example, HD 31295 has two candidate companions at 7.77'' and 
8.66'' seen only at epoch 2009-01-14, so all contrast curves at all epochs 
for this star are stopped at 7.77'', i.e. we assume we have zero data at 
larger separations.

\noindent 
 (2) If candidates are within the 100\% angular coverage region and lie 
between the contrast curves from different epochs, then any contrast curve 
that is more sensitive than the brightest of these candidates is replaced 
with the best contrast curve that could not have detected the brightest of 
these candidates.  For example, the target star HIP 50191 
has a candidate at 4.11'' that 
was detected in the 2010-01-05 ADI dataset but was undetected in the 
2008-12-16 ASDI dataset.  As a result, we use the 2008-12-16 ASDI contrast 
curve to compute completeness to planets at both epochs.

\noindent 
In principle, both changes could be applicable to a single star, but 
that scenario never arose for our NICI B and A stars.  In cases where 
candidate companions are within the 100\% coverage region, brighter than 
the contrast curve, 
and there is only one NICI epoch, the star is dropped from 
our analysis altogether.  This is the case for 7 of our 70 target stars, 
all of which have a large number of candidate companions, galactic 
latitude within 20$^\circ$ of the galactic plane and 
galactic longitude within 30$^\circ$ of the galactic center.  
The last column of 
Table~\ref{table_oneepoch} summarizes 
our adjustments for the contrast curves of stars with unconfirmed 
companions.

\subsubsection{Monte Carlo Simulations of Completeness to Planets}

To interpret the contrast curves as limits on planet 
fraction, we follow the Monte Carlo procedure developed 
by \citet{sdifinalpab} and \citet{sdimeta}.  Briefly, we place 
an ensemble of simulated planets around each target star 
and then determine the fraction that are detectable by comparing them to the 
contrast curve for that star's NICI dataset.  We begin by assigning orbital 
parameters to each simulated planet.  The position angle of nodes 
represents rotation on the sky and is generally not included in the 
simulations, as our 
contrast curves are one-dimensional (functions of separation only).  
The semi-major axis and mass are drawn from a regular grid (see below for 
details).  The eccentricity, inclination angle, longitude of periastron, and 
mean anomaly are randomly drawn from the 
appropriate probability distributions.  We have fit a linear function to 
the probability distribution for eccentricity based on known RV planets (as 
described in \citealt{sdimeta}), 
while the probability distributions of the other parameters are given by 
geometric considerations only.  The instantaneous projected separation is 
then calculated for each simulated planet from its orbital parameters.  In 
addition, the age of the star and mass of each simulated planet are used to 
produce an $H$-band flux using the \citet{cond} models, which is 
converted to a flux ratio using the known distance and $H$-band 
magnitude of the star.  Then, each simulated planet can be directly compared 
to the NICI contrast curve for that star; the detectable fraction is simply the 
fraction of simulated planets lying above the contrast curve.

In cases where 
a star was observed at multiple epochs, simulated planets are generated at 
the first epoch, advanced forward in their orbits to the second epoch, then 
compared to the contrast curve for the second epoch dataset.  
Similarly, the same set of simulated 
planets are compared separately to the ADI and ASDI contrast curves.  A 
simulated planet that remains below 
all contrast curves is considered undetected, while one that is above at 
least one contrast curve is detected.

An additional consideration is the use of age distributions, rather than a 
single defined age, to describe many of the B and A stars in this paper.  In 
\citet{sdimeta} each target star is assigned a single age, and the 
\citet{cond} models are used to assign absolute magnitudes 
corresponding to that age.  For stars with an external age measurement 
(i.e. stars belonging to a moving group) we use this same method.  For the 
other stars with a distribution of ages from our Bayesian analysis, 
we define ten points logarithmically spaced along the age axis, between 5 Myr 
(the minimum used in our 
luminosity grid) and the maximum age of the star, defined as 
where the probability of the 
star having that age or larger is 1\%.  The fraction of detectable planets is 
then calculated at each of those ten values of the age and then interpolated 
to the age grid used in Section~\ref{bayes}.  The final fraction of detectable 
planets is the weighted average of this array of interpolated fractions 
weighted by the relative probabilities of each age bin derived by our 
Bayesian method.

We also account for SDI self-subtraction when computing the flux of the 
simulated planets.  The effect is strongest close to the star: 
one narrowband methane image is demagnified to place 
the speckles at the same angular scale as in the other narrowband 
image, and so at 
smaller separations the on-methane and off-methane companions increasingly 
overlap 
each other and more self-subtraction occurs.  If the companion is strongly 
methanated, it will have much more flux in the off-methane filter compared to 
the on-methane filter, and so the amount of self-subtraction will be 
minimal.  However, if the companion has no methane absorption, it will have 
roughly equal flux between the two narrowband filters, and self-subtraction 
will be a strong effect.

To address this issue in the Monte Carlo simulations we modify 
the fluxes of the simulated planets to simulate the effect of ASDI 
self-subtraction.  
As in \citet{sdimeta}, we calculate the flux in the off-methane ($CH_4S$) 
filter by convolving the NICI filter transmission curve with the SpeX Prism 
Library of ultra-cool dwarfs and deriving a relation between ($CH_4S - H$) 
color and spectral type.  Temperatures are then derived from 
spectral type using the polynomial fit of \citet{sptyperef2}.  Then, we 
use the theoretical models of \citet{cond} to compute $H$-band magnitudes 
and temperatures as a function of age and planet mass.  We then apply 
our synthetic photometry results to 
obtain a grid of $CH_4S$ magnitudes for each combination of age and mass.  

To deal with self-subtraction, we use a similar procedure to produce 
grids of $CH_4S$ and $CH_4L$ magnitudes.  For small separation planets, the 
images of the planet in both bands overlap, and the final flux is 
simply the difference between the two fluxes.  At large separations the two 
images are completely offset from each other so no self-subtraction 
occurs, and the final flux is just the $CH_4S$ flux.  For intermediate 
separations, we calculate the amount of geometric self-subtraction (the 
degree of overlap in the two images) as a function of radius and reduce the 
flux for the $CH_4L$ image of the planet by this geometric factor before 
subtracting it from the $CH_4S$ flux.

We also account for nonuniform position angle 
coverage of our observations 
at large angular separations.  The NICI detector is square, with the focal 
plane mask and target star placed 
offset from the center.  As a result, while we image 360$^\circ$ 
in position angle at small separations, at larger separations ($\gtrsim$6.3'') 
our coverage 
declines as some position angles are off the edge of the detector.  We 
note the separations with reduced coverage in Table~\ref{tab:contrasts}.  In 
our Monte Carlo simulations we account for this effect by generating a 
uniform random variable between 0 and 1 for each simulated planet.  If 
that random variable is greater than the fractional angular coverage at 
the projected separation of the simulated planet, then that planet is 
considered undetectable even if it is brighter than the contrast curve.  
This parameter is similar to the position angle of 
nodes (rotation of the orbit on the plane of the sky), which follows a 
uniform distribution.  When multiple contrast curves 
are available for a single target star, this random variable is also 
preserved across all epochs so that the same set of 
simulated planets are compared to each contrast curve for the same star.

\subsubsection{Planet Frequency Analysis}

For each star we then compute the fraction of detectable planets 
with 10$^3$ simulated planets all having the same values of planet 
mass and orbital semi-major axis.  This is then repeated for a grid of mass 
and semi-major axis and for each target star.  As in \citet{sdimeta}, we 
consider the following quantity:

\begin{equation}
N(a,M) = \sum_{i=1}^{N_{obs}} f_p(a,M) P_i(a,M) C_{2.0} (M_{*,i})
\end{equation}

\noindent where $N(a,M)$ is the expected number of detected 
planets as a function of orbital 
semi-major axis ($a$) and planet mass ($M$); $N_{obs}$ is the number of 
stars observed; $f_p(a,M)$ is the frequency of planets for a specific 
combination 
of mass and semi-major axis; and $P_i(a,M)$ is the probability of detecting a 
planet with that mass and semi-major axis as determined 
from our Monte Carlo simulations.  Finally to account for the different 
masses of the stars in the sample, we define a mass correction that weights 
stars of different masses by their probability of harboring giant planets.  
This mass correction, $C_{2.0} (M_{*,i})$, is given by: 

\begin{equation}
C_{2.0} (M_{*}) = \frac{F_p(M_*)}{F_p(2.0 M_{\sun})}
\end{equation}

\noindent where $F_p(M_*)$ is the relative probability of hosting giant 
planets as a function of stellar mass.  We define $F_p(M_*)$ using RV results, 
specifically the power-law fit to planet frequency as a 
function of stellar mass from \citet{johnson_new}, where 
$F_p(M_*) \propto M_*^{1.0}$.  
Our approach assumes that the stellar mass dependence of close-in RV planets 
applies to wide-separation giant planets as well.  While the current 
paucity of 
wide-separation planets makes this assumption impossible to test, 
we use it as a starting point for our analysis of the frequency of these 
planets.  If wide-separation giant planets form a distinct population from 
close-in RV planets (as suggested by \citealt{crepp11}), then our 
adopted mass scaling might not be appropriate for long-period planets.  
We choose to normalize at 2 M$_{\sun}$, 
since that is the peak of the total distribution of stellar masses in our 
sample from our Bayesian analysis.  The total range of 
stellar mass in our 
sample is 1.5--4.6 M$_\sun$, as given by the median of our posterior mass 
PDFs for each star, with 60 of our 70 stars having a median between 
1.5--2.5~M$_{\sun}$, and only 4 having a median greater than 
3 M$_\sun$ (right panel of Figure~\ref{targets_fig}).  
As a result, the mass correction has a modest effect on the 
final results.

We are seeking to define 95\% confidence upper limits 
on the fraction of stars with planets as a function of semi-major axis and 
planet mass ($f_p(a,M)$).  We use the Poisson distribution: 

\begin{equation}
P(k) = \frac{\mu^k e^{-\mu}}{k!}
\end{equation}

\noindent with expectation value $\mu$ for all integers $k$.  In this case, 
$\mu$ is the predicted number of planets in a given semi-major axis and 
mass bin, and $k$ is the number of planets detected in that bin.  The 
95\% confidence level for a null result ($k=0$) is $\mu = 2.996$; that is, 
for an expectation value of 3 planets, the probability of a null result 
(i.e., detecting no planets) occurring by 
chance is 5\%.  So, solving for the fraction of stars with planets, 
$f_p(a,M)$, we obtain

\begin{equation}
f_p(a,M) = \frac{N(a,M)}{\sum_{i=1}^{N_{obs}} P_i(a,M) C_{2.0}(M_{*,i})}
\end{equation}

\noindent This is the upper limit on the 
fraction of stars with planets, as a function of semi-major axis and planet 
mass.  Setting $N(a,M)=3$ gives the 95\% confidence upper limit, allowing us 
to map out the upper limit that can be placed on the fraction of stars with 
planets at each combination of semi-major axis and planet mass.

\subsubsection{Binaries}

An additional concern is the presence of 
stellar and brown dwarf binaries around some 
of our target stars.  A target star with a 100 AU binary, for example, 
would not be expected to have a 100 AU planet from dynamical arguments, 
and so we cannot simply combine constraints on 100 AU planets 
from single stars with the 100 AU binary star for our statistical 
analysis.  \citet{holman99} suggest 
that a forming planet inside 20\% of the binary separation should see its 
parent star as a single star, and so we use this factor of 5 to define 
an ``exclusion zone'' around a stellar or binary brown dwarf companion to 
our target stars.  Between 20\% and 500\% of the binary semi-major 
axis (where a forming planet would be disrupted by the orbit of 
the binary), we exclude any constraints on planets from that star, i.e. we 
assume no information exists about planet frequency, since planets never 
could have formed in the first place.  We 
caution that this approach does not account for migration in either the 
binary system or its potential planets (e.g. \citealt{kratter12}).  
In addition, we do not change 
the analysis for stars with known planetary-mass companions ($\beta$ Pic and 
Fomalhaut), i.e. we do not consider any dynamical effects their known planets 
would have on other orbital separations.

We compile a list of known binaries among our target stars in 
Table~\ref{tab:binaries}, including the binary separation (or semi-major 
axis when available), spectral type, and references.  We also indicate 
the range of allowable semi-major axes, beyond the exclusion zone for that 
star which is excluded when deriving constraints on planet fraction.  As 
discussed in Section~\ref{hd31295}, we treat our target star 
HD~31295 as a single star despite it being listed in CCDM as a binary.

\subsubsection{Results}

\begin{figure}
\epsscale{0.8}
\plotone{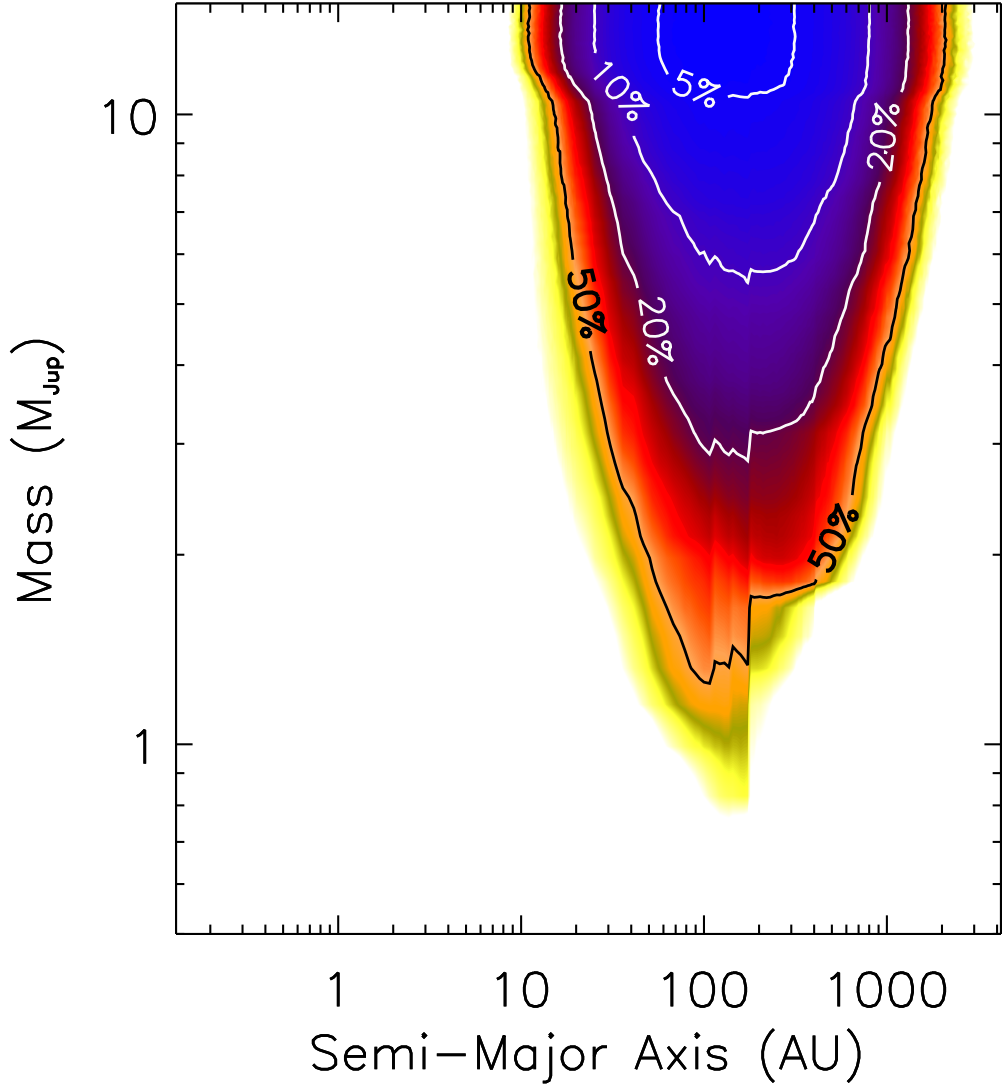}
\caption{\small{The 95\% confidence upper limit on planet frequency 
from our Monte Carlo analysis of 
our 70 B and A star sample.  We used the \citet{johnson_new} relation between 
stellar mass and RV planet frequency to scale the range of stellar host 
masses to a common mass of 2 M$_\sun$, the median for our sample.  For 
the 5\% frequency contour, for example, 
fewer than 1 in 20 high-mass stars can have giant planets 
with masses and semi-major axes inside the contour.  
The sawtooth pattern 
between 100 and 200 AU occurs as the constraints from the very young target 
stars HR~4796, HD~141569, and GJ~560~A are excluded at 110, 140, and 
180 AU due to the presence of stellar binaries around those stars.  The 
effect is most pronounced at contours for high upper limits on 
planet fraction, where the removal of a single star with large 
completeness to planets most impacts the statistics.}
\label{planet_frac_fig}}
\end{figure}

In Figure~\ref{planet_frac_fig} we show contours giving the upper limit on 
the fraction of 2 M$_{\sun}$ stars with planets.  
We also give the ranges of semi-major axis 
at which the upper limit on planet fraction drops below 5\%, 10\%, 20\%, and 
50\% for four planet masses in Table~\ref{table_pf}.  In general, the 
upper limits are modest, especially when compared to samples 
including solar type and later stars (e.g. \citealt{moving_groups} and 
\citealt{debris}): fewer than 20\% of B and A stars can have 
giant planets larger than 4 M$_{Jup}$ between 59 and 460 AU, with less than 
10\% of stars having giant planets above 10 M$_{Jup}$ between 38 and 650 AU.  

We can also determine limits on the fraction of A stars with a planet like 
HR~8799~b \citep{hr8799b}, by focusing on its corresponding values of mass 
(7 M$_{Jup}$) and semi-major axis (68 AU, assuming its 
semi-major axis is equal to its projected separation) in 
Figure~\ref{planet_frac_fig}.  Our survey's non-detection of planets 
states that fewer than 9.8\% of 2 M$_{\sun}$ stars can have a planet 
with the same properties as HR 8799 b at 95\% confidence.  
Planets like 
$\beta$ Pic b (8 M$_{Jup}$, 8 AU) are unconstrained by our results, since 
most of our target stars are older and more distant than $\beta$ Pic.

The weaker constraints on planet frequency for B and A stars compared to 
later-type stars are expected.  B and A stars are intrinsically brighter 
than solar-type and M stars, and so the same instrumental contrast 
corresponds to brighter (and more massive) 
detectable companions for higher-mass stars.  In addition, the stars here 
with ages determined by our Bayesian analysis are 
significantly older than the young moving group stars in 
\citet{moving_groups} and the debris disk hosts in \citet{debris}, 
with a median age of 250 Myr 
and 68\% confidence interval between 90 and 500 Myr (see 
Figure~\ref{ageout1_fig}).  In comparison, the median age for stars in the 
moving group sample is 12 Myr, while for the debris disk sample it is 
70 Myr.

\subsection{Comparison with Previous Studies}

We compare our constraints on the fraction of 
high-mass stars with planets to previous predictions and survey 
results.

\citet{crepp11} predicted the planet yield of the NICI 
survey with a simplified description of the NICI contrast curve, consisting 
of inner and outer working angles of 0.2'' and 5.5'' and a contrast floor 
within that range that increases monotonically from 11.4 mags (flux ratio of 
$2.9 \times 10^{-5}$) to 12.9 mags for the $H$-band.  This assumed contrast 
curve is similar to our median $CH_4$ contrast at 0.36'' of 11 mags, but 
is far too pessimistic at larger separations: we achieve a median 
contrast of 16.8 mags at 4'' compared to the 12.9 mags assumed by 
\citet{crepp11}.  Their calculations use 
simulated target stars corresponding to the age, distance, and stellar 
mass distributions of solar neighborhood stars.  Given their preferred input 
model of planet populations, they 
predict that a NICI-like campaign of 610 stars 
would detect 15.4 planets around target stars with masses between 1.5 and 
2.6 M$_{\sun}$.  They predict such a survey would observe 192 high-mass 
stars, compared to the 70 B and A stars observed by NICI, so 
we reduce this predicted yield by 70/192 to 5.6 planets\footnote{This is 
likely an underestimate.  If one were to design a sub-sample of 70 stars 
from an initial sample of 192 stars one would choose the 70 best stars, which 
would account for proportionally more planet detections than the 70 worst 
stars.}.

We note, however, that the null result of \citet{sdimeta} rules out 
essentially the \citet{crepp11} input model of planet populations (an 
extension of the 
RV planet population out to 120 AU) at $>$99\% confidence (e.g. Figure~15 
of \citealt{sdimeta}).  
The only difference is that \citet{crepp11} have a mild stellar mass 
dependence on their semi-major axis upper cut-off, 
$a_{max} \propto (M_*/2.5 M_\sun)^{4/9}$, while \citet{sdimeta} 
do not.  Using the more plausible \citet{crepp11} planet model with an upper 
cut-off of 35 AU for solar-type stars (only ruled 
out by \citealt{sdimeta} at the 68\% confidence level), a NICI-like survey 
of 610 stars should find only 4.1 planets around high-mass stars, or 1.5 
planets given the actual number of B and A stars observed by NICI.  This 
is consistent with our null result as a Poisson distribution rules it 
out at only the 78\% level.  Similarly, \citet{crepp11} predict that a 
NICI survey of 192 high-mass stars would 
detect only 0.1 ``cold-start'' planets (planets that form via the 
core-accretion scenario, based on the planet luminosity models of 
\citealt{fortney}).

\citet{janson11} conducted an adaptive optics 
survey of 18 high-mass stars with spectral types B2-A0, more massive than the 
stars discussed here.  Their median contrast at 1'' was 12.4 mags, compared to 
14.3 mags for our 
NICI B and A stars.  They did not detect any planets around their 
target stars and set limits on the fraction of planets, brown dwarfs, 
and low-mass stars that can form by the disk instability mechanism.  By 
applying formation model predictions of what combinations of semi-major 
axis and planet mass are allowed, they find an upper limit 
on companion fraction of 21\% (95\% confidence) for objects within 300 AU 
and smaller than 100~M$_{Jup}$.  We note that their assumed formation 
limits often preclude the presence of planetary-mass companions that 
are detectable by their contrast curve, so they are essentially comparing the 
theoretical prediction of brown dwarf frequency to their null result for 
brown dwarfs.  Without assuming a formation model and using the NICI 
B and A star results, we 
find that fewer than 7\% of high-mass stars can have a companion more massive 
than 8 M$_{Jup}$ between 85 and 300 AU, also at 95\% confidence.

\citet{vigan12} examined the frequency of giant planets 
around 42 high mass stars based on the results of IDPS, which included the 
detections of planets around HR 8799.  Their median detection limits were 
12.7 mags at 1'' and 15.3 mags at 5'', compared to our median 
NICI contrasts for B and A stars of 14.3 
mags at 1'' and 16.8 mags at 5''.  They use a Bayesian analysis assuming a 
fixed mass and semi-major axis distribution, where planets are uniformly 
distributed between 5 and 320 AU and 3 and 13 M$_{Jup}$, 
finding a planet frequency with 68\% confidence between 6 and 19\%.  
We find consistent results: at 95\% confidence, 
fewer than 25\% of B and A stars can have a giant planet more massive than 
3 M$_{Jup}$ between 70 and 375 AU.  And at 68\% confidence we find that less 
than 10\% of high-mass stars can have 3 M$_{Jup}$ planets in the same 
ranges of semi-major axis.  However, we note that the analysis 
of \citet{vigan12} 
differs significantly from ours, as we do not specify a particular 
distribution of planets but instead 
put upper limits on the frequency of wide-separation 
giant planets as a function of semi-major axis and planet mass.  In addition, 
as we illustrate in Figure~\ref{vigan_fig} and discuss in 
Section~\ref{ages_sec}, we 
find significantly older ages for many of their target stars than used 
in their analysis.  With these older ages, the 
actual IDPS constraints on planet fraction around high-mass stars 
would be significantly weaker, since fewer planets would be detectable 
around many of their stars.  

We note that our upper limits are similar to the 20\% frequency of giant RV 
planets ($\gtrsim$1~M$_{Jup}$) 
found around high-mass ($>$1.4 M$_{\sun}$) stars at small 
($a <$ 2.5 AU) separations 
\citep{johnson_new}.  So while our 10\% constraints on more massive 
($>$10 M$_{Jup}$), wide-separation (38--650 AU)
planets around B and A stars are a factor of two more restrictive 
than the RV frequency for all masses, 
directly comparing the frequencies between RV and directly imaged 
planets for a similar range of masses 
will require larger sample sizes and/or higher contrasts.

\section{Conclusions}

We have surveyed 70 young, nearby B and A for planets as part of the Gemini 
NICI Planet-Finding Campaign.  From this search we have discovered a brown 
dwarf and an M star orbiting HD~1160 and 
determined that the previously known brown dwarf companion HIP 79797 
B is actually a tight brown dwarf binary.  Deep images of 
$\beta$ Pic and Fomalhaut show no new companions to either star.  
The non-detection of any planets in the survey implies that giant planets are 
not common around high mass stars.  A Monte Carlo 
analysis shows that less than 10\% of 2~M$_\sun$ stars 
host a planet more massive than 
10~M$_{Jup}$ between 38 and 650 AU at 95\% confidence, and fewer than 
20\% of B and A stars harbor a hot-start giant planet more massive than 
4~M$_{Jup}$ between 59 and 460 AU.  Planets like HR~8799~b are 
similarly uncommon, with 
fewer than 9.8\% of 2~M$_{\sun}$ stars having such planets 
(7 M$_{Jup}$, 68 AU) at the 95\% confidence level.

The 70 B and A stars in the NICI Campaign represent the largest, deepest 
direct imaging survey around high mass stars conducted to date.  We 
reach similar upper limits on the fraction of massive stars 
with giant planets that \citet{sdimeta} reached for solar-type stars.  
We find that $<$20\% of 2 M$_{\sun}$ stars have a 4 M$_{Jup}$ planet between 
59 and 460 AU, compared to $<$20\% of 1~M$_{\sun}$ stars having a 4 M$_{Jup}$ 
planet between 30 and 466 AU.

If wide-separation giant planets are formed by outward 
scattering between giant planets close to the star, then our non-detection 
of planets from the NICI Campaign suggests that close 
scattering is not common in giant planet systems around B and A 
stars.  Such scattering is also proposed to be the origin of high 
eccentricities of small-separation giant planets detected by the RV 
method (e.g., \citealt{rasio96}, \citealt{weidenschilling96}, 
\citealt{lin97}).

The radial velocity observations for GK clump giant stars 
($<$1.5--3 M$_{\sun}$) that were main sequence B and A dwarfs show that the 
orbital distributions of giant planets around these ``retired A stars'' 
are quite different from those around solar-type stars 
\citep{johnson_new,sato10}.  One feature is that giant planets around GK 
clump giant stars may have smaller orbital eccentricities than 
intermediate separation RV planets ($\gtrsim$1 AU) around 
solar-type stars (e.g. \citealt{sato10}).  Both the RV result and our 
null result from the NICI Campaign B and A stars suggest that systems of giant 
planets may be more orbitally stable around high-mass stars than around 
solar-type stars.  Additionally, the paucity of hot Jupiters (orbital 
radii $\lesssim$0.1 AU) around high mass stars compared to solar-type 
stars could be due to less type II migration occurring around B and A 
stars \citep{burkert07,currie09}.  
The difference in hot Jupiter frequency cannot be fully accounted 
for by enhanced tidal decay during the RGB phase of the GK clump giant 
stars \citep{kunitomo11}.  If giant planets around B and A stars 
are more orbitally stable as well as less affected by type II migration, this 
presents a challenge to planet formation theory.

While wide-separation giant planets like HR 8799 b are relatively rare, 
with fewer than 9.8\% of high-mass stars having such a planet, we cannot place 
any constraints on planets like $\beta$ Pic b, (8~M$_{Jup}$, 8 AU).  
The star $\beta$ Pic is so much younger (12 Myr) and closer (19~pc) than 
any other B or A star in our sample, with the planet just at the edge of 
what is detectable by NICI.  To detect similar planets around the other A 
stars in our sample will require a planet-finding instrument coupled to 
an extreme adaptive optics system like GPI or SPHERE \citep{gpi07,sphere08}.  
While large-separation giant planets (like HR~8799~b) are rare around 
high mass stars, it is possible that giant planets in similar orbital 
distances as Jupiter are more common.  Observations of other B and A stars 
with future instruments will be able to answer this question definitively.

Nevertheless, our NICI sample of B and A stars is the one best suited 
for directly imaging planets, as it represents the youngest and closest high 
mass stars.  
The number of B and A stars in a volume-limited sample is significantly 
lower than for solar-type and M stars, so young 
high-mass stars within 100 pc will always be very limited.  With the 
exception of the handful of nearby B and A stars whose ages can be 
independently determined (e.g. 
members of a moving group or components of a wide multiple system with a 
lower-mass star that can be independently age-dated), searching 
for ``low CMD'' stars is the 
best technique for locating young B and A stars.  While we have shown 
that low position on the color-magnitude diagram 
does not indicate Pleiades 
age, this approach still does flag stars in the first 
third of their main-sequence lifetimes.  So the stars discussed here are 
likely to be the youngest B and A stars within 100~pc, with their ages 
most precisely determined from our Bayesian method.

To further refine the sample, 
spectroscopy of nearby B and A stars can clean out the low metallicity 
stars (which are likely to have larger ages) from the ``low CMD'' 
stars.  Evolutionary model predictions of log(g) and 
T$_{eff}$, which can also be determined from spectroscopy, may be more 
accurate than the photometry 
we use here, but the fundamental limitation of determining accurate ages 
will still remain.  B and A stars 
do not move significantly on either a color-magnitude diagram or an HR 
diagram in the first third of their main-sequence lifetimes.  While 
constraints on the fraction of high mass stars with giant planets can 
be improved with future instruments that can achieve higher 
contrasts at smaller separations, the optimal 
target list for these future surveys 
is unlikely to differ significantly from the stars presented here.

\acknowledgments

We thank Jessica Lu, Adam Kraus, Rolf Kudritzki, 
and Lisa Kewley for helpful discussions that 
greatly benefited this work.  We thank the anonymous referee for the 
constructive suggestions that have improved this work.  
B.A.B was supported by Hubble Fellowship grant HST-HF-01204.01-A awarded by 
the Space Telescope Science Institute, which is operated by AURA for 
NASA, under contract NAS 5-26555.  This work was supported in part by NSF 
grants AST-0713881 and AST-0709484 awarded to M. Liu, NASA Origins grant 
NNX11 AC31G awarded to M. Liu, and NSF grant AAG-1109114 awarded to 
L. Close.  
The Gemini Observatory is operated by the Association of Universities for
Research in Astronomy, Inc., under a cooperative agreement with the NSF on
behalf of the Gemini partnership: the National Science Foundation (United
States), the Science and Technology Facilities Council (United Kingdom), the
National Research Council (Canada), CONICYT (Chile), the Australian Research
Council (Australia), CNPq (Brazil), and CONICET (Argentina).  
Based on observations made with the European Southern Observatory
telescopes obtained from the ESO/ST-ECF Science Archive Facility.
This 
publication makes use of data products from the Two Micron All Sky Survey, 
which is a joint project of the University of Massachusetts and the Infrared 
Processing and Analysis Center/California Institute of Technology, funded by 
the National Aeronautics and Space Administration and the National Science 
Foundation.  This research has made use of the SIMBAD database,
operated at CDS, Strasbourg, France.  This research has made use of the 
VizieR catalogue access tool, CDS, Strasbourg, France.  The Digitized Sky 
Survey was produced at the Space Telescope Science Institute under U.S. 
Government grant NAG W-2166.  The images of these surveys are based on 
photographic data obtained using the Oschin Schmidt Telescope on Palomar 
Mountain and the UK Schmidt Telescope.  The plates were processed into the 
present compressed digital form with the permission of these institutions.

{\it Facilities:} \facility{Gemini:South (NICI)}. 

\bibliographystyle{apj}
\bibliography{apj-jour,astar}
\clearpage

\clearpage

\clearpage

\begin{deluxetable}{ccccccccccccccccc}
\rotate
\tabletypesize{\tiny}
\tablecaption{Properties of Target Stars\label{table1}}
\tablewidth{0pt}
\tablehead{ 
 & & & & & & & & & & & & \multicolumn{5}{c}{Age} \\
\cline{13-17}
\colhead{Name} & \colhead{HD} & \colhead{HIP} & \colhead{RA} 
& \colhead{Dec} & \colhead{SpT} & \colhead{Dist} 
& \colhead{$\sigma_{Dist}$} 
& \colhead{$B-V$} & \colhead{$V$} & \colhead{$H$} & \colhead{$K_S$}
& \colhead{Median} & \colhead{68\%} 
& \colhead{95\%} & \colhead{Adopted} & \colhead{Ref} \\ 
\colhead{} & \colhead{} & \colhead{} & \colhead{} 
& \colhead{} & \colhead{} & \colhead{(pc)} 
& \colhead{(pc)} & \colhead{(mag)} & \colhead{(mag)} 
& \colhead{(mag)} & \colhead{(mag)}
& \colhead{(Myr)} & \colhead{(Myr)} 
& \colhead{(Myr)} & \colhead{(Myr)} & \colhead{} 
}
\startdata

HD 1160 & 1160 & 1272 & 00:15:57.302 & +04:15:04.01 & A0 & 103 & 5 & 0.05 & 7.13 & 7.01 & 7.04 & 92 & 36--178 & 14--267 & \ldots & \ldots \\ 
HIP 2072 & 2262 & 2072 & 00:26:12.202 & -43:40:47.39 & A7 & 23.81 & 0.09 & 0.17 & 3.93 & 3.59 & 3.59 & 498 & 375--637 & 243--761 & \ldots & \ldots \\ 
49 Cet & 9672 & 7345 & 01:34:37.779 & -15:40:34.90 & A1 & 59.4 & 1.0 & 0.07 & 5.62 & 5.53 & 5.46 & 255 & 150--361 & 49--446 & 40 & Z12 \\ 
HD 10939 & 10939 & 8241 & 01:46:06.263 & -53:31:19.33 & A1 & 62.0 & 0.7 & 0.03 & 5.04 & 5.03 & 4.96 & 346 & 305--379 & 266--398 & \ldots & \ldots \\ 
HIP 12394 & 16978 & 12394 & 02:39:35.357 & -68:16:01.00 & B9 & 46.55 & 0.20 & -0.06 & 4.12 & 4.43 & 4.25 & 132 & 93--172 & 45--214 & 30 & B13 \\ 
HIP 12413 & 16754 & 12413 & 02:39:48.002 & -42:53:30.08 & A2 & 39 & 3 & 0.06 & 4.74 & 4.62 & 4.46 & 236 & 104--356 & 21--451 & \ldots & \ldots \\ 
HD 17848 & 17848 & 13141 & 02:49:01.487 & -62:48:23.48 & A2 & 50.5 & 0.5 & 0.10 & 5.25 & 5.16 & 4.97 & 372 & 269--467 & 159--580 & \ldots & \ldots \\ 
HIP 14146 & 18978 & 14146 & 03:02:23.499 & -23:37:28.09 & A4 & 27.17 & 0.13 & 0.16 & 4.08 & 3.54 & 3.57 & 567 & 419--670 & 261--757 & \ldots & \ldots \\
\enddata
\tablecomments{For most stars we use the results of our Bayesian age-dating.  
For stars with an independent age measurement in the literature, we use 
that value, given in the 
``Adopted'' column.  We only use literature ages for cases where there is 
solid evidence that the star belongs to a well-known moving group.  
The source of these external ages is given by 
B13: \citet{moving_groups}, K05: \citet{kouwenhoven05}, 
N12: \citet{nakajima12}, W00: \citet{weinberger00}, 
Z04: \citet{zs04}, Z11: \citet{zrsb11}, and Z12: \citet{zs12}.}
\end{deluxetable}

\begin{deluxetable}{lcrrrr}
\tabletypesize{\scriptsize}
\tablecaption{NICI Observation of Young B and A Stars}
\tablewidth{0pt}
\tablehead{ \colhead{Target}  &\colhead{UT Date} & \colhead{Obs Mode} & \colhead{N$_{images}$} & \colhead{Exp. Time (s)}             & \colhead{Rotation (deg)}}
\startdata
HD 1160 & 2010-10-31 & ADI &      20 &    1208 &     6.4 \\
HD 1160 & 2010-10-31 & ASDI &      45 &    2701 &    18.1 \\
HIP 2072 & 2011-11-03 & ADI &      40 &    2508 &    46.2 \\
49 Cet & 2009-12-02 & ASDI &      45 &    2736 &    40.6 \\
49 Cet & 2009-12-02 & ADI &      20 &    1208 &    16.4 \\
HD 10939 & 2010-12-26 & ASDI &      44 &    2708 &     3.5 \\
HIP 12394 & 2008-12-15 & ASDI &      42 &    2633 &    21.3 \\
HIP 12394 & 2008-12-15 & ADI &      20 &    1200 &     7.9 \\
\enddata
\label{tab:observing}
\end{deluxetable}

\begin{deluxetable}{ccccccccccc}
\tabletypesize{\scriptsize}
\tablecaption{Properties of Candidate Companions\label{table_comp2a}}
\tablewidth{0pt}
\tablehead{
\colhead{Name} & \colhead{\#} & \colhead{Sep} & 
\colhead{PA} & 
\colhead{$\Delta H$} & \colhead{$\Delta \tau$} & 
\colhead{Epochs} & \colhead{$\chi^2_{\nu}$(BG)} & 
\colhead{$\chi^2_{\nu}$(CPM)} & \colhead{dof} & \colhead{Comp?} \\
\colhead{} & \colhead{} & \colhead{($''$)} & \colhead{(deg)} & 
\colhead{(mag)} & \colhead{(years)} & \colhead{} & \colhead{} & 
\colhead{} & \colhead{} & \colhead{}
}
\startdata
HD 1160 & 1 & 0.780 & 244.3 & 9.6 & 9.28 & 12 & 11.22 & 0.47 & 22 & CPM \tablenotemark{1} \\ 
HD 1160 & 2 & 5.150 & 349.8 & 7.5 & 9.28 & 19 & 2.47 & 0.49 & 36 & CPM \tablenotemark{1} \\ 
HIP 14146 & 1 & 3.936 & 296.6 & 17.3 & 1.78 & 3 & 0.31 & 57.65 & 4 & BG  \\ 
HD 31295 & 1 & 6.261 & 271.6 & 13.3 & 1.95 & 2 & 0.09 & 69.71 & 2 & BG  \\ 
HD 31295 & 2 & 8.805 & 316.7 & 14.5 & 1.95 & 2 & 0.02 & 187.48 & 2 & BG  \\ 
HIP 25280 & 1 & 5.219 & 80.4 & 14.9 & 1.84 & 2 & 1.82 & 25.00 & 2 & BG  \\ 
$\zeta$ Lep & 1 & 5.295 & 110.0 & 18.0 & 2.24 & 2 & 0.44 & 28.80 & 2 & BG  \\ 
HD 54341 & 1 & 3.080 & 131.2 & 12.7 & 2.03 & 2 & 0.40 & 3.05 & 2 & BG  \\
\enddata
\tablenotetext{1}{\citet{hd1160}}
\tablenotetext{2}{Originally identified by \citet{hip79797}}
\tablenotetext{3}{Originally identified by \citet{lowrance00}}
\tablecomments{Summary of the candidate companions detected around 
each target star.  For each candidate we list the separation and 
position angle at the reference epoch, the contrast between 
candidate and host star, the time baseline for our astrometric 
data, the number of epochs, the reduced chi-square statistic for 
the candidate to be background (BG) or common proper motion (CPM), 
the number of degrees of freedom for the astrometric fit, 
and the final determination of 
each candidate: background or common proper motion.  Astrometry 
for individual epochs is in Table~\ref{table_comp2b}.
}
\end{deluxetable}

\begin{deluxetable}{cccccccccccccc}
\tabletypesize{\scriptsize}
\tablecaption{Astrometry of Candidate Companions\label{table_comp2b}}
\tablewidth{0pt}
\tablehead{
 & & & \multicolumn{4}{c}{Measured Position} & & 
\multicolumn{4}{c}{Background Position} & 
& \\
\cline{4-7}
\cline{9-12}
\colhead{Name} & \colhead{\#} & \colhead{Epoch} & \colhead{Sep} & 
\colhead{$\sigma_{Sep}$} & \colhead{PA} & \colhead{$\sigma_{PA}$} & 
\colhead{} & \colhead{Sep} & \colhead{$\sigma_{Sep}$} & 
\colhead{PA} & \colhead{$\sigma_{PA}$} & 
\colhead{Inst.} & \colhead{Comp?} \\
\colhead{} & \colhead{} & \colhead{} & \colhead{($''$)} & 
\colhead{} & \colhead{(deg)} & \colhead{} & \colhead{} & 
\colhead{($''$)} & \colhead{} & \colhead{(deg)} & \colhead{} & 
\colhead{} & \colhead{} \\
}
\startdata
HD 1160 & 1 & 2010.83 & 0.780 & 0.025 & 244.3 & 0.2 &   & \ldots & \ldots & \ldots & \ldots & N & CPM \\
  &   & 2002.57 & 0.767 & 0.030 & 246.2 & 1.0 &   & 0.711 & 0.025 & 229.528 & 0.629 & V & CPM \\
  &   & 2003.84 & 0.766 & 0.030 & 245.6 & 1.0 &   & 0.706 & 0.025 & 231.709 & 0.546 & V & CPM \\
  &   & 2005.98 & 0.760 & 0.030 & 244.7 & 1.0 &   & 0.721 & 0.025 & 235.770 & 0.392 & V & CPM \\
  &   & 2008.50 & 0.804 & 0.060 & 245.3 & 2.0 &   & 0.766 & 0.026 & 240.510 & 0.238 & V & CPM \\
  &   & 2010.71 & 0.769 & 0.060 & 242.8 & 2.0 &   & 0.785 & 0.026 & 244.094 & 0.183 & V & CPM \\
  &   & 2010.89 & 0.760 & 0.025 & 244.5 & 0.2 &   & 0.778 & 0.026 & 244.367 & 0.184 & N & CPM \\
  &   & 2010.90 & 0.773 & 0.020 & 244.9 & 0.5 &   & 0.778 & 0.026 & 244.390 & 0.184 & K & CPM \\
\enddata
\tablenotetext{1}{Astrometry from \citet{hr4796_1997}}
\tablenotetext{2}{Astrometry from \citet{hip79797}}
\tablenotetext{3}{Astrometry from \citet{lowrance00}}
\tablecomments{\tiny{Astrometry for each candidate companion detected around 
our target stars from NICI and archival observations.  At each 
epoch we give the measured separation, position angle, and 
uncertainties as well as the predicted separation and position 
angle for a background object based on the proper motion and 
parallax of the primary and the candidate position at the reference 
epoch, which is the first epoch listed for each candidate.  
Astrometry is taken from Gemini-South NICI (N), VLT NACO (V), 
Keck NIRC2 (K), HST NICMOS (H), 
VLT ISAAC (I), ESO 3.6m COME-ON-PLUS/ADONIS (E), and Gemini-North NIRI (G).
}}
\end{deluxetable}

\begin{deluxetable}{ccc}
\tablecaption{Relative Astrometry of HIP 79797 Bb with respect to Ba}
\tablewidth{0pt}
\tablehead{\colhead{UT Epoch} & \colhead{Sep (mas)} & \colhead{PA ($^\circ$)}}
\startdata
2010.2712 & 60 $\pm$ 20 & 331 $\pm$ 13 \\
2010.3534 & 60 $\pm$ 6 & 339 $\pm$ 3 \\
2011.3589 & 63 $\pm$ 15 & 340 $\pm$ 10 \\
2012.2623 & 63 $\pm$ 7 & 342 $\pm$ 3 \\
2012.6667 & 77 $\pm$ 14 & 338 $\pm$ 9 \\
\enddata
\label{hip79797_table}
\end{deluxetable}

\begin{deluxetable}{lccc}
\tablecaption{Properties of the HIP 79797 System}
\tablewidth{0pt}
\tablehead{\colhead{} & \colhead{HIP 79797 A} & \colhead{HIP 79797 Ba} & \colhead{HIP 79797 Bb}}
\startdata
Age (Myr) & 200$^{+170}_{-130}$ & & \\
Distance (pc) & 52.2 $\pm$ 1.1\tablenotemark{1} & & \\
Spectral Type & A4 & M9\tablenotemark{2} & M9\tablenotemark{2} \\
$\Delta J$ (Bab $-$ A, mag) & & \multicolumn{2}{c}{8.55 $\pm$ 0.14} \\
$\Delta H$ (Bab $-$ A, mag) & & \multicolumn{2}{c}{7.37 $\pm$ 0.17} \\
$\Delta K_S$ (Bab $-$ A, mag) & & \multicolumn{2}{c}{7.31 $\pm$ 0.19} \\
$\Delta J$ (Bb $-$ Ba, mag) & & \multicolumn{2}{c}{0.33 $\pm$ 0.07} \\
$\Delta H$ (Bb $-$ Ba, mag) & & \multicolumn{2}{c}{0.12 $\pm$ 0.06} \\
\enddata

\tablenotetext{1}{\citet{newhip}}
\tablenotetext{2}{\citet{hip79797}, and assuming the integrated light 
spectrum is similar to the spectra of the individual objects, which is 
consistent with the near-unity flux ratios.}
\tablenotetext{3}{\citet{2mass}}
\tablenotetext{4}{Astrometry is from epoch 2010.3534 and is the separation 
and position angle of the photocenter of the binary from HIP 79797 A.}
\label{hip79797_properties}
\end{deluxetable}

\begin{deluxetable}{cccccccc}
\tabletypesize{\scriptsize}
\tablecaption{Candidate Companions with One Epoch of Data\label{table_oneepoch}}
\tablewidth{0pt}
\tablehead{ 
\colhead{Star} & \colhead{\#} & \colhead{Sep} & \colhead{Sep} & \colhead{PA} 
& \colhead{$\Delta H$} & \colhead{Epoch} & \colhead{Action} \\ 
\colhead{} & \colhead{} & \colhead{(``)} & \colhead{(AU)} & \colhead{(deg)} 
& \colhead{(mag)} & \colhead{} & \colhead{} \\ 
}
\startdata
HD 31295 & 1 & 7.77 & 277 & 162.4 & 15.6 & 2009.0356 & Sep $<$ 7.77'' \\
    & 2 & 8.66 & 309 & 277.1 & 17.5 & 2009.0356 &     \\
HIP 36188 & 1 & 10.64 & 528 & 352.7 & 14.6 & 2011.8384 & Sep $<$ 10.64'' \\ 
HIP 40916 & 1 & 6.26 & 421 & 332.7 & 12.8 & 2009.0466 & Sep $<$ 6.26'' \\
    & 2 & 6.95 & 467 & 131.0 & 12.6 & 2009.0466 &     \\
    & 3 & 7.65 & 514 & 60.9 & 7.4 & 2009.0466 &     \\
    & 4 & 7.67 & 515 & 90.9 & 8.2 & 2009.0466 &     \\
HD 71155 & 1 & 9.93 & 372 & 221.0 & 15.2 & 2008.9563 & Sep $<$ 9.93'' \\
\enddata
\tablecomments{Target stars with candidate 
companions for which we only have a single epoch of data.  Since 
we cannot classify these as either CPM or background, we provide the 
properties of these candidates here and note the changes we 
make to the contrast curves in the Action column (see 
Section~\ref{concurve_sec} for details).  Contrast curves are 
edited either by reverting to a less sensitive contrast curve or 
restricting the contrast curve to within a given separation.  
For cases where we have 
only a single epoch and there are 
candidates inside the 100\% coverage region for 
position angle, we drop the star from our analysis.}
\end{deluxetable}

\begin{deluxetable}{lrrrrrrrrrrrrrrrrrr}
\rotate
\tabletypesize{\scriptsize}
\tablecaption{95\% completeness $CH_4S$ and $H$-band contrasts ($\Delta$mag)}
\tablewidth{0pt}
\tablehead{ \colhead{Target} &\colhead{0.36$''$} & \colhead{0.5$''$} & \colhead{0.75$''$} & \colhead{1$''$}&\colhead{1.5$''$} & \colhead{2$''$} & \colhead{3$''
$} & \colhead{4$''$}&\colhead{5$''$}&\colhead{7$''$} & \colhead{Cov.} & \colhead{9$''$} & \colhead{Cov.}&\colhead{12$''$} & \colhead{Cov.} & \colhead{14.8$''$}
 & \colhead{Cov.}}
\startdata
HD 1160, $CH_4$ & 10.8 & 12.2 & 12.9 & 13.6 & 14.0 & 13.8 & 14.0 & 13.9 & 13.9 & 13.8 & 0.98 & 13.3 & 0.72 & 12.2 & 0.29 & 11.8 & 0.02 \\
HD 1160, $H$-band &  7.5 &  9.6 & 11.5 & 13.0 & 14.3 & 14.6 & 14.6 & 14.7 & 14.5 & 14.5 & 0.88 & 14.3 & 0.61 & 13.8 & 0.26 & 13.7 & 0.03 \\
HIP 2072, $H$-band &  -  &  -  &  -  & 13.1 & 14.9 & 16.1 & 17.0 & 17.1 & 17.1 & 16.9 & 0.96 & 16.2 & 0.82 & 15.6 & 0.44 & 15.5 & 0.15 \\
49 Cet, $CH_4$ & 10.7 & 12.4 & 13.4 & 14.0 & 14.6 & 14.9 & 14.9 & 14.8 & 14.6 & 14.2 & 1.00 & 13.3 & 0.84 & 12.4 & 0.47 & 12.1 & 0.10 \\
49 Cet, $H$-band &  -  &  -  &  4.0 & 13.2 & 14.8 & 15.9 & 16.8 & 16.9 & 16.8 & 16.6 & 0.89 & 16.3 & 0.66 & 15.8 & 0.32 & 15.3 & 0.07 \\
HD 10939, $CH_4$ &  7.6 &  9.6 & 11.8 & 13.1 & 14.2 & 14.5 & 14.6 & 14.6 & 14.5 & 14.5 & 0.93 & 14.4 & 0.68 & 14.0 & 0.30 & 13.5 & 0.06 \\
HIP 12394, $CH_4$ & 11.5 & 12.9 & 14.2 & 15.1 & 15.3 & 15.5 & 15.5 & 15.4 & 15.2 & 15.2 & 0.98 & 15.0 & 0.74 & 14.3 & 0.32 & 13.4 & 0.03 \\
HIP 12394, $H$-band &  -  &  -  &  -  &  -  & 10.0 & 16.1 & 17.4 & 17.8 & 17.7 & 17.9 & 0.86 & 17.7 & 0.62 & 17.4 & 0.27 & 16.9 & 0.04 \\
\enddata
\label{tab:contrasts}
\tablecomments{Contrasts achieved in the $CH_4S$ and $H$-bands. For 
separations less than 1.5$''$, the $CH_4$ filter contrasts (obtained 
in ASDI mode) are usually better. For larger separations, the $H$-band 
contrasts (obtained in ADI mode) are usually 
better. Beyond 6.3$''$ each final image has data only at a fraction of the 
360$^\circ$ range in position angle.  This angular 
coverage fraction (ranging from 0 to 1) 
is given beside the contrast at these large separations. 
All the contrasts are 95\% completeness contrasts, except at separations 
beyond 6.3$''$ where the nominal 5$\sigma$ contrast curve is used. A 
constant is added to the 5$\sigma$ curve so that both curves match at 6.3$''$.}
\end{deluxetable}

\begin{deluxetable}{lcccc}
\tabletypesize{\scriptsize}
\tablecaption{Stellar and Brown Dwarf Binaries to Sample Stars}
\tablewidth{0pt}
\tablehead{ \colhead{Star} &\colhead{Projected Separation (AU)} & \colhead{Sp. Type} & 
\colhead{Reference} & \colhead{Allowable Planetary $a$ (AU)}}
\startdata
HD 1160 & 81, 530 & $\sim$L0, M3.5 & \citet{hd1160} & $<$20, $>$2700 \\
HIP 36188 & 1.2\tablenotemark{1} & $\sim$A3 & \citet{hip36188} & $>$6 \\
HIP 49669 & 3900, 4200, 4800 & K2, M5, M5 & \citet{ccdm} & $<$780 \\
HD 102647 & 720, 1100, 2600 & $\sim$M, $\sim$M, G2 & \citet{ccdm} & $<$140 \\
HIP 60965 & 630 & K0 & \citet{ccdm} & $<$130, $>$3200 \\
HR 4796 A & 560, 14000 & M2, M4.5 & \citet{jura95}, \citet{hr4796_mstar} & $<$110 \\
GJ 560 A & 260 & K5 & \citet{ccdm} & $<$52, $>$1300 \\
HD 141569 & 880, 1000 & M2, M4 & \citet{weinberger00} & $<$180 \\
HIP 78106 & 490 & A3 & \citet{ccdm} & $<$99, $>$2500 \\
HIP 79797 & 350, 350 & $\sim$M9, $\sim$M9 & \citet{hip79797}, this work & $<$70, $>$1800 \\
HIP 85038 & 580 & G8 & \citet{ccdm} & $<$120, $>$2900 \\
HD 172555 & 2000 & K5 & \citet{feigelson06} & $<$410 \\
HR 7329 & 200, 20000 & M7-8, F6 & \citet{hr7329} & $<$40, 1000$<a<$4000 \\
Fomalhaut & 57400 & K4 & \citet{fomalhaut_mamajek} & $<$11000 \\
\enddata
\tablenotetext{1}{Semi-major axis}
\label{tab:binaries}
\end{deluxetable}

\begin{deluxetable}{ccccc}
\tablecaption{Upper Limits on Planet Fraction (95\% Confidence Level)\label{table_pf}}
\tablewidth{0pt}
\tablehead{
 & \multicolumn{4}{c}{Planet Fraction} \\
\colhead{Mass (M${_{Jup}}$)}  & \colhead{$\le$ 5\%}  & \colhead{$\le$ 10\%}  & \colhead{$\le$ 20\%}  & \colhead{$\le$ 50\%}  \\
}
\startdata
1& \ldots & \ldots & \ldots & \ldots \\
2& \ldots & \ldots & \ldots & 50 -- 490 AU \\
4& \ldots & \ldots & 59 -- 460 AU & 25 -- 940 AU \\
10& \ldots & 38 -- 650 AU & 24 -- 1100 AU & 15 -- 1800 AU \\
\enddata
\tablecomments{This table gives the range in semi-major 
axis at which a given upper limit on planet fraction is reached for a given 
planet mass.  For example, between semi-major axis of 
38 and 650 AU fewer than 10\% of B and A stars can have giant planets 
more massive than 10 M$_{Jup}$.}
\end{deluxetable}

\clearpage

\end{document}